\RequirePackage{fix-cm}
\documentclass[a4paper, 12pt]{article} 
\usepackage[margin=1.07in]{geometry}
\usepackage{amsmath, amssymb, amsthm, thm-restate, mathtools, xcolor, todonotes, abstract, etoolbox}
\usepackage[inline]{enumitem}
\usepackage[hidelinks]{hyperref}
\usepackage{mathscinet, authblk}
\usepackage[capitalize, nameinlink]{cleveref}
\usepackage[T1]{fontenc}
\usepackage[utf8]{inputenc}
\usepackage{libertine} % text font
\usepackage{inconsolata} % texttt font
\usepackage{algorithm, algpseudocode, tocloft, titlesec, braket}
\newcommand{\T}{\mathcal{T}}
\usepackage{tikz, tkz-euclide, adjustbox}
\tikzset{MyNode/.style={circle, draw, inner sep=2,outer sep=0, fill=gray}}
\tikzset{MyRedArc/.style={line width=2.5pt, red}}
\tikzset{MyBlueArc/.style={line width=1.5pt, blue}}
\usetikzlibrary{hobby,arrows.meta}
\tikzset{
  arrRepeat/.style={
    postaction={
      decorate,
      decoration={
        markings,
        mark=between positions 0.1 and 0.95 step 134pt with {\arrow{>}},
}}}}
\definecolor{EulerColor}{RGB}{0,105,155}
\tikzset{
  EulerCurve/.style={draw=EulerColor,line width=1.1pt,
    line cap=round,line join=round},
  EulerArrow/.style={
    decoration={markings,
      mark=at position #1 with {\arrow{Stealth[length=2.1mm,width=1.5mm]}}},
    postaction={decorate}}
}

\makeatletter
\renewcommand\thmt@autorefsetup{%
  \@xa\def\csname\thmt@envname autorefname\@xa\endcsname\@xa{\thmt@thmname}%
}
\makeatother
\title{Finding full immersions is fixed-parameter tractable in labeled tour graphs}
\author[1]{Hadas Barabash}
\author[2]{Jim Geelen}
\author[3]{Rose McCarty}
\author[4]{Caleb McFarland}
\author[5]{Paul Wollan}
\affil[1,3,4]{Georgia Institute of Technology, USA.}
\affil[2]{University of Waterloo, Canada.}
\affil[5]{Sapienza Universit\`{a} di Roma, Italy.}
\date{\today}
\let\le\leqslant
\let\ge\geqslant
\let\leq\leqslant
\let\geq\geqslant

\newtheorem{theorem}{Theorem}
\newtheorem{corollary}[theorem]{Corollary}
\newtheorem{proposition}[theorem]{Proposition}
\newtheorem{assumption}[theorem]{Assumption}
\newtheorem{conjecture}[theorem]{Conjecture}
\crefname{conjecture}{conjecture}{conjectures}
\newtheorem{lemma}[theorem]{Lemma}
\newtheorem{claim}{Claim}[theorem]
\crefname{claim}{claim}{claims}
\newtheorem{observation}[theorem]{Observation}
\crefname{observation}{observation}{observations}
\theoremstyle{definition}

\newcommand{\N}{\mathbb{N}}

\newcommand{\Oh}{\mathcal{O}}
\newcommand{\Zplus}{\mathbb{Z}^+}
\DeclareMathOperator{\val}{value}

\newif\ifcomments
\commentstrue   %set to \commentsfalse to hide comments

\setlist[itemize]{topsep=4pt,itemsep=3pt,parsep=0pt} 
\setlist[enumerate]{topsep=4pt,itemsep=3pt,parsep=0pt}
\setlist[description]{topsep=4pt,itemsep=3pt,parsep=0pt} 

\newcommand{\zerodisplayskips}{%
  \setlength{\abovedisplayskip}{5pt}%
  \setlength{\belowdisplayskip}{5pt}%
  \setlength{\abovedisplayshortskip}{5pt}%
  \setlength{\belowdisplayshortskip}{5pt}}
\appto{\normalsize}{\zerodisplayskips}
\appto{\small}{\zerodisplayskips}
\appto{\footnotesize}{\zerodisplayskips}

\newlength{\NormalParindent}
\newlength{\NormalParskip}
\AtBeginDocument{%
  \setlength{\NormalParindent}{\parindent}%
  \setlength{\NormalParskip}{\parskip}%
}

\def\cqedsymbol{\ifmmode$\lrcorner$\else{\unskip\nobreak\hfil
\penalty50\hskip1em\null\nobreak\hfil$\lrcorner$
\parfillskip=0pt\finalhyphendemerits=0\endgraf}\fi}

\let\OLDthebibliography\thebibliography
\renewcommand\thebibliography[1]{
  \OLDthebibliography{#1}
  \setlength{\parskip}{0pt}
  \setlength{\itemsep}{0pt plus 0.3ex}
}%
\AtBeginDocument{%Remove MR numbers from bibliography
   \def\MR#1{}
}%

\titleformat{\part}[hang]
  {\normalfont\LARGE\bfseries} % Format of the entire line
  {\partname\ \thepart:}       % Label format (e.g., "Part I:")
  {.5em}                      % Space between label and title
  {}                          % Code preceding title
\titlespacing*{\part}{0pt}{0pt}{.75em}
\titlespacing*{\subsection}{0pt}{8pt}{3pt} %left, before, after

\begin{document}
\pagenumbering{roman}
\maketitle

%ABSTRACT
\begin{abstract}
    We give an $\Oh_{\mathcal{H}}(n^5)$-time algorithm to test whether an $n$-vertex labeled tour graph has a full immersion isomorphic to a labeled tour graph $\mathcal{H}$. A \emph{labeled tour graph} is a connected $4$-regular graph whose edges are labeled by binary vectors of fixed length. To obtain a \emph{full immersion}, a vertex may be removed only by pairing up its four incident edges and ``splitting off'' both pairs, summing the labels within each pair. Even for one-bit labels, no polynomial-time algorithm was previously known for arbitrary fixed $\mathcal{H}$. Our main tool is of independent interest: we give an algorithmic min-max theorem that, in particular, characterizes the maximum number of odd-length circuits in a rooted circuit decomposition of an Eulerian graph.

    Our results have implications for \emph{vertex-minors}, whose study is motivated by applications in quantum networking, connections with graph minor theory, and the goal of developing an irrelevant vertex technique for dense graphs. For vertex-minors, circle graphs play the same role as planar graphs do for graph minors. As a corollary of our main theorem, we obtain a fixed-parameter tractable algorithm for \emph{apex-vertex-minor} testing in \emph{$k$-apex circle graphs}. We also prove an irrelevant vertex theorem for such graphs of large rank-width. More broadly, our results are essential to forthcoming work of three of the authors proving a structure theorem for graphs with a forbidden vertex-minor. There, our min-max theorem plays the role of both the algorithmic $A$-paths and Two Paths Theorems.
\end{abstract}

\newpage
\tableofcontents
\newpage

\pagenumbering{arabic}
\setcounter{page}{1}

\part{Extended abstract}
\section{Introduction}
\label{sec:introExtendedAbs}

\titlespacing*{\part}{0pt}{1em}{.75em}

A classical algorithmic problem is testing whether a small graph $H$ is ``contained'' in a large $n$-vertex graph $G$. Many different notions of containment have been studied, including graph minors \cite{graphMinors13}, topological minors and immersions \cite{TopologicalMinorTesting}, and their parity variants \cite{huynhThesis, KRW2011, LiuYoo25}. For graph minors, which capture routing problems such as finding vertex-disjoint paths between fixed pairs of terminals, there is now an almost-linear time algorithm~\cite{korhonenPS2024}, thirty years after the original cubic algorithm \cite{graphMinors13}. These algorithms are \emph{fixed-parameter tractable} (FPT) in the size of the target graph $H$, meaning that they test whether $H$ is contained in $G$ in $\Oh_H(n^{d})$ time for a fixed constant~$d$. One cannot usually hope to find an algorithm which is polynomial in both $H$ and $G$ as NP-hardness is known for essentially every notion of containment \cite{planarHamCycle, vmTesting}. 

%Rose: I removed totallyOdd2013 because it does not study the testing problem

All of these FPT algorithms adapt the same technique, the \emph{irrelevant vertex} technique of Robertson and Seymour~\cite{graphMinors13}. Broadly speaking, this involves finding a vertex whose deletion does not destroy all the copies of $H$. Such vertices are found and removed until the host graph $G$ has ``width'' bounded by a function of $H$. Then a tree-like decomposition of $G$ is found, at which point we use dynamic programming on the tree structure. Although different notions of containment have motivated different width parameters, all of the algorithms mentioned here ultimately reduce to the case of bounded treewidth.

%(graphs with many edges relative to their number of vertices)
Treewidth, however, is not the right width parameter for \emph{dense} graphs. For instance, cliques and bicliques are very simple yet have arbitrarily large treewidth. For dense graphs like these, the functionally equivalent\footnote{Two graph parameters are \emph{functionally equivalent} if each is bounded as a function of the other.} notions of \emph{clique-width} and \emph{rank-width} are more natural: cliques and bicliques have rank-width~1. Rank-width is also the dense analog of treewidth in another sense: a graph has small treewidth if and only if it has small rank-width and forbids a clique and a biclique as induced subgraphs~\cite{GarsukiWanke}. These examples motivate the goal of developing an \emph{irrelevant vertex technique for dense graphs}. This goal is particularly compelling through the lens of \emph{logic}, which provides a general framework for obtaining ``dense versions'' of classic sparse properties; see  \cite{lowComplexityLinear2021, stableTwinWidth2022, logicSurvey2025}.

%Clique-width and rank-width are functionally equivalent because for any graph $G$ of rank-width $k$, its clique-width is between $k$ and $2^{k+1}-1$~\cite[Proposition~6.3]{approxCWBW2006}.

There is a standard notion of containment which arises in this dense setting. We say that a graph $H$ is a \emph{vertex-minor} of a graph $G$ if $H$ can be obtained from $G$ by locally complementing at vertices and by deleting vertices. \emph{Locally complementing} at a vertex $v$ replaces the induced subgraph on the neighborhood of $v$ by its complement. Taking vertex-minors does not increase rank-width. In fact, the rank-width of a graph is functionally equivalent to the largest integer $k$ such that it contains a certain graph called the \emph{$k \times k$ comparability grid} as a vertex-minor~\cite{GEELEN202393}. This parallels the relationship between graph minors and treewidth given by the Grid Theorem~\cite{graphMinors5Grid}. Vertex-minors also have applications in logic \cite{shallowVM,COURCELLE200791} and in quantum networking; they capture when one graph state can be prepared from another using only local Clifford gates, single-qubit Pauli measurements, and classical communication \cite{graphicalDescription2004, transformingStates}.

The central algorithmic question is whether vertex-minor testing is FPT in the size of the target graph. A natural folklore approach is to find and delete irrelevant vertices until the host graph has bounded rank-width, as we propose in \Cref{conj:irrelevant}.

\subsection*{Main theorem}\label{sec:tour-graph-intro}
%\label{subsec:Results}

In this paper we take a step towards developing an FPT algorithm for vertex-minor testing by giving an FPT algorithm for \emph{full immersions} in \emph{labeled tour graphs}. This notion of containment was recently introduced in \cite{fullImmersions2026}, motivated by the connection with vertex-minors.
Broadly speaking, we propose to reduce vertex-minor testing to testing for full immersions, which are back in the sparse regime. We first describe our theorem in the sparse setting, and then we explain its implications for vertex-minors.

\begin{figure}
\centering
\begin{tikzpicture}[line cap=round,line join=round,font=\small, scale=1]
  % Reserve space around the drawing, including the column-vector labels.
  \path[use as bounding box] (-1.02,-2.60) rectangle (9.60,3.22);

  % These definitions are local to this tikzpicture.
  \def\threeWaysVector#1#2{%
    \left[\begin{smallmatrix}#1\\#2\end{smallmatrix}\right]%
  }
  \def\threeWaysCoordinates{%
    % A and V are mirror images across the vertical line through B and C.
    \coordinate (A) at (-0.15,0.55);
    \coordinate (B) at (0.70,-0.13);
    \coordinate (C) at (0.70,-0.82);
  }
  \def\threeWaysCommonEdges{%
    \draw[thick] (A) -- (B);
    \draw[thick] (A)
      to[curve through={(-0.32,-0.03) .. (-0.02,-0.56)}] (C);
    \draw[thick] (B)
      to[curve through={(0.47,-0.46)}] (C);
    \draw[thick] (B)
      to[curve through={(0.93,-0.46)}] (C);
  }
  \def\threeWaysVertices{%
    \node[MyNode] at (A) {};
    \node[MyNode] at (B) {};
    \node[MyNode] at (C) {};
  }

  % Original graph, shifted left to separate its labels from the arrows.
  \begin{scope}[xshift=-0.55cm]
    \threeWaysCoordinates
    \coordinate (V) at (1.55,0.55);
    \threeWaysCommonEdges
    \draw[thick] (A)
      to[curve through={(0.30,0.89) .. (1.10,0.89)}] (V);
    \draw[thick] (V) -- (B);
    \draw[thick] (V)
      to[curve through={(1.72,-0.03) .. (1.42,-0.56)}] (C);
    \draw[MyRedArc] (A)
      to[curve through={(0.36,0.32) .. (1.04,0.32)}] (V);
    \threeWaysVertices
    \node[MyNode,label={[label distance=1pt]above right:$v$}] at (V) {};

    % Labels on the four edges incident to v.
    \node[inner sep=1pt] at (0.70,0.58) {};
    \node[inner sep=1pt,anchor=south] at (0.70,1.00)
      {$\threeWaysVector{0}{1}$};
    \node[inner sep=1pt,anchor=north west] at (1.17,0.18) {};
    \node[inner sep=1pt,anchor=west] at (1.78,-0.24)
      {$\threeWaysVector{1}{0}$};
  \end{scope}

  % Arrows indicating the three alternative splittings.
  \draw[thick,-{Stealth[length=2.8mm,width=2mm]}]
    (2.12,0.53) -- (4.17,1.54);
  \draw[thick,-{Stealth[length=2.8mm,width=2mm]}]
    (2.16,-0.02) -- (6.08,-0.02);
  \draw[thick,-{Stealth[length=2.8mm,width=2mm]}]
    (2.12,-0.56) -- (4.17,-1.62);

  % Top: pair a with c, and b with d.
  \begin{scope}[shift={(4.70,1.65)}]
    \threeWaysCoordinates
    \threeWaysCommonEdges
    \draw[thick] (A)
      to[curve through={(0.55,0.93) .. (1.32,0.76) .. (1.46,-0.13)}] (C);
    \draw[MyRedArc] (A)
      to[curve through={(0.55,0.30) .. (1.30,0.38) .. (1.19,0.12)}] (B);
    \threeWaysVertices

    % a+c = zero; b+d = [1;1].
    \node[inner sep=1pt] at (0.79,0.67) {};
    \node[inner sep=1pt,anchor=south] at (0.80,1.00)
      {$\threeWaysVector{1}{1}$};
  \end{scope}

  % Middle: pair a with b (creating a loop), and c with d.
  \begin{scope}[shift={(6.70,0)}]
    \threeWaysCoordinates
    \threeWaysCommonEdges
    \draw[thick] (B)
      to[curve through={(1.25,0.13) .. (1.78,0.45)
                        .. (1.83,-0.18) .. (1.28,-0.65)}] (C);
    % A shorter loop, symmetric about the horizontal line through A.
    \draw[MyRedArc] (A)
      to[out angle=15,in angle=-15,
         curve through={(0.42,0.84) .. (1.28,0.55)
                        .. (0.42,0.26)}] (A);
    \threeWaysVertices

    % a+b = [0;1]; c+d = [1;0].
    \node[inner sep=1pt,anchor=south] at (0.46,1.00)
      {$\threeWaysVector{0}{1}$};
    \node[inner sep=1pt,anchor=west] at (2.02,-0.16)
      {$\threeWaysVector{1}{0}$};
  \end{scope}

  % Bottom: pair a with d, and b with c.
  \begin{scope}[shift={(4.70,-1.65)}]
    \threeWaysCoordinates
    \threeWaysCommonEdges
    \draw[thick] (A)
      to[curve through={(0.58,0.90) .. (1.31,0.67) .. (1.18,0.19)}] (B);
    % The white under-stroke distinguishes the crossing from a vertex.
    \draw[MyRedArc,preaction={draw=white,line width=4pt}] (A)
      to[out angle=-15,
         curve through={(0.55,0.32) .. (1.44,0.50)
                        .. (1.70,-0.04) .. (1.43,-0.57)}] (C);
    \threeWaysVertices

    % a+d = [1;0]; b+c = [0;1].
    \node[inner sep=1pt,anchor=west] at (1.84,-0.22)
      {$\threeWaysVector{1}{0}$};
    \node[inner sep=1pt,anchor=south] at (0.79,1.00)
      {$\threeWaysVector{0}{1}$};
  \end{scope}
\end{tikzpicture}
\vspace{-.3cm}
\caption{The three ways to split off a vertex $v$ of a labeled tour graph; edges are labeled by $\vec{0}$ unless otherwise depicted. One edge is highlighted in bold red for convenience.}
\label{fig:threeWays}
\vspace{-.3cm}
\end{figure}
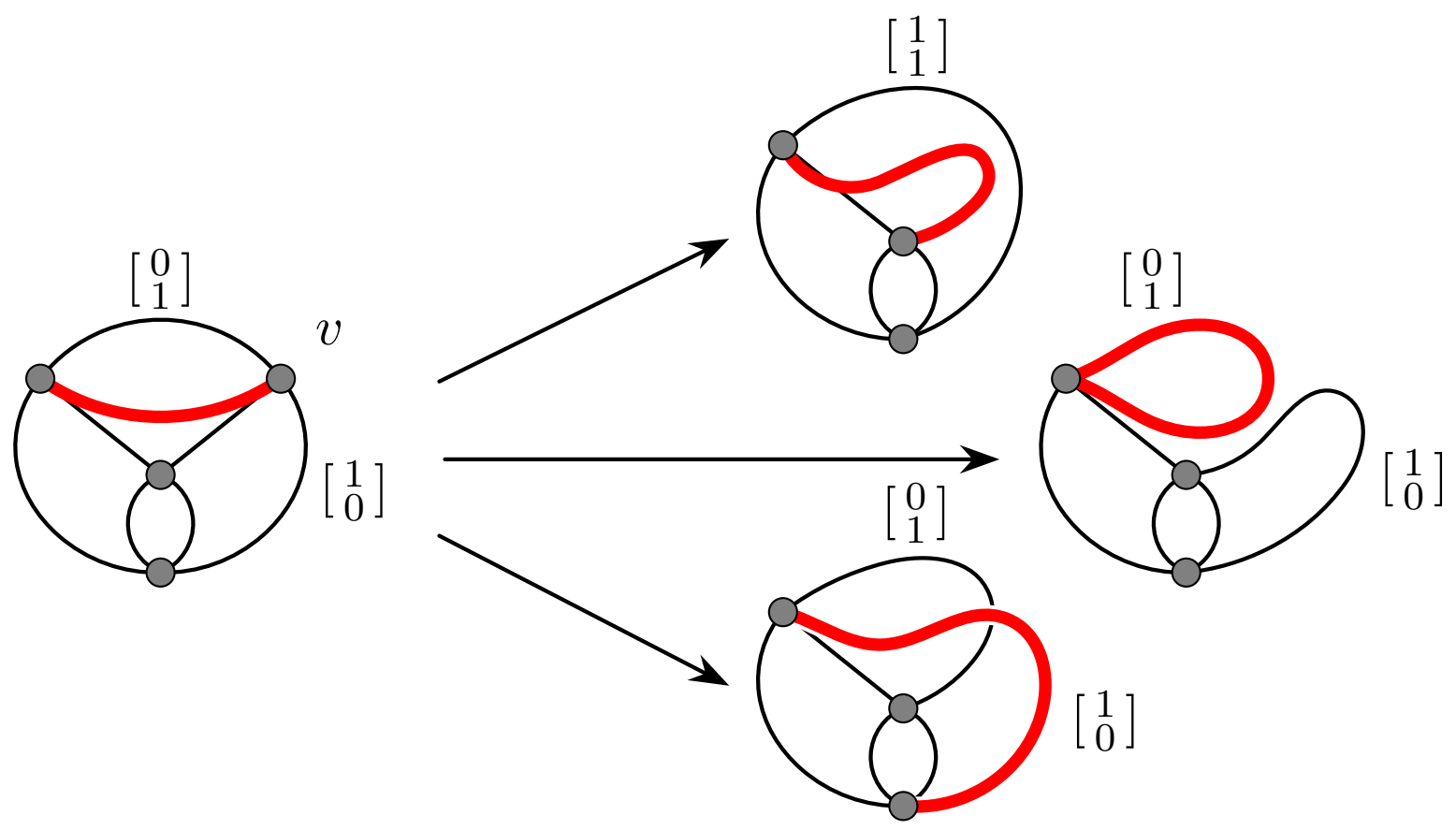

A \emph{tour graph} is a connected $4$-regular graph, and it is \emph{labeled} if its edges are labeled by binary vectors of a fixed length. We allow tour graphs to have loops and parallel edges. We consider labeled tour graphs up to the standard equivalence relation of \emph{shifting}, which adds any binary vector of the given length to every edge of a cut. Given a vertex $v$, there are three ways to pair its four incident edges into two parts of size two\footnote{Technically, we pair the half-edges incident to $v$ in order to handle loops. Also, we do not allow two half-edges to be paired if they are part of the same loop or $2$-edge cut.}. For each of these three ways, we can ``split off'' the two pairs of edges by combining them, thus removing $v$. We sum the labels of the paired edges to obtain the label of the new edge, as depicted in \Cref{fig:threeWays}. This operation is called \emph{splitting off $v$}. A labeled tour graph $\mathcal{H}$ is a \emph{full immersion} of a labeled tour graph $\mathcal{G}$ if $\mathcal{H}$ can be obtained from $\mathcal{G}$ by a sequence of shiftings and of splitting off vertices. It is always possible to perform the shiftings first.

Our algorithm is FPT in the \emph{size} $|\mathcal{H}|$ of the target graph $\mathcal{H}$, which is the number of vertices plus the dimension of the vector space in which it is labeled. Isomorphisms of labeled tour graphs are just graph isomorphisms that preserve the edge labels; we do not allow vector space isomorphism or coordinate projection.

\begin{restatable}{theorem}{mainTheorem}
\label{thm:main}
    There is an algorithm that tests whether an $n$-vertex labeled tour graph $\mathcal{G}$ has a full immersion isomorphic to a labeled tour graph $\mathcal{H}$ in time $\Oh_{|\mathcal{H}|}(n^{5})$.
\end{restatable}

\noindent We actually prove a more precise technical result (\Cref{prop:dynamic}) which allows for any number of the vertices of $\mathcal{H}$ to be mapped to prescribed vertices of~$\mathcal{G}$. Even when the vector space has dimension one, there was previously no known polynomial-time algorithm for arbitrary fixed $\mathcal{H}$, much less an FPT algorithm. 

Full immersions can equivalently be defined as immersions where the branch trails use all of the edges of the host graph\footnote{To obtain $\mathcal{H}$ as an \emph{immersion} of $\mathcal{G}$, we first shift in $\mathcal{G}$, and then we map vertices of $\mathcal{H}$ to distinct vertices of $\mathcal{G}$ called \emph{branch vertices}, and edges of $\mathcal{H}$ to edge-disjoint trails in $\mathcal{G}$ called \emph{branch trails}. (A \emph{trail} is a walk which can repeat vertices but not edges.) Each branch trail must join the right pair of vertices, and the sum of the labels of its edges must agree with the label of the corresponding edge of $\mathcal{H}$. We consider \emph{weak} immersions, where internal vertices of branch trails may be branch vertices, but for tour graphs this does not occur.}. Thus our methods draw on techniques for immersions \cite{TopologicalMinorTesting} and linkages with parity constraints \cite{LiuYoo25}. However, even for immersions that are not full and only have $1$-bit labels, the fastest known algorithm has running time $\Oh(n^{\ell+8})$, where $\ell$ is the number of vertices of the target graph \cite{LiuYoo25}. This algorithm tests all $n^\ell$ ways of mapping $\ell$ vertices into the host graph, applying an FPT algorithm in the rooted setting. To obtain an FPT algorithm in general, we  would need to develop the right notion of \emph{folios}, as mentioned in \cite{LiuYoo25}. For full immersions in tour graphs, it turns out that we only need folios to handle $2$-edge cuts, but this step is already subtle enough.

There is another major difference between \Cref{thm:main} and prior work. Suppose that a labeled tour graph $\mathcal{H}$ is an immersion of a labeled tour graph $\mathcal{G}$, and consider the edges of $\mathcal{G}$ which are unused by the branch trails. These edges can be grouped into components, and since both graphs are 4-regular, we can choose an Eulerian circuit of each component. (A \emph{circuit} is a closed walk which can have repeated vertices but not repeated edges; it is \emph{Eulerian} if it uses all of the edges.) Since $\mathcal{G}$ is connected, we may assume that this Eulerian circuit begins at a vertex which is on a branch trail. However, adding this Eulerian circuit onto the branch trail might change the label of the trail, thus destroying the immersion. So new techniques are required to handle full immersions, as we describe in \Cref{sec:techniquesTesting}. 

When the vector space has dimension zero, \Cref{thm:main} is equivalent to testing for immersions in tour graphs, and so an $\Oh_{\mathcal{H}}(n^3)$ algorithm was already known \cite{TopologicalMinorTesting}. Even this problem is NP-hard when $\mathcal{H}$ is part of the input and a doubled cycle, and, as a corollary, it is NP-hard to test if a graph $G$ has a vertex-minor isomorphic to~$H$ \cite{vmTesting}. 

Graphs whose edges are labeled by binary vectors arise in obstacle removal \cite{pointSeparation2022}, multiway cut \cite{earlierMulticut2016, cutCycleHitting2022}, and ``nearly graphic'' matroids \cite{GerardsGraphicMatroids, structureMatroidsSurvey}. When the vector space has dimension one, they are called \emph{signed graphs} and are usually specified by the set $\Sigma$ of edges labeled by $1$, called the \emph{signature}. So a labeling by vectors of length $k$ is the same as having $k$ signatures $\Sigma_1, \Sigma_2, \ldots, \Sigma_k$. Such labels encode several parity conditions simultaneously, and shifting preserves the sum of the labels around each circuit. 

\subsection*{Applications to vertex-minors}

Our interest in full immersions comes from circle graphs. A \emph{circle graph} is an intersection graph of a collection of chords on a circle: there is a vertex for each chord, and two vertices are adjacent if their chords intersect. Circle graphs play a similar role for vertex-minors as planar graphs do for graph minors. Indeed, Bouchet's characterization of circle graphs in terms of their forbidden vertex-minors \cite{BouchetCircleChar} is directly related \cite{pivotingCircle} to Wagner's characterization of planar graphs in terms of their forbidden minors. 

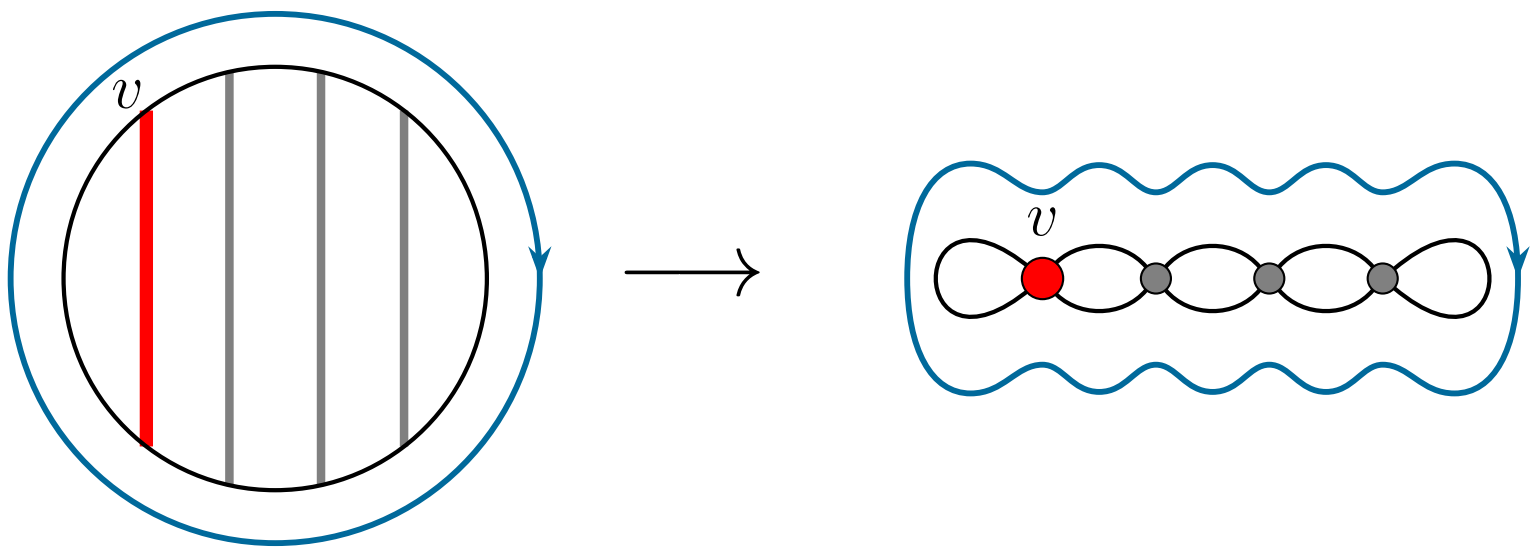
\begin{figure}
\centering
\begin{tikzpicture}[baseline=(current bounding box.center)]
  % Chord diagram.
  \begin{scope}[scale=.7,
    every node/.style={inner sep=0,outer sep=0,fill=none}]
    \def\dif{25}
    \draw[ultra thick,gray]
      (90-.5*\dif:2) -- (-90+.5*\dif:2);
    \draw[ultra thick,gray]
      (90-1.5*\dif:2) -- (-90+1.5*\dif:2);
    \draw[ultra thick,gray]
      (90+.5*\dif:2) -- (-90-.5*\dif:2);
    \draw[ultra thick,MyRedArc]
      (90+1.5*\dif:2) -- (-90-1.5*\dif:2);
    \draw[thick] (0,0) circle[radius=2];
    \node[label=above left:{$v$}] at (90+1.5*\dif:2) {};
    % Clockwise, offset from the circle and its label.
    \draw[EulerCurve,EulerArrow=.25] (90:2.5)
      arc[start angle=90,end angle=-270,radius=2.5] -- cycle;
  \end{scope}

  \node[inner sep=0] at (2.75,0) {\Large$\longrightarrow$};

  % Tour graph.
  \begin{scope}[xshift=6.2cm,every node/.style={MyNode}]
    \def\width{.75}
    \node (X0) at (-1.5*\width,0) {};
    \node (X1) at (-.5*\width,0) {};
    \node (X2) at ( .5*\width,0) {};
    \node (X3) at ( 1.5*\width,0) {};
    \foreach \i in {0,1,2}{
      \pgfmathtruncatemacro{\j}{\i+1}
      \draw[thick] (X\i) to[bend left=50] (X\j);
      \draw[thick] (X\i) to[bend right=50] (X\j);
    }
    \draw[thick] (X0) to[out=140,in=220,looseness=22] (X0);
    \draw[thick] (X3) to[out=-40,in=40,looseness=22] (X3);
    \node[label=above:{$v$},fill=red,inner sep=2.75pt]
      at (-1.5*\width,0) {};

    % A single smooth clockwise outline, following all five lobes.
    % Symmetric about the vertex line, with clearance for the label v.
    % The arrow is at the rightmost point, where the tangent is vertical.
    \draw[EulerCurve,EulerArrow=.5]
      (-2.02,0)
      .. controls (-2.02,.47) and (-1.86,.76) .. (-1.60,.76)
      .. controls (-1.38,.76) and (-1.30,.57) .. (-1.125,.57)
      .. controls (-.99,.57) and (-.93,.75) .. (-.75,.75)
      .. controls (-.57,.75) and (-.51,.57) .. (-.375,.57)
      .. controls (-.24,.57) and (-.18,.75) .. (0,.75)
      .. controls (.18,.75) and (.24,.57) .. (.375,.57)
      .. controls (.51,.57) and (.57,.75) .. (.75,.75)
      .. controls (.93,.75) and (.99,.57) .. (1.125,.57)
      .. controls (1.30,.57) and (1.38,.76) .. (1.60,.76)
      .. controls (1.86,.76) and (2.02,.47) .. (2.02,0)
      % Exact reflection of the upper half in the vertex line y=0.
      .. controls (2.02,-.47) and (1.86,-.76) .. (1.60,-.76)
      .. controls (1.38,-.76) and (1.30,-.57) .. (1.125,-.57)
      .. controls (.99,-.57) and (.93,-.75) .. (.75,-.75)
      .. controls (.57,-.75) and (.51,-.57) .. (.375,-.57)
      .. controls (.24,-.57) and (.18,-.75) .. (0,-.75)
      .. controls (-.18,-.75) and (-.24,-.57) .. (-.375,-.57)
      .. controls (-.51,-.57) and (-.57,-.75) .. (-.75,-.75)
      .. controls (-.93,-.75) and (-.99,-.57) .. (-1.125,-.57)
      .. controls (-1.30,-.57) and (-1.38,-.76) .. (-1.60,-.76)
      .. controls (-1.86,-.76) and (-2.02,-.47) .. (-2.02,0)
      -- cycle;
  \end{scope}
\end{tikzpicture}
\caption{Chords on a circle (left) and the tour graph (right) with its Eulerian circuit in blue.}
\label{fig:tourGraphCircleGraph}
\vspace{-.2cm}
\end{figure}

Kotzig \cite{KotzigImmersions} and de Fraysseix \cite{VMandInterlacement} showed that any circle graph can be encoded by an Eulerian circuit through an associated tour graph. To obtain this tour graph, we view the chords on the circle as a $3$-regular graph, and then we contract each chord into a vertex as depicted in \Cref{fig:tourGraphCircleGraph}; the circle becomes an Eulerian circuit. The vertex-minors of the circle graph then correspond to full immersions of the tour graph. Bouchet \cite{BouchetCircleChar} showed that the neighborhood within the circle graph of an additional apex vertex can be encoded by a signature on the tour graph. So $k$ apex vertices give rise to a labeling of the tour graph by vectors of length $k$. Thus, as a corollary of \cref{thm:main}, we obtain an FPT algorithm for \emph{apex-vertex-minor} testing in \emph{$k$-apex circle graphs} (\cref{cor:apexVM}). We postpone definitions until \Cref{sec:proposedVM}; essentially, we fix the apices and forget about adjacencies between them.

Our proof of \Cref{thm:main} yields an irrelevant vertex theorem in this dense setting. %The rank-width of the host graph must be sufficiently large as a function of both $\mathcal{H}$ and~$k$.

\begin{restatable}{theorem}{irrelevantCircle}
\label{thm:irrelevantCircle}
    For any $k$-apex circle graph $\mathcal{H}$, there is $c = c(\mathcal{H})$ so that every $k$-apex circle graph of rank-width at least $c$ contains an $\mathcal{H}$-irrelevant vertex with respect to apex-vertex-minors.
\end{restatable}

\noindent This is a vertex which can be removed in any of the three ways relevant for vertex-minors without destroying all copies of $\mathcal{H}$. We obtain this vertex from the labeled tour graph: it is a vertex which can be split off in any of the three ways depicted in \Cref{fig:threeWays} without destroying all copies of any small labeled tour graph. We need to handle multiple tour graphs at once to deal with inequivalent representations of circle graphs; see \Cref{thm:interpretation}.

We also propose a strategy to use \Cref{thm:main,thm:irrelevantCircle} to approach the full vertex-minor testing problem, as discussed in \Cref{sec:proposedVM}. In forthcoming work, three of the authors (Geelen, McCarty, and Wollan) prove a structure theorem for graphs with a forbidden vertex-minor, along the lines of the Graph Minors Structure Theorem \cite{graphMinors16Structure}. The work in this paper is an essential step in the upcoming proof; we require nearly all of the results in this paper in order to make our structure theorem algorithmic. For instance, our \Cref{thm:mainMinMax} plays the role of both the algorithmic $A$-paths \cite{packingApaths, algorithmApaths} and Two Paths \cite{Jung, Seymour2paths, Thomassen2linked, Shiloach2paths} theorems in the vertex-minors program. We are optimistic that this algorithmic structure theorem might then be used to find an irrelevant vertex when the rank-width is large, analogously to how the Flat Wall Theorem \cite{graphMinors13} is used to find an irrelevant vertex within an ``almost planar part'' of large treewidth.%\footnote{Departing from the analogy with graph minors, we only know how to find an ``almost circle-graphic part'' which is analogous to a flat wall by going through the full structure theorem.}

The results in this paper already have a concrete application in the vertex-minors program. The third author's Ph.D. thesis \cite{rosethesis} uses our \Cref{prop:allsignatures} to prove ``the Local Structure Theorem'' for graphs with a forbidden vertex-minor\footnote{Part of our proof of \Cref{prop:allsignatures} is based on Chapter~5 of the third author's Ph.D. thesis~\cite{rosethesis}; this work has not previously been published elsewhere. Here we also prove an algorithmic version.}. Very roughly, this proposition says that a $4$-edge-connected labeled tour graph with large treewidth contains both a full immersion of a grid-like graph, and a corresponding small edge cut separating off the signatures which do not appear richly on the grid. It is within the vertices of this grid-like full immersion that we ultimately find the irrelevant vertex for \Cref{thm:irrelevantCircle}.

\subsection*{Finding rooted circuit decompositions in Eulerian graphs}

We provide a new tool in order to prove \Cref{thm:main}. This tool is of independent interest and applies to every Eulerian graph. A graph $G$ is \emph{Eulerian} if it is connected and every vertex has even degree. A \emph{circuit decomposition} is a collection of circuits which partitions~$E(G)$.  

We begin with an illustrative example. Suppose that we have an Eulerian 
circuit beginning at a vertex $t$ in $G$. We traverse this Eulerian circuit, and every time we return to~$t$, we ``cut off a circuit''. This yields a circuit decomposition with exactly $\text{degree}(t)/2$-many circuits, each beginning and ending at $t$. We call such a circuit decomposition a \emph{flooding} \emph{rooted at} $t$. We give an $\Oh(|E(G)|^5)$-time algorithm to find a flooding rooted at $t$ with the maximum number of odd-length circuits. While it does not change the algorithmic complexity of the problem, it is convenient to frame this result in terms of a graph $G$ with a specified set of edges $\Sigma$ called the \emph{signature}. Then a circuit is \emph{odd} if it has an odd number of edges in the signature $\Sigma$. When $\Sigma = E(G)$, these are just the circuits of odd length.

Previously there was no known polynomial-time algorithm for this problem. The third author gave a combinatorial min-max theorem \cite{decomposingSigned}, but the proof was not constructive. That paper mentions that, for applications to vertex-minors, a generalization to graphs labeled by binary vectors is probably required. We avoid this issue by introducing ``free edges'': the signature stays, and we only add a second distinguished set of edges, yet this already suffices for the applications we have in mind. 

So, we say that a \emph{free graph} is a tuple $\mathcal{G}=(G, \Sigma, F, t)$ so that $G$ is an Eulerian graph, $\Sigma$ is a set of edges of $G$ called the \emph{signature}, $F$ is a set of edges of $G$ called the \emph{free edges}, and $t$ is a vertex of $G$ called the \emph{root}. (We have no restrictions on how $\Sigma$ and $F$ interact; for instance they may intersect. For readers familiar with group-labeled graphs, it may be helpful to consider $F$ as a labeling in a free group, but without shifting.) A circuit of $G$ is \emph{feasible} if it contains at least one free edge or an odd number of signature edges, or both. In a free graph, every flooding is rooted at $t$, and the \emph{value} of a flooding is the number of feasible circuits that it contains. We want to maximize this value. 

The dual objects are \emph{certificates}: tuples $(X, \Sigma')$ so that $X$ is a set of vertices containing $t$ and $\Sigma'$ is a signature which is equivalent to $\Sigma$ up to shifting. If $G$ is $4$-edge-connected, then, roughly, the \emph{value} of a certificate is small if and only if $\delta(X)$ is a small cut which separates almost all of the signature $\Sigma'$ and the free edges $F$ from the root. The full definition also takes parity into account. For instance, if the root has degree two and there are no free edges, then every flooding is an Eulerian circuit, and its value is equivalent to the number of signature edges modulo~$2$. In general, we save $1$ if the parity is ``wrong''; see \Cref{sec:techniquesMinMax}. %We prove in \Cref{lem:easyDir} that the maximum value of a flooding is at most the minimum value of a certificate.

Since we expect the target value $k$ of a flooding to be a constant in applications to vertex-minors, we state the theorem with $k$ as input. 

\begin{restatable}{theorem}{mainTheoremMM}
\label{thm:mainMinMax}
    For every free graph, the maximum value of a flooding equals the minimum value of a certificate. Moreover, given $k\in \mathbb{Z}^+$ and an $m$-edge free graph, we can in $\mathcal{O}(km^4)$ time find either a flooding of value at least $k$, or a flooding and a certificate of equal value.
\end{restatable}

\noindent This generalizes and independently proves the min-max theorem for signed graphs in \cite{decomposingSigned}. Taking $k=m+1$ yields an optimal flooding and certificate in $\mathcal{O}(m^5)$ time.

Free edges already arise in the proof of \Cref{thm:main}. We handle $2$-edge cuts by dynamic programming, recording the full immersions of the ``cut-off pieces'' (i.e., its \emph{folio}) as colors on the remaining edges. We then apply \Cref{thm:mainMinMax} with the edges of a given color as the free edges. We also apply \Cref{thm:mainMinMax} for signed graphs in order to obtain an irrelevant vertex. Intuitively, one can think of the $\Oh_{|\mathcal{H}|}(n^5)$ running time from \Cref{thm:main} as coming from $n$ applications of \Cref{thm:mainMinMax}, each time finding one irrelevant vertex. In particular, \Cref{thm:mainMinMax} is the barrier to improving the runtime of \Cref{thm:main}.

\section{Connection to vertex-minor testing}
\label{sec:proposedVM}

In this section we interpret our results in the context of vertex-minors.

\subsection*{Irrelevant vertices}

Many conjectures about vertex-minors are motivated by a direct connection with graph minors \cite{BOUCHET198858} as explained in \cite{RWAndVM}. Essentially, there is a special type of vertex-minor called a \emph{pivot-minor}, and results about pivot-minors often simultaneously imply results about both vertex-minors and graph minors. \emph{Pivoting} on an edge $uv$ means to locally complement at $u$, then $v$, then $u$ again (or, equivalently, on $v, u, v$). The \emph{pivot-minors} of a graph are obtained by pivoting on edges and by deleting vertices. Well-quasi-ordering (WQO) of graphs under pivot-minors would imply both WQO of graphs under vertex-minors and WQO of graphs under minors, which is the famous Graph Minors Theorem~\cite{graphMinors20WQO}.

Inspired by this connection, in 2005 Oum asked if there is a polynomial-time algorithm to test for any fixed graph $H$ as a vertex-minor up to isomorphism \cite{OumConjecture}. It is also natural to conjecture the following stronger FPT version.

\begin{conjecture}
\label{conj:vmTesting}
    There exists a constant $d$ so that we can test whether a graph $H$ is isomorphic to a vertex-minor of an $n$-vertex graph $G$ in time $\Oh_H(n^d)$.
\end{conjecture}

The natural approach to \Cref{conj:vmTesting} is to find irrelevant vertices, and ``irrelevance'' should be invariant under local complementation. We say that two graphs are \emph{locally equivalent} if each can be obtained from the other by locally complementing, and \emph{pivot equivalent} if each can be obtained from the other by pivoting. Given graphs $H$ and $G$, a vertex $v$ of $G$ is \emph{$H$-irrelevant} if for any graph $\widetilde{G}$ which is locally equivalent to $G$, the graph $\widetilde{G}-v$ contains a vertex-minor isomorphic to $H$ if and only if $G$ does. So, informally, locally complementing and then deleting $v$ cannot destroy the only copy of $H$ as a vertex-minor. 

Bouchet \cite{BOUCHET198858} and Fon-der-Flaas \cite{FdF1988} (see \cite[Lemma~1.4]{pivotingCircle})
showed that there are only ``three ways to remove a vertex''; this yields an equivalent definition of irrelevance. Let $G*v$ be the graph obtained from $G$ by locally complementing at $v$, and let $G/v$ be the graph obtained from $G$ by pivoting on an arbitrary edge incident to $v$ (if one exists) and then deleting $v$. The graph $G/v$ is defined up to pivot equivalence \cite[Proposition~2.5]{RWAndVM}. The ``three ways'' lemma says that if $H$ is a vertex-minor of $G$ and $v \notin V(H)$, then $H$ is also a vertex-minor of at least one of the three graphs $G-v$, $G*v-v$, and $G/v$. Thus a vertex $v$ of $G$ is \emph{$H$-irrelevant} if, when $G$ contains a vertex-minor isomorphic to $H$, then so do all three of the graphs $G-v$, $G*v-v$, and $G/v$. For circle graphs, these three ways of removing $v$ correspond to the three ways of splitting off $v$ depicted in \Cref{fig:threeWays}.

Now we can formalize the expected folklore approach to vertex-minor testing.

\begin{conjecture}
\label{conj:irrelevant}
    For any graph $H$, there exists an integer $c=c(H)$ so that every graph $G$ of rank-width at least $c$ contains an $H$-irrelevant vertex.
\end{conjecture}

\noindent This can be thought of as a vertex-minor version of the Unique Linkage Theorem \cite{uniqueLinkage}. 

An algorithmic irrelevant vertex theorem is the last missing piece needed to prove the \emph{non-uniform} version of \Cref{conj:vmTesting}, where there is a different algorithm for each fixed target graph $H$; see \Cref{sec:conclusion}. While we see no real obstacle to finding a uniform algorithm this way, one would need to verify that the bound in \cite{GEELEN202393} is computable.

\subsection*{Apex circle graphs}

We now formalize what our results say about vertex-minors. First, a \emph{$k$-apex circle graph} is a tuple $\mathcal{G}=(G, A)$ so that $G$ is a graph and $A$ is a set of exactly $k$ vertices in $G$ called \emph{apices} so that $G-A$ is a circle graph. An \emph{apex-vertex-minor} of $\mathcal{G}$ is any $k$-apex circle graph which can be obtained from $\mathcal{G}$ by locally complementing at and deleting only non-apex vertices, and by switching the adjacency between pairs of apex vertices. So we lose track of what is going on between the apex vertices, but we do keep track of their neighborhoods within the circle graph. We require \emph{isomorphisms} of $k$-apex circle graphs to fix each apex vertex.

We discussed in the introduction how Kotzig \cite{KotzigImmersions}, de Fraysseix \cite{VMandInterlacement}, and Bouchet \cite{BouchetCircleChar} showed that each circle graph has a corresponding tour graph and each apex vertex yields a corresponding signature on the tour graph. Thus any $k$-apex circle graph $\mathcal{G}=(G, A)$ has a corresponding labeled tour graph $\mathcal{T}_{\mathcal{G}}$. The vertex-set of $\mathcal{T}_{\mathcal{G}}$ is $V(G)\setminus A$, and we view its edges as being labeled over the binary vector space $\mathbb{F}_2^A$; thus each vertex $a \in A$ has a corresponding signature $\Sigma_a$ on $\mathcal{T}_{\mathcal{G}}$ given by the labeling. 

This labeled tour graph $\mathcal{T}_{\mathcal{G}}$ is not unique, however the work of Kotzig, de Fraysseix, and Bouchet immediately implies the following theorem (see \cite{rosethesis} and \cite{vmTesting}).

\begin{theorem}[\cite{KotzigImmersions,VMandInterlacement,BouchetCircleChar}]
\label{thm:interpretation}
Let $\mathcal{H}$ and $\mathcal{G}$ be apex circle graphs, and let $\mathcal{T}_{\mathcal{G}}$ be a labeled tour graph of $\mathcal{G}$. Then $\mathcal{H}$ is (respectively, is isomorphic to) an apex-vertex-minor of $\mathcal{G}$ if and only if there exists a labeled tour graph $\mathcal{T}_{\mathcal{H}}$ of $\mathcal{H}$ so that $\mathcal{T}_{\mathcal{H}}$ is (respectively, is isomorphic to) a full immersion of $\mathcal{T}_{\mathcal{G}}$.
\end{theorem}

We use \Cref{thm:interpretation} and an $\Oh(n^3)$-time algorithm for finding representations of circle graphs \cite{recognizingCircleGraphs} to prove the following corollary of \Cref{thm:main} in \Cref{sec:conclusion}. 

\begin{restatable}{corollary}{apexTestingCor}
\label{cor:apexVM}
    There is an algorithm that tests whether an $n$-vertex $k$-apex circle graph $\mathcal{G}$ has an apex-vertex-minor isomorphic to a $k$-apex circle graph $\mathcal{H}$ in time $\Oh_{\mathcal{H}}(n^{5})$.
\end{restatable}

To define irrelevant vertices, let $\mathcal{G}$ and $\mathcal{H}$ be $k$-apex circle graphs with $\mathcal{G}=(G, A)$. An \emph{$\mathcal{H}$-irrelevant vertex} in $\mathcal{G}$ is a vertex $v \in V(G)\setminus A$ such that $\mathcal{G}$ has an apex-vertex-minor isomorphic to $\mathcal{H}$ if and only if all of the $k$-apex circle graphs $(G-v, A)$, $(G*v-v, A)$, and $(G/v,A)$ also have an apex-vertex-minor isomorphic to $\mathcal{H}$. In \Cref{sec:conclusion} we show that the rank-width of a circle graph is functionally equivalent to the treewidth of its tour graph(s); a quadratic upper bound was already shown by Oum \cite[Lemma~9.2]{excludingBipCircle}.

\begin{restatable}{lemma}{widthTour}
\label{lem:widthTourFuncEquiv}
If $G$ is a circle graph with rank-width $r$ and $\mathcal{T}$ is a tour graph of $G$ with treewidth $t$, then $(t-5)/48\leq r \leq 2t+1$.
\end{restatable}

Then \Cref{thm:irrelevantCircle} (the irrelevant vertex theorem for apex circle graphs) immediately follows from \Cref{thm:interpretation}, \Cref{lem:widthTourFuncEquiv}, and the following theorem about labeled tour graphs proven in \Cref{sec:conclusion}. Given labeled tour graphs $\mathcal{H}$ and $\mathcal{G}$, a vertex $v$ of $\mathcal{G}$ is \emph{$\mathcal{H}$-irrelevant} if $\mathcal{G}$ contains a full immersion isomorphic to $\mathcal{H}$ if and only if all of the tour graphs which can be obtained from $\mathcal{G}$ by splitting off $v$ contain a full immersion isomorphic to $\mathcal{H}$.

\begin{restatable}{theorem}{irrTheorem}
\label{thm:irrelevantLabeled}
    For any integers $k$ and $\ell$, there exists an integer $c=c(k, \ell)$ so that every labeled tour graph of treewidth at least $c$ contains a vertex which is $\mathcal{H}$-irrelevant for every labeled tour graph $\mathcal{H}$ which has $\ell$ vertices and $k$ bits on the labels.
\end{restatable}

\subsection*{Vertex-minors program}

We now briefly sketch how the results in this paper will be used in upcoming work of the authors obtaining an algorithmic structure theorem for graphs with a forbidden vertex-minor. Very roughly, the theorem says that for any fixed graph $H$, every graph without a vertex-minor isomorphic to $H$ ``decomposes'' into parts which are ``almost'' circle graphs. The theorem is closely related\footnote{Our theorem is not exactly the same as the conjecture for technical reasons; it turns out that we have to decompose on sets of cut-rank $2$ one at a time rather than all simultaneously.} to the conjecture of the second author described in \cite{rosethesis}. The ``decomposition'' has to do with rank-width, and ``almost'' means that we can add a bounded number of vertices (as a function of $H$), perform local complementations, and then delete them. This latter operation is called a \emph{perturbation}.

Suppose that we are given a desired target graph $H$, an $n$-vertex host graph $G$, and a set of vertices $T$ which induces a large comparability grid in $G$. We grow a circle graph around $T$ one vertex at a time, performing perturbations when necessary. Ultimately, at each stage of the algorithm we have an apex circle graph $\mathcal{G}$ which is not quite the same as our original graph $G$ but is closely related to it. The apex vertices of $\mathcal{G}$ are split into two sets: a small set $A$ which (intuitively) specifies the perturbation, and a set $V$ of vertices that we have not yet understood. We use \Cref{thm:mainMinMax} and \Cref{prop:allsignatures} to determine when a vertex $v\in V$ can either be added to the circle graph, or ``specifies a necessary perturbation'' and thus can be moved into $A$. The relevant new signature under consideration is $\Sigma_v$, and we need the free edges in order to very precisely understand how $\Sigma_v$ interacts with certain ``fragile'' edges of the previous signatures $(\Sigma_a: a\in A)$. If $A$ ever grows too large, then we use \Cref{thm:mainMinMax} and \Cref{prop:allsignatures} to find a full immersion which yields a vertex-minor of $G$ isomorphic to $H$. Otherwise, the procedure ends when every vertex has been added to the circle graph\footnote{At various points throughout the procedure we may also need to decompose on a set of cut-rank at most~$2$.}. 

We view \Cref{thm:mainMinMax} as being analogous to both the $A$-paths and Two Paths Theorems because it is how we ``localize non-circle-graphic signatures'' in $(\Sigma_v: v \in V)$ to one ``small part'' of the tour graph. We use \Cref{thm:mainMinMax} to obtain a small edge cut so that on the side containing $T$, all of the signatures ``look circle-graphic''. To obtain this edge cut, we identify the vertices in $T$ to a single root vertex, apply \Cref{thm:mainMinMax}, and use the resulting certificate.

\section{Finding a flooding and certificate}
\label{sec:techniquesMinMax}

In this section we outline our strategy for finding an optimal flooding and corresponding certificate in a free graph, that is, for proving \Cref{thm:mainMinMax}. 

First we give the precise definition of the \emph{value} of a certificate $(X, \Sigma')$ of a free graph $\mathcal{G}=(G, \Sigma, F, t)$. We say that a component $H$ of $G-X$ is \emph{$\Sigma'$-odd} if $V(H)$ is not incident to any free edge of $G$, and the number of edges of $G$ which are incident to $V(H)$ and in $\Sigma'$ is different from $|\delta(H)|/2$ modulo~2. We write $odd_{\Sigma'}(G-X)$ for the number of components of $G-X$ which are $\Sigma'$-odd, and $E_{F, \Sigma'}(X)$ for the set of all edges of $G$ with both ends in $X$ which are either free or in $\Sigma'$ or both. The \emph{value} of a certificate $(X, \Sigma')$ is then
\begin{align*}
 \val(X, \Sigma') = |E_{F, \Sigma'}(X)|+|\delta(X)|/2-odd_{\Sigma'}(G-X).
\end{align*}

Now, for the rest of the proof overview, it may be instructive to consider the case where we are given only an Eulerian graph $G$ with a root vertex $t$; then a \emph{flooding} is a circuit decomposition ``rooted'' at $t$, and its \emph{value} is the number of odd-length circuits that it contains (we call these circuits \emph{feasible}). A flooding is \emph{optimal} if its value is maximum.

We prove a stronger theorem than just the min-max theorem, analogous to how the Gallai--Edmonds decomposition is a stronger version of the Tutte--Berge formula for the maximum size of a matching. In this analogy, instead of considering a vertex to be inessential if there exists a maximum matching that avoids it, we consider an edge $e$ to be \emph{inessential} if there exists an optimal flooding which puts $e$ in an infeasible circuit. We then want to find a certificate so that every edge which ``looks like it could be inessential according to the certificate'' actually is. In \Cref{sec:prelim} we word things the other way around and say that the ``obviously essential'' edges are \emph{forced by} the certificate.

Using this stronger type of certificate, we can handle the free edges by ``greedily absorbing them into the signature'' (\Cref{thm:reduction}). From now on we have no free edges, but the tradeoff is that we are committed to finding this stronger type of certificate.

At a high level, the rest of our approach is inspired by the $A$-paths algorithm in \cite{algorithmApaths}, although the details are entirely different. The main commonality is that we grow a ``dictionary displaying the inessential parts''. For us, a \emph{dictionary} is a collection of ``not too many'' floodings which all have the same value. It \emph{displays} an edge $e$ if there is a flooding in the dictionary which puts $e$ in an infeasible circuit. Consider the instructive case where a circuit is infeasible if it has an even number of edges. There are actually two ways this even circuit could look relative to the ends $u$ and $v$ of $e$. Beginning at the root, it could first reach $u$ by an odd-length trail, traverse $e$, and then end with an even-length trail from $v$. Alternatively, the parity of these two trails could be switched. Something similar holds if there is a signature $\Sigma$. So there are actually two \emph{ways} to display each edge. 

Beginning with a dictionary $\mathcal{D}$, we find one of three outcomes: 1) a dictionary $\mathcal{D}'$ so that the sum over all edges $e$ of the number of ways $e$ is displayed is larger than in $\mathcal{D}$, 2) a dictionary $\mathcal{D}'$ consisting of a single flooding of larger value than $\mathcal{D}$, or 3) a certificate of the same value as $\mathcal{D}$ so that all of the non-forced edges are displayed by $\mathcal{D}$. If we find the certificate then we are done; it satisfies the stronger property that we require. Otherwise, we continue with the new dictionary until after at most $2m$ steps the second outcome occurs: we get a flooding of larger value. Our main technical result, \Cref{prop:algSigned}, says that we can find one of these three outcomes in $\Oh(m^3)$ time. Thus repeated applications of \Cref{prop:algSigned} immediately give the $\Oh(km^4)$ running time of \Cref{thm:mainMinMax}. 

So, how do we actually find the certificate or new dictionary $\mathcal{D}'$ in \Cref{prop:algSigned}? Sometimes we can combine two floodings from $\mathcal{D}$ directly in order to obtain a single new, helpful flooding (see \Cref{sec:dictionary}). This brings us to a new technique for Eulerian graphs: if $e$ is an edge incident to a vertex $v$, then there are an odd number of other edges incident to $v$ with which we can pair $e$. In \Cref{prop:mostWork} we show that a majority of these pairings ``maintain'' certain desirable properties of a flooding. Thus, given two floodings with different desirable properties, we can find a pairing which simultaneously maintains \emph{both} properties. Ultimately, we find a single flooding which has both desirable properties. 

When we cannot find these two different floodings in $\mathcal{D}$ directly, we instead repeat the whole process above recursively. First we ``pull down'' an edge which is displayed exactly one way into two new edges through the root, as in \cite{decomposingSigned}. Then, depending on the outcome of this recursive call, we may have to again combine two different floodings into one. As a crucial intermediate step, we show that every edge that is displayed in two ways gives rise to either a flooding of larger value, or a loose circuit (\Cref{lem:findLoose}). A circuit $L$ is \emph{loose} in a given flooding if $L$ is a feasible subcircuit of an infeasible circuit in the flooding: intuitively, removing $L$ from the graph yields a flooding of higher value.

\section{Testing for a full immersion}
\label{sec:techniquesTesting}

Now we discuss our FPT algorithm for testing if an $n$-vertex labeled tour graph $\mathcal{G}$ has a full immersion isomorphic to a labeled tour graph $\mathcal{H}$. Our core algorithm requires $\mathcal{G}$ to be 4-edge-connected, but we are only guaranteed that it is connected. So we adapt the \textit{folio} machinery from \cite{TopologicalMinorTesting} to be able to focus on the 4-edge-connected case.

\subsection*{Folios}
Since tour graphs are Eulerian, every edge cut has even size, so the only obstruction to $\mathcal{G}$ being 4-edge-connected is a 2-edge cut. Informally, we deal with this by decomposing $\mathcal{G}$ along 2-edge cuts into 4-edge-connected components using the algorithm of~\cite{cactus}.

\begin{figure}
    \centering
    \includegraphics[width=\textwidth]{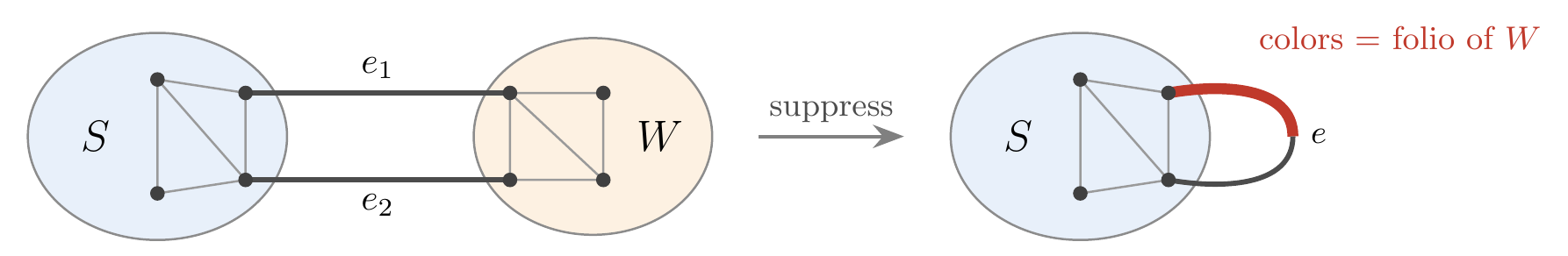}
    \vspace{-.5cm}
    \caption{Suppressing a 2-edge cut; we use a half-edge color to represent the folio of~$W$.}
    \label{fig:virtual}
    \vspace{-.2cm}
\end{figure}

We examine each component separately, beginning at the leaf components, using the algorithm for 4-edge-connected graphs. Then we combine the solutions using dynamic programming. To explain how this method works, consider a 2-edge cut $\delta(W)=\{e_1, e_2\}$ as depicted in \Cref{fig:virtual} (left). Imagine ``cutting down the middle'' so that $e_1$ and $e_2$ are cut in half, and suppose that we already know all of the small graphs that the $W$ side fully immerses. This side has ``dangling'' half-edges corresponding to $e_1$ and $e_2$, but we can combine these into one new edge and keep track of how each full immersion treats this special edge. Then, roughly speaking, the \textit{folio} of the labeled tour graph on $W$ is the set of small labeled tour graphs that it fully immerses.

\begin{figure}[ht]
    \centering
    \includegraphics[width=\textwidth]{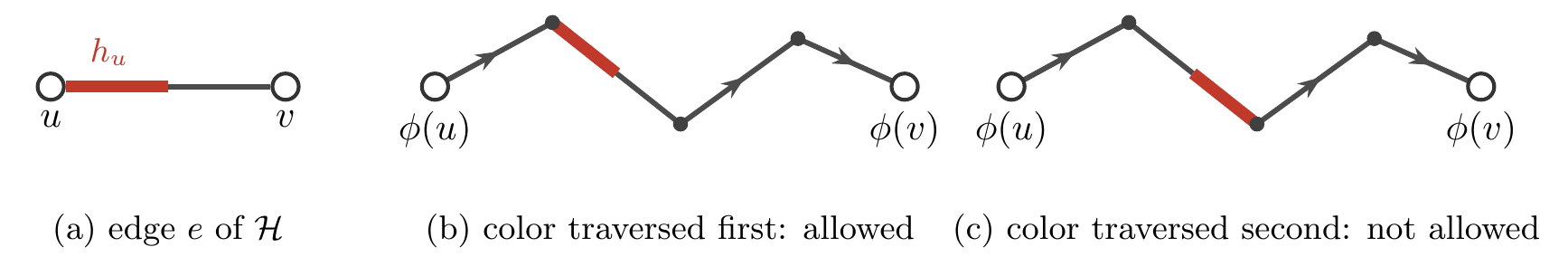}
    \vspace{-1cm}
    \caption{The two ways that a trail can traverse a colored half-edge.}
    \label{fig:halfEdgeColor}
    \vspace{-.2cm}
\end{figure}

In order to find the folio of the labeled tour graph on the other side of the cut (which may itself have two ``dangling'' half-edges if it is not the root of the decomposition tree), we suppress $\delta(W)$ into a \emph{virtual edge} and encode the folio of $W$ using half-edge colors as depicted in \Cref{fig:virtual} (right). Each full immersion in the folio gives rise to one color, and half-edges may receive multiple colors. The label on this virtual edge is the sum of the labels of all edges incident to $W$. Morally, half-edge colors are required because there are two ways to insert a full immersion, corresponding to Whitney flips. When our target graph has a color on a half-edge $h_u$ incident to $u$, the corresponding branch trail beginning at the branch vertex $\phi(u)$ must traverse a half-edge of that color ``first'', as depicted in \Cref{fig:halfEdgeColor}.

In general, there may be many $2$-edge cuts $\delta(W_1), \delta(W_2),\ldots, \delta(W_r)$ attached to the next $4$-edge-connected component that we want to understand. Thus there may be many colored half-edges; we require the corresponding branch trail to contain at least one of the right color. The number of colors (i.e. the folio size) is always bounded by a function of~$|\mathcal{H}|$.

\subsection*{Finding a clean grid in $\mathcal{G}$}
Now we see how to handle the 4-edge-connected case. From now on $\mathcal{H}$ and $\mathcal{G}$ additionally have half-edge colors and $\mathcal{G}$ is $4$-edge-connected\footnote{In the full statement we consider \emph{decorated graphs}, which are also allowed to have \emph{root vertices}; see \Cref{sec:prelimTesting}. Root vertices allow us to map the vertices of $\mathcal{H}$ to prescribed vertices of~$\mathcal{G}$.}. 

In this case we use the irrelevant vertex technique to reduce to the bounded treewidth case. If $\mathcal{G}$ has small treewidth, we compute a tree decomposition of small width \cite{Bodlaender96, Korhonen2021}, express ``$\mathcal{G}$ fully immerses $\mathcal{H}$'' in counting monadic second-order logic, and apply Courcelle's theorem~\cite{Courcelle90} to test for a full immersion in $\Oh_{|\mathcal{H}|}(n)$ time; see \Cref{lem:using-courcelles}.
 
If, on the other hand, the treewidth of $\mathcal{G}$ is large, then the idea is roughly as follows. By \cite{chudnovsky2013immersionfouredgeconnectedgraphs}, large treewidth guarantees that $\mathcal{G}$ immerses a $t \times t$ grid\footnote{Since we require a computable bound, we end up using \cite{immersionEdgeConn} and \cite{edgeDisjOdd} instead; see \Cref{lem:gridImmersion}.}. Here, $t$ is an integer which is sufficiently large relative to $|\mathcal{H}|$, to be chosen later. We can use the algorithm of \cite{TopologicalMinorTesting}  to attain this immersion in $\Oh_{|\mathcal{H}|}(n^4)$ time\footnote{We split off each pair of edges in turn, applying the $\Oh_{|\mathcal{H}|}(n^3)$ decision algorithm in \cite{TopologicalMinorTesting}.}. To transform this immersion into a full immersion, we greedily connect pairs of branch vertices on the boundary of the grid by a trail among the unused edges. Ultimately this yields a full immersion of a tour graph which has a spanning $t \times t$ grid subgraph; we call any such graph a \emph{grid-like graph}, and we call the $t\times t$ grid its \emph{underlying grid}. Its \emph{outside edges} are the edges which are not part of its underlying grid. We call a full immersion of a grid-like graph in $\mathcal{G}$ a \textit{grid in $\mathcal{G}$}.

We use the structure of this grid in $\mathcal{G}$ to find an irrelevant vertex.
Informally, if some signature or half-edge color appears ``often'' and is sufficiently ``spread out'' in this grid, we are easily able to route trails in a full immersion such that they have this signature or color. Roughly speaking, this is robust under splitting off a branch vertex of the grid arbitrarily. If this condition holds for a set of signatures $A$ and a set of colors $S$ for a grid in $\mathcal{G}$, we call this grid ``$A$-rich and $S$-saturated''. Hence, starting with this initial grid in $\mathcal{G}$, we attempt to ``enrich'' and ``saturate'' the grid with as many signatures and colors as possible. Each time we add a signature or a color the grid will become somewhat smaller, but this is fine since we only need it to be large relative to $|\mathcal{H}|$.

If two different signatures $\Sigma_1$ and $\Sigma_2$ contain exactly the same grid edges, then, in the $2$-bit case, this would mean that the only vector labels we can get from the grid are $\begin{bmatrix} 0 & 0 \end{bmatrix}$ and $\begin{bmatrix} 1 & 1 \end{bmatrix}$. So, more precisely, a grid in $\mathcal{G}$ is \emph{$A$-rich} if, when we restrict to the $k$ bits corresponding to the $k$ signatures in $A$, all edges of the underlying grid are labeled by $\vec{0}$, and there is a basis of the vector space $\mathbb{F}_2^k$ which appears ``often'' on the outside edges.

We have roughly explained what it means for a grid in $\mathcal{G}$ to be $A$-rich and $S$-saturated. Our goal is now to use \Cref{thm:mainMinMax} to either grow $S$ or $A$ (using the flooding), or to find a good reason why we cannot (using the certificate). For this step, we first \emph{clean} the grid in $\mathcal{G}$. For the purposes of this overview, we think of this as meaning that each signature outside of $A$ appears nowhere on the grid in $\mathcal{G}$. However the actual condition is more subtle. We show how to clean a grid in~$\mathcal{G}$ in \Cref{sec:clean-grid}.

\subsection*{Enriching and saturating a clean grid in $\mathcal{G}$}
At this stage we have found a clean grid in $\mathcal{G}$ that is $A$-rich and $S$-saturated. 

To attempt to enrich this grid with a signature $\Sigma$, we obtain a free graph by identifying the branch vertices of the grid to a single new root vertex, and apply the flooding algorithm from \Cref{thm:mainMinMax}, considering only the signature $\Sigma$. If we find a flooding of large value, then we are able to enrich the grid with signature $\Sigma$. An analogous statement holds for colors, except that each color $c$ gives rise to a set of free edges: the edges with a half-edge of color $c$. We apply this algorithm iteratively over all signatures and colors. If we ever obtain the flooding, we enrich/saturate further and repeat the process; see \Cref{sec:enrich grid,sec:colors}. 
 
So suppose, on the other hand, that for each remaining signature and color, \Cref{thm:mainMinMax} yields a certificate of small value. Since $\mathcal{G}$ is $4$-edge-connected, each certificate gives rise to a small edge cut separating off almost all of that signature or color from the branch vertices of the grid in $\mathcal{G}$. In general, given a small edge cut, almost all of the branch vertices of the grid must be on one side (\Cref{lem:connGrid}); we call this the \emph{big side} and the other side the \emph{small side}. (This terminology comes from tangles.) Since $\mathcal{G}$ is $4$-regular, we can move the few remaining signature/color edges out of the big side without growing the cut by much. Finally, by taking the union of all small sides, we obtain a small edge cut separating all signatures not in $A$ and all colors not in $S$ from almost all of the branch vertices.

\subsection*{Finding an irrelevant vertex}
Consider the small edge cut we have found which separates all signatures not in $A$ and all colors not in $S$ from almost all of the branch vertices of the $A$-rich $S$-saturated grid in $\mathcal{G}$. 

\begin{figure}
    \centering
    \includegraphics[width=0.65\textwidth]{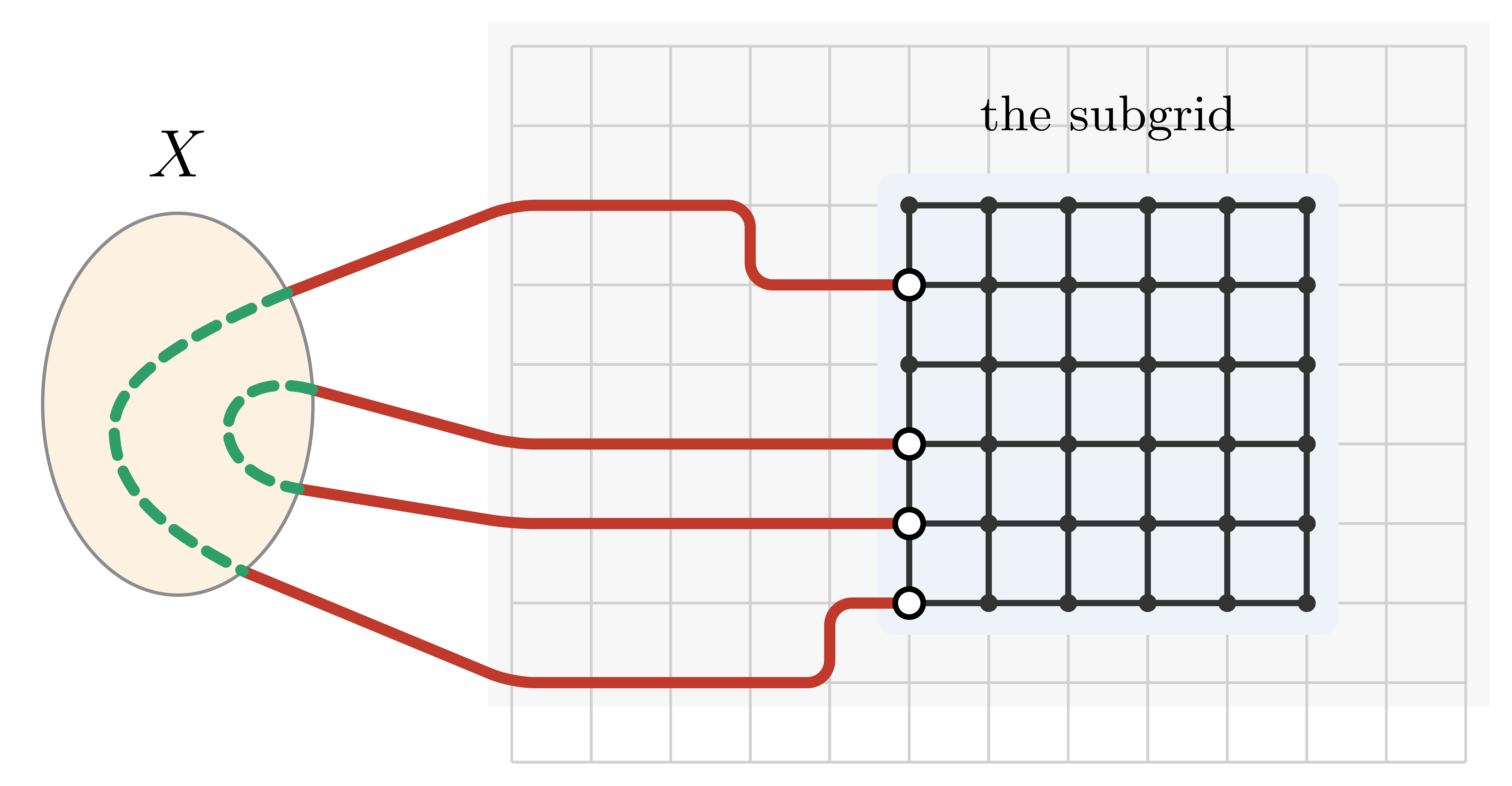}
    \vspace{-.4cm}
    \caption{The trails $P_1, P_2 \ldots, P_{\kappa}$ (red) and the untouched branch trails of $\mathcal{H}$ (dashed green).}
    \label{fig:route}
    \vspace{-.35cm}
\end{figure}

To see how we find an irrelevant vertex, suppose that $\mathcal{H}$ is a full immersion of $\mathcal{G}$ and consider the branch trails of this full immersion. These trails can be broken up into subtrails whose vertices are on the small side of the cut (we leave these untouched), and subtrails whose internal vertices are in the big side (we re-route these -- they only carry labels and colors that the grid can reproduce). So, if $\kappa$ is the number of edges in the cut, then our goal is to find $\kappa$ pairwise edge-disjoint trails $P_1, P_2, \ldots, P_{\kappa}$ from the cut to a large subgrid of our underlying grid. We require that $P_1, P_2, \ldots, P_{\kappa}$ are edge-disjoint from the branch trails of this subgrid; thus in particular they must end at vertices on the boundary of the subgrid as depicted in \Cref{fig:route}. Furthermore, we require that the subgrid is ``universal enough'' to connect these endpoints in any way, with any labels in $A$ and colors in $S$, while still using every edge on its side (i.e. the big side). In fact, we prove that it is universal enough even after a chosen vertex is split off arbitrarily, thus obtaining an irrelevant vertex. 

Finding these trails $P_1, P_2, \ldots, P_{\kappa}$ is a key new technique which we now elaborate on. Even if we grow the small side of the cut such that the cut size is minimal, it is not immediately clear that such trails exist, since Menger's Theorem might yield trails that traverse edges on branch trails of the subgrid. However, ``pushing the cut in towards the grid'' is still a good first step. So, consider a maximal set $X$ that contains the small side of the original cut and is \textit{$\kappa$-connected} to the grid in $\mathcal{G}$. This means that there is no edge cut of size less than $\kappa=|\delta(X)|$ whose small side contains $X$. Roughly, we show that for almost every vertex outside of $X$, no matter how we split it off, we still have that $|\delta(X)|=\kappa$ and $X$ is $\kappa$-connected to the grid in $\mathcal{G}$. We find such a vertex which is not a branch vertex of the grid in $\mathcal{G}$ and split it off to maintain the full immersion of the grid. Once we can no longer repeat this process, there must be subgrid which is actually just a subgraph of our host graph. At this point Menger's Theorem yields the desired trails.

\newpage

\part{The algorithmic min-max theorem}

\section{Preliminaries}
\label{sec:prelim}

This part of the paper is dedicated to proving \Cref{thm:mainMinMax}. In this part, for convenience, a \emph{signed graph} is a tuple $(G, \Sigma, t)$ so that $G$ is an Eulerian graph, $\Sigma$ is a set of edges in $G$ called the \emph{signature}, and $t$ is a vertex in $G$ called the \emph{root}. Recall that a \emph{free graph} is a tuple $(G, \Sigma, F, t)$ so that $(G, \Sigma, t)$ is a signed graph and $F$ is a set of edges in $G$ which we call \emph{free}. So signed graphs are just free graphs with no free edges. We quickly reduce from free graphs to signed graphs in \Cref{thm:reduction}. Then most of this part is dedicated to proving \Cref{thm:mainMinMax} for signed graphs. At a high level, our algorithm for signed graphs follows the same general strategy used by Chudnovsky, Cunningham, and Geelen \cite{algorithmApaths} to prove an algorithmic version of the min-max theorem for packing non-zero $A$-paths in group-labeled graphs \cite{packingApaths}. In particular, we maintain ``a dictionary which displays the irrelevant edges''.

All graphs are undirected and may have loops and parallel edges. We consider each edge $e$ to be an unordered set $\{h_1, h_2\}$ of two \emph{half-edges} $h_1$ and $h_2$, where $h_1$ is only incident to one end of $e$ and $h_2$ is incident to the other. (If $e$ is a loop, then $h_1$ and $h_2$ are incident to the same vertex.) We sometimes wish to consider orientations; each edge $e$ has two distinct orientations which we denote by $\vec{e}$ and $\vec{e}^{-1}$. Equivalently, the two orientations of $e=\{h_1, h_2\}$ can be identified with the two tuples $(h_1, h_2)$ and $(h_2, h_1)$. Each oriented edge $\vec{e}$ (also called an \emph{arc}) has a \emph{tail} and a \emph{head}, and we say that $\vec{e}$ is oriented \emph{from} its tail \emph{to} its head. We call $e$ the \textit{underlying} edge of~$\vec{e}$.

A \textit{trail} $T$ in a graph $G$ is a sequence of oriented edges $\vec{e}_1\vec{e}_2\dots\vec{e}_n$ such that the underlying edges are all distinct, and for each $i \in [n-1]$, the head of $\vec{e}_i$ equals the tail of $\vec{e}_{i+1}$. (The endpoints need not be distinct; the same vertex may be visited multiple times.) The \emph{reversal} of $T$ is the trail $T^{-1}$ which is the sequence of edges in the reverse order and with the reverse orientation. The \textit{tail} of $T$ is the tail of $\vec{e}_1$ and the \textit{head} of $T$ is the head of $\vec{e}_n$. We call the head and tail of $T$ its \textit{ends}. A \emph{subtrail} of $T$ is any trail obtained by deleting zero or more arcs from the beginning and end of $T$. A \emph{circuit} is a trail $C$ whose head and tail are equal. Given a vertex $t$, a \emph{$t$-circuit} is a circuit which begins and ends at $t$ but does not otherwise visit $t$. A \emph{subcircuit} of a trail is a subtrail which is also a circuit. 

We say that a circuit in a free graph is \emph{feasible} if it contains at least one free edge or an odd number of signature edges or both. Otherwise it is \emph{infeasible}. It is \emph{odd/even} if it contains an odd/even (respectively) number of signature edges; we also say that its \emph{label} is 1/0, respectively. We use similar terminology for edges, arcs, and trails.

A \emph{shifting} of $\Sigma$ is any set $\Sigma'$ of the form $\Sigma' = \Sigma \Delta R$ for some cut $R$. (A \emph{cut} is a set of edges of the form $\delta(S)$ for a set of vertices $S$. We write $\delta(S)$ for the set of all edges with exactly one end in $S$ and $\Delta$ for the symmetric difference of two sets.) Note that shifting does not change whether a circuit is feasible/infeasible or odd/even. Shifting is a standard notion of equivalence for graphs with signatures, although it is more often called ``switching''; see for instance~\cite{HararyBalance, zaslavskySignedGraphs}.

\subsection*{Floodings and certificates}

Let $\mathcal{G}=(G, \Sigma, F, t)$ be a free graph. Recall that a \emph{flooding} is a collection $\mathcal{C}$ of $\textrm{degree}(t)/2$-many $t$-circuits which partition the edge-set of $G$. We view any flooding obtained from $\mathcal{C}$ by reversing some of its circuits as being \emph{equivalent} to $\mathcal{C}$. The \emph{value} of a flooding is the number of feasible circuits that it contains. The \emph{flooding number} $\nu(\mathcal{G})$ is the maximum value of a flooding. Any flooding attaining this value is called \emph{optimal}.

\begin{figure}
\centering
\begin{tikzpicture}[scale=1, every node/.style={MyNode}]
    \def \w {2}
    \def \boxH {1.6}
    \def \boxW {4}
    \draw[dashed] (1,-.6) --++ (\boxW,0) --++ (0,-\boxH) --++ (-\boxW,0) -- cycle;
    \node[label=center:$X$, draw=none, fill=none] (A) at (1.3,-.6-\boxH+.3) {};
    \node (A1) at (.4*\w,0) {};
    \node (A2) at (1.5*\w,0) {};
    \node (A3) at (2.6*\w,0) {};
    \node[label=below:$t$] (a) at (1.5*\w,-1.5) {};
    \draw[thick] (A1) to (A2);
    \draw[thick] (A2) to (A3);
    \draw[MyRedArc] (A1) to [bend left=25] (A3);
    \foreach \i in {1,3}{
        \draw[MyRedArc] (a) to [bend left=15] (A\i);
        \draw[thick] (a) to [bend right=15] (A\i);
    }\foreach \i in {2}{
        \draw[MyRedArc] (a) to [bend left=22] (A\i);
        \draw[thick] (a) to [bend right=22] (A\i);
    }
    \draw[MyBlueArc] (a) to [out=-125,in=-185,looseness=25] (a);
\end{tikzpicture}
\vspace{-.3cm}
\caption{A depiction of a free graph with one free edge (the loop at $t$) and a certificate $(X,  \Sigma')$ of value $3$. The edges in $\Sigma'$ are depicted in bold red, and $G-X$ is $\Sigma'$-odd.}
\label{fig:certificate}
\end{figure}
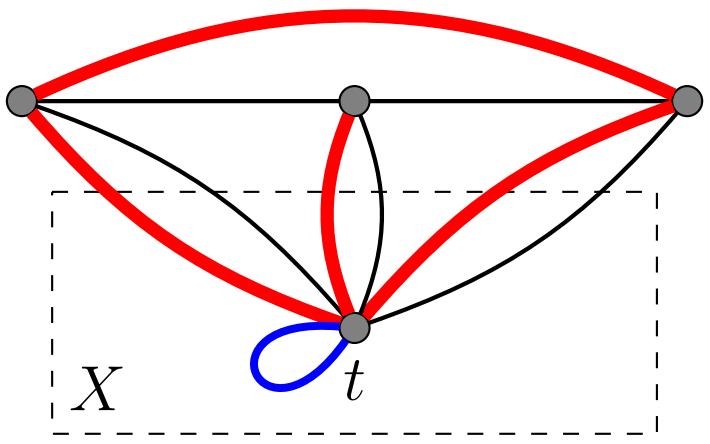

Recall that a \emph{certificate} of $\mathcal{G}$ is a tuple $(X, \Sigma')$ so that $X$ is a set of vertices which contains $t$ and $\Sigma'$ is a shifting of $\Sigma$. A component $H$ of $G-X$ is \emph{$\Sigma'$-odd} if $V(H)$ is not incident to any free edge of $G$, and the number of edges of $G$ which are incident to $V(H)$ and in $\Sigma'$ is different from $|\delta(H)|/2$ modulo~2. (We write $\delta(H)$ as shorthand for $\delta(V(H))$.) We write $odd_{\Sigma'}(G-X)$ for the number of components of $G-X$ which are $\Sigma'$-odd. We write $E_{F, \Sigma'}(X)$ for the set of all edges of $G$ with both ends in $X$ which are either free or in $\Sigma'$ or both. The \emph{value} of a certificate $(X, \Sigma')$ is then
\begin{align*}
 \val(X, \Sigma') = |E_{F, \Sigma'}(X)|+|\delta(X)|/2-odd_{\Sigma'}(G-X).
\end{align*}
\noindent See \Cref{fig:certificate} for an example. We say that an edge $e$ of $G$ is \emph{forced by} $(X, \Sigma')$ if either it is in $E_{F, \Sigma'}(X)$ or it is incident to a component of $G-X$ which is not $\Sigma'$-odd. Finally, note that all of these definitions also apply to signed graphs by setting $F \coloneqq \emptyset$.

We go ahead and prove the easier direction of the min-max theorem now.

\begin{lemma}
\label{lem:easyDir}
    For any free graph $\mathcal{G}=(G, \Sigma, F, t)$, the maximum value of a flooding is at most the minimum value of a certificate.
\end{lemma}
\begin{proof}
    Let $(X, \Sigma')$ be a certificate, and let $\mathcal{C}$ be a flooding. Since shifting preserves the feasibility of circuits, we may work with the signature \(\Sigma'\) throughout the proof. We now ``split up'' the circuits in $\mathcal{C}$ into a collection $\mathcal{T}$ of trails whose edge-sets partition $E(G)$. Imagine traversing each circuit $C \in \mathcal{C}$ beginning at $t$ until immediately before the first arc in $\delta(X)$ is about to be traversed; add this trail to $\mathcal{T}$. Then continue traversing $C$ until immediately after the second arc in $\delta(X)$ has been traversed. By repeating this process until we return to $t$, we find a partition of $C$ into subtrails  which are ``inside $X$'' and subtrails which ``leave and then return to $X$''. Then the collection $\mathcal{T}$ is obtained by repeating this procedure for each circuit $C \in\mathcal{C}$.
    
    Now, every feasible trail contained in $G[X]$ contains an edge of $E_{F,\Sigma'}(X)$. For the other trails, note that each component $H$ of $G-X$ is incident to exactly $|\delta(H)|/2$ trails in $\mathcal{T}$. Moreover, if $H$ is $\Sigma'$-odd, then at least one of these trails is infeasible due to the parity constraint. (Note that these trails partition the set of all edges of $G$ which are incident to $V(H)$.) It follows that at most $\val(X, \Sigma')$ many trails in $\mathcal{T}$ are feasible. Since each feasible circuit in $\mathcal{C}$ contains at least one feasible trail in $\mathcal{T}$, it follows that at most $\val(X, \Sigma')$ many circuits in $\mathcal{C}$ are feasible. 
\end{proof}

\subsection*{Algorithmic aspects}

Let us briefly discuss how the algorithm stores a free graph $\mathcal{G}=(G, \Sigma, F, t)$, a flooding $\mathcal{C}$ of $\mathcal{G}$, and a certificate $(X, \Sigma')$ of $\mathcal{G}$. First of all, we assume that graphs are stored as adjacency lists of half-edges, with pointers between the two half-edges of each edge. We assume that $\Sigma$ and $F$ are stored by variables on the edges which can be accessed in constant time. Similarly, when an algorithm returns a certificate $(X, \Sigma')$, it labels each edge by whether it is in $\Sigma'$ or not, and it labels each vertex by whether it is in $X$, a component of $G-X$ which is $\Sigma'$-odd, or neither. A flooding $\mathcal{C}$ is stored by labeling each edge $e$ by the orientation $\vec{e}$ which is in a circuit $C \in \mathcal{C}$, and by the edges and half-edges which come before/after $\vec{e}$ in $C$ (or whether $\vec{e}$ is the first/last arc of~$C$).

\section{Reducing to the signed case}
\label{sec:reduction}

In this section we show that it is enough to find an algorithm for signed graphs. 

First we show how to find an arbitrary flooding to start with.

\begin{lemma}
\label{lem:findFlooding}
    We can find a flooding of a free graph with $m$ edges in $\mathcal{O}(m)$ time. 
\end{lemma}
\begin{proof}
    We can find an Eulerian circuit in $\mathcal{O}(m)$ time using Hierholzer’s algorithm, for instance. Then we can convert this Eulerian circuit into a flooding by traversing it and ``stopping'' each time we see an arc whose head is the root. 
\end{proof}

Our reduction will assume that we have an algorithm of the following form for signed graphs. We note that the floodings $\mathcal{C}_e$ below need not be returned by the algorithm; we are just promised that they exist.

\begin{assumption}
\label{assumption1}
Given a signed graph $(G, \Sigma, t)$ and a flooding $\mathcal{C}$, we can find in time $\phi(|E(G)|)$ either a flooding $\mathcal{C}'$ of value greater than the value of $\mathcal{C}$, or a certificate $(X, \Sigma')$ of the same value as $\mathcal{C}$ with the additional guarantee that: \begin{itemize}
\item for each edge $e$ of $G$ which is not forced by $(X, \Sigma')$, there exists an optimal flooding $\mathcal{C}_e$ so that $e$ is in an infeasible circuit in~$\mathcal{C}_e$.
\end{itemize}
\end{assumption}

We now prove that such an algorithm yields an algorithm for free graphs.

\begin{theorem}
\label{thm:reduction}
Suppose that Assumption~\ref{assumption1} holds. Then, given a free graph $\mathcal{G}=(G, \Sigma, F, t)$ and an integer $k$, we can find either a flooding of value at least $k$ or a flooding and a certificate of the same value in time $\mathcal{O}(k \cdot \phi( |E(G)|))$. 
\end{theorem}
\begin{proof}
    Let $\mathcal{A}$ be the algorithm from Assumption~\ref{assumption1}, and set $m\coloneqq |E(G)|$. Since $\phi(m)\geq m$, we can make $\mathcal{O}(k)$ calls to $\mathcal{A}$ and perform $\mathcal{O}(km)$ additional work.

    Set $F_0\coloneqq\emptyset$ and $\Sigma_0\coloneqq\Sigma$. Starting with a flooding from \Cref{lem:findFlooding}, repeatedly apply $\mathcal{A}$ to $(G,\Sigma_0,t)$ and the current flooding until either its value is at least $k$ or a matching certificate is returned. This requires at most $k$ calls. In the first case, we return the flooding; it also has value at least $k$ in $\mathcal{G}$. Otherwise, we obtain an optimal flooding $\mathcal{C}_0$ of value $k_0<k$ and a matching certificate $(X_0, \Sigma_0')$ of value $k_0$ satisfying the guarantee of Assumption~\ref{assumption1}.

    We maintain $F_i\subseteq F$ with $|F_i|=i$, $\Sigma_i=\Sigma\Delta F_i$, and
    \begin{equation*}
        \nu(G,\Sigma_i,F_i,t)=\nu(G,\Sigma_i,t)=k_0+i<k
    \end{equation*}
    together with an optimal flooding $\mathcal{C}_i$ of $(G,\Sigma_i,t)$ and a matching certificate $(X_i,\Sigma_i')$ satisfying the guarantee of Assumption~\ref{assumption1}. These conditions hold for $i=0$. Since $\Sigma_i=\Sigma\Delta F_i$ and $F_i \subseteq F$, every circuit which is odd with respect to $\Sigma_i$ is feasible in~$\mathcal{G}$.

    First check in $\mathcal{O}(m)$ time whether every edge of $F$ is forced by $(X_i,\Sigma_i')$ in $(G,\Sigma_i,t)$. If so, return $\mathcal{C}_i$ and $(X_i,\Sigma_i'\Delta F_i)$. The latter is a certificate of $\mathcal{G}$ since $\Sigma_i'\Delta F_i$ is a shifting of $\Sigma$. Moreover, every edge of $F$ with both ends in $X_i$ belongs to $\Sigma_i'$, and no edge of $F$ is incident to a $\Sigma_i'$-odd component of $G-X_i$. Thus making the edges of $F$ free and replacing $\Sigma_i'$ by $\Sigma_i'\Delta F_i$ does not increase the value of the certificate. Since $\mathcal{C}_i$ has value at least $k_0+i$ in $\mathcal{G}$, \Cref{lem:easyDir} gives equality.

    Otherwise, choose an edge $e\in F$ which is not forced by $(X_i,\Sigma_i')$. By Assumption~\ref{assumption1}, there exists an optimal flooding $\mathcal{C}_e$ of $(G,\Sigma_i,t)$ in which $e$ is in an even circuit. Thus $e\notin F_i$, since otherwise $\mathcal{C}_e$ would have value greater than $k_0+i$ in $(G,\Sigma_i,F_i,t)$. Set $F_{i+1}\coloneqq F_i\cup\{e\}$ and $\Sigma_{i+1}\coloneqq\Sigma_i\Delta\{e\}$. The flooding $\mathcal{C}_e$ has value $k_0+i+1$ with respect to $\Sigma_{i+1}$. On the other hand, passing from $(G,\Sigma_i,F_i,t)$ to $(G,\Sigma_{i+1},F_{i+1},t)$ can change feasibility only for the circuit containing $e$, so it increases the value of any flooding by at most one. It follows that $\nu(G,\Sigma_{i+1},F_{i+1},t)=\nu(G,\Sigma_{i+1},t)=k_0+i+1$.

    If $\mathcal{C}_i$ has value $k_0+i+1$ with respect to $\Sigma_{i+1}$, set $\mathcal{C}_{i+1}\coloneqq\mathcal{C}_i$. Otherwise, its value is $k_0+i-1$, so applying $\mathcal{A}$ to $(G,\Sigma_{i+1},t)$ and $\mathcal{C}_i$ must return a flooding $\mathcal{C}_{i+1}$ of larger value. Every flooding of this signed graph has value congruent to $|\Sigma_{i+1}|$ modulo $2$, so $\mathcal{C}_{i+1}$ has value exactly $k_0+i+1$. If this value is $k$, then return $\mathcal{C}_{i+1}$; it also has value at least $k$ in $\mathcal{G}$. Otherwise, apply $\mathcal{A}$ to this optimal flooding to obtain the certificate for the next iteration. The total running time of this algorithm is $\mathcal{O}(k\phi(m))$ since it concludes after at most $k$ iterations.
\end{proof}

\section{Setting up the signed case}
\label{sec:setupMainAlg}

The rest of this part is dedicated to finding an algorithm for signed graphs which satisfies Assumption~\ref{assumption1}. In this section we give an overview of the approach.

\subsection*{Pulling down edges}

There is a key ``reduction move'' which was used to prove the min-max theorem in~\cite{decomposingSigned}. Let $\mathcal{G}=(G, \Sigma, t)$ be a signed graph, and let $\vec{e}$ be an arc of $G$ which is not incident to $t$. The graph obtained from $G$ by \emph{pulling $\vec{e}$ down into $\vec{e}_1, \vec{e}_2$} is the graph where we delete $e$ and add new edges $e_1$ and $e_2$ so that $\vec{e}_1$ is oriented from the tail of $\vec{e}$ into $t$ and $\vec{e}_2$ is oriented from $t$ into the head of $\vec{e}$. 

A signed graph $\widehat{\mathcal{G}}=(\widehat{G}, \widehat{\Sigma}, t)$ is \emph{obtained from $\mathcal{G}$ by pulling $\vec{e}$ down into $\vec{e}_1, \vec{e}_2$} if:\begin{itemize}
\item $\widehat{G}$ is obtained from $G$ by pulling $\vec{e}$ down into $\vec{e}_1, \vec{e}_2$, 
\item $\widehat{\Sigma}$ and $\Sigma$ are identical on all edges except possibly $e, e_1, e_2$, and
\item $e$ is in $\Sigma$ if and only if exactly one of $e_1, e_2$ is in $\widehat{\Sigma}$.
\end{itemize}\noindent There are exactly two signed graphs which can be obtained from $\mathcal{G}$ by pulling $\vec{e}$ down into $\vec{e}_1, \vec{e}_2$. First we choose the label of $e_1$, and that determines the label of~$e_2$.

We wrote the definition above for an arc $\vec{e}$ instead of its underlying edge $e$ to emphasize the fact that any trail $T$ of $G$ yields a trail $\widehat{T}$ of $\widehat{G}$; if $\vec{e}$ or $\vec{e}^{-1}$ is in $T$, then we replace it by $\vec{e}_1\vec{e}_2$ or $\vec{e}_2^{-1}\vec{e}_1^{-1}$, respectively. Similarly, any $t$-circuit $C$ of $G$ containing $\vec{e}$ or $\vec{e}^{-1}$ yields two $t$-circuits $\widehat{C}_1$ and $\widehat{C}_2$ of $\widehat{G}$. Thus a flooding $\mathcal{C}$ of $\mathcal{G}$ yields a flooding $\widehat{\mathcal{C}}$ of $\widehat{\mathcal{G}}$. The value of $\widehat{\mathcal{C}}$ is either equal to the value of $\mathcal{C}$ or exactly two higher; the latter occurs when the circuit $C \in \mathcal{C}$ containing $e$ is even and both $\widehat{C}_1$ and $\widehat{C}_2$ are odd. We often use this correspondence. Since pulling $\vec{e}$ down is the same thing as pulling $\vec{e}^{-1}$ down, we often talk about pulling an (unoriented) edge $e$ down as well.

The paper~\cite{decomposingSigned} says that if $\mathcal{G}=(G, \Sigma, t)$ is a signed graph so that every way of pulling down an edge increases the flooding number, then there is a certain simple certificate which satisfies the guarantee of Assumption~\ref{assumption1}. So, in order to obtain an efficient algorithm, the key idea is to grow a collection of floodings which shows, for more and more ways of pulling down edges, that the flooding number increases. To explain this more carefully we need to introduce some definitions. Essentially, we store a collection of floodings of the same value; for each of these floodings, we are interested in ``representatives'' of its even circuits.

\subsection*{Representatives, candidates, and loose circuits}

Let $\mathcal{G}=(G, \Sigma, t)$ be a signed graph. A \emph{candidate} is a tuple $(\vec{f}, \alpha)$ so that $\vec{f}$ is an arc, $\alpha \in \{0,1\}$, and:\begin{itemize}
\item if the tail of $\vec{f}$ is $t$ then $\alpha=0$, and 
\item if the head of $\vec{f}$ is $t$ then $\alpha$ equals the label of $f$.
\end{itemize}
Recall that the \emph{label} of $f$ is 1 or 0, respectively, depending on whether $f$ is in $\Sigma$ or not. So for instance, the definition of a candidate does not allow $\vec{f}$ to be an odd loop at $t$. We view a candidate $(\vec{f}, \alpha)$ as being \emph{equivalent} to the candidate $(\vec{f}^{-1}, \alpha')$, where $\alpha'$ equals $\alpha$ plus the label of $f$. (Arithmetic on labels is performed modulo 2.) In this manner, each edge $e$ of $G-t$ corresponds to exactly two candidates up to equivalence: $(\vec{e},0)$ and $(\vec{e},1)$. Similarly, each edge $e$ which is incident to $t$ and not an odd loop corresponds to exactly one candidate, up to equivalence.

Given a trail $T=\vec{e}_1\vec{e}_2 \ldots \vec{e}_n$ and an arc $\vec{e}_{i}$ of $T$, the \emph{prefix trail at $\vec{e}_{i}$ in $T$} is $\vec{e}_1\vec{e}_2\dots\vec{e}_{i-1}$. (This trail can be empty if $i=1$.) The \emph{prefix label at $\vec{e}_{i}$ in $T$} is then the label of this trail. A candidate $(\vec{f}, \alpha)$ \emph{represents} a circuit $C$ if the circuit $C$ is even and, after possibly reversing $C$, it contains the arc $\vec{f}$ and the prefix label at $\vec{f}$ in $C$ is $\alpha$. Note that two equivalent candidates represent the same circuits. We say that $(\vec{f}, \alpha)$ \emph{represents} a flooding $\mathcal{C}$ if it represents a circuit in $\mathcal{C}$. 

Sometimes we would like a single flooding that is represented by both $(\vec{f}, 0)$ and $(\vec{f}, 1)$ at once, but this -- by definition -- is not possible. However, we often use the following type of flooding to achieve something similar. We say that a circuit $L$ is \emph{loose} in a flooding $\mathcal{C}$ if $L$ is odd and there exists an even circuit $C \in \mathcal{C}$ so that $L$ is a subcircuit of $C$. We also say that $L$ is \emph{loose in} that circuit~$C$. Note that if $\mathcal{C}$ is represented by $(\vec{f},0)$ and $L$ contains $\vec{f}$ or $\vec{f}^{-1}$, then by ``reversing $L$'' we can obtain a flooding of the same value as $\mathcal{C}$ which is represented by $(\vec{f},1)$.

\subsection*{The dictionary}

Let $\mathcal{G}=(G, \Sigma, t)$ be a signed graph. Given a set of floodings $\mathcal{D}$ of $\mathcal{G}$, we say that a candidate $(\vec{f}, \alpha)$ is \emph{displayed by} $\mathcal{D}$ if there exists a flooding $\mathcal{C} \in \mathcal{D}$ which is represented by $(\vec{f}, \alpha)$. Our algorithm always maintains a \emph{dictionary}, which is a non-empty set of floodings $\mathcal{D}$ so that\begin{itemize} 
\item every flooding in $\mathcal{D}$ has the same value, and 
\item the size of $\mathcal{D}$ is at most the number of candidates (up to equivalence) displayed by $\mathcal{D}$ plus one.
\end{itemize}
This implies that every dictionary contains at most $2|E(G)|$-many floodings. We call the value of the floodings in $\mathcal{D}$ the \emph{value} of the dictionary, denoted by $\val(\mathcal{D})$. A dictionary is \emph{full} if it displays every candidate. A dictionary $\mathcal{D}'$ \emph{improves} $\mathcal{D}$ if it has greater value than $\mathcal{D}$, or if it has the same value and displays more candidates.

Note that for each edge $e$, a dictionary displays either zero, one, or two candidates containing $\vec{e}$, and this number is the same for $\vec{e}^{-1}$; we refer to this number as \emph{how many ways $e$ is displayed by $\mathcal{D}$}. We say that a certificate $(X, \Sigma')$ \emph{respects $\mathcal{D}$} if it has the same value as $\mathcal{D}$ and each edge which is not forced by $(X, \Sigma')$ is displayed by $\mathcal{D}$. Any such certificate satisfies the guarantee of Assumption~\ref{assumption1} by \Cref{lem:easyDir}.

\subsection*{Splitting off transitions}

We require one more graph operation for Section~\ref{sec:lemmas}.

Let $\mathcal{G}=(G, \Sigma, t)$ be a signed graph. The \textit{transitions} of $\mathcal{G}$ are the sets $\{h_1, h_2\}$ of two distinct half-edges which are incident to the same non-root vertex. We say the transition is \emph{at} that vertex. Let $T=\vec{e}_1\vec{e}_2\dots\vec{e}_n$ be a trail so that $t$ is not the tail of any arc except possibly $\vec{e}_1$ and not the head of any arc except possibly $\vec{e}_n$. The \emph{transitions of} $T$ are the transitions obtained by taking the last half-edge at $\vec{e}_i$ and the first half-edge at $\vec{e}_{i+1}$ for $i \in [n-1]$. This transition \emph{splits $T$ up} into the two trails $\vec{e}_1\vec{e}_2\dots\vec{e}_i$ and $\vec{e}_{i+1}\vec{e}_{i+2}\dots\vec{e}_n$. Note that $T$ and $T^{-1}$ have the same transitions. The \emph{transitions of a flooding} $\mathcal{C}$ of $\mathcal{G}$ are the transitions of the circuits in $\mathcal{C}$.

\begin{figure}
\centering
\begin{tikzpicture}[scale=1.75, every node/.style={MyNode}]
    \def \w {.7}
    \node (A1) at (-\w,0) {};
    \node[label=below:$v$] (v) at (0,0) {};
    \node (A2) at (\w,0) {};
    \draw[thick] (A1) -- (v) -- (A2);
    \draw[thick, dotted] (A1) --++ (-.28, 0);
    \draw[thick, dotted] (A1) --++ (-.2, -.2);
    \draw[thick, dotted] (A1) --++ (0, -.28);
    \draw[thick, dotted] (A2) --++ (.28, 0);
    \draw[thick, dotted] (A2) --++ (.2, -.2);
    \draw[thick, dotted] (A2) --++ (0, -.28);
    \draw[MyRedArc] (v) to [out=60,in=120,looseness=40] (v);
    \draw[ultra thick, dashed] (-\w*.5, .05) to (-.1,.05) to [bend left=10] (-.22, .45);
\end{tikzpicture}\hskip .25cm
\begin{tikzpicture}[scale=1, every node/.style={MyNode}]
    \def \w {1}
    \node[draw=none, fill=none] (v) at (0,0) {};
    \node[rectangle, draw=none, fill=none] (center) at (0, .5*\w) {\Large$\longrightarrow$};
\end{tikzpicture}\hskip .25cm
\begin{tikzpicture}[scale=1.75, every node/.style={MyNode}]
    \def \w {.7}
    \node (A1) at (-\w,0) {};
    \node[label=below:$v$] (v) at (0,0) {};
    \node (A2) at (\w,0) {};
    \draw[MyRedArc] (A1) -- (v);
    \draw[thick] (v) -- (A2);
    \draw[thick, dotted] (A1) --++ (-.28, 0);
    \draw[thick, dotted] (A1) --++ (-.2, -.2);
    \draw[thick, dotted] (A1) --++ (0, -.28);
    \draw[thick, dotted] (A2) --++ (.28, 0);
    \draw[thick, dotted] (A2) --++ (.2, -.2);
    \draw[thick, dotted] (A2) --++ (0, -.28);
    \draw[draw=none] (v) to [out=60,in=120,looseness=40] (v);
    \draw[ultra thick, dashed] (-.1, .05) -- (-\w*.6, .05);
    \draw[ultra thick, dashed] (.1, .05) -- (\w*.6, .05);
    \node at (v) {};
\end{tikzpicture}\hskip .25cm
\begin{tikzpicture}[scale=1, every node/.style={MyNode}]
    \def \w {1}
    \node[draw=none, fill=none] (v) at (0,0) {};
    \node[rectangle, draw=none, fill=none] (center) at (0, .5*\w) {\Large$\longrightarrow$};
\end{tikzpicture}\hskip .25cm
\begin{tikzpicture}[scale=1.75, every node/.style={MyNode}]
    \def \w {.7}
    \node (A1) at (-\w,0) {};
    \node[label=below:{\textcolor{white}{$v$}}, draw=none, fill=none] (v) at (0,0) {};
    \node (A2) at (\w,0) {};
    \draw[MyRedArc] (A1) -- (A2);
    \draw[thick, dotted] (A1) --++ (-.28, 0);
    \draw[thick, dotted] (A1) --++ (-.2, -.2);
    \draw[thick, dotted] (A1) --++ (0, -.28);
    \draw[thick, dotted] (A2) --++ (.28, 0);
    \draw[thick, dotted] (A2) --++ (.2, -.2);
    \draw[thick, dotted] (A2) --++ (0, -.28);
    \draw[draw=none] (v) to [out=60,in=120,looseness=40] (v);
\end{tikzpicture}
\caption{Splitting off two transitions at $v$. The edges in $\Sigma$ are depicted in bold red, and the half-edges of the transitions are dashed.}
\label{fig:splittingOff}
\end{figure}
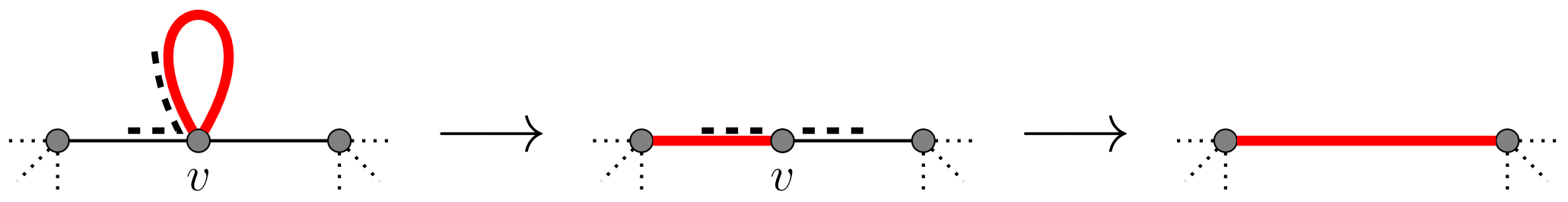

Let $\mathcal{G}=(G, \Sigma, t)$ be a signed graph, and let $\{h_1, h_2\}$ be a transition of some flooding of $\mathcal{G}$. \emph{Splitting off} $\{h_1, h_2\}$ means to delete the underlying edges $e_1$ and $e_2$ that contain $h_1$ and $h_2$, respectively, and to add a new edge $e$ whose ends are the other ends of $e_1$ and $e_2$ (not corresponding to $h_1$ and $h_2$). We delete any isolated vertex that is created by this operation, and we put $e$ into the new signature if and only if exactly one of $e_1, e_2$ is in $\Sigma$. Splitting off a vertex of degree four, as in \Cref{sec:introExtendedAbs}, amounts to splitting off two transitions at it; see \Cref{fig:splittingOff}. We require $\{h_1, h_2\}$ to be a transition of some flooding in order to ensure that the resulting graph is connected. Some of our algorithms involve successively splitting off transitions; in this case, each arc is associated with a trail in the original graph. It is enough to store the first and last arc of this trail, updating the pointers as we split off more transitions.

Sometimes we need to keep track of an edge $f$ or a candidate $(\vec{f}, \alpha)$ while splitting off transitions. If this transition contains a half-edge of $f$, then we need to be careful. This is more of a technical issue than a conceptual one. In particular, using the symmetry given by the equivalence relation on candidates, it suffices to understand the behavior when the transition contains the last half-edge of $\vec{f}$. Splitting off such a transition does not affect the prefix label at the new arc obtained from $\vec{f}$. (Also note that we never split off a transition of the form $\{h_1, h_2\}$ where $\{h_1, h_2\}$ is a loop edge, because there is no flooding which contains this transition.)

\section{Combining floodings}
\label{sec:lemmas}

In this section we prove some helpful lemmas. Essentially, we show how to go from two floodings $\mathcal{C}_1$ and $\mathcal{C}_2$ which respectively satisfy two properties $P_1$ and $P_2$, to a single flooding $\mathcal{C}$ which simultaneously satisfies both $P_1$ and~$P_2$. These properties are generally about how edges ``interact'' with even circuits in a flooding.

We now show, roughly, that for any flooding $\mathcal{C}$ and any half-edge $h_1$ which is not incident to $t$, most of the transitions containing $h_1$ are transitions of some flooding with similar properties to~$\mathcal{C}$. We have two stages: there is one algorithm that lists the admissible transitions, and a second algorithm that constructs a corresponding flooding for a selected transition. This avoids constructing a separate flooding for every transition in the list. Note that there are an odd number of transitions containing $h_1$ since every vertex has even degree. Thus, if most of these transitions ``work'' for each of the two properties $P_1$ and $P_2$, we can find one transition that ``works for both''.

We require the following technical definition. Given a subcircuit $L$ of a circuit $C$ and a half-edge $h_1$, we say that $L$ is \emph{$h_1$-dangerous} if the edge $e_1$ that contains $h_1$ is either the first or last edge of $L$ or comes immediately before or after $L$ in~$C$.

\begin{proposition}
\label{prop:mostWork}
    There are algorithms $\mathcal{A}_1$ and $\mathcal{A}_2$ which run in time $\mathcal{O}(|E(G)|)$ and behave as follows. Algorithm $\mathcal{A}_1$ takes as input a signed graph $\mathcal{G}=(G, \Sigma, t)$, a flooding $\mathcal{C}$, a half-edge $h_1$ not incident to $t$, and, optionally, one of the following:\begin{enumerate}
        \item an edge $f$ which is in an even circuit in $\mathcal{C}$,
        \item distinct edges $f_1$ and $f_2$ which are in distinct even circuits in $\mathcal{C}$, or
        \item a candidate $(\vec{f}, \alpha)$ which represents a circuit $C \in \mathcal{C}$ so that there is no loose circuit in $C$ which contains $f$ and is $h_1$-dangerous.
    \end{enumerate}Algorithm $\mathcal{A}_1$ returns either a flooding of larger value than $\mathcal{C}$, or a list of more than half of the transitions containing $h_1$. Algorithm $\mathcal{A}_2$ takes the same input as $\mathcal{A}_1$ plus a transition $\{h_1, h_2\}$ returned by $\mathcal{A}_1$. It returns a flooding $\mathcal{C}'$ of the same value as $\mathcal{C}$ so that $\{h_1, h_2\}$ is a transition of $\mathcal{C}'$, and if there was optional input, then respectively:
    \begin{enumerate}
        \item $f$ is in an even circuit in $\mathcal{C}'$,
        \item $f_1$ and $f_2$ are in distinct even circuits in $\mathcal{C}'$, and
        \item $(\vec{f}, \alpha)$ represents $\mathcal{C}'$.
    \end{enumerate}
\end{proposition}
\begin{proof}
    For each algorithm, we first traverse every circuit of $\mathcal C$, storing at each arc its prefix label, which circuit of $\mathcal{C}$ it is in, whether that circuit is even or odd, and the position of the arc in the circuit. This takes $\mathcal{O}(|E(G)|)$ time. Now, note that there is exactly one transition $\{h_1, h_2\}$ of $\mathcal{C}$ which contains $h_1$; this is one of the transitions in the list returned by algorithm $\mathcal{A}_1$. Algorithm $\mathcal{A}_2$ can check whether this transition is the input in constant time, in which case it just returns~$\mathcal{C}$.
    
    Suppose that the vertex incident to $h_1$ has degree $2d+2$. Then there are exactly $d$ transitions $\{g_1, g_2\}$ of $\mathcal{C}$ which are at this vertex and do not contain $h_1$. By traversing the circuits in $\mathcal{C}$, we encounter every such transition in $\mathcal{O}(|E(G)|)$ time. For algorithm $\mathcal{A}_1$, we show how to decide in constant time whether to add $\{h_1, g_1\}$, $\{h_1, g_2\}$, or neither to the list; if we cannot add either, then we will in $\mathcal{O}(|E(G)|)$ time find a flooding of larger value than $\mathcal{C}$ to return. For algorithm $\mathcal{A}_2$, we perform the same process until we would add the input transition to the list; then we stop traversing and in $\mathcal{O}(|E(G)|)$ time find the corresponding flooding to return. This process adds at least $d$ transitions to the list. Along with the transition $\{h_1, h_2\}$, this means that, in total, the list contains at least $d+1$ of the $2d+1$ transitions containing $h_1$, as desired.

    So it suffices to show what to do once such a transition $\{g_1, g_2\}$ is encountered. Due to the pre-processing, we can access the following information in constant time: 
    \begin{itemize}
        \item whether each of the optional input edges $f, f_1, f_2$ is on the same circuit as $h_1$ and/or $g_1$, and if so, whether it comes before or after $\{h_1, h_2\}$ and/or $\{g_1, g_2\}$.
    \end{itemize}
    
    \noindent Given access to all of this information, we now show how to decide in constant time whether to add $\{h_1, g_1\}$, $\{h_1, g_2\}$, or neither to the list.

    \smallskip\smallskip

    \noindent \emph{Case 1:} $\{h_1, h_2\}$ and $\{g_1, g_2\}$ are transitions of the same circuit $C \in \mathcal{C}$.

    \smallskip

    Then there exist subtrails $T_1$ and $T_2$ of $C$ and a subcircuit $L$ of $C$ so that $C=T_1 L T_2$, and the transitions $\{h_1, h_2\}$ and $\{g_1, g_2\}$ split $C$ into the two trails $T_1$ and $L T_2$, and $T_1 L$ and $T_2$ (or vice-versa). We claim that the flooding $\mathcal{C}'$ obtained by replacing $T_1LT_2$ with $T_1 L^{-1}T_2$ satisfies the desired conditions. This will complete \emph{Case~1}; we can figure out which transition to append to the list in constant time, and we can ``reverse $L$'' in linear time to obtain the new flooding. The algorithmic aspects in the other cases will follow similarly, and so we will omit them.
    
    So, to prove the claim, notice that $\mathcal{C}'$ partitions the edge-set of $G$ in the same way as $\mathcal{C}$. Thus we can assume that we have the third optional input; we just need to show that $(\vec{f}, \alpha)$ represents $\mathcal{C}'$. If $f$ is in $T_1$, $T_2$, or a circuit of $\mathcal{C}$ other than $C$, then this is true. So we may assume that $f$ is in $L$. If $L$ is even, then reversing it preserves the representative. Otherwise, $L$ is loose in $C$ and $h_1$-dangerous, a contradiction.

    \smallskip\smallskip

    \noindent \emph{Case 2:} $\{h_1, h_2\}$ and $\{g_1, g_2\}$ are transitions of different circuits in $\mathcal{C}$.

    \smallskip

    Suppose that $\{h_1, h_2\}$ splits up its circuit into trails $T_1$ and $T_2$, and $\{g_1, g_2\}$ splits up its circuit into trails $R_1$ and $R_2$. Consider the floodings $\mathcal{C}_1$ and $\mathcal{C}_2$, respectively, obtained from $\mathcal{C}$ by replacing the circuits $T_1T_2$ and $R_1R_2$ with 1) $T_1R_1^{-1}$ and $T_2^{-1}R_2$, and 2) $T_1R_2$ and $R_1T_2$. It suffices to show that at least one of these floodings $\mathcal{C}_1,\mathcal{C}_2$ satisfies the desired conditions. If both of the original circuits $T_1T_2$ and $R_1R_2$ are odd, then we can pair $T_1$ with whichever of $R_1^{-1}$ and $R_2$ has different label from $T_1$; both of the two new circuits of the new flooding are still odd.
    
    Now suppose that one of the circuits $T_1T_2$ and $R_1R_2$ is odd and the other is even. There is symmetry between $\{h_1, h_2\}$ and $\{g_1, g_2\}$, so we may assume that $T_1T_2$ is even. Notice that at most one of the trails $T_1$ and $T_2$ can contain any of the optional edges $f, f_1, f_2$. We pair whichever trail $T_i$ contains an optional edge with whichever of $R_1$ and $R_2$ has the same label as $T_i$. This flooding has the desired properties.

    In the final case, both of the circuits $T_1T_2$ and $R_1R_2$ are even. If $T_1$ and $R_1$ have different labels, then pairing them up yields a flooding of strictly larger value than $\mathcal{C}$. So we may assume that all of the trails $T_1,T_2,R_1,R_2$ have the same label. At most one of the trails $T_1$ and $T_2$ contains any of the optional input edges $f, f_1, f_2$, and similarly for $R_1$ and $R_2$. So we may pair whichever of $T_1$ and $T_2$ contains an optional input edge with whichever of $R_1$ and $R_2$ does not. This was the final case and completes the proof of \Cref{prop:mostWork}.
\end{proof}

We now obtain several useful corollaries of \Cref{prop:mostWork}. First we show that if an arc $\vec{f}$ is in two candidates $(\vec{f}, 0)$ and $(\vec{f}, 1)$, then we can transform two floodings represented by those candidates into one flooding with a loose circuit containing~$f$.

\begin{lemma}
\label{lem:findLoose}
    Given a signed graph $\mathcal{G}=(G, \Sigma, t)$, floodings $\mathcal{C}_0$ and $\mathcal{C}_1$ of value $k$, and an arc $\vec{f}$ so that\begin{enumerate}
        \item $(\vec{f}, 0)$ represents $\mathcal{C}_0$, and
        \item $(\vec{f}, 1)$ represents $\mathcal{C}_1$,
    \end{enumerate}
    we can find in $\mathcal{O}(|E(G)|^2)$ time either a flooding of value greater than $k$, or a flooding $\mathcal{C}$ of value $k$ and a circuit $L$ containing $f$ so that $L$ is loose in $\mathcal{C}$.
\end{lemma}
\begin{proof}
    Choose an arbitrary non-loop edge $e_1$ which is incident to $t$, and let $h_1$ be the half-edge of $e_1$ which is not incident to $t$. For $i=0,1$, let $C_i$ be the circuit in $\mathcal{C}_i$ which contains $f$. First we show how to test in $\mathcal{O}(|E(G)|)$ time whether $C_i$ has a loose circuit $L_i$ which contains $f$ and is $h_1$-dangerous. We may assume that $e_1$ is in $C_i$ and, after possibly reversing $C_i$, that $\vec{e}_1$ is the first arc of $C_i$. We traverse $C_i$ beginning right after $\vec{e}_1$. We keep track of whether we have seen $f$ and whether we have seen an odd or even number of edges in $\Sigma$. Each time we again reach the head of $\vec{e}_1$, if we have seen $f$ and the tabulated value is odd, then we return $\mathcal{C}_i$ and the circuit $L$ we have found. 
    
    Now suppose that the procedure above did not find any such loose circuit. Then apply algorithm $\mathcal{A}_1$ from \Cref{prop:mostWork} to the flooding $\mathcal{C}_0$ and candidate $(\vec{f}, 0)$. Also apply it to the flooding $\mathcal{C}_1$ and candidate $(\vec{f}, 1)$. If either time this yields a flooding of value larger than $k$, then return it. So we may assume that both times, $\mathcal{A}_1$ returns a list of over half the transitions containing $h_1$. Find a transition $\{h_1, h_2\}$ which is in both of the lists. Now, for $i=0,1$, apply algorithm $\mathcal{A}_2$ to the flooding $\mathcal{C}_i$, candidate $(\vec{f}, i)$, and transition $\{h_1, h_2\}$. Thus we obtain, for $i=0,1$, a flooding $\mathcal{C}_i'$ of value $k$ so that $\{h_1, h_2\}$ is a transition of $\mathcal{C}_i'$ and $(\vec{f}, i)$ represents $\mathcal{C}_i'$.

    We now split off the transition $\{h_1, h_2\}$ and repeat the whole procedure above on the resulting signed graph, which has one less edge. Thus, by repeating the procedure $\mathcal{O}(|E(G)|)$ times, we eventually find either a flooding of value larger than $k$ or a loose circuit containing $f$ in a flooding of value $k$. Note that in general, each arc $\vec{e}$ of the smaller graph corresponds to a trail in the larger graph of the same label as $\vec{e}$. So if we find a flooding/circuit in the smaller graph, we can immediately ``read off'' the corresponding flooding/circuit in $\mathcal{G}$. 
\end{proof}

The next lemma is similar in spirit to \Cref{lem:findLoose}. 

\begin{lemma}
\label{lem:threeInEven}
    Given a signed graph $\mathcal{G}=(G, \Sigma, t)$, floodings $\mathcal{C}_0$ and $\mathcal{C}_1$ of value $k$, an arc $\vec{f}$, edges $f_1, f_2$, and $\alpha \in \{0,1\}$ so that\begin{enumerate}
        \item $(\vec{f}, \alpha)$ represents $\mathcal{C}_0$, and
        \item $f_1$ and $f_2$ are in distinct even circuits in $\mathcal{C}_1$,
    \end{enumerate}
    we can find in $\mathcal{O}(|E(G)|^2)$ time either a flooding of value greater than $k$, or a flooding of value $k$ represented by $(\vec{f}, \alpha)$ in which $f_1$ and $f_2$ are in distinct even circuits.
\end{lemma}
\begin{proof}
    If $t$ is incident to at most two non-loop edges, then we return the flooding $\mathcal{C}_0$. Note that $\mathcal{C}_0$ and $\mathcal{C}_1$ partition the edge-set of $G$ in the same way. Thus $f_1$ and $f_2$ are in distinct even circuits in $\mathcal{C}_0$, as desired.

    Now, by iterating through the edges incident to $t$, we can find a non-loop edge $e_1$ which is incident to $t$, and which is not in the same circuit as $f$ in the flooding $\mathcal{C}_0$. Let $h_1$ be the half-edge of $e_1$ which is not incident to $t$. Apply algorithm $\mathcal{A}_1$ from \Cref{prop:mostWork} to the flooding $\mathcal{C}_0$ and candidate $(\vec{f}, \alpha)$. Also apply it to the flooding $\mathcal{C}_1$ and edges $f_1$ and $f_2$. If either time this yields a flooding of value larger than $k$, then return it. So we may assume that both times, $\mathcal{A}_1$ returns a list of over half the transitions containing $h_1$. We find a transition $\{h_1, h_2\}$ which is in both lists, and then we call algorithm $\mathcal{A}_2$ twice to obtain two new floodings $\mathcal{C}_0'$ and $\mathcal{C}_1'$. Then we split off $\{h_1, h_2\}$ and continue as in the proof of \Cref{lem:findLoose}.
\end{proof}

We have one final related lemma to show. This lemma, unlike the others, is guaranteed to return a flooding of larger value.

\begin{lemma}
\label{lem:improvingByCombining}
    Given a signed graph $\mathcal{G}=(G, \Sigma, t)$, floodings $\mathcal{C}_0$ and $\mathcal{C}_1$ of value $k$, edges $f, f_1, f_2$ with a common endpoint $x \neq t$, and a circuit $L$ so that\begin{enumerate}
        \item $f$ is in $L$ and $L$ is loose in $\mathcal{C}_0$, and
        \item $f_1$ and $f_2$ are in distinct even circuits in $\mathcal{C}_1$,
    \end{enumerate}
    we can find in $\mathcal{O}(|E(G)|^2)$ time a flooding of value greater than $k$.
\end{lemma}
\begin{proof}
    It suffices to find either a flooding of value greater than $k$, or a flooding $\mathcal{C}$ of value $k$ with a loose circuit $L'$ containing $f$ and with $f_1,f_2$ in distinct even circuits $C_1,C_2$. In the latter case, choose $i\in\{1,2\}$ so that $C_i$ does not contain $L'$. Both $C_i$ and $L'$ contain $x$, so we can remove $L'$ from its even circuit, cyclically reorder it to begin at $x$, and insert it into $C_i$. Both affected circuits become odd, increasing the value of the flooding by two in $\mathcal{O}(|E(G)|)$ time. We now show how to obtain one of these outcomes without any assumption about $f, f_1, f_2$ having a common endpoint.

    As in the previous lemmas, we find in linear time either a desired outcome or a transition which can be split off while preserving the two input properties. In fact, splitting off this transition will leave $L$ unchanged. First, check in $\mathcal{O}(|E(G)|)$ time whether every vertex of $G-t$ is incident to an edge of $L$. Suppose so. If $\mathcal{C}_0$ contains an even circuit $C$ other than the one containing $L$, and $C$ is not a loop at $t$, then $C$ meets $L$ at a non-root vertex. Moving $L$ onto $C$ as above increases the value of the flooding by two. We can find such a circuit and perform the move in linear time. Otherwise, if no such $C \in \mathcal{C}_0$ exists, then since $\mathcal{C}_0$ and $\mathcal{C}_1$ have the same value, at least one of $f_1,f_2$, say $f_1$, is an even loop at $t$. If $f_2$ is in an even circuit of $\mathcal{C}_0$, then $\mathcal{C}_0$ and $L$ give the second outcome. Otherwise, $f_2$ is in an odd circuit $C'\in\mathcal{C}_0$, which is not a loop at $t$. Move $L$ onto $C'$ at a common non-root vertex, cyclically reordering $L$ if necessary. Its original circuit becomes odd and $C'$ becomes even, giving a flooding of value $k$ and a loose circuit $L'$ with the required properties.

    We may therefore assume that some vertex $v\neq t$ is not incident to any edge of $L$. Choose a half-edge $h_1$ incident to $v$, and apply algorithm $\mathcal{A}_1$ from \Cref{prop:mostWork} to $\mathcal{C}_1$, $h_1$, and $f_1,f_2$. If it returns a flooding of value greater than $k$, return it. Otherwise, we obtain a list of more than half of the transitions containing $h_1$.

    To obtain a corresponding list for $\mathcal{C}_0$, let $\vec{e}_1$ and $\vec{e}_2$ be the arcs immediately before and after $L$ in its even circuit. Split off all transitions of the trail $\vec{e}_1L\vec{e}_2$ to obtain an arc $\vec{f}'$ in a signed graph $\mathcal{G}'$. The flooding $\mathcal{C}_0$ yields a flooding of $\mathcal{G}'$ of value $k$ in which $f'$ is in an even circuit. Every transition split off is at a vertex incident to $L$, and hence not at $v$. Thus the half-edges incident to $v$ are preserved, giving a bijection between the transitions containing $h_1$ in $\mathcal{G}$ and $\mathcal{G}'$. Apply $\mathcal{A}_1$ to the flooding of $\mathcal{G}'$, with $h_1$ and $f'$. If it returns a flooding of value greater than $k$, expand the arc $\vec{f}'$ into $\vec{e}_1L\vec{e}_2$ and return the resulting flooding of $\mathcal{G}$. Otherwise, identify its list with transitions in $\mathcal{G}$ and find a transition $\{h_1,h_2\}$ in both lists. Applying $\mathcal{A}_2$ to the respective inputs and this transition, and expanding the flooding obtained in $\mathcal{G}'$, yields floodings $\mathcal{C}_0'$ and $\mathcal{C}_1'$ of value $k$ which both contain $\{h_1,h_2\}$, with $L$ loose in $\mathcal{C}_0'$ and $f_1,f_2$ in distinct even circuits of $\mathcal{C}_1'$. All of these steps take $\mathcal{O}(|E(G)|)$ time.

    Now split off $\{h_1,h_2\}$ and repeat on the smaller graph, updating the edges $f_1,f_2$. As in the proof of \Cref{lem:findLoose}, there are at most $|E(G)|$ reductions, and so the total running time is $\mathcal{O}(|E(G)|^2)$.
\end{proof}

\section{Improving a dictionary}
\label{sec:dictionary}

In this section we show several helpful lemmas about how to improve a dictionary. Note that any signed graph $\mathcal{G}=(G, \Sigma, t)$ can be written uniquely as a union of signed graphs $\mathcal{G}_1, \mathcal{G}_2, \ldots, \mathcal{G}_r$ whose only shared vertex is $t$; there is one signed graph for each loop at $t$ and one for each non-empty component of $G-t$. We call $\mathcal{G}_1, \mathcal{G}_2, \ldots, \mathcal{G}_r$ the \emph{$t$-components} of $\mathcal{G}$. If $r=1$, then we say that $\mathcal{G}$ is \emph{root-connected}.

We begin by handling full dictionaries. The first two lemmas assume that $\mathcal{G}$ is root-connected; then we handle the general case in \Cref{prop:endingState}.

\begin{lemma}
\label{lem:improveFull}
    Given a root-connected signed graph $\mathcal{G}=(G, \Sigma, t)$ and a full dictionary $\mathcal{D}$ of value $k$, we can in time $\mathcal{O}(|E(G)|^3)$ find either a certificate $(X, \Sigma')$ which respects $\mathcal{D}$, or a dictionary $\mathcal{D}'$ which improves~$\mathcal{D}$.
\end{lemma}
\begin{proof}
    All floodings in $\mathcal{D}$ have the same number of even circuits. If this number is at most one, then we return the certificate $(\{t\}, \Sigma)$, which respects $\mathcal{D}$. 
    
    If $G-t$ consists of just an isolated vertex, then let $\Sigma'$ be obtained from $\Sigma$ by shifting at $\delta(t)$, if necessary, so that at most half of the edges incident to $t$ are in $\Sigma'$. If the value of $\mathcal{D}$ is strictly less than the size of $\Sigma'$, then we find and return a flooding of larger value than $\mathcal{D}$. Otherwise, the value of $\mathcal{D}$ equals the size of $\Sigma'$. In this case $\Sigma'= \emptyset$ since $\mathcal{D}$ is full and the edges in $\Sigma'$ are only ever in odd circuits in floodings of value $|\Sigma'|$. So we return the certificate $(V(G), \emptyset)$, which respects $\mathcal{D}$.

    We may now assume that $G-t$ is connected and every flooding in $\mathcal{D}$ has at least two even circuits. We show how to find in $\mathcal{O}(|V(G)|\cdot |E(G)|^2)$ time either a flooding of value greater than $k$, or a flooding $\mathcal{C}$ of value $k$ containing two distinct even circuits which meet at a vertex other than $t$. Choose an arbitrary flooding $\mathcal{C}_1\in\mathcal{D}$, distinct even circuits $C_1,C_2\in\mathcal{C}_1$, and edges $f_i\in E(C_i)$ for $i=1,2$. Orient these edges as $\vec{f}_1,\vec{f}_2$ so that their heads are different from $t$, and find a shortest path $P$ in $G-t$ joining their heads. If $P$ is empty, then the two circuits already meet outside~$t$.

    Otherwise, let $\vec{f}$ be the first arc of $P$, oriented from the head of $\vec{f}_1$ towards the head of $\vec{f}_2$. Since $\mathcal{D}$ is full, we can find in $\mathcal{O}(|E(G)|^2)$ time a flooding $\mathcal{C}_0\in\mathcal{D}$ and $\alpha\in\{0,1\}$ so that $(\vec{f},\alpha)$ represents $\mathcal{C}_0$. Apply \Cref{lem:threeInEven} to $\mathcal{C}_0$, $\mathcal{C}_1$, and these distinguished edges and candidate. In $\mathcal{O}(|E(G)|^2)$ time, it returns either a flooding of value greater than $k$, or a flooding $\mathcal{C}_1'$ of value $k$ in which $f$ is in an even circuit and $f_1,f_2$ are in distinct even circuits. If $f$ and $f_1$ are in distinct circuits, then those circuits meet at the head of $\vec{f}_1$. Otherwise, $f$ and $f_2$ are in distinct even circuits, and replacing $\vec{f}_1$ by $\vec{f}$ strictly decreases the distance between the chosen heads in $G-t$. After repeating at most $|V(G)|$ times, we obtain either the desired flooding $\mathcal{C}$ or an improvement to~$\mathcal{D}$.

    Now, let $C_1,C_2\in\mathcal{C}$ be distinct even circuits which contain respective edges $f_1, f_2$ that meet at a vertex $x\neq t$. Since $G-t$ is connected and has an edge, it has an arc $\vec{f}_0$ incident to $x$. Since the dictionary is full, we can find in $\mathcal{O}(|E(G)|^2)$ time two floodings in $\mathcal{D}$ represented by $(\vec{f}_0,0)$ and $(\vec{f}_0,1)$, respectively. Apply \Cref{lem:findLoose}. It returns either a flooding of value greater than $k$, or a flooding $\mathcal{C}'$ of value $k$ with a loose circuit $L$ containing $f_0$. In the latter case, apply \Cref{lem:improvingByCombining} to $\mathcal{C}'$, $\mathcal{C}$, $f_0,f_1,f_2$, and $L$ to obtain a flooding of value greater than $k$. These final steps take $\mathcal{O}(|E(G)|^2)$ time. Since $|V(G)|\leq |E(G)|$, the total running time is $\mathcal{O}(|E(G)|^3)$.
\end{proof}

Next we show how to handle another case that is roughly the opposite of \Cref{lem:improveFull}.

\begin{lemma}
\label{lem:allZero}
    Given a root-connected signed graph $\mathcal{G}=(G, \Sigma, t)$ and a dictionary $\mathcal{D}$ of value $k$ which displays every edge of $G-t$ in zero ways, we can in time $\mathcal{O}(|E(G)|^3)$ find either a certificate $(X, \Sigma')$ which respects $\mathcal{D}$, or a dictionary $\mathcal{D}'$ which improves~$\mathcal{D}$.
\end{lemma}
\begin{proof}
    All floodings in $\mathcal{D}$ have the same number of even circuits. If this number is zero, or if the graph consists of just one loop at $t$, then we return the certificate $(\{t\}, \Sigma)$, which respects $\mathcal{D}$. 

    Thus we may assume that $G$ has no loops at $t$, and every even circuit in a flooding in $\mathcal{D}$ consists of two parallel edges joining $t$ to another vertex. Choose an arbitrary flooding $\mathcal{C}\in\mathcal{D}$. For each even circuit $C\in\mathcal{C}$, write $C=\vec{e}_1\vec{e}_2$, where $\vec{e}_1$ is oriented from $t$ to some vertex $x \neq t$ and both edges have label $\alpha$. If another even circuit at $x$ has both edges of label different from $\alpha$, then pairing each edge of $C$ with an edge of that circuit gives two odd circuits, increasing the value of the flooding. So we may assume that this value $\alpha$ is the same for all even circuits in $\mathcal{C}$ meeting $x$. Let $X$ consist of $t$ and all such vertices $x$. Let $\Sigma'$ be the shifting of $\Sigma$ obtained by shifting on the cuts $\delta(x)$ where $x \in X\setminus \{t\}$ has a corresponding label $\alpha=1$.

    We work with $\Sigma'$ for the rest of the proof. So every even circuit in $\mathcal{C}$ now consists of two even edges. Consider an odd circuit $C'\in\mathcal{C}$ which visits a vertex in $X\setminus \{t\}$. Then we can write $C'=R_1R_2$, where $R_1$ ends in $X\setminus \{t\}$. Note that whichever of $R_1$ and $R_2$ has label $0$ must consist of only a single even edge which is incident to $t$ and already displayed by $\mathcal{D}$; otherwise we could improve $\mathcal{D}$. It follows that each odd circuit in $\mathcal{C}$ has at most two transitions which are at vertices in $X\setminus \{t\}$. 

    Remove from each odd circuit in $\mathcal{C}$ its even edges joining $t$ to $X\setminus \{t\}$. The remaining trail is an odd trail with ends in $X$ and all internal vertices in a component of $G-X$. It follows that no component of $G-X$ is $\Sigma'$-odd, and that $(X, \Sigma')$ is a certificate of value $k$. Every edge not forced by this certificate is an even edge with both ends in $X$. Such an edge either belongs to an even circuit of $\mathcal{C}$ or was removed from an odd circuit above, so it is displayed by $\mathcal{D}$. Hence $(X,\Sigma')$ respects $\mathcal{D}$, and we return it. The displayed edges can be identified in $\mathcal{O}(|E(G)|^2)$ time. There are $\mathcal{O}(|E(G)|^2)$ choices of an even circuit in $\mathcal{C}$ and a transition in another circuit, and each exchange can be tested and performed in $\mathcal{O}(|E(G)|)$ time. Constructing the certificate takes $\mathcal{O}(|E(G)|)$ time. The total running time is therefore $\mathcal{O}(|E(G)|^3)$.
\end{proof}

Next we show a circumstance under which a dictionary can be improved. 

\begin{lemma}
\label{lem:improveZeroTwo}
    Given a signed graph $\mathcal{G}=(G, \Sigma, t)$, a dictionary $\mathcal{D}$, and edges $f$ and $f'$ which are incident to a common vertex $x \neq t$ and displayed by $\mathcal{D}$ in two and zero ways, respectively, we can in time $\mathcal{O}(|E(G)|^2)$ find a dictionary $\mathcal{D}'$ which improves~$\mathcal{D}$.
\end{lemma}
\begin{proof}
    Let $\vec{f}$ be the orientation of $f$ away from $x$. In $\mathcal{O}(|E(G)|^2)$ time, we can find floodings $\mathcal{C}_0, \mathcal{C}_1 \in \mathcal{D}$ represented by $(\vec{f}, 0)$ and $(\vec{f}, 1)$, respectively. By \Cref{lem:findLoose}, we can then in $\mathcal{O}(|E(G)|^2)$ time find either a flooding of value greater than the value of $\mathcal{D}$, or a flooding $\mathcal{C}$ of the same value as $\mathcal{D}$ and a circuit $L$ containing $f$ so that $L$ is loose in $\mathcal{C}$. We may assume that we have found $\mathcal{C}$ and $L$. 
    
    If $f'$ is in an even circuit in $\mathcal{C}$, then we add $\mathcal{C}$ to the dictionary and return. So we may assume that $f'$ is in an odd circuit $C \in \mathcal{C}$. Then we find trails $T_1, T_2, R_1, R_2$ so that $x$ is the head of $R_1$, the circuit $C$ equals $R_1 R_2$, and the circuit in $\mathcal{C}$ containing $f$ equals $T_1 L T_2$. Let $L'$ be any circuit obtained from $L$ by cyclically reordering its arcs to begin and end at $x$ (note that $L$ is incident to $x$ since it contains $f$). Then the flooding $\mathcal{C}'$ obtained from $\mathcal{C}$ by removing the circuits containing $f$ and $f'$ and adding $T_1 T_2$ and $R_1 L' R_2$ still has the same value, and has $f'$ being in an even circuit. We obtain this flooding in $\mathcal{O}(|E(G)|^2)$ time, add it to the dictionary, and return.
\end{proof}

Now we use the other lemmas from this section to handle the case that each edge of $G-t$ is displayed in either two ways or zero ways. 

\begin{proposition}
\label{prop:endingState}
    Given a signed graph $\mathcal{G}=(G, \Sigma, t)$ and a dictionary $\mathcal{D}$ so that each edge of $G-t$ is displayed by $\mathcal{D}$ in either two ways or zero ways, we can in time $\mathcal{O}(|E(G)|^3)$ find either a certificate $(X, \Sigma')$ which respects $\mathcal{D}$, or a dictionary $\mathcal{D}'$ which improves~$\mathcal{D}$.
\end{proposition}
\begin{proof}
    Write $k$ for the value of $\mathcal{D}$. First determine in $\mathcal{O}(|E(G)|^2)$ time how many ways each edge is displayed by $\mathcal{D}$. If there are edges $f,f'$ incident to a common vertex $x\neq t$ which are displayed in two and zero ways, respectively, then apply \Cref{lem:improveZeroTwo} and return the improved dictionary. We may therefore assume that no such pair exists.

    Let $\mathcal{G}_1,\ldots,\mathcal{G}_r$ be the $t$-components of $\mathcal{G}$. Each flooding $\mathcal{C}\in\mathcal{D}$ partitions into floodings $\mathcal{C}_i$ of $\mathcal{G}_i$ for $i=1,\ldots,r$. If some $\mathcal{C},\mathcal{C}'\in\mathcal{D}$ satisfy $\val(\mathcal{C}_i)>\val(\mathcal{C}_i')$, then $\mathcal{C}_i\cup(\mathcal{C}'\setminus\mathcal{C}_i')$ is a flooding of value greater than $k$, and we return it. So we may assume that for each $i$, all restrictions to $\mathcal{G}_i$ have the same value $k_i$, and $\sum_{i=1}^r k_i=k$. For each $i$, retain one arbitrary restriction and one restriction witnessing each displayed candidate, up to equivalence. This gives a dictionary $\mathcal{D}_i$ of value $k_i$ displaying exactly the same candidates on $\mathcal{G}_i$ as $\mathcal{D}$. Computing the restrictions, comparing their values, and selecting these witnesses take $\mathcal{O}(|E(G)|^2)$ time in total.

    If $\mathcal{G}_i=(G_i, \Sigma_i, t)$ contains an edge displayed in two ways, then every edge of $G_i-t$ is displayed in two ways, and every edge of $G_i$ incident to $t$ is displayed in one way. Thus $\mathcal{D}_i$ is full, and we apply \Cref{lem:improveFull}. Otherwise, every edge of $G_i-t$ is displayed in zero ways, and we apply \Cref{lem:allZero}. If this returns an improved dictionary $\mathcal{D}_i'$ of $\mathcal{G}_i$, then we can obtain an improved dictionary $\mathcal{D}'$ of $\mathcal{G}$. We may therefore assume that each call returns a certificate $(X_i,\Sigma_i')$ of value $k_i$ in $\mathcal{G}_i$ which respects $\mathcal{D}_i$. 
    
    Set $X\coloneqq\bigcup_{i=1}^r X_i$ and $\Sigma'\coloneqq\bigcup_{i=1}^r\Sigma_i'$. Since the union of cuts in $G_1, \ldots, G_r$ yields a cut in $G$, we have that $\Sigma'$ is a shifting of $\Sigma$. The components of $G-X$ are precisely the components of the graphs $G_i-X_i$, and hence the value of $(X, \Sigma')$ is $k$. An edge of $G_i$ is forced by $(X,\Sigma')$ if and only if it is forced by $(X_i,\Sigma_i')$. Since $\mathcal{D}_i$ and $\mathcal{D}$ display the same candidates on $\mathcal{G}_i$, the certificate $(X,\Sigma')$ respects $\mathcal{D}$, and we return it.

    Applying the lemmas for all of $\mathcal{G}_1, \ldots, \mathcal{G}_r$ takes $\mathcal{O}(|E(G)|^3)$ time since the edge-sets of $G_1, \ldots, G_r$ partition the edge-set of $G$. All other steps take $\mathcal{O}(|E(G)|^2)$ time in total, giving the required running time.
\end{proof}

We handle edges displayed in exactly one way by pulling them down. We need to introduce some notation for this. Let $\mathcal{G}=(G, \Sigma, t)$ be a signed graph, $\mathcal{D}$ be a dictionary of $\mathcal{G}$, and $e$ be an edge of $G-t$ that is displayed in exactly one way. Say that $(\vec{e},\alpha)$ is displayed by $\mathcal{D}$. Let $\widehat{\mathcal{G}}=(\widehat{G},\widehat{\Sigma},t)$ be obtained by pulling $\vec{e}$ down into $\vec{e}_1,\vec{e}_2$ so that $e_1$ has label $\alpha$. (This determines the label of $e_2$.) Note that each flooding $\mathcal{C} \in \mathcal{D}$ yields a flooding $\widehat{\mathcal{C}}$ of $\widehat{\mathcal{G}}$ of the same value as $\mathcal{C}$; an even circuit containing $e$ splits into two even circuits, while an odd circuit containing $e$ splits into one even circuit and one odd circuit. Thus the dictionary $\mathcal{D}$ yields a dictionary $\widehat{\mathcal{D}}$ of $\widehat{\mathcal{G}}$. Every displayed candidate on an edge other than $e$ remains displayed, and both new edges $e_1$ and $e_2$ are displayed. Note that $\widehat{\mathcal{D}}$ also satisfies the dictionary size condition. We say that $(\widehat{\mathcal{G}}, \widehat{\mathcal{D}})$ is \emph{obtained from $(\mathcal{G}, \mathcal{D})$ by pulling $e$ down}.

\begin{lemma}
\label{lem:pullingUpOneEdge}
    Let $\mathcal{G}=(G, \Sigma, t)$ be a signed graph, $\mathcal{D}$ be a dictionary of $\mathcal{G}$, and $e$ be an edge of $G-t$ that is displayed exactly one way. Let $(\widehat{\mathcal{G}}, \widehat{\mathcal{D}})$ be obtained from $(\mathcal{G}, \mathcal{D})$ by pulling $e$ down. Then, given these objects as well as a dictionary $\widehat{\mathcal{D}}'$ of $\widehat{\mathcal{G}}$ which improves $\widehat{\mathcal{D}}$, we can find a dictionary $\mathcal{D}'$ of $\mathcal{G}$ which improves $\mathcal{D}$ in $\mathcal{O}(|E(G)|^2)$ time.
\end{lemma}
\begin{proof}
    Let $k$ be the value of $\mathcal{D}$, and say that $(\vec{e},\alpha)$ is displayed by $\mathcal{D}$. Also write $\vec{e}_1, \vec{e}_2$ for the arcs obtained by pulling $\vec{e}$ down; recall that $\vec{e}_1$ has label $\alpha$.
    
    First suppose that $\widehat{\mathcal{D}}'$ contains a flooding $\widehat{\mathcal{C}}$ of value $q>k$. If $e_1,e_2$ are in distinct circuits, then join those circuits by replacing $\vec{e}_1\vec{e}_2$ with $\vec{e}$, reversing the circuits if necessary. If they are in the same circuit, then restoring $e$ instead gives a circuit $L$ which avoids $t$. Since $G$ is connected, $L$ meets another circuit at a non-root vertex; cyclically reorder $L$ and insert it into that circuit. In either case, the resulting flooding $\mathcal{C}$ of $\mathcal{G}$ has value at least $q-2$. Since $q\equiv k\pmod 2$, either $\mathcal{C}$ has value greater than $k$, or $q=k+2$ and $\mathcal{C}$ has value $k$. In the latter case, if $e_1,e_2$ were in distinct circuits, then both circuits were odd, so $(\vec{e},\alpha+1)$ represents $\mathcal{C}$. Otherwise, $L$ and its receiving circuit were both odd, so $L$ is loose in $\mathcal{C}$ and contains $e$. Reversing $L$ if necessary gives a flooding represented by $(\vec{e},\alpha+1)$. This candidate is not displayed by $\mathcal{D}$.

    Now suppose that $\widehat{\mathcal{D}}'$ contains a flooding $\widehat{\mathcal{C}}$ of value $k$ which is represented by a candidate $(\vec{f},\beta)$ of $\widehat{\mathcal{G}}$ which is not displayed by $\widehat{\mathcal{D}}$. By the definition of $\widehat{\mathcal{D}}$ and since it displays $e_1$ and $e_2$ in one way, this means that $f$ is also an edge of $G$ and $(\vec{f},\beta)$ is not displayed by $\mathcal{D}$. Let $\widehat{\mathcal{C}}_e$ be the flooding in $\widehat{\mathcal{D}}$ which is obtained from the flooding $\mathcal{C}_e \in \mathcal{D}$ which displays $(\vec{e},\alpha)$. So $e_1,e_2$ are in distinct even circuits in $\widehat{\mathcal{C}}_e$. Apply \Cref{lem:threeInEven} to $\widehat{\mathcal{C}}$, $\widehat{\mathcal{C}}_e$, $(\vec{f},\beta)$, and $e_1,e_2$. If it returns a flooding of greater value, use the preceding argument. Otherwise, it returns a flooding of value $k$ represented by $(\vec{f},\beta)$ in which $e_1,e_2$ are in distinct even circuits. Joining these circuits and restoring $e$ preserves the value and the representative $(\vec{f},\beta)$. The lemma takes $\mathcal{O}(|E(G)|^2)$ time, and all other operations take $\mathcal{O}(|E(G)|)$ time.
\end{proof}

\section{Completing the signed case}
\label{sec:algorithm}

We are ready to prove the following key proposition. Afterwards, we quickly show how \Cref{thm:mainMinMax} follows from \Cref{prop:algSigned}. 

\begin{proposition}
\label{prop:algSigned}
    Given a signed graph $\mathcal{G} = (G, \Sigma, t)$ and a dictionary $\mathcal{D}$, we can find in $\mathcal{O}(|E(G)|^3)$ time either a dictionary $\mathcal{D}'$ which improves $\mathcal{D}$, or a certificate $(X, \Sigma')$ which respects $\mathcal{D}$.
\end{proposition}
\begin{proof}
    If every edge of $G-t$ is displayed in zero or two ways, then apply \Cref{prop:endingState} and return its output. Otherwise, choose an edge $e\in E(G-t)$ which is displayed in exactly one way, and let $(\widehat{\mathcal{G}},\widehat{\mathcal{D}})$ be obtained from $(\mathcal{G},\mathcal{D})$ by pulling $e$ down. Apply the same procedure recursively to $(\widehat{\mathcal{G}},\widehat{\mathcal{D}})$. If it returns an improved dictionary, then apply \Cref{lem:pullingUpOneEdge} to obtain an improved dictionary of $\mathcal{G}$ and return it.
        
    We may therefore assume that the recursive call returns a certificate $(X,\widehat{\Sigma}')$ respecting $\widehat{\mathcal{D}}$. Let us write $k$ for the value of $\mathcal{D}$ and $\vec{e}_1, \vec{e}_2$ for the arcs obtained by pulling $\vec{e}$ down. Let us also write $\widehat{\mathcal{G}}=(\widehat{G}, \widehat{\Sigma}, t)$ so that $\widehat{\Sigma}'$ is a shifting of $\widehat{\Sigma}$. Apply the same shifting to $\Sigma$ to obtain $\Sigma'$. (That is, if $\widehat{\Sigma}' = \widehat{\Sigma} \Delta \delta_{\widehat{G}}(S)$ for some set of vertices $S$, then we consider the shifting $\Sigma \Delta \delta_G(S)$.) Thus $\Sigma'$ agrees with $\widehat{\Sigma}'$ on retained edges, and the label of $e$ is the sum of the labels of $e_1,e_2$.

    We claim that the old certificate $(X, \widehat{\Sigma}')$ of $\widehat{\mathcal{G}}$ and the new certificate $(X,{\Sigma}')$ of $\mathcal{G}$ have the same value of $k$. By exhaustively considering the cases, we can see that the value of the new certificate is at most the value of the old certificate. Moreover, by \Cref{lem:easyDir}, the value of the new certificate $(X,{\Sigma}')$ is at least the value of $\mathcal{D}$, which is $k$. Since $\widehat{\mathcal{D}}$ has the same value as $\mathcal{D}$ and $(X, \widehat{\Sigma}')$ respects $\widehat{\mathcal{D}}$, this means that the value of the new certificate $(X,{\Sigma}')$ in fact equals~$k$.
    
    Now we evaluate in more detail what the new certificate can look like. Notice that since $e$ is displayed by $\mathcal{D}$ and $(X,{\Sigma}')$ has value $k$, the edge $e$ is not forced by $(X,{\Sigma}')$. (If it was forced, then $(X,{\Sigma}' \Delta \{e\})$ would have value less than the flooding number of $(G, {\Sigma}' \Delta \{e\}, t)$, contradicting \Cref{lem:easyDir}.) Since $e$ is not forced and the two certificates have the same value, one of the following cases occurs. \begin{enumerate}
    \item Both ends of $e$ are in $X$ and both of the new edges $e_1, e_2$ are even.
    \item Exactly one end of $e$ is in $X$, the new edge inside $X$ is even, and the component containing the other end of $e$ is $\widehat{\Sigma}'$-odd.
    \item Neither end of $e$ is in $X$, and the ends of $e$ lie in distinct $\widehat{\Sigma}'$-odd components.
    \end{enumerate} 
    Here, the labels are taken in $\widehat{\Sigma}'$ and the components are taken in $\widehat{G}-X$.
    
    Now check whether every edge $f$ of $G$ which is not forced by $(X,{\Sigma}')$ is displayed by $\mathcal{D}$. If so, return $(X,{\Sigma}')$. We can check this condition in $\mathcal{O}(|E(G)|^2)$ time by marking which edges are forced and traversing the floodings in $\mathcal{D}$. Now suppose we find an edge $f$ of $G$ which is not forced by $(X,{\Sigma}')$ and is not displayed by $\mathcal{D}$. So $f \neq e$ since $e$ is displayed. Moreover, in each of the cases above, $f$ is also not forced by $(X,\widehat{\Sigma}')$.

    So there is a flooding $\widehat{\mathcal{C}}_0 \in \widehat{\mathcal{D}}$ represented by a candidate $(\vec{f},\alpha)$. Similarly, since $e$ is displayed by $\mathcal{D}$, there exists a flooding $\widehat{\mathcal{C}}_1 \in \widehat{\mathcal{D}}$ so that $e_1$ and $e_2$ are in distinct even circuits in $\widehat{\mathcal{C}}_1$. Apply \Cref{lem:threeInEven} to $\widehat{\mathcal{C}}_0$, $\widehat{\mathcal{C}}_1$, $(\vec{f},\alpha)$, and $e_1,e_2$. Since $(X,\widehat{\Sigma}')$ has value $k$, we obtain a flooding of value $k$ represented by $(\vec{f},\alpha)$ in which $e_1,e_2$ are in distinct even circuits. Joining these circuits and restoring $e$ gives a flooding $\mathcal{C}$ of $\mathcal{G}$ of value $k$ with $f$ in an even circuit. Return $\mathcal{D}\cup\{\mathcal{C}\}$. This is a dictionary which improves $\mathcal{D}$. 

    Each recursive call decreases the number of edges not incident to $t$ by one and increases the total number of edges by one. Thus, starting with $m$ edges, there are at most $m$ pull-downs and every intermediate graph has at most $2m$ edges. Constructing each pulled-down dictionary, checking the displayed edges, and lifting each output take $\mathcal{O}(m^2)$ time. \Cref{prop:endingState} is called only once and takes $\mathcal{O}(m^3)$ time. The total running time is therefore $\mathcal{O}(m^3)$, as desired.
\end{proof}

We now prove \Cref{thm:mainMinMax}, which is restated below for convenience. 

\mainTheoremMM*
\begin{proof}
    Recall from \Cref{lem:easyDir} that the maximum value of a flooding is at most the minimum value of a certificate. By \Cref{thm:reduction}, it suffices to show that there is an algorithm which, given a signed graph $\mathcal{G}=(G, \Sigma, t)$ and a flooding $\mathcal{C}$ of $\mathcal{G}$, returns in $\mathcal{O}(|E(G)|^4)$ time either a flooding $\mathcal{C}'$ of value strictly more than the value of $\mathcal{C}$, or a certificate $(X, \Sigma')$ of the same value as $\mathcal{C}$ with the additional guarantee that: \begin{itemize}
    \item for each edge $e$ of $G$ which is not forced by $(X, \Sigma')$, there exists an optimal flooding $\mathcal{C}_e$ so that $e$ is in an infeasible circuit in~$\mathcal{C}_e$.
    \end{itemize} 
    Such an algorithm exists since we can initialize $\mathcal{D}$ to be the dictionary consisting only of $\mathcal{C}$ and then apply \Cref{prop:algSigned} at most $2|E(G)|$-many times. Within this many iterations, it returns either a dictionary of greater value than $\mathcal{D}$, in which case we obtain the flooding $\mathcal{C}'$, or a certificate $(X, \Sigma')$ which respects the current dictionary, in which case we have found the desired certificate.
\end{proof}

\section{A more general conjecture}
\label{sec:conclusionMinMax}

We conclude this part by briefly discussing a conjecture which would generalize \Cref{thm:mainMinMax} to group-labeled graphs.

In this setting, we have an Eulerian graph $G$ with a fixed root vertex $t$ and a labeling $\gamma$ of $G$ over an arbitrary group $\Gamma$. (In the case of vertex-minors, the group $\mathbb{Z}_2^k$ naturally arises.) That is, $\gamma$ is a function which assigns to each arc $\vec{e}$ a group element $\alpha \in \Gamma$ such that $\gamma(\vec{e}^{-1})=\alpha^{-1}$. The \emph{label} of a trail $\vec{e}_1\vec{e}_2\ldots \vec{e}_n$ is then $\gamma(\vec{e}_1) \cdot \gamma(\vec{e}_2)\cdot \ldots \cdot \gamma(\vec{e}_n)$. A trail is called \emph{feasible} (or often in the literature, \emph{non-zero}, but we write groups multiplicatively to emphasize that they need not be abelian) if its label is not the identity element. \emph{Floodings}, their \emph{value}, and the \emph{flooding number} are defined similarly to before.

We conjecture that there is a polynomial-time algorithm to compute the flooding number of a group-labeled graph. (To be more precise, we conjecture that there is a polynomial-time algorithm for each fixed finite group $\Gamma$. However, we suspect that even an oracle model like the one in~\cite{EulerianUnique} -- which uses queries to the word problem over $\Gamma$ -- would suffice.) In particular, say that a \emph{certificate} is a pair $(X, \gamma')$ so that $X$ is a set of vertices which contains the root and $\gamma'$ is a shifting of $\gamma$ (see~\cite{packingApaths}). A component $H$ of $G-X$ is \emph{$\gamma'$-odd} if there exists a group element $\delta$ of order 2 so that every edge incident to $V(H)$ is labeled by 1 or $\delta$ in $\gamma'$, and the parity of the number of edges labeled by $\delta$ is different from $|\delta(H)|/2$. The \emph{value} of a certificate is then the number of edges which have both ends in $X$ and have non-identity label according to $\gamma'$, plus $|\delta(X)|/2$, minus the number of components of $G-X$ which are $\gamma'$-odd.

There are several reasons not to expect the maximum value of a flooding to equal the minimum value of a certificate. However, if $G$ is obtained from a 4-edge-connected graph by subdividing every edge once, then we do conjecture this to be true. We note that the theorems in~\cite{EulerianUnique} are relevant to handling 2-edge cuts and that subdividing (while changing the labeling appropriately) does not change the flooding number; it is mainly just a convenient way to talk about half-edges.

\part{Full immersion testing}

\section{Introduction}

In this part we prove that full immersion testing in labeled tour graphs is fixed-parameter-tractable; we restate this theorem below for convenience.

\mainTheorem*

Broadly, our algorithm consists of two steps. The core algorithm depends on the input graph $\mathcal{G}$ being 4-edge-connected, so in Section \ref{sec:2-cuts}, we reduce to this case by ``decomposing'' $\mathcal{G}$ along 2-edge cuts into maximally 4-edge-connected components. Section \ref{sec:2-cuts} then uses dynamic programming to test which full immersions can be constructed in all of $\mathcal{G}$. So, we first show, in Sections \ref{sec:overview} to \ref{sec:irrelevant}, how to test for full immersions in each of these 4-edge-connected components. However, using the decomposition found in Section \ref{sec:2-cuts}, we encode the full immersions possible in neighboring components with half-edge colors. The idea is that a color on a half-edge indicates which small full immersions this edge can be replaced with, and the choice of half-edge indicates the ``orientation'' that a small immersion can be inserted with along a 2-edge cut. Hence, our algorithm for full immersion testing in 4-edge-connected graphs also accounts for these half-edge colors, for which we define \textit{decorated graphs}. Further, we call the set of full immersions in a decorated graph, including these colors, its \textit{color-folio}. Roughly, the algorithm for the 4-edge-connected case works by reducing to the bounded treewidth case, in which we invoke Courcelle's theorem (see Lemma \ref{lem:using-courcelles}). We reduce to this case by finding ``irrelevant for the color-folio'' vertices to split off, until the treewidth is small enough to test directly.

\section{Preliminaries}
\label{sec:prelimTesting}

Recall from Section \ref{sec:tour-graph-intro} that the edges of a labeled tour graph are labeled by binary vectors of fixed length. For all remaining sections, we consider all labeled tour graphs to be labeled over the binary vector space $\mathbb{F}_2^U$, where $|U|$ is fixed. Formally, for a graph $G$, we define a function $\lambda$, called an \textit{$\mathbb{F}_2^U$-labeling on $G$}, from its edge set to $\mathbb{F}_2^U$ for some set $U$. We call $\lambda(e)$, $e$'s \textit{weight}. Then, for a trail $T$, $T$'s weight is the sum of the weights of its edges: $\lambda(T) = \sum_{e \in T}\lambda(e)$. Note that this is a natural extension of signed graphs, which is the case where $|U|=1$ and there is a root vertex. These are introduced in Section \ref{sec:tour-graph-intro} and formally defined in Section \ref{sec:prelim}.

The following is a type of augmented labeled tour graph, which we will use for full immersion testing in 4-edge-connected graphs.
\subsection {Decorated graphs}
A \textit{decorated graph} is a tuple $\mathcal{G} = (G, \lambda, R, \rho, \alpha)$ so that \begin{enumerate}
    \item $G$ is a connected 4-regular graph,
    \item $G$ has $\mathbb{F}_2^U$-labeling $\lambda$,
    \item $G$ has roots $R \subseteq V(G)$ with an injective mapping $\rho: R \rightarrow \mathbb{N}$, and
    \item $G$ has $q$-coloring $\alpha$ (a $q$-coloring maps half-edges of G to a collection of 0 or more colors in $\{1,....,q\}$, see below for more).
\end{enumerate}

Note that we add root vertices to handle the case where we are given a mapping from vertices of the target graph $\mathcal{H}$ to vertices of the input graph $\mathcal{G}$, specifying branch vertices in the target full immersion.

As a slight abuse of notation, we will consider the vertices and edges of $G$ to be the vertices and edges of $\mathcal{G}$, and may refer to these interchangeably. Also, unless otherwise stated we will use the convention that the 4-regular graph associated with a decorated graph $\mathcal{H}$ is always $H$: the same letter is used. To fully understand decorated graphs, we elaborate on the definition of \textit{$q$-colored graphs}.

\paragraph{\texorpdfstring{$q$}{q}-colored graphs.}

We define $q$-colored graphs, which we layer onto $\mathbb{F}_2^U$-labeled graphs to take into account colored half-edges (the motivation for this is in \Cref{sec:2-cuts}). Unlike with $\mathbb{F}_2^U$ labelings, we will not care about the parity of the number of colored half-edges on a trail, only that at least one half-edge of a color intersects the trail, and we will care if this half-edge lies at the tail or head of an edge oriented by a trail. We let $\{1,...,q\}$ be the set of colors, and we define a function $\alpha$ from the set of half-edges of a graph to $\mathbb{F}_2^q$, a binary vector representing the subset of colors in $\{1,...,q\}$ on each half-edge. We call a graph colored in this way \textit{$q$-colored}, and we call $\alpha$ a \textit{$q$-coloring} of $G$. Then, consider a trail $T$, and orient each edge of $T$ in the direction $T$ traverses it, letting $T_1$ be the set of tail half-edges of these arcs and $T_2$ the set of head half-edges.
Then, we define the \textit{forward color} of a trail $T$ as the union of the colors of half-edges in $T_1$, that is, $\overrightarrow{\alpha}(T) = \bigvee_{h \in T_1}\alpha(h)$, where $\alpha(h) \in \{0,1\}^q$ and $\alpha_i(h) = 1$ if and only if $h$ has color $i$. Symmetrically, we define the \textit{backward color} of a trail $T$ as the union of the colors of its half-edges that are in $T_2$, that is, $\overleftarrow{\alpha}(T) = \bigvee_{h \in T_2}\alpha(h)$, where $\alpha(h) \in \{0,1\}^q$ and $\alpha_i(h) = 1$ if and only if $h$ has color $i$. As a shorthand, we will also say that an edge $e$ has color $i$ if at least one of its half-edges has color $i$. Further, we will define the color of an edge this way; given an edge-coloring $\alpha$, if $e = \{h_1,h_2\}$, then say $\alpha(e) = \alpha(h_1) \vee \alpha(h_2)$. This is done to simplify arguments where we only care if an edge carries a certain color, which applies in \Cref{sec:enrich} and \Cref{sec:clean-grid}.

\paragraph{Colorless decorated graphs.} 
We can think of labeled tour graphs as decorated graphs without colors, so we define such a term. Let a \textit{colorless decorated graph} be a decorated graph $\mathcal{G}$ with a coloring $\alpha$ such that $\alpha(h) = \{0\}^q$ for all half-edges in $\mathcal{G}$. 
\newline

The final definition required before defining full immersions of decorated graphs is \textit{shifting}, which is discussed in section \ref{sec:prelim}, but in our setting we also specify how shifting generalizes to a labeling instead of just one signature.

\paragraph{Shifting in decorated graphs.}
Shifting in an $\mathbb{F}_2^U$-labeled graph is a natural extension of shifting in a signed graph. Given a decorated graph $\mathcal{G}$ and a set $X \subseteq V(G)$, \textit{shifting at $X$} is defined as adding any vector in $\mathbb{F}_2^U$ to the labelings $\lambda$ of all edges in $\delta(X)$. In particular, we typically consider shifting at a vertex $v$, which adds a vector in $\mathbb{F}_2^U$ to $\lambda(e)$ for every non-loop edge $e$ incident to $v$. If a decorated graph $\mathcal{G'}$ can be obtained from $\mathcal{G}$ by a sequence of shiftings, up to isomorphism (preserving $\rho$-values of root vertices and colors) we call $\mathcal{G}$ and $\mathcal{G'}$ \textit{equivalent}.

\subsection{Full immersions of decorated graphs}
We define full immersions very similarly to how rooted topological minors are defined in ~\cite{TopologicalMinorTesting}. 
Let $\mathcal{G} = (G, \lambda, R_G, \rho_G, \alpha)$ and $\mathcal{H} = (H, \mu, R_H, \rho_H, \beta)$ be two decorated graphs. We say that $\mathcal{H}$ is a \textit{full immersion} of $\mathcal{G}$ if there exist shiftings $\lambda'$ of $\lambda$ and $\mu'$ of $\mu$ such that \begin{enumerate}
\item $\rho_H(R_H) = \rho_G(R_G)$,
    \item there is a mapping $\phi: V(H) \rightarrow V(G)$ that maps each vertex $v \in V(H)$ to a distinct vertex of $V(G)$ so that for all $v \in R_H$, $\rho_G(\phi(v)) = \rho_H(v)$, and
    \item $\phi$ maps each edge $uv \in E(H)$ with half-edges $h_1$ at $u$ and $h_2$ at $v$ to a $\phi(u)\phi(v)$-trail $T$ in $G$, such that $\lambda'(T) = \mu'(uv)$, $\overrightarrow{\alpha}_i(T) \ge \beta_i(h_1)$ and $\overleftarrow{\alpha}_i(T) \ge \beta_i(h_2)$  for all colors $i \in [q]$, all trails in the image of $\phi$ are edge-disjoint, and
    \item all edges in $E(G)$ belong to exactly one trail in the image of $\phi$.
\end{enumerate} 

In other words, the rooted immersion of $H$ in $G$ must use all edges in $E(G)$, and a trail $T$ in $G$ corresponding to an edge $e=\{h_1,h_2\}$ in $E(H)$ must have the same $\mathbb{F}_2^U$-labeling as $e$, and its forward and backward colors must be supersets of $h_1$ and $h_2$'s colors, respectively. Note that we can also only shift the labeling of one of $\mathcal{H}$ and $\mathcal{G}$, and the definition is equivalent.

\subsubsection{Full immersion model}

Given decorated graphs $\mathcal{H}$, $\mathcal{G}$ such that $\mathcal{H}$ is a full immersion of $\mathcal{G}$, a full immersion model of $\mathcal{H}$ in $\mathcal{G}$ is a mapping $\phi$ from the vertices and edges of $H$ to the vertices and a set of trails in $G$ that certify the full immersion conditions given above.

We are ready now to elaborate on our algorithm for 4-edge-connected decorated graphs.

\section{Algorithm for 4-edge-connected decorated graphs}\label{sec:overview}

In the following four sections, we will show how, through an irrelevant vertex technique, to test if a decorated graph $\mathcal{H}$ is fully immersed by a 4-edge-connected decorated graph $\mathcal{G}$, in time $\Oh_{|V(H)|, |U|, q}(n^5)$. However, since our aim is to be able to do full immersion testing without this connectivity condition, we go slightly further and use irrelevant vertices to compute the \textit{color-folio} of a 4-edge-connected graph. Then, \Cref{sec:2-cuts} shows that we can decompose $\mathcal{G}$ into 4-edge-connected components, and use their color-folios to determine which colorless decorated graphs $\mathcal{G}$ fully immerses. So, we need the following definition.

\paragraph{Color-folios.} Let $\mathcal{G}$ be a decorated graph. The \textit{color-folio} of $\mathcal{G}$ is the set of all decorated graphs $\mathcal{F}$, up to equivalence, fully immersed by $\mathcal{G}$, such that every edge of $F$ has at most one half-edge with a non-empty set of colors, and this set consists of a single color. Then, for any integer $\delta \ge 0$, the $\delta$-color-folio $\mathcal{C}_\delta(\mathcal{G})$ of $\mathcal{G}$ is the set of such graphs with $|V(F)| \le \delta$.

So, the following four sections are dedicated to proving the following:

\begin{restatable}{proposition}{propcolfolio}\label{prop:col-folio-alg}
    For any 4-edge-connected decorated graph $\mathcal{G}$ with $r_G$ roots and any $\tau \in \mathbb{N}$, the $\tau$-color-folio of $\mathcal{G}$ can be computed in time $\Oh_{\tau, r_G, |U|, q}(n^5)$.
\end{restatable}

To give an overview, at each step of the algorithm, we find an \textit{irrelevant vertex} $v \in V(G)$, meaning for every way of splitting off $v$, the $\tau$-color-folio of the resulting graph contains a decorated graph $\mathcal{H}$ if and only if the $\tau$-color-folio of $\mathcal{G}$ contains $\mathcal{H}$. This irrelevant vertex will be found by exploiting the structure of a grid-like full immersion in $\mathcal{G}$, that will be found if $\mathcal{G}$ has sufficiently high treewidth. So, we focus on proving:

\begin{restatable}{proposition}{irrelevant}\label{prop:inessential-vertex}
    There exists a function $f_{\ref{prop:inessential-vertex}}: \mathbb{N}^4 \to \mathbb{N}$ such that the following holds. Let $\mathcal{G}= (G, \lambda, R_G, \rho_G, \alpha_G)$ be a 4-edge-connected decorated graph, let $\tau \in \N$ and $r_G \coloneqq |R_G|$, and suppose that $G$ has treewidth at least $f_{\ref{prop:inessential-vertex}}(\tau, r_G, |U|, q)$. Then, there exists a vertex $v \in V(G)$ such that for any of the three decorated graphs $\mathcal{G'}$ obtained from splitting off at $v$ in $\mathcal{G}$, a decorated graph $\mathcal{F}$ is in the $\tau$-color-folio of $\mathcal{G}$ if and only if $\mathcal{F}$ is in the $\tau$-color-folio of $\mathcal{G}'$. Furthermore, such a vertex can be found in time $\Oh_{\tau, r_G, |U|, q}(n^4)$.
\end{restatable}

Once we have this proposition, the last piece required to prove \Cref{prop:col-folio-alg} is the following:

\begin{lemma}\label{lem:using-courcelles}
    For any $\tau \in \mathbb{N}$, suppose there exists a function $f_{\ref{lem:using-courcelles}}: \mathbb{N}^4 \to \mathbb{N}$ such that the treewidth of a decorated graph $\mathcal{G}$ is at most $f_{\ref{lem:using-courcelles}}(\tau, r_G, |U|, q)$. Then, the $\tau$-color-folio of $\mathcal{G}$ can be found in time $\Oh_{\tau, r_G, |U|, q}(n)$.
\end{lemma}
\begin{proof}
    To compute the $\tau$-color-folio of $\mathcal{G}$, its candidate members have at most $\tau$ vertices, so up to equivalence there are at most $g(\tau, |U|, q)$ of them for some function $g$, so we test each one. To apply Courcelle's theorem, we have the following claim:
    \begin{claim}\label{clm:c2mso2}
        For every fixed decorated graph $\mathcal{F}$, the property that a decorated graph $\mathcal{G}$ fully immerses $\mathcal{F}$ is definable in $C_2MSO_2$.
    \end{claim}
    \begin{proof}
        Suppose we are testing whether $\mathcal{F}$ is fully immersed by $\mathcal{G}$. We build the following auxiliary graph $\mathcal{G}'$. Subdivide every edge $e = xy$ of $\mathcal{G}$ once, so that the half-edges of $e$ become edges, and give each new edge the colors of the corresponding half-edge. Arbitrarily give one of the two new edges the labeling of $e$, and the other weight zero; the choice does not matter, since every trail using one of the new edges must use both. This does not increase the treewidth. Labels and colors in $\mathcal{G}'$ become unary predicates on edges, and roots with their $\rho$-values become unary predicates on vertices. We quantify the images of the vertices of $F$, an edge set $X_f$ for each edge $f$ of $F$, and a set of vertices $Y_u$ for each $u \in U$ encoding the shifting. Note that we only need one set of vertices for the shifting of each label, since if we must shift at multiple sets of vertices, this is equivalent to shifting at their symmetric difference. We require that the images of the vertices of $F$ are distinct, that roots are mapped to roots with the same $\rho$-value, and that the sets $X_f$ partition $E(\mathcal{G}')$.

        That $X_f$ is the edge set of a trail between the images of the ends of $f$ is expressible by connectivity of $X_f$ which can be done in $C_2MSO_2$~\cite{CourcelleEngelfriet2012}, together with the condition that the ends of the trail have odd degree and all other vertices have even degree. The $\mathbb{F}_2^U$-labeling of this trail, after shifting, is constrained by, for each $u \in U$, checking the parity of the number of edges in $X_f$ with $\lambda_u$-weight one, and taking into account the parity of shifted edges in $X_f \cap \delta(Y_u)$, which $C_2MSO_2$ allows.

        For colors, let $f = ab$ be an edge of $F$ whose half-edge at $a$ has color $i$. Enumerate the edges of the trail $\phi(f)$ in order, starting at $\phi(a)$. Since every edge of $\mathcal{G}$ is subdivided, the edges in odd positions are exactly the half-edges that the trail traverses first. So $i$ is a forward color of the trail in $\mathcal{G}'$ if and only if $i$ appears on an odd edge of $\phi(f)$ in $\mathcal{G}'$. We express this by requiring that $X_f$ can be partitioned into an ``odd'' set $X_1$ and an ``even'' set $X_0$ such that:
        \begin{enumerate}
            \item the edge of $X_f$ incident to $\phi(a)$ is in $X_1$, and the edge of $X_f$ incident to $\phi(b)$ is in $X_0$,
            \item every other vertex incident to an edge of $X_f$ is incident to equally many edges of $X_0$ and of $X_1$, that is, one of each or two of each, and
            \item all colors on the half-edge of $f$ incident to $a$ appear on some edge of $X_1$, and all colors on the half-edge of $f$ incident to $b$ appear on some edge of $X_0$.
        \end{enumerate}
        Every trail satisfies (1) and (2) with $X_1$ its set of odd edges, since consecutive edges have different parities, and since every trail in $\mathcal{G}'$ between original vertices has even length. Further, if the corresponding trail in the original graph $\mathcal{G}$ has $i$ as a forward color, then an edge in $X_1$ will have color $i$, with an analogous statement holding for backward colors. Conversely, suppose $X_f$ is connected and $(X_0, X_1)$ satisfies (1), (2), and (3). At each vertex other than the ends, pair every edge of $X_1$ with an edge of $X_0$. Following these pairs decomposes $X_f$ into an alternating trail from $\phi(a)$ to $\phi(b)$ and a set of alternating closed trails. However, it is clear that two alternating trails sharing a vertex $v$, where at least one of the two trails is closed, can be combined by following one trail until $v$, then following the other, closed trail until reaching $v$ again, choosing to start with the edge incident to $v$ on the closed trail that keeps the trail alternating, and continuing with the first trail from $v$. So, this trail can be mapped to a trail in $\mathcal{G}$ with the correct forward and backward colors. Since every edge of $F$ has finitely many colors, the color conditions are a finite conjunction over $[q]$.
    \end{proof}
    By the above claim, for a fixed candidate $\mathcal{F}$, the statement ``$\mathcal{G}$ fully immerses $\mathcal{F}$'' is expressible in $C_2MSO_2$ by a sentence $\varphi_{\mathcal{F}}$ whose length depends only on $\tau$, $|U|$ and $q$.
    So by Courcelle's theorem~\cite{Courcelle90}, each $\varphi_{\mathcal{F}}$ can be decided in time $\Oh_{\tau, |U|, q}(n)$, given the tree decomposition. 
    Then, since the treewidth of $\mathcal{G}$ is bounded by a function of $\tau, r_G, |U|, q$, by \cite{Bodlaender96}, the tree decomposition can be found in time $\Oh_{\tau, r_G, |U|, q}(n)$.
\end{proof}

Then, we prove \Cref{prop:col-folio-alg} as follows:
\begin{proof}
    Let $w \coloneqq f_{\ref{prop:inessential-vertex}}(\tau, r_G, |U|, q)$. While $G$ has treewidth at least $w$, find the vertex $v$ of \Cref{prop:inessential-vertex} and split it off. By a theorem of Lov\'asz~\cite{LovaszSplitting} (see also~\cite{Mader78}), one of the three ways of splitting off a vertex of degree 4 in a 4-edge-connected Eulerian graph keeps the graph 4-edge-connected; we choose that one, so \Cref{prop:inessential-vertex} applies again. Each step removes a vertex and preserves the $\tau$-color-folio, so after at most $n$ steps we obtain a 4-edge-connected decorated graph $\mathcal{G}'$ with the same $\tau$-color-folio and treewidth less than $w$. Testing whether the treewidth is at least $w$, and otherwise finding a tree decomposition of width $\mathcal{O}(w)$, takes time $2^{\mathcal{O}(w^3)} n$~\cite{Bodlaender96}. Then, by \Cref{lem:using-courcelles}, the $\tau$-color-folio of $\mathcal{G}'$ can be found in $\Oh_{\tau, r_G, |U|, q}(n)$ time.
    Hence, the reduction takes $\Oh_{\tau, r_G, |U|, q}(n^5)$ time, since there are at most $n$ steps, each taking $\Oh_{\tau, r_G, |U|, q}(n^4)$ time by \Cref{prop:inessential-vertex}, and the final step takes $\Oh_{\tau, r_G, |U|, q}(n)$ time, so the total is $\Oh_{\tau, r_G, |U|, q}(n^5)$.
\end{proof}

The next four sections show how we can use a grid-like full immersion (defined in Section \ref{sec:clean-summary}) in $\mathcal{G}$ to find an irrelevant vertex in $\mathcal{G}$. In broad strokes, we are promised a large grid-like full immersion if the treewidth of $\mathcal{G}$ is sufficiently large, then we show: \begin{enumerate}
    \item we may obtain a ``clean'' grid-like full immersion from a grid-like full immersion,
    \item we may either ``enrich'' a grid-like full immersion with a signature or find a small cut separating edges with ``problematic'' labelings,
    \item we may either ``saturate'' a grid-like full immersion with a color or find a small cut separating edges with ``problematic'' colorings,
    \item we can route trails in a full immersion model from this small cut to a sufficiently large grid, and
    \item there exists a vertex that can be split off in this grid such that all full immersions survive.
\end{enumerate}
\section{Cleaning the grid}\label{sec:clean-summary}

We need to define a few terms surrounding grid-like graphs to proceed.

\subsection{Grid-like graphs}
We will need the following notation on grids.

\begin{figure}
\centering
    \begin{tikzpicture}[scale=1, every node/.style={MyNode, inner sep=1.5}]
    \def \m {3}
    \def \n {6}
    \def \width {1cm}
    \foreach \i in {0,...,\n} %Draw nodes
    	{\foreach \j in {0,...,\m}{
      	\node[draw,circle] (A\i\j) at (\i*\width,-\j*\width) {};
    }}
    \pgfmathparse{int(\n-1)}
    \foreach \i in {0,...,\pgfmathresult}%half-graph edges
    	{\foreach \j in {0,...,\m}{
        	\pgfmathparse{int(\i+1)}\edef\next{\pgfmathresult}
            \draw (A\i\j) -- (A\next\j);
    }}
    \pgfmathparse{int(\m-1)}
    \foreach \i in {0,...,\n}%half-graph edges
    	{\foreach \j in {0,...,\pgfmathresult}{
        	\pgfmathparse{int(\j+1)}\edef\next{\pgfmathresult}
            \draw (A\i\j) -- (A\i\next);
    }}
    \draw[MyRedArc] (A10) -- (A20) -- (A21) -- (A11) -- (A10);
    %\draw[MyRedArc] (A21) -- (A22);
    %\draw[MyRedArc] (A42) -- (A43);
    \draw[MyRedArc] (A62) -- (A63);
    \node[draw, label=left:{$(1,1)$}] at (A00) {};
    \node[draw, label=left:{$(4,1)$}] at (A03) {};
    \node[draw, label=right:{$(4,7)$}] at (A63) {};
    \node[draw, label=right:{$(1,7)$}] at (A60) {};
    \end{tikzpicture}
\caption{A $4 \times 7$ grid with a face on row~$1$ and column~$2$ (bold red), its four corner vertices (labelled), and a vertical boundary edge (also bold red).}
\label{fig:grid}
\end{figure}
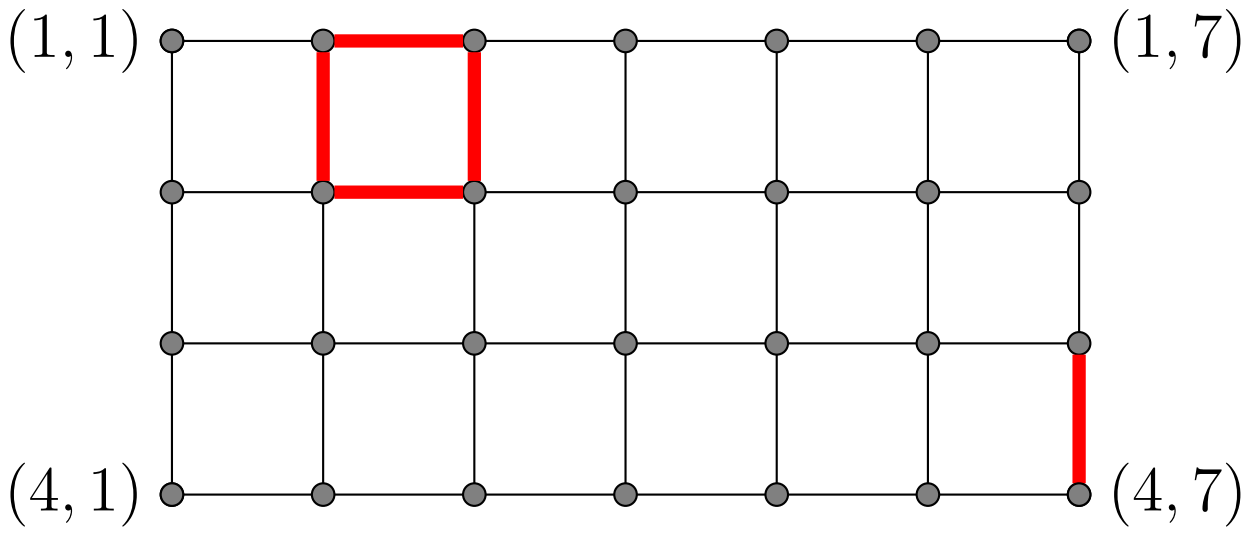

For positive integers $m$ and $n$, an \emph{$m \times n$ grid} is the graph with vertex set $\{(i,j): i\in \{1, \ldots, m\}, j \in \{1, \ldots, n\}\}$, where there is an edge between vertices $(i,j)$ and $(i',j')$ whenever $|i-i'|+|j-j'|=1$ (see \Cref{fig:grid}). We think of the vertex set as having $m$ rows and $n$ columns. Thus we call $m$ the \emph{height} and $n$ the \emph{width}, and we will refer to the \emph{row} and \emph{column} of a vertex. A \emph{face} is a cycle of length~$4$; its \emph{row} (respectively, \emph{column}) is the smallest row (respectively, column) of one of its vertices. 

As in the figure, we consider the vertex $(1,1)$ to be the \emph{top-left vertex}. We will view the columns as being ordered from left to right and the rows as being ordered from top to bottom (or equivalently, from smallest to largest). The \emph{corner vertices} are the top-left, top-right, bottom-left, and bottom-right vertices. The \emph{boundary vertices} are the vertices that are on the same row or column as a corner vertex, and the \emph{boundary edges} are the edges that are in only one face. An edge between vertices $(i,j)$ and $(i+1,j)$ is \emph{vertical}; other edges are \emph{horizontal}. Finally, a \emph{subgrid} of a grid is just a subgraph that is isomorphic to a grid. We will use all of this same terminology for subgrids.

\begin{figure}
\centering
    \begin{tikzpicture}[scale=1, every node/.style={MyNode, inner sep=1.7}]
    \def \m {3}
    \def \n {4}
    \def \width {1cm}
    \foreach \i in {0,...,\n} %Draw nodes
    	{\foreach \j in {0,...,\m}{
      	\node[draw,circle] (A\i\j) at (\i*\width,-\j*\width) {};
    }}
    \pgfmathparse{int(\n-1)}
    \foreach \i in {0,...,\pgfmathresult}%horizontal edges
    	{\foreach \j in {0,...,\m}{
        	\pgfmathparse{int(\i+1)}\edef\next{\pgfmathresult}
            \draw[ultra thick] (A\i\j) -- (A\next\j);
    }}
    \pgfmathparse{int(\m-1)}
    \foreach \i in {0,...,\n}%vertical edges
    	{\foreach \j in {0,...,\pgfmathresult}{
        	\pgfmathparse{int(\j+1)}\edef\next{\pgfmathresult}
            \draw[ultra thick]  (A\i\j) -- (A\i\next);
    }}
    \draw (A03) to [bend left=60] (A00);
    \draw (A23) to [bend left=24] (A20);
    \draw (A43) to [bend right=40] (A40);
    \draw (A03) to [bend left=24] (A00);
    \draw (A01) to (A42);
    \draw (A02) to (A41);
    \draw (A30) to (A43);
    \draw (A33) to [bend left=40] (A13);
    \draw (A10) to [bend left=30] (A40);
    \end{tikzpicture}
\caption{A $4 \times 5$ grid-like graph and its underlying grid (bold).}
\label{fig:grid-like}
\end{figure}
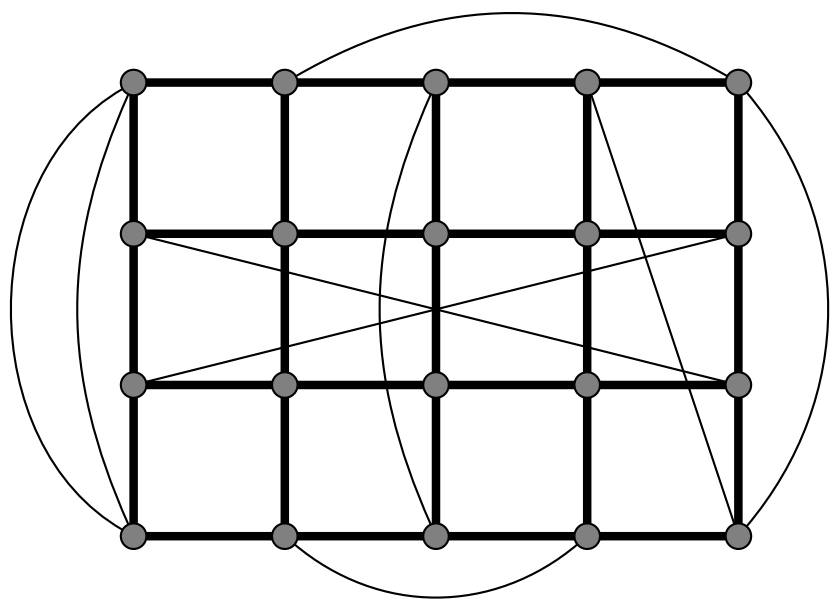

Then, we say that a 4-regular graph $G$ is \emph{grid-like} if there exists $F\subseteq E(G)$ so that $G-F$ is isomorphic to a grid (see \Cref{fig:grid-like}). Thus every end of an edge in $F$ is a boundary vertex of that grid. We call $G-F$ the \emph{underlying grid of $G$} and use all of the prior definitions on grids for $G$ as well (by implicitly referring to the underlying grid). With this set-up, we are implicitly fixing $F$ as well as an isomorphism from $G-F$ to an actual grid.

Finally, let $G$ be a grid-like graph. An \emph{outside edge} is one that is not part of the underlying grid. \textit{Vertical outside edges} are ones where both endpoints lie on the same column; \textit{horizontal outside edges} are ones where both endpoints lie on the same row. 

We will discuss full immersions of grid-like graphs often, so as a short-hand we call these \textit{grids in $\mathcal{G}$}, as below.
\subsection{Grids in decorated graphs}
Let $\mathcal{G} = (G, \lambda, R, \rho, \alpha)$ be a decorated graph. Then, a \textit{grid in $\mathcal{G}$} is a tuple $(\mathcal{H}, \phi)$ so that \begin{enumerate}
    \item $\mathcal{H} = (H, \mu, \emptyset, \emptyset, \beta) $ is a decorated graph with an empty set of roots where $H$ is a grid-like graph,
    \item $\mathcal{G}$ fully immerses $\mathcal{H}$,
    \item $\phi$ maps each vertex of $H$ to a distinct vertex of $G$, and maps the edges $uv$ of $H$ to edge-disjoint $\phi(u)\phi(v)$-trails such that these trails form a full immersion of $\mathcal{H}$ in $\mathcal{G}$.  
\end{enumerate}

Next, it will be useful in finding an irrelevant vertex to be able to manipulate how labelings and colors ``interact'' with a full immersion of a grid-like graph, so we need the following.

\subsection{Color-saturated and rich grid-like graphs}
Let $\mathcal{G} = (G, \lambda, R, \rho, \alpha)$ be a decorated graph, and let $A \subseteq U$. We say $\mathcal{G}$ is \textit{$A$-rich} \textit{with richness $m$}, if \begin{enumerate}
\item $G$ is a grid-like graph,
\item each edge of $G$'s underlying grid has $\lambda_A$-weight zero, and
\item there exist $m$ pairwise disjoint sets of outside edges of $G$ whose $\lambda_A$ weights generate $\mathbb{F}_2^A$, called the \textit{batches} of $\mathcal{G}$.
\end{enumerate} 

Then, let $S \subseteq \{1,...,q\}$. We say a decorated graph $\mathcal{G}$ that is $A$-rich is \textit{$S$-saturated}, if each batch of $\mathcal{G}$ contains at least one half-edge of each color in $S$, and the maximum size of a batch is $|A| + |S|$.
Note that we do not require distinct half-edges for each color.

Then, roughly speaking, our algorithm will identify the branch vertices of a full grid immersion to a single root vertex to obtain a free graph on which we will run the flooding algorithm from Theorem \ref{thm:mainMinMax}. So, it is helpful to define ``clean'' grids, where the underlying grid does not have signatures or colors that may be relevant in the associated free graph.

\subsection{\texorpdfstring{$(A,B)$}{(A,B)}-clean grids}

An $(A,B)$-clean grid is a type of $A$-rich decorated graph that has properties that will be useful in finding a ``maximally rich'' full immersion. Let $\mathcal{G} = (G, \lambda, R, \rho, \alpha)$ be a decorated grid-like graph, and let $A$ and $B$ be disjoint sets with $A, B \subseteq U$. Then $\mathcal{G}$ is \emph{$(A,B)$-clean} with \emph{richness $m$} if each edge of $G$'s underlying grid has $\lambda_A \times \lambda_B$-weight zero, and there exist pairwise disjoint sets of outside edges of $G$, $F_1, \ldots, F_m$ so that\begin{enumerate}
    \item $\mathcal{G}$ is $A$-rich with batches $F_1, \ldots, F_m$,
    \item there exist disjoint subgrids $H_1, \ldots, H_m$ of the same height as $G$ so that for each $i \in \{1, \ldots, m\}$, every edge in $F_i$ is an outside-vertical edge with its ends in $V(H_i)$, and
    \item the $\lambda_A \times \lambda_B$-weights of each of $F_1, \ldots, F_m$ generate the same subspace of $\mathbb{F}_2^A \times \mathbb{F}_2^B$, $N$, where $\dim(N) = |A|$.
\end{enumerate} We call $N$ the \emph{subspace} of $\mathcal{G}$. It is convenient to also consider the subspaces generated by the projection $\pi_v: \mathbb{F}_2^A \times \mathbb{F}_2^B \rightarrow \mathbb{F}_2^A \times \mathbb{F}_2^v$ of $N$ onto just the $\lambda_A\times \lambda_v$-weights, for any $v \in B$; let each such subspace be denoted $N_v$. 

\subsection{Algorithm for cleaning a grid}\label{sec:clean-grid}

Finally, let $\mathcal{G}$ be a decorated graph, and let $(\mathcal{H}, \phi)$ be a grid in $\mathcal{G}$, with $\mathcal{H} = (H, \mu, \emptyset, \emptyset, \beta)$. Note that trivially, $\mathcal{H}$ is $\emptyset$-rich and $\emptyset$-saturated.
We claim that $\mathcal{H}$ fully immerses an $(A,U \setminus A)$-clean $\emptyset$-saturated decorated graph $\mathcal{H}'$ for some $A \subseteq U$, which implies $\mathcal{H}'$ is fully immersed by $\mathcal{G}$. Formally,
\begin{restatable}{proposition}{controlledprop}
\label{prop:controlled}
There is a function $f_{\ref{prop:controlled}}: \N^4 \rightarrow \N$, non-decreasing in every argument, so that for any $m \in \N$, disjoint sets $A,B \subseteq U$, and $S \subseteq [q]$, every $A$-rich $S$-saturated decorated graph $\mathcal{G}$ of richness, height, and width at least $f_{\ref{prop:controlled}}(m, |A|, |B|, q)$ fully immerses an $(A',B')$-clean $S$-saturated decorated graph $\mathcal{H}$ of richness, height, and width at least~$m$, where $A' \supseteq A$ and $B' = B \setminus (A' \setminus A)$. Further, a full immersion model can be recovered in time $\Oh_{m, |U|, q}(n)$.
\end{restatable}

Intuitively, this proposition says we can sacrifice some amount of height, width, and richness to ``clean'' a grid. We will use this proposition often to make sure the graph we are working with is ``clean'', which will simplify our proofs. The remainder of this section proves the above proposition, throughout which, $n \coloneqq |V(G)|$; when $\mathcal{G}$ is itself given by a full immersion model in a larger graph, composing models costs time linear in the size of that graph.

For ``cleaning up'' signatures, we repeatedly use the following proposition that says, given a sufficiently  large, rich, saturated grid, we can get any rich and saturated grid as a full immersion:

\begin{proposition}[Grid universality for decorated graphs]
\label{prop:universality}
There is a function $f_{\ref{prop:universality}}:\N^3 \rightarrow \mathbb{N}$ so that the following holds for any $S \subseteq [q]$. If $\mathcal{G} = (G, \lambda, R_G, \rho_G, \alpha)$ and $\mathcal{H} = (H, \mu, R_H, \rho_H, \beta)$ are two decorated grid-like graphs such that $\lambda,\mu$ sum to the same element, and \begin{enumerate}
    \item $\mathcal{G}$ is $A$-rich and $S$-saturated with height, width, and richness at least $f_{\ref{prop:universality}}(|V(H)|, |A|, |S|)$, and
    \item $\mathcal{H}$ is $A'$-rich and $S'$-saturated for some $A' \subseteq A$ and $S' \subseteq S$.
\end{enumerate}
Then $\mathcal{G}$ fully immerses $\mathcal{H}$. Furthermore, a model of this full immersion can be found in time $\Oh_{|\mathcal{H}|, |A|, |S|}(n)$.
\end{proposition}
\begin{proof}
    Let $t \in \Zplus$, let $A$ be a set, and define\begin{align*}
        d &\coloneqq 2t(|A|+2|S|+1)+1,\\
        k &\coloneqq 2t(2|S|+1)+4t+ 8d,\textrm{ and}\\
        f_{\ref{prop:universality}}(t, |A|, |S|) & \coloneqq k.
    \end{align*}
Now, let $\mathcal{G} = (G, \lambda, R_G, \rho_G, \alpha)$ and $\mathcal{H} = (H, \mu, \emptyset, \emptyset, \beta)$ be decorated graphs that satisfy the conditions of the proposition, where $t$ is the number of vertices of $H$. Let $\hat{H}$ denote the underlying grid of $H$. We map the underlying grid of $H$ to a subgrid of $G$ so that the branch vertices of $H$ each have distance at least $d$ from the boundary vertices of $G$.

First of all, we claim that it suffices to find a collection $\T = (T_e: e$ is an outside edge of $H)$ of pairwise edge-disjoint trails of $G$ so that\begin{enumerate}
    \item for each outside edge $e=uv$ of $H$, the trail $T_e$ has ends $u$ and $v$, has $\lambda(T_e) =\mu(e)$, and $\alpha(T_e) \supseteq \beta(e)$, and
    \item no trail in $\T$ uses any edge of $\hat{H}$ or any boundary edge of $G$.
\end{enumerate}So, suppose that such a collection $\T$ exists; we will show that then the proposition holds.

Consider the components of the graph obtained from $\mathcal{G}$ by deleting the edges of $\hat{H}$ and the edges of each trail in $\T$. Since there is a component that includes all boundary edges of $\mathcal{G}$, and only the outside edges of $\mathcal{G}$ can be non-zero, it follows that there is at most one component with a non-zero edge. Furthermore, since the sum of the weights of the edges is the same in the two graphs, the weights of the edges of that component sum to zero. So each component is an Eulerian graph whose edge-weights sum to zero; thus we can add on each component to a trail in $\T$ without changing its weight. In this manner, we obtain another collection of trails with the appropriate weights so that the edge-sets of the trails partition $E(G)-E(\hat{H})$. Thus $\mathcal{G}$ fully immerses $\mathcal{H}$, as desired.

Now it just remains to show that such a collection $\T$ exists. Here is our overall approach. First of all, $H$ has at most $2t$ edges and therefore at most $2t$ outside edges. For each of those outside edges, we will use one of the $k$ batches of $(G, \lambda)$ to build the corresponding trail in $\T$. We have plenty of batches to choose from. Each batch has size at most $|A|+|S|$, and we will use $|A|+2|S|+1$ ``consecutive concentric cycles around $\hat{H}$'' to build the corresponding trail.

To state the definition of a concentric cycle, let $(m,n)$ be the bottom-right vertex of $G$. For each $i \in \{1, \ldots, d-1\}$, the \emph{concentric cycle at distance $i$} is the cycle consisting of the boundary edges of the subgrid with top-left vertex $(1+i,1+i)$ and bottom-right vertex $(m-i,n-i)$; see \Cref{fig:rerouting1}. These cycles exist and are disjoint from the boundary of $\hat{H}$. Since $d=2t(|A|+2|S|+1)+1$, we can partition these concentric cycles into $2t$ parts of size $|A|+2|S|+1$. Furthermore, we can assume that the distances of each part are of the form $i, i+1, \ldots, i+|A|+2|S|$. Now fix an arbitrary injection from the outside edges of $H$ to these $2t$ parts.

Next we show how to choose batches. Consider just those batches whose ends have distance at least $d$ from the corners of $G$ and are not on a row or column of $\hat{H}$. There are at least $2t(2|S|+1)$ such batches; at most $8d$ batches are too close to a corner of $G$, and at most $4t$ outside edges are on a row or column of $\hat{H}$. So we can fix an injection from the outside edges of $H$ to $2t$ sets of size $2|S|+1$ of these batches. We will need one batch for the labeling, one batch for each forward color of a trail and one batch for each backward color of a trail (the colorings of the half-edges of an outside edge of $H$).

Now we show how to obtain $\T$; see \Cref{fig:rerouting2}. Consider an outside edge $e=uv$ of $H$ with at most $1+2|S|$ associated batches and associated concentric cycles at distances $i, i+1, \ldots, i+|A|+2|S|$. The corresponding trail $T_e$ begins at $u$, and its first edge is an edge of $E(G)-E(\hat{H})$ that is incident to $u$. Then it proceeds along this direction until hitting the first concentric cycle, at distance $i+|A|+2|S|$. If $\mu(e) = 0$ and $\beta(e) = 0$, then it proceeds around this concentric cycle and then returns to $v$ in the same manner.

\begin{figure}
\centering
    \begin{tikzpicture}[scale=1, every node/.style={MyNode, inner sep=.75}]
    \def \m {7}
    \def \n {10}
    \def \width {.75cm}
    \foreach \i in {0,...,\n} %Draw nodes
    	{\foreach \j in {0,...,\m}{
      	\node[draw,circle] (A\i\j) at (\i*\width,-\j*\width) {};
    }}
    \pgfmathparse{int(\n-1)}
    \foreach \i in {0,...,\pgfmathresult}%half-graph edges
    	{\foreach \j in {0,...,\m}{
        	\pgfmathparse{int(\i+1)}\edef\next{\pgfmathresult}
            \draw (A\i\j) -- (A\next\j);
    }}
    \pgfmathparse{int(\m-1)}
    \foreach \i in {0,...,\n}%half-graph edges
    	{\foreach \j in {0,...,\pgfmathresult}{
        	\pgfmathparse{int(\j+1)}\edef\next{\pgfmathresult}
            \draw (A\i\j) -- (A\i\next);
    }}
    \draw[MyRedArc, very thick] (A11) -- (A91) -- (A96) -- (A16) -- (A11);
    \draw[MyRedArc, very thick] (A22) -- (A82) -- (A85) -- (A25) -- (A22);
    \draw[line width={2pt}, dashed, arrRepeat] (A33) -- (A32) -- (A82) -- (A85) -- (A25) -- (A20) to [bend left=25] (A70) -- (A71) -- (A91) -- (A94) -- (A44);
    \draw (A20) to [bend left=25] (A70);
    %draw bold grid edges
    \foreach \i in {3}%horizontal edges
    	{\foreach \j in {3,4}{
        	\pgfmathparse{int(\i+1)}\edef\next{\pgfmathresult}
            \draw[ultra thick] (A\i\j) -- (A\next\j);
    }}
    \foreach \i in {3,...,4} %vertical edges
    	{\foreach \j in {3}{
      	       \pgfmathparse{int(\j+1)}\edef\next{\pgfmathresult}
            \draw[ultra thick] (A\i\j) -- (A\i\next);
    }}
    %re-draw regular nodes
    \foreach \i in {0,...,\n} %Draw nodes
    	{\foreach \j in {0,...,\m}{
      	\node[draw,circle] at (A\i\j) {};
    }}
    %draw subgrid nodes
    \foreach \i in {3,...,4} %Draw nodes
    	{\foreach \j in {3,...,4}{
      	\node[draw,circle, inner sep=1.7]  at (\i*\width,-\j*\width) {};
    }}
    %\node[draw,circle, inner sep=1.7]  at (A20) {};
    %\node[draw,circle, inner sep=1.7]  at (A70) {};
    \end{tikzpicture}
\caption{A depiction of a possible trail $T_e$ (bold dashed) where the ends of $e$ are the top-left and bottom-right vertices of $\hat{H}$.}
\label{fig:rerouting2}
\end{figure}
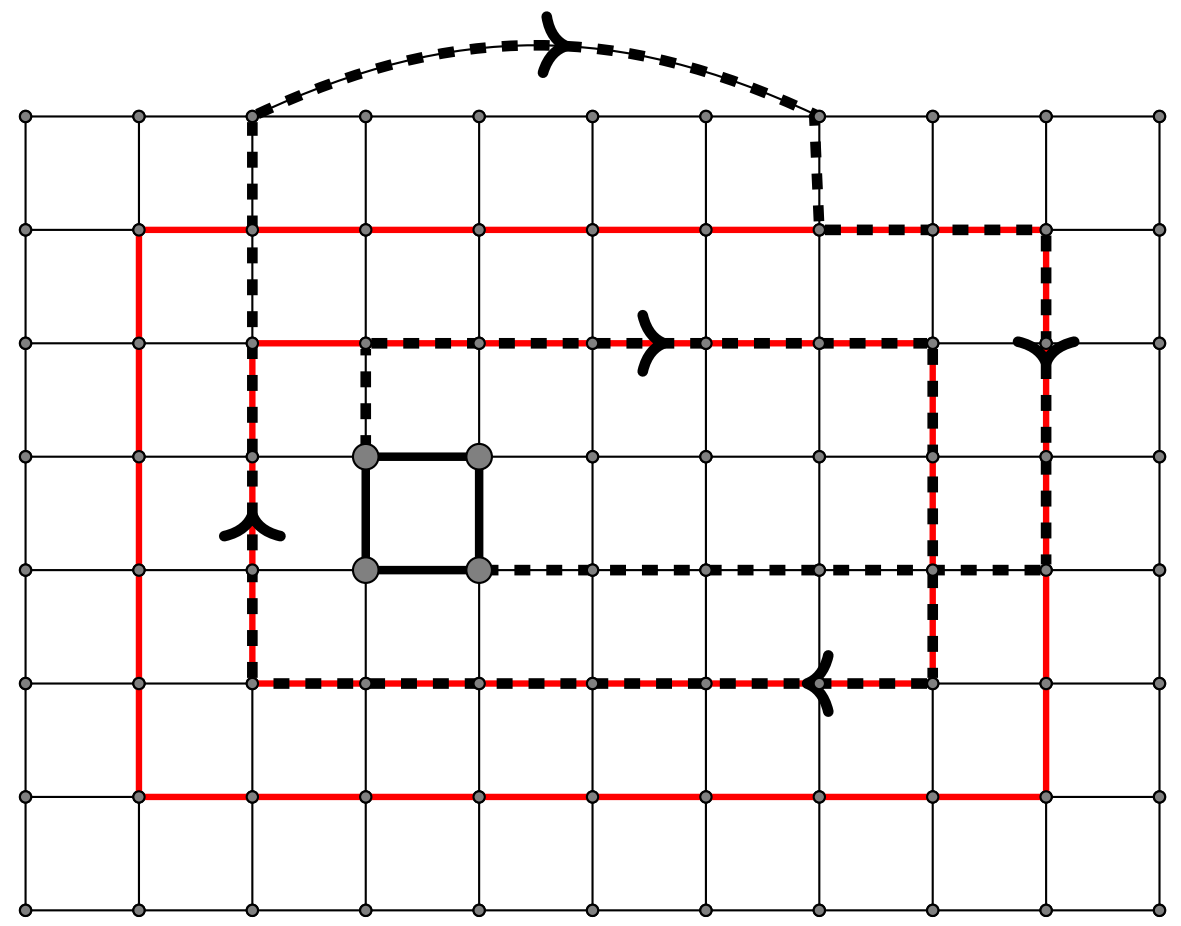

So now assume that $\mu(e) \neq 0$, $\beta(h_u) \neq 0$ or $\beta(h_v) \neq 0$. We show how to give its trail in $G$ the correct coloring and labeling; we start with its coloring. Let the half-edges of $e=uv$ be $\{h_u,h_v\}$, with $h_u$ incident to $u$, and $h_v$ incident to $v$. Our trail begins at $u$, so we show how to ensure that the forward colors of this trail contain all colors in $\beta(h_u)$. See that there are distinct half-edges $g_1, \ldots, g_p$ contained in edges of at most $|S|$ batches so that $\beta(h_u) \subseteq \alpha(g_1)\vee\ldots \vee \alpha(g_p)$. Then the trail proceeds along the concentric cycle, and then directly out to the end of $g_1$ lying on the boundary of $G$, then along $g_1$, and finally back to the next concentric cycle, at distance $i+|A|+2|S|-1$. This process repeats until the trail has used all of the half-edges $g_1, \ldots, g_p$, of which there are at most $|S|$, and returned to the concentric cycle at distance $i+|A|+2|S|-p$. Our constructed trail so far has all required colors for $h_u$. It is clear that the same process can be repeated for the colors of $h_v$ using $p' \le |S|$ outside-edges of remaining batches, where instead the trail enters an outside edge containing the required half-edge from its other end, so the required color is a backward color of the trail. See that at the end of this process, the trail has returned to the concentric cycle at distance $i+|A|+2|S|-p-p' \ge i+|A|$.
Now, we show how to obtain the required labeling. The trail constructed so far may have a non-zero labeling picked up from the outside-edges used so far, so let the labeling of the trail constructed so far be $r$.
Then there are distinct edges $f_1, \ldots, f_\ell \in F_3$ so that $\mu(e) = r + \lambda(f_1)+\ldots+\lambda(f_{\ell})$. Finding these edges takes time $\Oh((|A|+|S|)^3)$ using Gaussian elimination. Then the trail proceeds along the concentric cycle, and then directly out to an end of $f_1$, then along $f_1$, and finally back to the next concentric cycle, at distance at least $i+|A|-1$. This process repeats until the trail has used all of the edges $f_1, \ldots, f_\ell$ and returned to the concentric cycle at distance at least $i+|A|-\ell$. Finally, the trail proceeds along that concentric cycle, and then back to $v$ in the same manner. Due to the choice of the batches, these trails exist and satisfy all of the conditions that were outlined earlier.

For the runtime, finding the edges for each of the at most $2t$ trails takes $\Oh((|A|+|S|)^3)$ time by Gaussian elimination, the trails are edge-disjoint so constructing them takes $\Oh(|E(G)|)$ time, and inserting the leftover components takes $\Oh(|E(G)|)$ time using Hierholzer's algorithm. So the total is $\Oh(t(|A|+|S|)^3 + n)$.
So this completes the proof of \Cref{prop:universality}.
\end{proof}

Now that we have this universality tool, we first focus on how to deal with signatures. Since signatures are mutable through shiftings, the following lemma on word decompositions will be helpful.
\begin{lemma}
\label{lem:RamseyWords}
There is a function $f_{\ref{lem:RamseyWords}}:(\Zplus)^2\rightarrow \Zplus$ so that for any $m,k\in \Zplus$, every word over $\mathbb{F}_2^m$ of length $f_{\ref{lem:RamseyWords}}(m,k)$ has a decomposition $w_0, z_1, \ldots, z_k, w_{k+1}$ so that $z_i$ evaluates to zero for each $i \in \{1, \ldots, k\}$. Moreover, this decomposition can be found in $\Oh(m(k+1)2^m)$ time.
\end{lemma}

\begin{proof}
    Let $f_{\ref{lem:RamseyWords}}(m,k) = (k+1)2^m$, and let $W=w_1,...,w_{(k+1)2^m}$ be a word over $\mathbb{F}_2^m$. Then, it is enough to find $k+1$ repeated prefix values in $W$, since if $w_i,w_j$ have the same prefix value and $i<j$, then the subword $w_{i+1},...,w_j$ evaluates to 0. Since there are $2^m$ possible prefix values for a word over $\mathbb{F}_2^m$, and $W$ has $(k+1)2^m$ prefixes, there must be some prefix value that repeats at least $k+1$ times. This proves the decomposition exists. To recover the decomposition, we can read the word left to right, evaluating each prefix by adding the current letter to the value of the previous prefix, and maintaining a dictionary storing the frequency of each prefix value. It is easy to see that this can be done in $\Oh(m(k+1)2^m)$ time.
\end{proof}

Broadly, given a decorated graph $\mathcal{G} = (G, \lambda, R, \rho, \alpha)$, we will use this lemma to combine the $\lambda_v$ weights of its edges to obtain $\lambda_v$ weight zero trails, for $v \in B$, to make up edges of the inner grid of the $(A,B)$-clean decorated graph we are after. We do this by considering the set of faces with vertical edges on columns $i$ and $i+1$; we consider this to be the set of faces on column $i$. We let each of the $n$ columns with $m$ faces of an $(m+1) \times (n+1)$ grid be a letter $z_i$ over $\mathbb{F}_2^m$. Then, by \Cref{lem:RamseyWords}, there's a way to ``combine'' columns so that the faces each have $\lambda_v$ weight zero. The faces having $\lambda_v$ weight zero is enough, by the following lemma:

\begin{lemma}
\label{lem:facesEnough}
If $\mathcal{G} = (G, \lambda, R, \rho, \alpha)$ is a decorated grid-like graph so that, only considering the underlying grid, each face has weight zero, then there exists a shifting $\lambda'$ of $\lambda$ so that for each edge $e$, $\lambda'(e)=0$. Further, this shifting can be performed in $\Oh(|V(G)|+|E(G)|)$ time.
\end{lemma}
\begin{proof}
    Consider the labeled graph $G'$ with labeling $\lambda$ obtained by deleting the outside edges of $\mathcal{G}$ and restricting $\lambda$ to the remaining edges. The set of facial cycles is a basis for the cycle space of $G'$. So every circuit has weight zero since, if $C_1$ and $C_2$ are the edge-sets of two circuits, then their symmetric difference has weight\begin{align*}
        \lambda(C_1)+\lambda(C_2)-2\lambda(C_1 \cap C_2) = \lambda(C_1)+\lambda(C_2).
    \end{align*} So Harary's theorem~\cite{HararyBalance} on balanced signed graphs says that the desired shifting exists.

    To perform the shifting, we consider the following algorithm. First, we find a spanning tree $\mathcal{T}$ of $G'$ in $\Oh(|V(G')| + |E(G')|)$ time. Then, we traverse the spanning tree, starting at an arbitrary root, using $BFS$. At each new vertex, do the necessary shifting such that the edge traversed has weight 0. So, all edges on the spanning tree have weight 0. Let this new labeling be $\lambda'$. Note that these shiftings preserve the total weight on each circuit in $G'$, so all circuits in $G'$ still have weight 0 under $\lambda'$. Consider any edge $e \notin E(\mathcal{T})$. Then, the weight of $e$'s fundamental cycle $C$ with respect to $\mathcal{T}$ is 0. Since the weight of $C \setminus \{e\}$ is also 0, $e$ must have weight 0.
\end{proof}

With these lemmas, we are ready to see how to ``clean up'' labelings in our immersion:
\controlledprop*

\begin{proof}
Write $s \coloneqq q$; since $|S| \le s$, every batch has at most $a+s$ edges. Define\begin{align*}
        k &\coloneqq (a+s+1)m2^{(a+s)b},\\
        t &\coloneqq (m+1)(k+1),\\
        L &\coloneqq (a+s+2)f_{\ref{prop:universality}}(t, a, s)+f^b_{\ref{lem:RamseyWords}}(m,k)+b+m+1,\textrm{ and}\\
        f_{\ref{prop:controlled}}(m,a,b,s) &\coloneqq   2f_{\ref{prop:universality}}(L^2, a, s)+L.
\end{align*}
Here $f^b_{\ref{lem:RamseyWords}}$ denotes the $b$-fold iterate of $f_{\ref{lem:RamseyWords}}$ in its second argument; that is, $f^0_{\ref{lem:RamseyWords}}(m,k) \coloneqq k$ and $f^{j}_{\ref{lem:RamseyWords}}(m,k) \coloneqq f_{\ref{lem:RamseyWords}}(m, f^{j-1}_{\ref{lem:RamseyWords}}(m,k))$ for $j \ge 1$.
Suppose that the conditions of the proposition hold for disjoint sets $A$ and $B$ of sizes $a$ and $b$ respectively, and some $A$-rich $S$-saturated decorated graph $\mathcal{G}$, with $|S|$ denoted by $s$. Note that the size of each batch of $\mathcal{G}$ is at most $a+s$, by definition.

Our aim is to apply \Cref{lem:RamseyWords}, the Ramsey-type theorem on words. Since each letter of the word will come from a column of faces, we will want to ``combine columns''. So it is helpful to have the following claim.

\begin{claim}
\label{clm:controlled1}
The graph $\mathcal{G}$ fully immerses an $A$-rich $S$-saturated decorated graph $\mathcal{H}$ of richness $f_{\ref{prop:universality}}(t, a, s)$ and height and width $L$ so that for each column $c$ of $H$, there is an edge of $\mu_A$-weight zero whose ends are the two boundary vertices on column~$c$. Further, each batch of $\mathcal{H}$ contains only horizontal outside edges, and this immersion can be recovered in $\Oh(|E(G)|+(a+s)f_{\ref{prop:universality}}(t, a,s))$ time.
\end{claim}
\begin{proof}
    Notice that $G$ has an $L \times L$ subgrid whose vertex-set has distance at least $f_{\ref{prop:universality}}(L^2, a,s)$ from the boundary vertices of $G$. This is simply the ``middle'' subgrid of $G$.
    So by \Cref{prop:universality} on universality (applied with $A$ in place of $U$, that is, to the labelings $\lambda_A$ and $\mu_A$), we only need to find a decorated graph $\mathcal{H}$ which satisfies the claim so that the $\mathbb{F}_2^A$-weights of the edges sum to the same element in both graphs. Such a graph exists since $L \geq (a+s)f_{\ref{prop:universality}}(t, a,s)+1$; we can add $f_{\ref{prop:universality}}(t, a,s)$ batches of size at most $a+s$ and then correct the sum of the weights with one additional edge. Computationally, we take the middle subgrid as the vertices of the inner grid of $H$, then calculate the difference between the sum of the labels in $\mathcal{G}$ and $f_{\ref{prop:universality}}(t, a,s)$ batches of size $a+s$ to obtain the signature of the additional edge, which can clearly be done in $\Oh(|E(G)|+(a+s)f_{\ref{prop:universality}}(t, a,s))$ time. Then, to recover the trails of the outside edges of $\mathcal{H}$, we may do so using the concentric cycle method as in \Cref{prop:universality} on universality, which is done in time $\Oh(|E(G)|)$.
\end{proof}

Now let $\mathcal{H} = (H, \mu, \emptyset, \emptyset, \beta)$ be the $L \times L$ decorated graph from the claim. By the definition of $L$, $H$ has a subgrid $\hat{H}$ of height $m+1$ and width $f^b_{\ref{lem:RamseyWords}}(m,k)+b$ so that $V(\hat{H})$ has distance at least $f_{\ref{prop:universality}}(t, a,s)$ from the boundary vertices of $H$ (this is the middle subgrid of $H$).

Then, let $v \in B$ be arbitrary. For each column $c$ of faces of $\hat{H}$, there is a corresponding element of $\mathbb{F}_2^m$ which represents the $\mu_v$-weights of the $m$ faces on column $c$. This word can be recovered in $\Oh(|E(H)|)$ time. Thus we obtain a word over $\mathbb{F}_2^m$ of length $f^b_{\ref{lem:RamseyWords}}(m,k)$. Let $r = f^{b-1}_{\ref{lem:RamseyWords}}(m,k)+1$. By \Cref{lem:RamseyWords}, this word has a decomposition $w_0, z_1, \ldots, z_{r-1}, w_{r}$, found in $\Oh(m(f^{b-1}_{\ref{lem:RamseyWords}}(m,k)+1)2^m)$ time, so that for each $i \in \{1, \ldots, f^{b-1}_{\ref{lem:RamseyWords}}(m,k)\}$, the subword $z_i$ evaluates to zero.

We will ``combine'' columns according to these subwords, thereby finding a full immersion of an $A$-rich $S$-saturated decorated graph in which the faces of a large subgrid have $\mu_v$-weight zero. We will combine columns again according to the corresponding subwords of the next $u \in B$, until we get a full immersion of an $(A,B)$-clean $S$-saturated decorated graph.

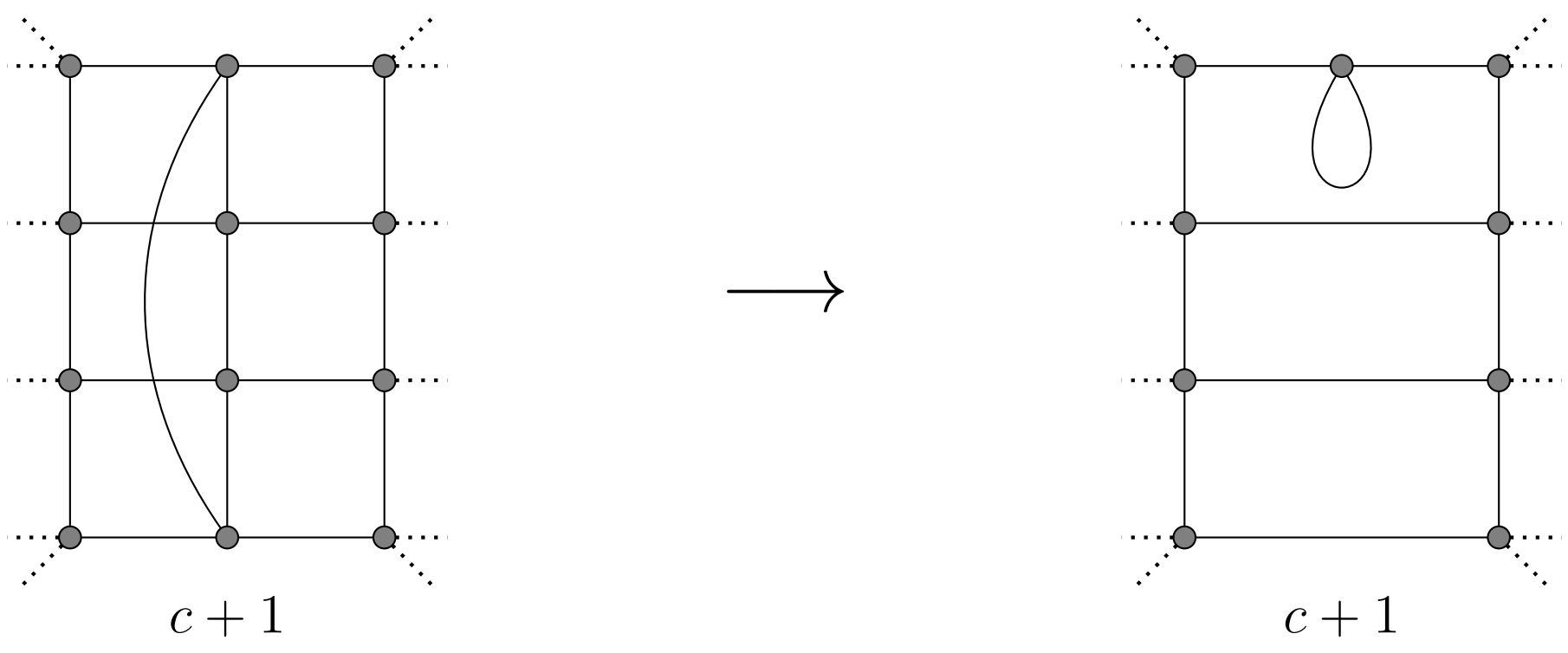
\begin{figure}
\centering
    \begin{tikzpicture}[scale=1, every node/.style={MyNode, inner sep=1.7}]
    \def \m {3}
    \def \n {2}
    \def \width {1.2cm}
    \foreach \i in {0,...,\n} %Draw nodes
    	{\foreach \j in {0,...,\m}{
      	\node[draw,circle] (A\i\j) at (\i*\width,-\j*\width) {};
    }}
    \pgfmathparse{int(\n-1)}
    \foreach \i in {0,...,\pgfmathresult}%half-graph edges
    	{\foreach \j in {0,...,\m}{
        	\pgfmathparse{int(\i+1)}\edef\next{\pgfmathresult}
            \draw (A\i\j) -- (A\next\j);
    }}
    \pgfmathparse{int(\m-1)}
    \foreach \i in {0,...,\n}%half-graph edges
    	{\foreach \j in {0,...,\pgfmathresult}{
        	\pgfmathparse{int(\j+1)}\edef\next{\pgfmathresult}
            \draw  (A\i\j) -- (A\i\next);
    }}
    %dotted lines
    \foreach \j in {0,...,\m}{
      	\draw[thick, dotted] (A0\j) --++ (-.4*\width, 0);
      	\draw[thick, dotted] (A2\j) --++ (.4*\width, 0);
    }
    \draw[thick, dotted] (A20) --++ (.3*\width, .3*\width);
    \draw[thick, dotted] (A23) --++ (.3*\width, -.3*\width);
    \draw[thick, dotted] (A00) --++ (-.3*\width, .3*\width);
    \draw[thick, dotted] (A03) --++ (-.3*\width, -.3*\width);
    %other arcs
    \draw (A10) to [bend right=35] (A13);
    \node[draw=none, fill=none, rectangle, label=below:{$c+1$}] at (A13) {};
    \end{tikzpicture}\hskip 2cm
\begin{tikzpicture}[scale=1, every node/.style={MyNode}]
    \def \width {1.2cm}
    \node[draw=none, fill=none] at (0,.3*\width) {};
    \node[draw=none, fill=none, rectangle, label=below:{\textcolor{white}{$c+1$}}] at (0, -3*\width) {};
    \node[rectangle, draw=none, fill=none] (center) at (0, -1.5*\width) {\Large$\longrightarrow$};
\end{tikzpicture}\hskip 2cm
    \begin{tikzpicture}[scale=1, every node/.style={MyNode, inner sep=1.7}]
    \def \m {3}
    \def \n {2}
    \def \width {1.2cm}
    \foreach \i in {0, 2} %Draw nodes
    	{\foreach \j in {0,...,\m}{
      	\node[draw,circle] (A\i\j) at (\i*\width,-\j*\width) {};
    }}
    \node[draw,circle] (A10) at (1*\width,-0*\width) {};
    \node[draw=none, fill=none, circle] (A11) at (1*\width,-1*\width) {};
    \node[draw=none, fill=none, circle] (A12) at (1*\width,-2*\width) {};
    \node[draw=none, fill=none, circle] (A13) at (1*\width,-3*\width) {};
    \pgfmathparse{int(\n-1)}
    \foreach \i in {0,...,\pgfmathresult}%horizontal edges
    	{\foreach \j in {0}{
        	\pgfmathparse{int(\i+1)}\edef\next{\pgfmathresult}
            \draw (A\i\j) -- (A\next\j);
    }}
    \pgfmathparse{int(\m-1)}
    \foreach \i in {0, 2}%vertical edges
    	{\foreach \j in {0,...,\pgfmathresult}{
        	\pgfmathparse{int(\j+1)}\edef\next{\pgfmathresult}
            \draw  (A\i\j) -- (A\i\next);
    }}
    %fill in holes
    \draw  (A01) -- (A21);
    \draw  (A02) -- (A22);
    \draw  (A03) -- (A23);
    %dotted lines
    \foreach \j in {0,...,\m}{
      	\draw[thick, dotted] (A0\j) --++ (-.4*\width, 0);
      	\draw[thick, dotted] (A2\j) --++ (.4*\width, 0);
    }
    \draw[thick, dotted] (A20) --++ (.3*\width, .3*\width);
    \draw[thick, dotted] (A23) --++ (.3*\width, -.3*\width);
    \draw[thick, dotted] (A00) --++ (-.3*\width, .3*\width);
    \draw[thick, dotted] (A03) --++ (-.3*\width, -.3*\width);
    %other arcs
    \node[draw=none, fill=none, rectangle, label=below:{$c+1$}] at (A13) {};
    \draw (A10) to [out=240,in=300,looseness=40] (A10);
    \end{tikzpicture}
\caption{Combining two columns, before removing the final vertex.}
\label{fig:combiningColumns}
\end{figure}

We will say how to remove each vertex of $\mathcal{H}$ one at a time by specifying which incident edges to ``combine''. Suppose that we wish to ``combine'' columns $c$ and $c+1$ of $(H, \mu)$. The vertices on column $c+1$ are incident to one face on column $c$ and one face on column $c+1$; these are the vertices that we will remove. So, first of all, remove every vertex on column $c+1$ except for the vertex on the first row by ``combining'' the two incident horizontal edges as in \Cref{fig:combiningColumns}. By Claim~\ref{clm:controlled1}, the final remaining vertex on column $c+1$ has a loop with $\mu_A$-weight zero. Thus any way of removing this vertex results in an $A$-rich decorated graph of the same richness and one smaller width. Moreover, since $f_{\ref{prop:universality}}(t, a,s)\geq 1$, none of the faces of $\hat{H}$ contain boundary edges of $H$; so the weights of the faces of $\hat{H}$ sum.

Let us update the names of these graphs $\mathcal{H}$ and $\hat{H}$ as we combine columns according to $w_0, z_1, \ldots, z_{f^{b-1}_{\ref{lem:RamseyWords}}(m,k)}, w_{f^{b-1}_{\ref{lem:RamseyWords}}(m,k)+1}$. We do not lose any richness as we ``combine columns'' since the batches use horizontal outside edges; so in the end we still have richness at least $f_{\ref{prop:universality}}(t, a,s)$ and are $S$-saturated. Furthermore, in the end, every face of $\hat{H}$ that is not on the first or last column has $\mu_v$-weight zero. Let $\hat{H}'$ be the subgrid obtained from $\hat{H}$ by deleting every vertex on the first or last column. Then $\hat{H}'$ is an $(m+1) \times (f^{b-1}_{\ref{lem:RamseyWords}}(m,k) + 1)$ subgrid of $H$ so that $V(\hat{H}')$ has distance at least $f_{\ref{prop:universality}}(t, a,s)$ from the boundary vertices of $H$.

Now, we repeat the same ``combining columns'' process for all $v \in B$, recursively on $\hat{H}'$. Suppose at a given iteration, we have ``combined columns'' for a set $B' \subseteq B$ and are considering a new $v \in B \setminus B'$. See that on this iteration, since we combine columns of the current $\hat{H}'$, whose faces have $\mu_u$-weight zero for all $u \in B'$, the faces of the new subgrid have $\mu_u$-weight zero for all $u \in B' \cup \{v\}$. So, after applying this process for all of $B$, we end up with an $(m+1) \times (k+1)$ subgrid $\hat{H}'$ of $H$ so that $V(\hat{H}')$ has distance at least $f_{\ref{prop:universality}}(t, a, s)$ from the boundary vertices of $H$, and all faces of the subgrid have $\mu_B$-weight zero.

So, since $t = (m+1)(k+1)$, we can apply \Cref{prop:universality} on universality to an $A$-rich decorated graph with underlying grid $\hat{H}'$.

So, since $k+1 = (a+s+1)m2^{(a+s)b}+1 > (a+s)m2^{(a+s)b}$, $\mathcal{H}$ fully immerses a decorated graph $\widetilde{\mathcal{H}}$ with labeling $\tilde{\mu}$ so that\begin{enumerate}
    \item $\widetilde{H}$ is $A$-rich and $S$-saturated with $m2^{(a+s)b}$ batches,
    \item each batch consists of outside-vertical edges on consecutive columns,
    \item $\hat{H}'$ is the underlying subgrid of $\widetilde{H}$, and
    \item each batch has the same set of $\widetilde{\mu}_A$-weights.
\end{enumerate}Furthermore, by applying \Cref{lem:facesEnough} and shifting $\widetilde{\mu}_v$ for each $v \in B$, we may assume that each edge of the underlying grid of $\hat{H}'$ has $\widetilde{\mu}_B$-weight zero. Now by the pigeonhole principle, there exist $m$ batches which each have the same set of $\widetilde{\mu}_A\times \widetilde{\mu}_B$-weights. 

Let $N \subseteq \mathbb{F}_2^A \times \mathbb{F}_2^B$ be the subspace generated by the $\widetilde{\mu}_A \times \widetilde{\mu}_B$-weights of any one of these $m$ batches. Notice that $\dim(N) \ge |A|$, and $N$ restricted to coordinates in $A$ spans $\mathbb{F}_2^A$, so there exists an $A' \supseteq A$ such that $\widetilde{H}$ is $A'$-rich, $\dim(N) = |A'|$, and each batch has the same set of $\widetilde{\mu}_{A'}$-weights. Let $B' = B \setminus (A' \setminus A)$. Then $\widetilde{\mathcal{H}}$ is an $(A',B')$-clean $S$-saturated decorated graph with subspace $N$ and height, width, and richness at least $m$ fully immersed by $\mathcal{G}$.

\emph{Running time.} Only the construction of $\mathcal{H}$ in Claim~\ref{clm:controlled1} and the final composition of models involve $G$; both take $\Oh(|E(G)|) = \Oh(n)$ time, since the trails of a full immersion model partition $E(G)$. Every other step operates on $\mathcal{H}$, and their runtimes only depend on $|\mathcal{H}|, |U|, |S|$. Note that $|\mathcal{H}|$ is bounded by a function of $m, |U|, |S|$. So $\widetilde{\mathcal{H}}$ and its model in $\mathcal{G}$ can be found in time $\Oh_{m, |U|, |S|}(n)$.

This completes the proof of \Cref{prop:controlled}.
\end{proof}

\section{Enriching and saturating the grid or finding a small cut}\label{sec:enrich}
Informally, this section gives two very similar algorithms that try to either \begin{enumerate}
    \item ``enrich'' a grid immersion with another signature $v \in U$ (\Cref{prop:main}), or
    \item further saturate a grid immersion with another color $s \in [q]$ (\Cref{prop:main2}).
\end{enumerate} 
And if this is not possible, both algorithms return a a small cut separating all ``problematic'' edges, where problematic either refers to edge weights or half-edge colors.

Both algorithms primarily rely on the algorithm given by \Cref{thm:mainMinMax}, that tests in $\Oh(k|E(G)|^4)$ time if a free graph $G$ contains a flooding of size $k$. To use this flooding algorithm for rerouting in a grid immersion, the following two lemmas will be helpful, so we state and prove them here before using them in both algorithms.

\begin{lemma}[Rerouting Lemma]
\label{lem:splittingOffFlooding}
If $(G, \Sigma, F, t)$ is a $4$-edge-connected Eulerian free graph with $\nu(G, \Sigma, F, t) \geq \ell$ and $|\delta(t)|/2>\ell+1$, then there exists a vertex $x \neq t$ and a half-edge $h$ incident to $x$ so that every transition at $x$ which includes $h$ is a transition of a flooding of value at least $\ell$. Further, given a flooding of $(G, \Sigma, F, t)$ of value at least $\ell$, we can find this vertex $x$ in $\Oh(n)$ time. 
\end{lemma}
\begin{proof}
Let $\mathcal{C}$ be a flooding of $(G, \Sigma, F, t)$ of value at least $\ell$. Since $|\delta(t)|/2>\ell+1$, there are more than $\ell+1$ circuits in $\mathcal{C}$ with more than one edge. Choose such a circuit $C \in \mathcal{C}$; if possible, choose $C$ to be an infeasible circuit. Scanning through the circuits takes time at most $\Oh(|\delta(t)|) = \Oh(n)$. Let $x \neq t$ be a vertex which is incident to a half-edge $h$ in the first arc of $C$ (thus, informally, $x$ is the first vertex after $t$ which is ``hit'' by $C$). Let $h'$ denote the half-edge so that $\{h,h'\}$ is a transition of $C$. We consider each transition $\{h_1, h_2\} \neq \{h,h'\}$ of $\mathcal{C}$ at $x$ one at a time. To prove the lemma, we need to show that both of the transitions $\{h, h_1\}$ and $\{h, h_2\}$ work. We break into two cases. 

\smallskip
\noindent\emph{Case 1:} The transition $\{h_1, h_2\}$ is a transition of $C$.
\smallskip

Then there are trails $T_1, L, T_2$ so that $C=(T_1, L, T_2)$ and exactly one half-edge from each of $\{h_1, h_2\}$ and $\{h,h'\}$ is contained in $L$, which is an $x$-circuit. Replacing $C$ by $(T_1, L^{-1}, T_2)$ does not change the number of feasible circuits in the flooding. Furthermore, since $G$ is $4$-edge-connected, it is possible to ``move $L$ off of $C$'' and ``onto'' a transition of $\mathcal{C}$ which does not include any of $h, h', h_1, h_2$. The number of feasible circuits in this new flooding either stays the same or goes down by at most~$2$. Furthermore, it can only go down by~$2$ if $C$ is feasible, in which case there were at least $\ell+2$ feasible circuits in $\mathcal{C}$ to begin with. So in this case, both $\{h, h_1\}$ and $\{h, h_2\}$ work.

\smallskip
\noindent\emph{Case 2:} The transition $\{h_1, h_2\}$ is a transition of a circuit $C' \in \mathcal{C}-\{C\}$.
\smallskip
    
Then there are trails $T_1, T_2, R_1, R_2$ so that $C=(T_1, T_2)$, $C'=(R_1, R_2)$, exactly one half-edge from $\{h,h'\}$ is in $T_1$, and exactly one half-edge from $\{h_1, h_2\}$ is in $R_1$. Again we have two new floodings to consider; we can replace $C$ and $C'$ by either $(T_1, R_2)$ and $(R_1, T_2)$, or by $(T_1, R_1^{-1})$ and $(R_2^{-1}, T_2)$. Again, the number of feasible circuits either stays the same or goes down by at most~$2$; the latter case can only occur when both $C$ and $C'$ are feasible. So again, both $\{h, h_1\}$ and $\{h, h_2\}$ work.

In each of the cases, both $\{h, h_1\}$ and $\{h, h_2\}$ work, and so \Cref{lem:splittingOffFlooding} holds.
\end{proof}

Since we will obtain a free graph by identifying vertices in a grid to a single vertex, we will need the following:
\begin{lemma}
\label{lem:reroute}
    Suppose $\mathcal{H}, \mathcal{G}$ are decorated graphs such that $G$ fully immerses $H$, and $\phi$ is a full immersion model of $\mathcal{H}$ in $\mathcal{G}$. Consider a free graph $(\hat{G}, \hat{\Sigma}, \hat{F}, b)$ obtained from $G$ by identifying the branch vertices of $H$ to a single vertex $b$, with an arbitrary signature $\hat{\Sigma}$ and set of free edges $\hat{F}$. If $\nu(\hat{G}, \hat{\Sigma}, \hat{F}, b) \ge \ell$ and more than $\ell+1$ edges of $H$ are not edges of $G$, then we can split off a vertex of $G$ so that the trails of $H$ are maintained and the flooding number of the resulting free graph is still at least $\ell$. Further, given a flooding of $(\hat{G}, \hat{\Sigma}, \hat{F}, b)$ of value at least $\ell$, the resulting graph and a flooding of value at least $\ell$ of its free graph can be found in $\Oh(n^2)$ time.
\end{lemma}

\begin{proof}
    Suppose more than $\ell+1$ edges of $H$ are not edges of $G$. This gives us that $|\delta_{\hat{G}}(b)|/2 > \ell + 1$. Then, if $G$ and therefore $(\hat{G}, \hat{\Sigma}, \hat{F}, b)$ are 4-edge-connected, we may apply the Rerouting Lemma. This gives us a vertex $x \neq b$ such that any way we split off at $x$ maintains that the flooding number of the resulting free graph is at least $\ell$. 
    Hence, we can split off at $x$ such that the trails of $H$ are preserved, and the resulting free graph has flooding number at least $\ell$. This splitting-off process gives rise to a new graph $G'$ with a new labeling $\lambda'$ and coloring $\alpha'$. We see that, if 4-edge-connectivity is preserved, we may continue this splitting off process until at most $\ell+1$ edges of $H$ are not edges of $G'$. If some splitting-off operation makes the resulting graph $G'$ not 4-edge-connected, we may correct this in the following way.

    If at some step $G'$ is not 4-edge-connected, there must be a 2-edge cut partitioning the vertex set of $G$, $(X, \bar{X})$. Since $H$ is a grid-like graph, $V(H)$ must lie in one side of the cut. Without loss of generality, $V(H) \subseteq \bar{X}$. Since the size of the cut is 2, there must be a subtrail of $H$ that uses both edges in the cut and traverses all edges in $E(X)$. Splitting off vertices along this subtrail gets rid of the 2-edge cut, while maintaining $\mathcal{H}$ as a full immersion of $\mathcal{G}$. Further, the associated free graph has the same flooding number as before splitting off along this subtrail.

    So, we can continue to split off in $G$ to obtain a new graph and labeling $G', \lambda'$, such that at most $\ell+1$ edges of $H$ are not edges of $G'$, and the flooding number of the associated free graph is at least $\ell$. 

    For the runtime, there are at most $n$ splitting-off steps. In each, the vertex $x$ is found in $\Oh(n)$ time by \Cref{lem:splittingOffFlooding}, and the flooding of value at least $\ell$ after splitting off is obtained from the current one in $\Oh(n)$ time by the construction in its proof. A 2-edge cut can be detected in $\Oh(n)$ time~\cite{cactus}, and splitting off along the subtrail carries the flooding over, since exactly one circuit crosses the cut and its segment inside $X$ can be replaced by that subtrail, which has the same edge set and ends.
    \end{proof}

Finally, before showing both algorithms, we need the following two lemmas for the case when we cannot enrich or further saturate a given grid immersion. Roughly, this will happen when the flooding number of an associated free graph is small, and in which case we show that there exists a ``small cut'' separating ``problematic edges''.

\begin{lemma}[Small signature-separating cut]
\label{lem:smallcut}
    If $(G, \Sigma, F, t)$ is a $4$-edge-connected Eulerian free graph with $\nu(G, \Sigma, F, t) \le \ell$, then there exists a shifting $\Sigma'$ of $\Sigma$ and a set of vertices $X$ containing $t$ such that \begin{enumerate}
        \item there are at most $4\ell$ edges in $\delta(X)$, and
        \item there are at most $\ell$ edges with both ends in $X$ that are free or in $\Sigma'$ (that is, $|E_{F,\Sigma'}(X)| \le \ell$).
    \end{enumerate}
    Further, $(X, \Sigma')$ are given by the certificate of value at most $\ell$ returned by the algorithm of \Cref{thm:mainMinMax}.
\end{lemma}
\begin{proof}
    Let $(G, \Sigma, F, t)$ be a 4-edge-connected Eulerian free graph with $\nu(G, \Sigma, F, t) \le \ell$ and certificate $(X, \Sigma')$ returned by the algorithm from \Cref{thm:mainMinMax}. Recall that 
    \begin{align}
        \val(X, \Sigma') = |E_{F, \Sigma'}(X)|+|\delta(X)|/2-odd_{\Sigma'}(G-X)
    \end{align}
    Then, by Theorem $\ref{thm:mainMinMax}$, the value of $(G, \Sigma, F, t)$'s certificate is equal to the maximum size of a flooding in  $(G, \Sigma, F, t)$, so:
    \begin{align}
        \nu(G, \Sigma, F, t) = |E_{F, \Sigma'}(X)|+|\delta(X)|/2-odd_{\Sigma'}(G-X)
    \end{align}
    Then since $G$ is 4-edge-connected, $odd_{\Sigma'}(G-X) \le \frac{1}{4}|\delta(X)|$. This gives:
    \begin{align}
        \ell \ge \nu(G, \Sigma, F, t) \ge |E_{F, \Sigma'}(X)|+\frac{1}{4}|\delta(X)|
    \end{align}
    which gives the result.
\end{proof}

Then, the following will be used to show that the side of the ``small cut found with problematic edges'' is also ``small''.
\begin{lemma}[Cuts in a grid immersion]
\label{lem:connGrid}
    Let $\mathcal{G} = (G, \lambda, R_G, \rho_G, \alpha)$ and $\mathcal{H} = (H, \mu, R_H, \rho_H, \beta)$ be two decorated graphs such that $(\mathcal{H}, \phi)$ is a grid in $\mathcal{G}$, and $H$ has height and width at least $n$. Let $T = \phi(V(H))$; the set of $H$'s branch vertices in $G$. Then, for any $X \subseteq V(G)$ such that $|\delta(X)| < n$, either $X \cap T$ or $(V(G)-X) \cap T$ has size at most $|\delta(X)|^2$.
\end{lemma}
\begin{proof}
    Each row and each column of $\mathcal{H}$ that contains a vertex in $T$ on each side of the cut contributes at least one to $|\delta(X)|$. So by symmetry between the two sides, we can assume that there is a column with no branch vertices in $X$. Then there is also a row with no branch vertices in $X$. So $X \cap T$ is contained in at most $|\delta(X)|$ rows and columns, and $|X \cap T| \leq |\delta(X)|^2$, as desired. 
\end{proof}

\subsection{Enriching the grid}\label{sec:enrich grid}

We start with an algorithm that, given a full immersion model of an $(A,B)$-clean decorated graph $\mathcal{H}$ in $\mathcal{G}$, either ``enriches'' the full immersion with another signature $v \notin A$, or returns a small cut separating all edges with ``problematic'' $\lambda_A \times \lambda_v$-weights. We will use this algorithm iteratively for each signature that a given decorated graph $\mathcal{G}$ is labeled with, to get a grid-like immersion that is as rich as possible, and a small cut that informally separates all labels with which we cannot ``enrich'' the grid-like immersion.

\begin{proposition}
\label{prop:main}
There exist functions $f_{\ref{prop:main}}, g_{\ref{prop:main}}:\N^4 \rightarrow \Zplus$, non-decreasing in every argument, such that the following holds for any $m \in \mathbb{Z}^+$. Let $\mathcal{G} = (G, \lambda, R_G, \rho_G, \alpha)$ be a 4-edge-connected decorated graph, and let $(\mathcal{H}, \phi)$ be an $(A,B)$-clean $S$-saturated grid in $\mathcal{G}$ of richness, height, and width at least $f_{\ref{prop:main}}(m, |A|, |B|, q)$, for some disjoint $A,B \subseteq U$, and $S \subseteq [q]$. Let $N$ be $\mathcal{H}$'s subspace. Then, for any $v \in B$, either\begin{enumerate}
    \item there is an $(A', B')$-clean $S$-saturated grid in $\mathcal{G}$ of richness, height, and width at least $m$, for some $A' \supseteq A \cup \{v\}$ with $B' = B \setminus (A' \setminus A)$, or
    \item there exists an $X \subseteq V(G)$ such that $|\delta(X)| \le g_{\ref{prop:main}}(m, |A|, |B|, q)$, $|X \cap \phi(V(H))| \le g_{\ref{prop:main}}(m, |A|, |B|, q)$, and a shifting $\lambda_v'$ of $\lambda_v$ such that $\{e \in E(G): \lambda_A(e) \times \lambda_v'(e) \notin N_v\} \subseteq E(X) \cup \delta(X)$. \end{enumerate}
Further, we can find the trails of the immersion, or the set $X$ and $\lambda'_v$, in time $\Oh_{m, |U|, q}(n^4)$.

\end{proposition}
\begin{proof}
Let $m \in \mathbb{Z}^+$, and define:
$$p = f_{\ref{prop:controlled}}(m, |A|+1, |B|, q)$$
$$\ell := 100(p-1)^2+5(p-1)+2$$
% $$\ell' = t|U|(144(\ell-1)^2 + 12(\ell -1) + 4|R_F|+1)$$
$$f_{\ref{prop:main}}(m,|A|, |B|, q) := 5p(\ell+2)$$
$$g_{\ref{prop:main}}(m, |A|, |B|, q) := 8(\ell-1)$$

Since $f_{\ref{prop:controlled}}$ is non-decreasing in every argument, so are $p$, $\ell$, $f_{\ref{prop:main}}$ and $g_{\ref{prop:main}}$; this is used in \Cref{prop:allsignatures}.

The proof will be split into two cases based on the flooding number of an associated free graph. First, for the subspace $N$ and $v$, consider the projection $\pi_v: \mathbb{F}_2^A \times \mathbb{F}_2^B \rightarrow \mathbb{F}_2^A \times \mathbb{F}_2^v$ of the $\lambda_A\times \lambda_B$-weights of each batch onto the $\lambda_A\times \lambda_v$-weights. Recall this is denoted by $N_v$.

Consider the free graph $(\hat{G}, \hat{\Sigma}, \emptyset, b)$ that is obtained from $G$ by identifying all branch vertices of $H$ to a single new vertex $b$, with no free edges and with signature $\hat{\Sigma} := \{e \in E(G) : \lambda_A(e) \times \lambda_v(e) \notin N_v\}$. Note that by \Cref{thm:mainMinMax}, we can compute whether the flooding number of $(\hat{G}, \hat{\Sigma}, \emptyset, b)$ is at least $\ell$, and find a flooding of value at least $\ell$ if so, in $\Oh(\ell|E(\hat{G})|^4)$ time.

\bigskip
\noindent\emph{Case 1:} The flooding number of $(\hat{G}, \hat{\Sigma}, \emptyset, b)$ is at least $\ell$.
\bigskip

Broadly, we want to add each of these circuits to a batch of $\mathcal{H}$ as an outer edge, which would make $H$ an $A \cup \{v\}$-rich grid with sufficient richness. If we maintain that our grid immersion is $(A \cup \{v\}, B \setminus \{v\})$-clean and $S$-saturated, we would be done. However, we must ensure that we have enough circuits that avoid branch trails of $H$'s inner grid, which we want to preserve. For this, by \Cref{lem:reroute}, by splitting off vertices in $G$ along the trails given by $\phi$ we obtain a graph $G'$, with labeling $\lambda'$, such that at most $\ell+1$ edges of $H$ are not edges of $G'$, and the free graph obtained from $G'$ as above still has flooding number at least $\ell$. Since $\mathcal{G}$ fully immerses $(G', \lambda', \ldots)$, it suffices to work in $G'$, and from now on all free graphs in Case 1 are obtained from $G'$. Then, by a pigeonhole argument, there is a large grid in $\mathcal{G}$ with only outside-horizontal edges possibly not edges of $G'$:

    \begin{observation}\label{obs:untouchedsubgrid}
        $\mathcal{G}$ fully immerses an $(A,B)$-clean $S$-saturated decorated graph $\mathcal{\hat{H}}$ of richness and width at least $5p$, and height equal to that of $\mathcal{H}$, such that every edge of $\mathcal{\hat{H}}$ that is not outside-horizontal is also an edge of $G'$.
    \end{observation}
\begin{proof}
    Since $\mathcal{H}$ is $(A,B)$-clean, there are $5p(\ell+2)$ disjoint subgrids of $\mathcal{H}$ which each have the same height as $\mathcal{H}$ and contain the ends of a batch of $\mathcal{H}$. Then we can ``break up $\mathcal{H}$'' into $\ell+2$ disjoint grids, each of which contains $5p$ of those subgrids and $5p$ batches made up of vertical edges. By \Cref{lem:reroute}, the edge set of one of those grids and its vertical outside edges are edges of $G'$. Call this grid $\hat{H}$. Then, to get a full immersion of an $(A,B)$-clean $S$-saturated decorated graph in $\mathcal{G}$ with underlying grid $\hat{H}$ and height, width, richness at least $5p$, we can use the vertical outside edges incident to $\hat{H}$. So, it is left to find trails for the outside horizontal edges of $\hat{H}$ which together use every remaining edge exactly once. We can route these trails as in \Cref{fig:growingA3}. Let $\mathcal{\hat{H}}$ be the decorated graph given by this full immersion. So, $\mathcal{G}$ fully immerses the $(A,B)$-clean $S$-saturated decorated graph $\mathcal{\hat{H}}$ of richness and width at least $5p$.
    \end{proof}

        \begin{figure}
    \centering
    \begin{tikzpicture}[scale=1, every node/.style={MyNode, inner sep=1}]
    \def \m {3}
    \def \n {6}
    \def \width {1cm}
    %Draw nodes
    \foreach \i in {0,...,\n} 
    	{\foreach \j in {0,...,\m}{
      	\node[draw,circle] (A\i\j) at (\i*\width,-\j*\width) {};
    }}
    %Draw large nodes
    \foreach \i in {2,...,4}
    	{\foreach \j in {0,...,\m}{
      	\node[draw,circle, inner sep=1.7] (A\i\j) at (\i*\width,-\j*\width) {};
    }}
    \pgfmathparse{int(\n-1)}
    \foreach \i in {0,...,\pgfmathresult}%horizontal edges
    	{\pgfmathparse{int(\i+1)}\edef\next{\pgfmathresult}
    	\foreach \j in {1,...,\m}{
            \draw (A\i\j) -- (A\next\j);
    }
            \draw[red, thick, dash pattern=on \pgflinewidth off 1pt] (A\i0) -- (A\next0);
    }
    \pgfmathparse{int(\m-1)}
    \foreach \i in {0,...,\n}%vertical edges
    	{\foreach \j in {0,...,\pgfmathresult}{
        	\pgfmathparse{int(\j+1)}\edef\next{\pgfmathresult}
            \draw[red, thick, dash pattern=on \pgflinewidth off 1pt] (A\i\j) -- (A\i\next);
    }}
    %Subgrid edges
    \foreach \i in {2,3}%horizontal edges
    	{\foreach \j in {0,...,\m}{
        	\pgfmathparse{int(\i+1)}\edef\next{\pgfmathresult}
            \draw[ultra thick] (A\i\j) -- (A\next\j);
    }}
    \foreach \i in {2,3,4}%vertical edges
    	{\foreach \j in {0,...,2}{
        	\pgfmathparse{int(\j+1)}\edef\next{\pgfmathresult}
            \draw[ultra thick] (A\i\j) -- (A\i\next);
    }}
    \foreach \i in {0,1,5,6}%outside vertical edges
    	{
    	    \draw[red, thick, dash pattern=on \pgflinewidth off 1pt] (A\i\m) to [bend left=22] (A\i0);
    	}
    \foreach \i in {2,3,4}%outside vertical edges
    	{
    	    \draw[ultra thick] (A\i\m) to [bend left=22] (A\i0);
    	}
    \foreach \j in {1,...,\m}%outside horizontal edges
    	{
    	    \draw (A\n\j) to [bend left=10] (A0\j);
    	}
    \draw[red, thick, dash pattern=on \pgflinewidth off 1pt] (A\n0) to [bend left=10] (A00);
    \end{tikzpicture}
\caption{The subgraph of $H$ induced by $V(\hat{H})$ (bold) and a depiction of how to find the full immersion (the red dotted edges are the edge-set of a single trail).}
\label{fig:growingA3}
\end{figure}
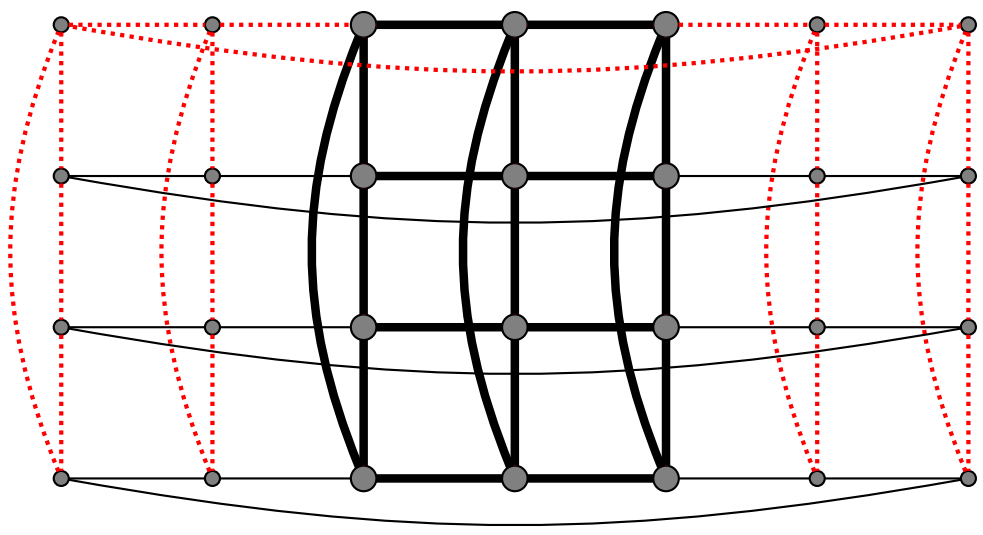
    Let $(\hat{G}', \hat{\Sigma}', \emptyset, \hat{b})$ be the free graph obtained from $G'$ by identifying the branch vertices of $\hat{H}$ to a single vertex $\hat{b}$, with no free edges and with signature $\hat{\Sigma}' := \{e \in E(G') : \lambda'_A(e) \times \lambda'_v(e) \notin N_v\}$. Roughly, if the flooding number of $(\hat{G}', \hat{\Sigma}', \emptyset, \hat{b})$ is at least $p$, then we can see that these $p$ circuits can be paired with $p$ batches made up of vertical outside edges of $\hat{H}$ to obtain an $A \cup \{v\}$-rich, $S$-saturated decorated graph as a full immersion of $\mathcal{G}$ with the same underlying grid as $\hat{H}$. Formally,
    \begin{lemma}\label{lem:circuit-enrich}
        If the flooding number of $(\hat{G}', \hat{\Sigma}', \emptyset, \hat{b})$ is at least $p$, then $\mathcal{G}$ fully immerses an $A \cup \{v\}$-rich $S$-saturated decorated graph of height, width and richness at least $p$.
    \end{lemma}
    \begin{proof}
        By \Cref{obs:untouchedsubgrid}, only the outside-horizontal edges of $\hat{H}$ are not edges of $G'$. So in any flooding of $(\hat{G}', \hat{\Sigma}', \emptyset, \hat{b})$, every circuit is disjoint from the underlying grid and vertical-outside edges of $\hat{H}$. Then, since $G'$ is 4-regular, a flooding of this free graph is equivalent to an alternative set of trails of horizontal outside edges for $\hat{H}$. Replacing $\hat{H}$'s set of horizontal outside edges with the ones provided by an optimal flooding gives at least $p$ horizontal outside edges $e$ such that $\lambda_A'(e) \times \lambda_v'(e) \notin N_v$. Since $\hat{H}$ has at least $p$ batches made up of vertical outside edges, we can add one horizontal outside edge $e$ with $\lambda_A'(e) \times \lambda_v'(e) \notin N_v$, that is, an odd circuit, to each of these batches.
        \begin{claim}
            The updated batches generate $\mathbb{F}_2^{A \cup \{v\}}$.
        \end{claim}
        \begin{proof}
            To see this, recall that the set of $\lambda_A'(e) \times \lambda_v'(e)$-weights of the edge set of each batch before the update spans $N_v$ and is in $N_v$, with $\dim(N_v)=|A|$. Since for each batch, the added edge's $\lambda_A'(e) \times \lambda_v'(e)$-weight is not in $N_v$, the dimension of the new subspace spanned by each batch's $\lambda_A'(e) \times \lambda_v'(e)$-weights increases by one to $|A|+1$. So, since $\dim(\mathbb{F}_2^{A \cup \{v\}}) = |A|+1$, each batch's $\lambda'_A \times \lambda'_v$ weights generate $\mathbb{F}_2^{A \cup \{v\}}$.
        \end{proof}
        Hence, with this outside-edge replacement, $\hat{H}$ has at least $p$ batches such that each batch's $\lambda'_A \times \lambda'_v$ weights generate $\mathbb{F}_2^{A \cup \{v\}}$. Let $\mathcal{H'}$ be the decorated graph given by this replacement of trails of $\mathcal{\hat{H}}$. Clearly, $\mathcal{H'}$ is $A \cup \{v\}$-rich and $S$-saturated with richness, height and width at least $p$, fully immersed by $\mathcal{G}$.
    \end{proof}
    
    Then, since $p = f_{\ref{prop:controlled}}(m, |A|+1, |B|, q) \ge f_{\ref{prop:controlled}}(m, |A \cup \{v\}|, |B \setminus \{v\}|, q)$, we can apply \Cref{prop:controlled} to find an $(A', B')$-clean $S$-saturated grid-like graph fully immersed by $\mathcal{G}$, of height, width, and richness at least $m$, where $A' \supseteq A \cup \{v\}$, and $B' = B \setminus (A' \setminus A)$.

    So, suppose for contradiction that the flooding number of $(\hat{G}', \hat{\Sigma}', \emptyset, \hat{b})$ is strictly less than $p$. Note that since $G'$ is 4-edge-connected, so is this free graph. So, we can apply \Cref{lem:smallcut} to obtain a subset $X \subseteq V(G')$ such that $X$ contains $V(\hat{H})$, and:
    \begin{enumerate}
        \item there are at most $4(p-1)$ edges of $G'$ with exactly one end in $X$, and
        \item there exists $V(\hat{H}) \subseteq X' \subseteq X$ such that there are at most $p-1$ edges with both ends in $X$ that either have $\lambda_A' \times \lambda'_v$-weight in $N_v$ and are in $\delta(X')$, or are not in $\delta(X')$ and have ${\lambda_A' \times \lambda'_v}$-weight not in $N_v$. 
    \end{enumerate}
Let an edge satisfying either of these two conditions be called $N_v$-bad. Hence, there are at most $5(p-1)$ $N_v$-bad edges. We will obtain a contradiction to the fact that the flooding number of the free graph obtained from $G'$ by identifying the branch vertices of $H$ (with the signature defined as above) is at least $\ell$. We need the following claim:

\begin{claim}
\label{clm:growingA4}
All but at most $25(p-1)^2$ vertices of $H$ are in $X'$.
\end{claim}
\begin{proof}
    Fix the full immersion model of $\mathcal{H}$ in $G'$. Let $Y$ be the set of all $y \in V(H)$ so that there is a path from $y$ to $V(\hat{H})$ in the underlying grid of $H$ whose corresponding trails, in $G'$, do not use any $N_v$-bad edge. Thus there is a trail from $y$ to $V(\hat{H})$ in $G'$ which does not use any $N_v$-bad edge and has ${\lambda_A' \times \lambda'_v}$-weight in $N_v$, since every edge of the underlying grid of $H$ has $\lambda_A \times \lambda_B$-weight zero. An edge of the trail has ${\lambda_A' \times \lambda'_v}$-weight not in $N_v$ if and only if it is in $\delta(X')$. Since $\dim(N_v) = |A|$, the quotient $(\mathbb{F}_2^A \times \mathbb{F}_2^v)/N_v$ is isomorphic to $\mathbb{F}_2$, so a sum of vectors lies in $N_v$ if and only if an even number of them lie outside $N_v$. Hence the trail crosses $\delta(X')$ an even number of times, and since $V(\hat{H}) \subseteq X'$, we have $y \in X'$. So in fact $Y \subseteq X'$.
    
    We complete the proof by showing that $|V(H)-Y| \leq 25(p-1)^2$. Observe that $H$ and $\hat{H}$ have height and width at least $5p >5(p-1)$, and the edge-boundary of $Y$ within the underlying grid of $H$ has size at most $5(p-1)$ (each edge of this boundary corresponds to a branch trail using an $N_v$-bad edge, each $N_v$-bad edge lies on at most one branch trail, and there are at most $5(p-1)$ $N_v$-bad edges). So by \Cref{lem:connGrid} about cuts in a grid, either $Y$ or $V(H)-Y$ has size at most $25(p-1)^2$. So the claim follows since $|Y|\geq |V(\hat{H})|\geq 25p^2>25(p-1)^2$.
\end{proof}

Now consider a collection of trails of $G'$ which corresponds to an optimal flooding of this free graph. By Claim~\ref{clm:growingA4}, at most $100(p-1)^2$ of the trails have an end outside of $X'$. Each of the other trails which is odd must use an $N_v$-bad edge. So \begin{align*}
    100(p-1)^2+5(p-1)+2 = \ell \leq \nu(\hat{G}, \hat{\Sigma}, \emptyset, b) \leq 100(p-1)^2+5(p-1),
\end{align*} a contradiction. This completes the first case. 

\bigskip
\noindent\emph{Case 2:} The flooding number of $(\hat{G}, \hat{\Sigma}, \emptyset, b)$ is less than $\ell$.
\bigskip

Here, since $G$ and therefore $(\hat{G}, \hat{\Sigma}, \emptyset, b)$ is 4-edge-connected, we can apply \Cref{lem:smallcut}. 

First, however, consider what it means to shift the signature $\hat{\Sigma}$ of $(\hat{G}, \hat{\Sigma}, \emptyset, b)$, which is obtained from $\lambda$ and the subspace $N_v$. Recall that the subspace $N_v$ has dimension $|A|$, and that its dimension does not change upon restricting from $\mathbb{F}_2^A\times \mathbb{F}_2^v$ to $\mathbb{F}_2^A$. So the vector which is zero everywhere except for the $v$-entry is not in $N_v$. Thus, equivalently, we can just shift $\lambda_v$. 

So, \Cref{lem:smallcut} gives a shifting $\lambda^{'}_v$ of $\lambda_v$ and a set $V(H) \subseteq X \subseteq V(G)$ such that \begin{enumerate}
    \item there are at most $4(\ell-1)$ edges of $G$ with exactly one end in $X$, and
    \item there are at most $\ell-1$ edges of $G$ with both ends in $X$ that have $\lambda_A \times \lambda^{'}_v$-weight not in $N_v$.
\end{enumerate}

This gives the following easy corollary:
\begin{corollary}\label{cor:cleancut}
    There exists a set $X' \subseteq V(G)$ such that \begin{enumerate}
        \item there are at most $8(\ell-1)$ edges of $G$ with exactly one end in $X'$,
        \item at most $\ell-1$ branch vertices of $H$ are not in $X'$, and
    \item all edges with both ends in $X'$ have $\lambda_A \times\lambda^{'}_v$-weight in $N_v$.
    \end{enumerate}
\end{corollary}
\begin{proof}
    Let $F$ be the set of edges with both ends in $X$ and $\lambda_A \times\lambda^{'}_v$-weight not in $N_v$; so $|F| \le \ell - 1$. For each edge of $F$, we move one of its endpoints to $V(G)-X$, choosing an endpoint which is not a branch vertex of $H$ if possible, and let $X'$ be the resulting set. Then every edge with both ends in $X'$ has $\lambda_A \times\lambda^{'}_v$-weight in $N_v$. Since $X$ contains every branch vertex of $H$ and at most $\ell-1$ vertices are moved, at most $\ell-1$ branch vertices of $H$ are not in $X'$. Since $G$ is 4-regular, each vertex moved across the cut increases the number of edges in the cut by at most 4. So, since there are at most $\ell - 1$ edges in $F$ and $\delta(X)$ had starting size at most $4(\ell-1)$, we have $|\delta(X')| \le 4(\ell-1) + 4(\ell-1) = 8(\ell-1)$.
    \end{proof}

So, $\lambda'_v$ and the set $X'' := V(G) - X'$ complete the second case: $X''$ contains at most $\ell - 1 \le g_{\ref{prop:main}}(m,|A|,|B|,q)$ branch vertices of $H$, $|\delta(X'')| = |\delta(X')| \le 8(\ell-1) = g_{\ref{prop:main}}(m,|A|,|B|,q)$, and every edge with $\lambda_A \times \lambda'_v$-weight not in $N_v$ has an end in $X''$, so it lies in $E(X'') \cup \delta(X'')$.

\emph{Running time.} Since $G$ is 4-regular, $|E(\hat{G})| = 2n$, so by \Cref{thm:mainMinMax} we decide which case holds, and find a flooding of value at least $\ell$ or a certificate of value less than $\ell$, in $\Oh(\ell n^4)$ time. In Case~1, \Cref{lem:reroute} takes $\Oh(n^2)$ time, Observation~\ref{obs:untouchedsubgrid} and \Cref{lem:circuit-enrich} take $\Oh(n)$ time, \Cref{thm:mainMinMax} applied to $(\hat{G}', \hat{\Sigma}', \emptyset, \hat{b})$ with $k = p$ takes $\Oh(p n^4)$ time, and \Cref{prop:controlled} takes $\Oh_{m,|U|,q}(n)$ time. In Case~2, the certificate gives $X$ and $\lambda'_v$ as in \Cref{lem:smallcut}, and \Cref{cor:cleancut} takes $\Oh(n)$ time. Since $\ell$ and $p$ depend only on $m$, $|U|$ and $q$, the total is $\Oh_{m,|U|,q}(n^4)$.
\end{proof}

\subsection{Saturating the grid}
\label{sec:colors}

Similarly, we want to ``saturate'' our grid immersion with as many colors as possible, or otherwise find a small cut containing half-edges with ``problematic'' colors. This algorithm will follow the structure of \Cref{prop:main} almost exactly, except the free graph chosen is different, and since we are concerned with colors and not labelings, any reasoning about shiftings is dropped.

\begin{proposition}
\label{prop:main2}
The following holds for any $m \in \mathbb{Z}^+$. Let $\mathcal{G} = (G, \lambda, R_G, \rho_G, \alpha)$ be a 4-edge-connected decorated graph, and let $(\mathcal{H}, \phi)$ be an $(A,B)$-clean $S$-saturated grid in $\mathcal{G}$ of richness, height, and width at least $f_{\ref{prop:main}}(m,|A|,|B|,q)$, for some disjoint $A,B \subseteq U$, and $S \subseteq [q]$. Then, for any $i \in [q]\setminus S$, either\begin{enumerate}
    \item there is an $(A', B')$-clean $S \cup \{i\}$-saturated grid in $\mathcal{G}$ of richness, height, and width at least $m$ for some $A' \supseteq A$ and $B' = B \setminus (A' \setminus A)$, or
    \item there exists $X \subseteq V(G)$ such that $|\delta(X)| \le g_{\ref{prop:main}}(m,|A|,|B|,q)$, $|X \cap \phi(V(H))| \le g_{\ref{prop:main}}(m,|A|,|B|,q)$, and $\{e \in E(G): \alpha_i(e)=1\} \subseteq E(X) \cup \delta(X)$. \end{enumerate}
Further, we can find the trails of the immersion or the set $X$ in time $\Oh_{m, |U|, q}(n^4)$.
\end{proposition}
\begin{proof}
Let $m \in \mathbb{Z}^+$, and define, as in the proof of \Cref{prop:main}:
$$p = f_{\ref{prop:controlled}}(m, |A|+1, |B|, q)$$
$$\ell := 100(p-1)^2+5(p-1)+2$$

Recall that:
$$f_{\ref{prop:main}}(m,|A|, |B|, q) := 5p(\ell+2)$$
$$g_{\ref{prop:main}}(m, |A|, |B|, q) := 8(\ell-1)$$

The proof will be split into two cases based on the flooding number of the free graph $(\hat{G}, \emptyset, F_i, b)$ defined as follows. Let $i \in [q] \setminus S$, and let $(\hat{G}, \emptyset, F_i, b)$ be the free graph obtained from $G$ by identifying all branch vertices of $H$ to a single new vertex $b$, with empty signature and free edges $F_i \coloneqq \{e \in E(G) : \alpha_i(e) = 1\}$. So a circuit is feasible if and only if it contains an edge of color $i$. Note that by \Cref{thm:mainMinMax}, we can compute whether the flooding number of $(\hat{G}, \emptyset, F_i, b)$ is at least $\ell$, and find a flooding of value at least $\ell$ if so, in $\Oh(\ell|E(\hat{G})|^4)$ time.

\bigskip
\noindent\emph{Case 1:} The flooding number of $(\hat{G}, \emptyset, F_i, b)$ is at least $\ell$.
\bigskip

Broadly, we want to add each of these circuits to a batch of $\mathcal{H}$ as an outer edge, which would make $H$ an $A$-rich grid that is $S \cup \{i\}$-saturated with sufficient richness. If we maintain that our grid immersion is $A$-rich, we would be done. However, we must ensure that we have enough circuits that avoid branch trails of $H$'s inner grid, which we want to preserve. For this, by \Cref{lem:reroute}, by splitting off vertices in $G$ along the trails given by $\phi$ we obtain a graph $G'$, with labeling $\lambda'$ and coloring $\alpha'$, such that at most $\ell+1$ edges of $H$ are not edges of $G'$, and the free graph obtained from $G'$ as above, with free edges $\{e \in E(G') : \alpha'_i(e) = 1\}$, still has flooding number at least $\ell$. Since $\mathcal{G}$ fully immerses $(G', \lambda', \ldots, \alpha')$, it suffices to work in $G'$, and from now on all free graphs in Case 1 are obtained from $G'$. Then, by a pigeonhole argument, there is a large grid in $\mathcal{G}$ with only outside-horizontal edges possibly not edges of $G'$:

    \begin{observation}\label{obs:untouchedsubgrid-colors}
        $\mathcal{G}$ fully immerses an $(A,B)$-clean $S$-saturated decorated graph $\mathcal{\hat{H}}$ of richness and width at least $5p$, and height equal to that of $\mathcal{H}$, such that every edge of $\mathcal{\hat{H}}$ that is not outside-horizontal is also an edge of $G'$.
    \end{observation}
\begin{proof}
    Since $\mathcal{H}$ is $(A,B)$-clean, there are $5p(\ell+2)$ disjoint subgrids of $\mathcal{H}$ which each have the same height as $\mathcal{H}$ and contain the ends of a batch of $\mathcal{H}$. Then we can ``break up $\mathcal{H}$'' into $\ell+2$ disjoint grids, each of which contains $5p$ of those subgrids and $5p$ batches made up of vertical edges. By \Cref{lem:reroute}, the edge set of one of those grids and its vertical outside edges are edges of $G'$. Call this grid $\hat{H}$. Then, to get a full immersion of an $(A,B)$-clean $S$-saturated decorated graph in $\mathcal{G}$ with underlying grid $\hat{H}$ and height, width, richness at least $5p$, we can use the vertical outside edges incident to $\hat{H}$. So, it is left to find trails for the outside-horizontal edges of $\hat{H}$ which together use every remaining edge exactly once. We can route these trails as in \Cref{fig:growingA3}. Let $\mathcal{\hat{H}}$ be the decorated graph given by this full immersion. So, $\mathcal{G}$ fully immerses the $(A,B)$-clean $S$-saturated decorated graph $\mathcal{\hat{H}}$ of richness and width at least $5p$.
\end{proof}

    Let $(\hat{G}', \emptyset, \hat{F}'_i, \hat{b})$ be the free graph obtained from $G'$ by identifying the branch vertices of $\hat{H}$ to a single vertex $\hat{b}$, with empty signature and with set of free edges $\hat{F}'_i := \{e \in E(G') : \alpha'_i(e) = 1\}$. Roughly, if the flooding number of $(\hat{G}', \emptyset, \hat{F}'_i, \hat{b})$ is at least $p$, then we can see that these $p$ circuits can be paired with $p$ batches made up of vertical outside edges of $\hat{H}$ to obtain an $A$-rich, $S \cup \{i\}$-saturated decorated graph as a full immersion of $\mathcal{G}$ with the same underlying grid as $\hat{H}$. Formally,
    \begin{lemma}\label{lem:circuit-saturate}
        If the flooding number of $(\hat{G}', \emptyset, \hat{F}'_i, \hat{b})$ is at least $p$, then $\mathcal{G}$ fully immerses an $A$-rich $S \cup \{i\}$-saturated decorated graph of height, width and richness at least $p$.
    \end{lemma}
    \begin{proof}
        By \Cref{obs:untouchedsubgrid-colors}, only the outside-horizontal edges of $\hat{H}$ are not edges of $G'$. So in any flooding of $(\hat{G}', \emptyset, \hat{F}'_i, \hat{b})$, every circuit is disjoint from the underlying grid and vertical-outside edges of $\hat{H}$. Then, since $G'$ is 4-regular, a flooding of this free graph is equivalent to an alternative set of trails of horizontal outside edges for $\hat{H}$. Replacing $\hat{H}$'s set of horizontal outside edges with the ones provided by an optimal flooding gives at least $p$ horizontal outside edges $e$ such that $\alpha'_i(e) = 1$; that is, at least $p$ feasible circuits. Since $\hat{H}$ has at least $p$ batches made up of vertical outside edges, we can add one horizontal outside edge $e$ with $\alpha'_i(e) = 1$ to each of these batches.
        \begin{claim}
            The updated batches generate $\mathbb{F}_2^{A}$, and each of them contains at least one edge of each color in $S \cup \{i\}$.
        \end{claim}
        \begin{proof}
            To see this, recall that before the update, the $\lambda'_A$-weights of the edge set of each batch generate $\mathbb{F}_2^A$, and each batch contains at least one edge of each color in $S$. Adding an edge to a batch can only enlarge the subspace generated by its $\lambda'_A$-weights, so each updated batch still generates $\mathbb{F}_2^A$, and still contains an edge of each color in $S$. Since the added edge has color $i$, each updated batch also contains an edge of color $i$.
        \end{proof}
        Hence, with this outside-edge replacement, $\hat{H}$ has at least $p$ batches such that each batch's $\lambda'_A$ weights generate $\mathbb{F}_2^{A}$ and each batch contains an edge of each color in $S \cup \{i\}$. Let $\mathcal{H'}$ be the decorated graph given by this replacement of trails of $\mathcal{\hat{H}}$. The edges of the underlying grid of $\mathcal{H'}$ are unchanged, so they still have $\lambda_A$-weight zero. Clearly, $\mathcal{H'}$ is $A$-rich and $S \cup \{i\}$-saturated with richness, height and width at least $p$, fully immersed by $\mathcal{G}$.
    \end{proof}

    Then, since $p \ge f_{\ref{prop:controlled}}(m, |A|, |B|, q)$, we can apply \Cref{prop:controlled} to find an $(A', B')$-clean $S \cup \{i\}$-saturated grid fully immersed by $\mathcal{G}$, of height, width, and richness at least $m$, where $A' \supseteq A$ and $B' = B \setminus (A' \setminus A)$.

    So, suppose for contradiction that the flooding number of $(\hat{G}', \emptyset, \hat{F}'_i, \hat{b})$ is strictly less than $p$. Note that since $G'$ is 4-edge-connected, so is this free graph. So, we can apply \Cref{lem:smallcut} to obtain a subset $X \subseteq V(G')$ such that $X$ contains $V(\hat{H})$, and:
    \begin{enumerate}
        \item there are at most $4(p-1)$ edges of $G'$ with exactly one end in $X$, and
        \item there are at most $p-1$ edges with both ends in $X$ that have color $i$.
    \end{enumerate}
Let an edge satisfying either of these two conditions be called $i$-bad; that is, an edge is $i$-bad if it is in $\delta(X)$, or it has color $i$ and both ends in $X$. Hence, there are at most $5(p-1)$ $i$-bad edges. We will obtain a contradiction to the fact that the flooding number of the free graph obtained from $G'$ by identifying the branch vertices of $H$ (with the free edges defined as above) is at least $\ell$. We need the following claim:

\begin{claim}
\label{clm:saturate-cut}
All but at most $25(p-1)^2$ vertices of $H$ are in $X$.
\end{claim}
\begin{proof}
    Fix the full immersion model of $\mathcal{H}$ in $G'$. Let $Y$ be the set of all $y \in V(H)$ so that there is a path from $y$ to $V(\hat{H})$ in the underlying grid of $H$ whose corresponding trails, in $G'$, do not use any $i$-bad edge. Thus there is a trail from $y$ to $V(\hat{H})$ in $G'$ which does not use any edge of $\delta(X)$. Since $V(\hat{H}) \subseteq X$, this trail stays in $X$, so $y \in X$. So in fact $Y \subseteq X$.

    We complete the proof by showing that $|V(H)-Y| \leq 25(p-1)^2$. Observe that $H$ and $\hat{H}$ have height and width at least $5p >5(p-1)$, and the edge-boundary of $Y$ within the underlying grid of $H$ has size at most $5(p-1)$ (each edge of this boundary corresponds to a branch trail using an $i$-bad edge, each $i$-bad edge lies on at most one branch trail, and there are at most $5(p-1)$ $i$-bad edges). So by \Cref{lem:connGrid} about cuts in a grid, either $Y$ or $V(H)-Y$ has size at most $25(p-1)^2$. So the claim follows since $|Y|\geq |V(\hat{H})|\geq 25p^2>25(p-1)^2$.
\end{proof}

Now consider a collection of trails of $G'$ which corresponds to an optimal flooding of this free graph; each trail joins two branch vertices of $H$. By Claim~\ref{clm:saturate-cut}, and since every branch vertex has degree 4, at most $100(p-1)^2$ of the trails have an end outside of $X$. Each of the other trails which is feasible contains an edge of color $i$. If this edge has both ends in $X$, it is $i$-bad; otherwise it has an end outside $X$, and since both ends of the trail are in $X$, the trail uses an edge of $\delta(X)$, which is also $i$-bad. Since the trails are edge-disjoint, each $i$-bad edge lies on at most one of them, so at most $5(p-1)$ of these trails are feasible. So \begin{align*}
    100(p-1)^2+5(p-1)+2 = \ell \leq \nu(\hat{G}, \emptyset, F_i, b) \leq 100(p-1)^2+5(p-1),
\end{align*} a contradiction. This completes the first case.

\bigskip
\noindent\emph{Case 2:} The flooding number of $(\hat{G}, \emptyset, F_i, b)$ is less than $\ell$.
\bigskip

Here, since $G$ and therefore $(\hat{G}, \emptyset, F_i, b)$ is 4-edge-connected, we can apply \Cref{lem:smallcut}, which gives a set $V(H) \subseteq X \subseteq V(G)$ such that \begin{enumerate}
    \item there are at most $4(\ell-1)$ edges of $G$ with exactly one end in $X$, and
    \item there are at most $\ell-1$ edges of $G$ with both ends in $X$ that have color $i$.
\end{enumerate}

This gives the following easy corollary:
\begin{corollary}\label{cor:cleancut-colors}
    There exists a set $X' \subseteq V(G)$ such that \begin{enumerate}
        \item there are at most $8(\ell-1)$ edges of $G$ with exactly one end in $X'$,
        \item at most $\ell-1$ branch vertices of $H$ are not in $X'$, and
    \item no edge with both ends in $X'$ has color $i$.
    \end{enumerate}
\end{corollary}
\begin{proof}
    Let $F$ be the set of edges with both ends in $X$ and color $i$; so $|F| \le \ell - 1$. For each edge of $F$, we move one of its endpoints to $V(G)-X$, choosing an endpoint which is not a branch vertex of $H$ if possible, and let $X'$ be the resulting set. Then no edge with both ends in $X'$ has color $i$. Since $X$ contains every branch vertex of $H$ and at most $\ell-1$ vertices are moved, at most $\ell-1$ branch vertices of $H$ are not in $X'$. Since $G$ is 4-regular, each vertex moved across the cut increases the number of edges in the cut by at most 4. So, since there are at most $\ell - 1$ edges in $F$ and $\delta(X)$ had starting size at most $4(\ell-1)$, we have $|\delta(X')| \le 4(\ell-1) + 4(\ell-1) = 8(\ell-1)$.
    \end{proof}

So, the set $X'' := V(G) - X'$ completes the second case: $X''$ contains at most $\ell - 1 \le g_{\ref{prop:main}}(m,|A|,|B|,q)$ branch vertices of $H$, $|\delta(X'')| = |\delta(X')| \le 8(\ell-1) = g_{\ref{prop:main}}(m,|A|,|B|,q)$, and every edge with color $i$ has an end in $X''$, so it lies in $E(X'') \cup \delta(X'')$.

\emph{Running time.} The runtime follows just as in \Cref{prop:main}, but we repeat it here for clarity. Since $G$ is 4-regular, $|E(\hat{G})| = 2n$, so by \Cref{thm:mainMinMax} we decide which case holds, and find a flooding of value at least $\ell$ or a certificate of value less than $\ell$, in $\Oh(\ell n^4)$ time. In Case~1, \Cref{lem:reroute} takes $\Oh(n^2)$ time, Observation~\ref{obs:untouchedsubgrid-colors} and \Cref{lem:circuit-saturate} take $\Oh(n)$ time, \Cref{thm:mainMinMax} applied to $(\hat{G}', \emptyset, \hat{F}_i', \hat{b})$ with $k = p$ takes $\Oh(p n^4)$ time, and \Cref{prop:controlled} takes $\Oh_{m,|U|,q}(n)$ time. In Case~2, the certificate gives $X$ as in \Cref{lem:smallcut}, and \Cref{cor:cleancut-colors} takes $\Oh(n)$ time. Since $\ell$ and $p$ depend only on $m$, $|U|$ and $q$, the total is $\Oh_{m,|U|,q}(n^4)$.
\end{proof}

\section{Dealing with multiple signatures}

The above section roughly shows that either we can ``enrich'' or ``saturate'' a decorated graph immersing a large grid-like graph, or we can find a small cut separating ``problematic'' edges.
The next proposition shows that these results compound nicely when we apply \Cref{prop:main,prop:main2} iteratively on all signatures and colors.

\begin{restatable}{proposition}{allsignatures}\label{prop:allsignatures}
   For every non-decreasing function $\psi: \N \to \N$ with $\psi(x) > x$ for all $x \in \N$, there exist $f_{\ref{prop:allsignatures}}(\psi, |U|, q), g_{\ref{prop:allsignatures}}(\psi, |U|, q) \in \N$ such that the following holds. Let $\mathcal{G} = (G, \lambda, R_G, \rho_G, \alpha_G)$ be a 4-edge-connected decorated graph, and let $(\mathcal{H}, \phi)$ be an $(A_0, U \setminus A_0)$-clean $S'$-saturated grid in $\mathcal{G}$ of richness, height and width at least $f_{\ref{prop:allsignatures}}(\psi, |U|, q)$, for some $A_0 \subseteq U$ and $S' \subseteq [q]$. Then there exist $A$ with $A_0 \subseteq A \subseteq U$, $S$ with $S' \subseteq S \subseteq [q]$, an $(A, U \setminus A)$-clean $S$-saturated grid $(\mathcal{\tilde{H}}, \tilde\phi)$ in $\mathcal{G}$ with subspace $N$, a shifting $\lambda'_{U \setminus A}$ of $\lambda_{U \setminus A}$, and a set $Z \subseteq V(G)$ such that
   \begin{enumerate}
        \item $\mathcal{\tilde{H}}$ has richness, height and width at least $\psi(|\delta(Z)|)$,
        \item $\{e \in E(G): \lambda_A(e) \times \lambda'_{U \setminus A}(e)\notin N\} \subseteq E(Z) \cup \delta(Z)$,
        \item $\{e \in E(G): \alpha_i(e) = 1 \text{ for some } i \in [q] \setminus S\} \subseteq E(Z) \cup \delta(Z)$,
        \item $|\delta(Z)| \le g_{\ref{prop:allsignatures}}(\psi, |U|, q)$, and
        \item $|Z \cap \tilde\phi(V(\tilde{H}))| \le |\delta(Z)|^2$.
    \end{enumerate}
    Further, given the full immersion model of $(\mathcal{H}, \phi)$, we can find $A$, $S$, the full immersion model of $(\mathcal{\tilde{H}}, \tilde\phi)$, $\lambda'_{U \setminus A}$ and $Z$ in time $\Oh_{\psi, |U|, q}(n^4)$.
\end{restatable}

\begin{proof}
By \Cref{prop:main}, both of $f_{\ref{prop:main}}$ and $g_{\ref{prop:main}}$ are non-decreasing in every argument. For $m \in \Zplus$, let
\begin{align*}
f^*(m) &:= \max\{f_{\ref{prop:main}}(m,a,b,q) : a + b \le |U|\}, \\
g^*(m) &:= \max\{g_{\ref{prop:main}}(m,a,b,q) : a + b \le |U|\},
\end{align*}
which are again non-decreasing. Define $F_0 := \psi(0)$, and for $j \ge 1$,
$$F_j := \max\big\{F_{j-1},\; f^*(F_{j-1}),\; \psi\big(j \cdot g^*(F_{j-1})\big)\big\}.$$
So $F_0 \le F_1 \le \cdots$. Set $f_{\ref{prop:allsignatures}}(\psi, |U|, q) := F_{|U|+q}$ and $g_{\ref{prop:allsignatures}}(\psi, |U|, q) := (|U|+q)\, g^*(F_{|U|+q-1})$ (and $g_{\ref{prop:allsignatures}} := 0$ if $|U| + q = 0$).

For an $(A, U\setminus A)$-clean $S$-saturated grid-like graph, call $j := |U \setminus A| + |[q] \setminus S|$ its \emph{deficiency}; it is the number of signatures and colors that could still be added, and $0 \le j \le |U| + q$.

\begin{claim}\label{clm:iterate}
    There is an $(A, U \setminus A)$-clean $S$-saturated grid $(\mathcal{\tilde{H}}, \tilde\phi)$ in $\mathcal{G}$, for some $A \supseteq A_0$ and $S \supseteq S'$, of deficiency $j$ and richness, height and width at least $F_j$, such that either $j = 0$, or applying \Cref{prop:main} with $m = F_{j-1}$ to every $v \in U \setminus A$, and \Cref{prop:main2} with $m = F_{j-1}$ to every $i \in [q] \setminus S$, always returns outcome (ii).
\end{claim}
\begin{proof}
    We maintain a current grid $(\mathcal{H}_c, \phi_c)$ which is $(A, U \setminus A)$-clean and $S$-saturated with $A \supseteq A_0$ and $S \supseteq S'$, has deficiency $j$, and has richness, height and width at least $F_j$. Initially $(\mathcal{H}_c, \phi_c) = (\mathcal{H}, \phi)$, $A = A_0$ and $S = S'$; this satisfies the invariant since the deficiency is at most $|U| + q$ and the $F_j$ are non-decreasing in $j$.

    If $j = 0$ we stop. Otherwise, since $F_j \ge f^*(F_{j-1}) \ge f_{\ref{prop:main}}(F_{j-1}, |A|, |U \setminus A|, q)$, we may apply \Cref{prop:main} with $m = F_{j-1}$ to $(\mathcal{H}_c, \phi_c)$ and each $v \in U \setminus A$, and \Cref{prop:main2} with $m = F_{j-1}$ to $(\mathcal{H}_c, \phi_c)$ and each $i \in [q] \setminus S$. If one of these applications returns outcome (i), we obtain an $(A'', U \setminus A'')$-clean $S''$-saturated grid for some $A'' \supseteq A \cup\{v\}$ and $S'' \supseteq S$, or for some $A'' \supseteq A$ and $S'' \supseteq S \cup \{i\}$, of richness, height and width at least $F_{j-1}$. Its deficiency $j'$ is at most $j - 1$, and since $F_{j-1} \ge F_{j'}$, it satisfies the invariant. We make it the current grid and repeat. If all applications return outcome (ii), we stop.

    Each repetition decreases the deficiency by at least one, so there are at most $|U| + q$ repetitions, each using at most $|U| + q$ applications of Propositions~\ref{prop:main} and~\ref{prop:main2}. So the process stops after at most $(|U|+q)^2$ applications, and the final current grid is as required.
\end{proof}

Let $(\mathcal{\tilde{H}}, \tilde\phi)$, $A$, $S$ and $j$ be as in Claim~\ref{clm:iterate}, and let $N$ be the subspace of $\mathcal{\tilde{H}}$. For each $v \in U \setminus A$, let $N_{A,v}$ be the projection of $N$ onto $\mathbb{F}_2^A \times \mathbb{F}_2^v$ (this is $N_v$ in the notation of \Cref{prop:main}).

If $j = 0$, then $A = U$ and $S = [q]$; we take $Z := \emptyset$ and the empty shifting. Items (ii), (iii) and (v) hold trivially, $|\delta(Z)| = 0$, and $\mathcal{\tilde{H}}$ has richness, height and width at least $F_0 = \psi(0)$, so (i) holds.

So suppose $j \ge 1$. By Claim~\ref{clm:iterate}, for each $v \in U\setminus A$, \Cref{prop:main} applied to $(\mathcal{\tilde{H}}, \tilde\phi)$ with $m = F_{j-1}$ gives a set $X_v \subseteq V(G)$ and a shifting $\lambda'_v$ of $\lambda_v$ such that
\begin{enumerate}
    \item $|\delta(X_v)| \le g_{\ref{prop:main}}(F_{j-1}, |A|, |U \setminus A|, q) \le g^*(F_{j-1})$,
    \item $|X_v \cap \tilde\phi(V(\tilde{H}))| \le g^*(F_{j-1})$, and
    \item $\{e \in E(G): \lambda_A(e) \times \lambda'_{v}(e) \notin N_{A,v}\} \subseteq E(X_v) \cup \delta(X_v)$.
\end{enumerate}
Let $\lambda'_{U \setminus A} := (\lambda'_v)_{v \in U \setminus A}$; this is a shifting of $\lambda_{U \setminus A}$, since each $\lambda'_v$ only changes the $v$-coordinate. Similarly, for each $i \in [q] \setminus S$, \Cref{prop:main2} applied to $(\mathcal{\tilde{H}}, \tilde\phi)$ with $m = F_{j-1}$ gives a set $Y_i \subseteq V(G)$ with $|\delta(Y_i)| \le g^*(F_{j-1})$, $|Y_i \cap \tilde\phi(V(\tilde{H}))| \le g^*(F_{j-1})$, and $\{e \in E(G): \alpha_i(e) = 1\} \subseteq E(Y_i) \cup \delta(Y_i)$. Note that all of these sets are obtained with respect to the \emph{same} grid $(\mathcal{\tilde{H}}, \tilde\phi)$, and hence the same subspace $N$.

\begin{claim}
    Let $Z := (\bigcup_{v \in U \setminus A} X_v) \cup (\bigcup_{i \in [q]\setminus S} Y_i)$. Then, \begin{enumerate}
        \item $|\delta(Z)| \le j \cdot g^*(F_{j-1})$,
        \item $\{e \in E(G): \lambda_A(e) \times \lambda'_{U \setminus A}(e) \notin N\} \subseteq E(Z) \cup \delta(Z)$, and
        \item $\{e \in E(G): \alpha_i(e) = 1\} \subseteq E(Z) \cup \delta(Z)$ for all $i \in [q] \setminus S$.
    \end{enumerate}
\end{claim}
\begin{proof}
    Every edge of $\delta(Z)$ has exactly one end in $Z$, so it has exactly one end in at least one of the sets $X_v$, $Y_i$; that is, $\delta(Z) \subseteq \bigcup_v \delta(X_v) \cup \bigcup_i \delta(Y_i)$. There are exactly $j$ such sets, each with boundary of size at most $g^*(F_{j-1})$, which gives the first property. For the second property, it suffices to show that if $\lambda_A(e) \times \lambda'_{U \setminus A}(e) \notin N$, then for some $v \in U \setminus A$, $\lambda_A(e) \times \lambda'_v(e) \notin N_{A,v}$: then $e$ has an end in $X_v \subseteq Z$, so $e \in E(Z) \cup \delta(Z)$. To see this, suppose for some $e \in E(G)$, $\lambda_A(e) \times \lambda'_{U \setminus A}(e) \notin N$. Since $\dim(N)=|A|$ and the dimension does not decrease when restricted to $\mathbb{F}_2^A$, there exists a unique labeling $\lambda''_{U \setminus A}(e)$ such that $\lambda_A(e) \times \lambda''_{U \setminus A}(e) \in N$. So, for some $v \in U \setminus A$, it must be that $\lambda'_v(e) \neq \lambda''_v(e)$. Hence, for this $v$, $\lambda_A(e) \times \lambda_v'(e) \notin N_{A,v}$. Finally, the third property holds since every edge of $E(Y_i) \cup \delta(Y_i)$ has an end in $Y_i \subseteq Z$, and so lies in $E(Z) \cup \delta(Z)$.
\end{proof}

Now we check the five items. Items (ii) and (iii) are the last two properties of the claim. For (iv), since $j \le |U|+q$ and $g^*$ is non-decreasing, $|\delta(Z)| \le j \cdot g^*(F_{j-1}) \le (|U|+q)\,g^*(F_{|U|+q-1}) = g_{\ref{prop:allsignatures}}(\psi,|U|,q)$. For (i), since $\psi$ is non-decreasing,
$$\psi(|\delta(Z)|) \le \psi\big(j \cdot g^*(F_{j-1})\big) \le F_j,$$
and $\mathcal{\tilde{H}}$ has richness, height and width at least $F_j$ by Claim~\ref{clm:iterate}. It remains to prove (v).

\begin{claim}\label{clm:Zsmallside}
    Let $T := \tilde\phi(V(\tilde{H}))$ be the set of branch vertices of $\tilde{H}$. Then $|Z \cap T| \le |\delta(Z)|^2$.
\end{claim}
\begin{proof}
    Let $c := j \cdot g^*(F_{j-1})$, and let $L$ be the minimum of the height and width of $\tilde{H}$. By Claim~\ref{clm:iterate} and the definition of $F_j$, we have $L \ge F_j \ge \psi(c) > c \ge |\delta(Z)|$. So $|\delta(Z)| < L$, and by \Cref{lem:connGrid} about cuts in a grid immersion, either $|Z \cap T| \le |\delta(Z)|^2$ or $|(V(G) - Z) \cap T| \le |\delta(Z)|^2$. We show the second option is impossible. Since $Z$ is the union of the $j$ sets $X_v$ and $Y_i$, each containing at most $g^*(F_{j-1})$ vertices of $T$, we have $|Z \cap T| \le c$. Also, $|T| = |V(\tilde{H})| \ge L^2$. Hence
    $$|(V(G) - Z) \cap T| \ge L^2 - c \ge (c+1)^2 - c = c^2 + c + 1 > c^2 \ge |\delta(Z)|^2.$$
    So $|Z \cap T| \le |\delta(Z)|^2$, as desired.
\end{proof}
This proves (v).

Note that under the new labeling, $\mathcal{G}' = (G, \lambda_A \times \lambda'_{U \setminus A}, R, \rho, \alpha)$, the full immersion model of $\mathcal{\tilde{H}}$ in $\mathcal{G}$ still gives an $(A, U \setminus A)$-clean $S$-saturated decorated graph with the same subspace. This is because shifting at internal vertices of a branch trail of $\tilde{H}$ does not impact the label of the branch trail itself. Since the shiftings are performed in an associated free graph where branch vertices of $\tilde{H}$ are identified to a single vertex, we may also assume that no shiftings are performed at branch vertices of $\tilde{H}$.

\emph{Running time.} The numbers $F_0, \dots, F_{|U|+q}$ can be computed in time depending only on $\psi$, $|U|$ and $q$. The process in Claim~\ref{clm:iterate} makes at most $(|U|+q)^2$ applications of Propositions~\ref{prop:main} and~\ref{prop:main2}, each with $m \le F_{|U|+q}$, so it takes time $\Oh_{\psi,|U|,q}(n^4)$. The sets $X_v$, $Y_i$ and shiftings $\lambda'_v$ are the outputs of the last round of these applications, and $Z$ and $\lambda'_{U \setminus A}$ are then obtained in $\Oh((|U|+q)\,n)$ time.
\end{proof}
\section{Irrelevant vertex}\label{sec:irrelevant}

We would like to, given a 4-edge-connected decorated graph $\mathcal{G}$, compute its $\tau$-color-folio (defined in Section \ref{sec:overview}). Since this can be done efficiently when the treewidth of $\mathcal{G}$ is small (see \Cref{sec:overview}), it is helpful to find a vertex $v \in V(G)$ that can be split off in any way to obtain a decorated graph $\mathcal{G'}$, such that, a decorated graph $\mathcal{F}$ is in $\mathcal{G}$'s $\tau$-color-folio if and only if $\mathcal{F}$ is in $\mathcal{G'}$'s $\tau$-color-folio. So, this section is dedicated to proving exactly that:

\begin{restatable}{proposition}{irrelevantprop}\label{prop:inessential-vertex-decorated}
    There is a function $f_{\ref{prop:inessential-vertex-decorated}}: \N^4 \to \N$ such that the following holds. Let $\mathcal{G}= (G, \lambda, R_G, \rho_G, \alpha_G)$ be a 4-edge-connected decorated graph, let $\tau \in \N$ and $r_G \coloneqq |R_G|$, and suppose that there is an $(A_0,U \setminus A_0)$-clean $S'$-saturated grid in $\mathcal{G}$ of height, width, and richness at least $f_{\ref{prop:inessential-vertex-decorated}}(\tau, r_G, |U|, q)$, for some $A_0 \subseteq U$ and $S' \subseteq [q]$. Then, there exists a vertex $v \in V(G)$ such that for any of the three decorated graphs $\mathcal{G'}$ obtained from splitting off at $v$ in $\mathcal{G}$, a decorated graph $\mathcal{F}$ is in $\mathcal{G}$'s $\tau$-color-folio if and only if $\mathcal{F}$ is in $\mathcal{G}'$'s $\tau$-color-folio. Further, given the full immersion model of the clean grid, this vertex can be found in time $\Oh_{\tau, r_G, |U|, q}(n^4)$.
\end{restatable}

First, we must clarify how splitting off a vertex impacts half-edge colors.
\subsection{Splitting off a vertex}
\begin{figure}[ht]
\centering
\includegraphics[width=\textwidth]{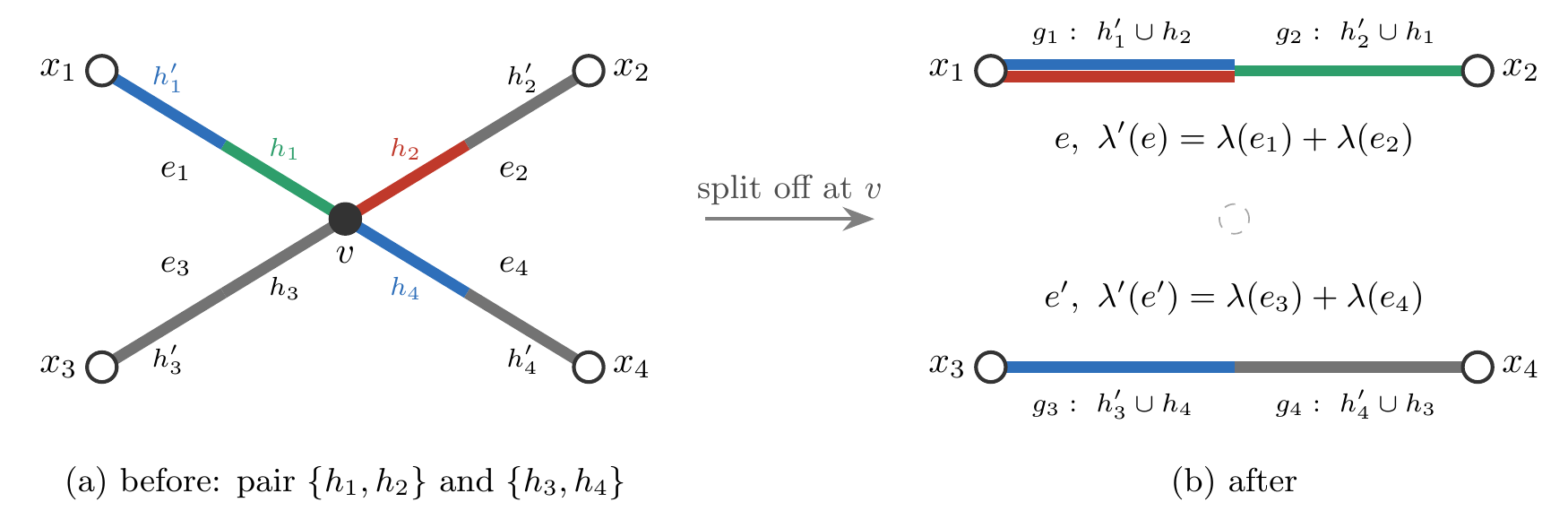}
\caption{Splitting off at $v$ with the pairing $\{h_1,h_2\}$, $\{h_3,h_4\}$. Colored half-edges are drawn in color and colorless ones in gray; each new half-edge carries the union of the color sets shown in its label.}
\label{fig:splitoffcolor}
\end{figure}

{

Suppose a non-root vertex $v$ in a decorated graph $\mathcal{G} = (G, \lambda, R, \rho, \alpha)$ is incident to edges $e_1,e_2,e_3,e_4$, such that $e_i = \{h_i, h_i'\}$ and $v$ is incident to the half-edge $h_i$ for each $i \in [4]$. Then, splitting off at $v$ means to choose a pairing of $\{h_1,h_2,h_3,h_4\}$, say $\{h_1,h_2\}$ and $\{h_3,h_4\}$, and obtain a new decorated graph $\mathcal{G'} = (G', \lambda', R, \rho, \alpha')$ by \begin{enumerate}
    \item deleting $e_1,e_2,e_3,e_4$,
    \item adding a new edge $e = \{g_1,g_2\}$ whose ends are the other ends of $e_1$ and $e_2$ (not corresponding to $h_1$ and $h_2$), where $g_1$ takes the place of $h'_1$ and $g_2$ takes the place of $h'_2$,
    \item adding a new edge $e' = \{g_3,g_4\}$ whose ends are the other ends of $e_3$ and $e_4$ (not corresponding to $h_3$ and $h_4$), where $g_3$ takes the place of $h'_3$ and $g_4$ takes the place of $h'_4$,
    \item deleting $v$,
    \item setting $\lambda'(e)=\lambda(e_1) + \lambda(e_2)$ and $\lambda'(e')=\lambda(e_3) + \lambda(e_4)$, and keeping $\lambda'(f) = \lambda(f)$ for all other edges in $G$,
    \item setting $\alpha'(g_1) = \alpha(h'_1) \cup \alpha(h_2)$, $\alpha'(g_2) = \alpha(h'_2) \cup \alpha(h_1)$,
      $\alpha'(g_3) = \alpha(h'_3) \cup \alpha(h_4)$, $\alpha'(g_4) = \alpha(h'_4) \cup \alpha(h_3)$,
      and keeping $\alpha'(h) = \alpha(h)$ for all other half-edges in $G$.
\end{enumerate}

In other words, when combining $e_1$ and $e_2$ for example, the forward colors of a trail passing through $e_1$ followed by $e_2$ remain forward colors in $e$, and the same for backward colors (see Figure \ref{fig:splitoffcolor}). If some $e_i$ is a loop at $v$, then both of its half-edges are among $h_1, \ldots, h_4$, and we require that they are not paired with each other. In this case the two new edges described above are joined through the loop into a single edge; for instance, if $e_2 = e_3$ is a loop with half-edges $h_2, h_3$, the result is a single edge from the other end of $e_1$ to the other end of $e_4$, with half-edges $g_1$ and $g_4$, weight $\lambda(e_1) + \lambda(e_2) + \lambda(e_4)$, and
$$\alpha'(g_1) = \alpha(h'_1) \cup \alpha(h_2) \cup \alpha(h_4), \qquad \alpha'(g_4) = \alpha(h'_4) \cup \alpha(h_3) \cup \alpha(h_1)$$}

To apply Proposition \ref{prop:inessential-vertex-decorated}, we must find a large clean grid in $\mathcal{G}$. Note that any grid-like decorated graph is automatically $\emptyset$-rich and $\emptyset$-saturated for any richness, using empty batches. So, if a decorated graph $\mathcal{G}$ fully immerses any large enough grid-like graph, by Proposition $\ref{prop:controlled}$ about ``cleaning'' a grid, $\mathcal{G}$ also immerses a large $(A,U \setminus A)$-clean $\emptyset$-saturated decorated graph of any desired richness, for some $A \subseteq U$. So, the following section shows that large treewidth guarantees a full immersion of a large grid-like graph.

\subsection{Computable bounds}
Chudnovsky, Dvo{\v{r}}{\'a}k, Klimo{\v{s}}ov{\'a}, and Seymour \cite{chudnovsky2013immersionfouredgeconnectedgraphs} proved that every $4$-edge-connected graph of sufficiently large treewidth contains a $k\times k$ grid as an immersion. It is not hard to show as a corollary that the same holds for $4$-regular graphs. However, we require a version of this theorem which has computable bounds in order to obtain an FPT algorithm, and it is not entirely straightforward to verify that their proof yields a computable bound. So we obtain such a lemma by applying other known results.

\begin{lemma}
\label{lem:gridImmersion}
    There is a computable function $f_{\ref{lem:gridImmersion}}\colon \mathbb{N}\to\mathbb{N}$ such that every $4$-regular graph $G$ with treewidth at least $f_{\ref{lem:gridImmersion}}(t)$ contains the $t\times t$ grid as an immersion.
\end{lemma}
\begin{proof}
    First of all, by \cite[Theorem~3.1]{edgeDisjOdd}, there is a computable function $h$ so that for every graph $G$ with treewidth more than $h(t)$ that has no $3$-edge cut, the extended line graph of $G$ contains a $t$-vertex clique as a minor. Thus, since the extended line graph is obtained from the line graph by subdividing some of its edges, the line graph of $G$ also contains a $t$-vertex clique as a minor for $t \geq 3$.

    Now we apply \cite[Theorem~6.3]{immersionEdgeConn} with $k=D=4$ and $n=t$. Inside the proof, there is an integer $\ell = \Oh(t^5)$, and we set $W = h(\ell)$. In this manner, we find that there is a computable function $g$ so that every $4$-edge-connected $4$-regular graph with tree-cut width at least $g(t)$ contains the $t \times t$ grid as an immersion. By \cite[Proposition~2.6]{applicationsTreeCut}, every graph has tree-cut width at least order square-root of its treewidth. So there is a computable function $f$ so that every $4$-regular $4$-edge-connected graph with treewidth at least $f(t)$ contains the $t \times t$ grid as an immersion. 

    Now, let $G$ be a $4$-regular graph with treewidth at least $\max (4, f(t))$. We now show that $G$ contains the $t \times t$ grid as an immersion. We may assume that $G$ is connected since its treewidth is the maximum treewidth of one of its components. Similarly, if $(X_1,X_2)$ is a partition of the vertex-set of $G$ so that there are exactly two edges going between $X_1$ and $X_2$, then for $i=1,2$, let $G_i$ be obtained from the subgraph of $G$ induced on $X_i$ by adding one edge between the two vertices incident to cut edges. Then the treewidth of $G$ equals the maximum of the treewidth of $G_1$ and $G_2$. Since $G_1$ and $G_2$ are both immersions of $G$, repeating this process until we obtain a $4$-edge-connected $4$-regular graph completes the proof.
\end{proof}

This gives the following:
\begin{lemma}\label{lem:find-grid}
    Every decorated graph $\mathcal{G}$ with treewidth at least $f_{\ref{lem:gridImmersion}}(5t)$ fully immerses a grid-like graph of height and width $t$, whose model in $\mathcal{G}$ can be found in time $\Oh_t(n^4)$.
\end{lemma}
\begin{proof}
    Let $\mathcal{G}$ satisfy the conditions of the statement. By lemma \ref{lem:gridImmersion}, $\mathcal{G}$ immerses a $5t \times 5t$-grid. Here, an immersion is a full immersion that disregards labels, colors, and roots, and does not require that every edge of $\mathcal{G}$ belongs to a trail. Consider a model of this immersion in $\mathcal{G}$. We construct an immersion model of a grid-like graph as follows. Map the vertices and trails of the grid-like graph to the middle $t \times t$ subgrid of the $5t \times 5t$-grid. To construct outside-edges, see that a $t \times t$ grid-like graph has $2t$ outside edges, so we can route these in an edge-disjoint way by using $2t$ concentric cycles around the middle subgrid. Here, a concentric cycle is a cycle formed by two columns and two rows at equal distance from the boundary of the $t \times t$ subgrid; see Figure \ref{fig:rerouting1}. Hence, $\mathcal{G}$ immerses a grid-like graph of height and width $t$.
    
    Then, by the algorithm given in~\cite{TopologicalMinorTesting} such an immersion can be found in time $\Oh_t(n^4)$.  Since $\mathcal{G}$ is Eulerian, an immersion implies the existence of a full immersion, which can be found from the immersion in linear time using Hierholzer's algorithm. So, given that its treewidth is at least $f_{\ref{lem:gridImmersion}}(5t)$, $\mathcal{G}$ fully immerses a grid-like decorated graph of height and width $t$, found in time $\Oh_t(n^4)$.
\end{proof}

Combining Lemma \ref{lem:find-grid} with \Cref{prop:controlled} (starting with $A = \emptyset$, $B = U$ and $S = \emptyset$), every 4-edge-connected decorated graph $\mathcal{G}$ of treewidth at least
$$f'(t, |U|, q) \coloneqq f_{\ref{lem:gridImmersion}}\big(5f_{\ref{prop:controlled}}(t, 0, |U|, q)\big)$$
fully immerses an $(A,U \setminus A)$-clean $\emptyset$-saturated decorated graph of height, width and richness at least $t$, for some $A \subseteq U$. So, \Cref{prop:inessential-vertex-decorated} implies \Cref{prop:inessential-vertex} (restated below) with
$$f_{\ref{prop:inessential-vertex}}(\tau, r_G, |U|, q) \coloneqq f'\big(f_{\ref{prop:inessential-vertex-decorated}}(\tau, r_G, |U|, q), |U|, q\big).$$

\irrelevant*

We return to finding an irrelevant vertex. Informally, this vertex will be a branch vertex in a large grid in $\mathcal{G}$, since a grid gives us many ways to reroute trails. We use \Cref{prop:allsignatures} to get a small cut in $\mathcal{G}$ separating all ``problematic'' edges from most branch vertices of a large ``clean'' grid in $\mathcal{G}$. Roughly, we will split off a vertex on the side of this cut with most branch vertices of a large grid, since any subtrails in a full immersion model of a decorated graph $\mathcal{F}$ in $\mathcal{G}$ on this side have labelings in the subspace of this grid, and colors that this grid is saturated for, and thus can be rerouted using the grid branch trails.

\subsection{Finding a small cut}

Since the roots of any decorated graph fully immersed by $\mathcal{G}$ must be mapped to the roots of $\mathcal{G}$ with the same $\rho$-values, we also want all roots of $\mathcal{G}$ to be on the ``problematic'' side of the cut. Note that this choice does not depend on any specific target full immersion $\mathcal{F}$, so the irrelevant vertex we find will be irrelevant for every decorated graph $\mathcal{F}$ on at most $\tau$ vertices fully immersed by $\mathcal{G}$. The following corollary of \Cref{prop:allsignatures} shows that the cut stays small, relative to the grid, when we add the roots of $\mathcal{G}$ to one side.

\begin{corollary}\label{cor:allsignaturesroots}
    Let $\psi: \N \to \N$ be non-decreasing with $\psi(x) > x$ for all $x \in \N$, let $r_G \coloneqq |R_G|$, and let $\psi_{R}(x) \coloneqq \psi(2x + 8r_G)$. Let $\mathcal{G} = (G, \lambda, R_G, \rho_G, \alpha_G)$ be a 4-edge-connected decorated graph, and let $(\mathcal{H}, \phi)$ be an $(A_0, U \setminus A_0)$-clean $S'$-saturated grid in $\mathcal{G}$ of richness, height and width at least $f_{\ref{prop:allsignatures}}(\psi_{R}, |U|, q)$, for some $A_0 \subseteq U$ and $S' \subseteq [q]$. Then there exist $A$ with $A_0 \subseteq A \subseteq U$, $S$ with $S' \subseteq S \subseteq [q]$, an $(A, U \setminus A)$-clean $S$-saturated grid $(\mathcal{\tilde{H}}, \tilde\phi)$ in $\mathcal{G}$ with subspace $N$, a shifting $\lambda'_{U \setminus A}$ of $\lambda_{U \setminus A}$, and a set $Z' \subseteq V(G)$ with $R_G \subseteq Z'$ such that
   \begin{enumerate}
        \item $\mathcal{\tilde{H}}$ has richness, height and width at least $\psi(|\delta(Z')|)$,
        \item $\{e \in E(G): \lambda_A(e) \times \lambda'_{U \setminus A}(e)\notin N\} \subseteq E(Z') \cup \delta(Z')$,
        \item $\{e \in E(G): \alpha_i(e) = 1 \text{ for some } i \in [q] \setminus S\} \subseteq E(Z') \cup \delta(Z')$,
        \item $|\delta(Z')| \le g_{\ref{prop:allsignatures}}(\psi_{R}, |U|, q) + 4r_G$, and
        \item $|Z' \cap \tilde\phi(V(\tilde{H}))| \le |\delta(Z')|^2$.
    \end{enumerate}
    Further, given the full immersion model of $(\mathcal{H}, \phi)$, these can be found in time $\Oh_{\psi, r_G, |U|, q}(n^4)$.
\end{corollary}
\begin{proof}
    The function $\psi_{R}$ is non-decreasing and $\psi_{R}(x) \ge \psi(x) > x$, so we may apply \Cref{prop:allsignatures} with $\psi_{R}$ in place of $\psi$. Let $A$, $S$, $(\mathcal{\tilde{H}}, \tilde\phi)$, $N$, $\lambda'_{U \setminus A}$ and $Z$ be as given, and let $Z' \coloneqq Z \cup R_G$. Write $c \coloneqq |\delta(Z)| + 4r_G$. Since $G$ is 4-regular, we have $|\delta(Z')| \le |\delta(Z)| + 4|R_G| = c$, which gives (iv). Then, (ii) and (iii) follow because $(E(Z) \cup \delta(Z)) \subseteq (E(Z') \cup \delta(Z'))$. For (i), $\mathcal{\tilde{H}}$ has richness, height and width at least $\psi_{R}(|\delta(Z)|) = \psi(2|\delta(Z)| + 8r_G) = \psi(2c) \ge \psi(|\delta(Z')|)$, since $\psi$ is non-decreasing and $2c \ge |\delta(Z')|$.

    Finally, for (v), let $T \coloneqq \tilde\phi(V(\tilde{H}))$, and let $L$ be the minimum of the height and width of $\tilde{H}$; so $L \ge \psi(2c) \ge 2c+1 > |\delta(Z')|$. By \Cref{lem:connGrid} about cuts in a grid immersion, either $|Z' \cap T| \le |\delta(Z')|^2$ or $|(V(G) - Z') \cap T| \le |\delta(Z')|^2$. By \Cref{prop:allsignatures}(v), $|Z' \cap T| \le |Z \cap T| + |R_G| \le |\delta(Z)|^2 + r_G \le c^2 + c$. So, since $L \ge 2c+1$,
    $$|(V(G) - Z') \cap T| \ge L^2 - c^2 - c \ge (2c+1)^2-c^2-c = 3c^2+3c+1 > c^2 \ge  |\delta(Z')|^2,$$
    and hence $|Z' \cap T| \le |\delta(Z')|^2$.
\end{proof}

Let $Z'$ be as given in \Cref{cor:allsignaturesroots}. Note that, for any full immersion model of a decorated graph $\mathcal{F}$ on at most $\tau$ vertices in $\mathcal{G}$, its trails can be partitioned into subtrails completely in $E(Z') \cup \delta(Z')$, and subtrails completely in $E(G) \setminus (E(Z') \cup \delta(Z'))$. We show that we can reroute the latter subtrails, even after splitting off at a vertex $v$ of $G$ that does not depend on $\mathcal{F}$. Working with the labeling $\lambda_A \times \lambda'_{U \setminus A}$ given by \Cref{cor:allsignaturesroots}, each subtrail of the second kind has weight in $N$ and only colors in $S$, since it avoids $E(Z') \cup \delta(Z')$. Moreover, each such subtrail ends either at an end of an edge of $\delta(Z')$, or at a branch vertex of the full immersion model of $\mathcal{F}$ outside of $Z'$. Since $Z'$ contains every root of $\mathcal{G}$, each branch vertex outside of $Z'$ is the image of a non-root vertex of $\mathcal{F}$, so there are at most $\tau$ of them. Hence, it suffices to show that, for any way of pairing up the edges of $\delta(Z')$ and up to $\tau$ new branch vertices, with any prescribed labelings in $N$ and colorings in $S$, we can find edge-disjoint trails realizing this pairing which together use every edge of $G$ not in $E(Z')$, and that we can do so even after splitting off at $v$. We do this in two steps: first, we route from $\delta(Z')$ to a large clean subgrid without using its edges (Lemmas~\ref{lem:route} and~\ref{lem:route-to-diagonal}), and then we route inside this subgrid using the concentric cycle method, choosing $v$ to be a branch vertex of the subgrid that none of these trails use (\Cref{prop:universal-part}).

\subsection{Routing from a small cut to a subgrid}\label{subsec:route-to-subgrid}
This subsection proves that we can always construct trails from a small cut to a large subgrid without using edges in the target subgrid.
First, we need to define a few new terms. Suppose $\mathcal{G} = (G, \lambda, R, \rho, \alpha)$ is a decorated graph and $(\mathcal{H}, \phi)$ is a grid in $\mathcal{G}$, with branch vertices $T = \phi(V(H))$. We say that $X \subseteq V(G)$ is \textit{$k$-connected} to $T \subseteq V(G)$ if for all $X \subseteq W \subseteq V(G)$ such that $|W \cap T| \le k^2$, we have $|\delta(W)| \ge k$. Call $\mathcal{G}$ \textit{$(\mathcal{H},X,T)$-minimal} if for all $v \in V(G)\setminus( T \cup X)$, if we split off at $v$ such that the branch trails of $\mathcal{H}$ specified by $\phi$ are maintained to obtain a decorated graph $\mathcal{G}'$ from $\mathcal{G}$, then $X$ is no longer $k$-connected to $T$ in $\mathcal{G}'$, where $k=|\delta_G(X)|$.

\begin{proposition}\label{prop:route-to-subgrid}
     Let $k \in \N$ and $b \in \mathbb{Z}^+$ with $b > k$. Let $\mathcal{G} = (G, \lambda, R, \rho, \alpha)$ be a decorated graph, and let $(\mathcal{H}, \phi)$ be an $(A,B)$-clean $S_c$-saturated grid in $\mathcal{G}$ with subspace $N$ and richness, height, and width at least $b(2k^2+5k+1)$, for some disjoint $A,B \subseteq U$ and $S_c \subseteq [q]$. Let $T = \phi(V(H))$ be its set of branch vertices. Let $X \subseteq V(G)$ be $k$-connected to $T$, and let $|\delta(X)| = k$. Then, there is an $(A,B)$-clean $S_c$-saturated grid $(\mathcal{\tilde{H}}, \tilde\phi)$ in $\mathcal{G}$ with subspace $N$ and richness, height, and width at least $b$, whose batches are batches of $\mathcal{H}$, such that \begin{enumerate}
        \item $X$ is disjoint from $\tilde\phi(V(\tilde{H}))$,
        \item there exist $k$ pairwise edge-disjoint trails $P_1,\ldots, P_k$ in $G$ from $X$ to boundary branch vertices of $\tilde{H}$, and
        \item $P_1,\ldots,P_k$ are pairwise edge-disjoint from the branch trails of the underlying grid and vertical-outside edges of $\tilde{H}$.
    \end{enumerate}
    Further, given the full immersion model of $(\mathcal{H}, \phi)$, we can find $(\mathcal{\tilde{H}}, \tilde\phi)$ and $P_1, \ldots, P_k$ in time $\Oh_{|T|, k}(n^2)$.
\end{proposition}

\begin{proof}
We assume that $\mathcal{G}$ is $(\mathcal{H},X,T)$-minimal. If not, we can keep splitting off at vertices $v \in V(G)\setminus( T \cup X)$, maintaining the branch trails of $\mathcal{H}$, until we achieve minimality. Since splitting off does not increase the size of any cut, and splittings at distinct vertices commute, a vertex at which splitting off destroys $k$-connectivity still does so after further splittings. So one pass over $V(G) \setminus (T \cup X)$ suffices.
% Further, let $Z$ be a maximum-sized set such that $X \subseteq Z$, $|Z \cap T| \le k^2$, and $|\delta_G(Z)| = k$.
This gives us the following:

\begin{claim}\label{lem:splitting-off}
  There exists a set $Y \supseteq X$ such that $|Y \cap T| \le k^2$, and $|\delta_G(Y)| = k$, and every vertex not in $Y \cup N(Y)$ is in $T$. Further, we can find $Y$ in $\Oh_{|T|,k}(n^2)$ time.
\end{claim}
\begin{proof}
    If every vertex not in $X \cup N(X)$ is in $T$, we are done, so suppose not. Let $Y \supseteq X$ be a candidate set for satisfying the claim, where $|Y \cap T| \le k^2$, and $|\delta_G(Y)| = k$; we start with $Y = X$. Then, there exists some $v \in V(G) \setminus (Y \cup N(Y) \cup T)$. Consider the graph $G'$ obtained by splitting off at $v$ such that the branch trails of $H$ are maintained. By the minimality of $G$, we have that in $G'$, $X$ is not $k$-connected to $T$. This means that for some $W \supseteq X$ such that $|W\cap T| \le k^2$, $|\delta_{G'}(W)| < k$. By $X$ being $k$-connected to $T$ in $G$, we have $|\delta_G(W)| \ge k$. So, by the 4-regularity of $G$, it must be that exactly 2 edges $e,f$ incident to $v$ are in $\delta_G(W)$, splitting off at $v$ to maintain $H$'s branch trails removes $e,f$ from the cut, and $|\delta_G(W)| = k$. Note that we can find this set $W$ in $\Oh(|T|^{k^2}kn)$ time: it is enough to check, for every set $T_0$ of at most $k^2$ vertices of $T$, whether a minimum cut in $G'$ separating $X \cup T_0$ from $T \setminus T_0$ has size less than $k$. There are $\Oh(|T|^{k^2})$-many such sets, and running at most $k$ iterations of the Ford-Fulkerson max-flow algorithm takes $\Oh(km) = \Oh(kn)$ time, since $G$ is 4-regular \cite{max-flow}.

    If $W \subseteq Y$, then $v \in W \cup N(W) \subseteq Y \cup N(Y)$, a contradiction. So, assume $W \setminus Y \neq \emptyset$.
    By the submodularity of the cut function, we have:
    \begin{align*}
        |\delta(Y)| + |\delta(W)| \ge |\delta(W \cap Y)| + |\delta(W \cup Y)| \\
        2k \ge |\delta(W \cap Y)| + |\delta(W \cup Y)|
    \end{align*}
    By $X$ being $k$-connected to $T$, we have $|\delta(W \cap Y)| \ge k$, so $|\delta(W \cup Y)| \le k$. If $|(W \cup Y) \cap T| \le k^2$, then $|\delta(W \cup Y)| = k$. In this case, we may consider $W \cup Y$ as a candidate set for the claim, and restart. Note that this growing process must terminate in $\Oh(n)$ steps, each taking $\Oh(|T|^{k^2} k n)$ time. Otherwise, $k^2 < |(W \cup Y) \cap T| \le 2k^2$. By \Cref{lem:connGrid} about cuts in a graph immersing a large grid-like graph, since $|\delta(W \cup Y)| \le k$, we have that either $|(W\cup Y) \cap T| \le k^2$, or $|(V(G) \setminus (W\cup Y)) \cap T| \le k^2$, a contradiction, since $|(V(G) \setminus (W\cup Y)) \cap T| \ge |T| -2k^2 > k^2$.
\end{proof}

Now, we will find a large enough ``untouched subgrid'' outside of $Y \cup N(Y)$ to which we can route the edges leaving $X$.

\begin{lemma}\label{lem:smallsideofcut2}
    All but at most $2k^2 + 5k$ branch trails of $H$ are fully contained in $V(G) - (Y \cup N(Y))$.
\end{lemma}
\begin{proof}
Since $|Y \cap T| \le k^2 $, and $|\delta(Y)|=k$, we have that $|(Y \cup N(Y)) \cap T| \le k^2 + k$. So, at most $2k^2+ 2k$ branch trails are fully contained in $Y \cup N(Y)$.
Further, by the 4-regularity of $G$, $|\delta(Y \cup N(Y))| \le 3k$. So, at most $3k$ branch trails of $H$
cross $\delta(Y \cup N(Y))$. Summing these, all but at most $2k^2 + 5k$ branch trails are fully contained in $V(G) - (Y \cup N(Y))$.
 \end{proof}

Most of the branch trails being in $V(G) - (Y \cup N(Y))$ gives us that a large contiguous part of $H$ is contained in $V(G) - (Y \cup N(Y))$:

\begin{lemma}\label{lem:preservedgrid2}
    There is an $(A,B)$-clean $S_c$-saturated grid $(\mathcal{\tilde{H}}, \tilde\phi)$ in $\mathcal{G}$ with subspace $N$ and richness, height, and width at least $b$, whose batches are batches of $\mathcal{H}$, such that all branch trails of $\tilde H$'s underlying grid and vertical outside edges are fully contained in $V(G) - (Y \cup N(Y))$.
\end{lemma}

\begin{proof}
    Let $H_1, \ldots, H_r$, with $r \ge b(2k^2+5k+1)$, be the disjoint subgrids of $H$ containing the ends of the batches $F_1, \ldots, F_r$ of $\mathcal{H}$, ordered from left to right. We cut $H$ into $2k^2+5k+1$ disjoint subgrids of the same height as $H$, which is at least $b$, each consisting of consecutive columns and containing at least $b$ of the subgrids $H_i$; so each of these has width at least $b$. Each branch trail of the underlying grid or of a vertical outside edge of $H$ lies in at most one of these subgrids, so by \Cref{lem:smallsideofcut2}, one of them must have all edges of its underlying grid and outside vertical edges in $V(G) - (Y \cup N(Y))$. Let this grid be $\tilde{H}$, with horizontal outside edges not yet specified. Just as in \Cref{obs:untouchedsubgrid}, we can find trails for the outside-horizontal edges as shown in \Cref{fig:growingA3}, using the remaining branch trails of $H$; this gives $\tilde\phi$. We take as the batches of $\mathcal{\tilde{H}}$ the batches $F_i$ with $H_i \subseteq \tilde{H}$. The branch trails of the underlying grid of $\tilde{H}$ and of these batches are unchanged, so $(\mathcal{\tilde{H}}, \tilde\phi)$ is an $(A,B)$-clean $S_c$-saturated grid in $\mathcal{G}$ with subspace $N$ and richness at least $b$.
\end{proof}

Finally, we can route from $X$ to $\tilde{H}$ with the following:
\begin{lemma}\label{lem:kconn}
    There exist $k$ edge-disjoint trails $P_1,\ldots, P_k$ from $X$ to boundary vertices of $\tilde{H}$ such that $P_1,\ldots,P_k$ are pairwise edge-disjoint from the edges of the underlying grid and vertical edges of $\tilde{H}$.
\end{lemma}
\begin{proof}
By $X$ being $k$-connected to $T$, we have that for all $X \subseteq W \subseteq V(G)-\tilde\phi(V(\tilde{H}))$, such that $|W \cap T| \le k^2$, $|\delta(W)| \ge k$. Then, consider a set $X \subseteq W \subseteq V(G)-\tilde\phi(V(\tilde{H}))$ such that $|W \cap T| > k^2$. Taking the contrapositive of \Cref{lem:connGrid}, since $|W \cap T| > k^2$ and $|\bar{W} \cap T| \ge |V(\tilde{H})| > k^2$, we have that $|\delta(W)| \ge k$, as required. By Menger's theorem, there exist $k$ edge-disjoint trails $P_1,\ldots, P_k$ from $X$ to $\tilde\phi(V(\tilde{H}))$. Shorten each $P_i$ to end at the first vertex in $\tilde\phi(V(\tilde{H}))$ it reaches, for $i \in [k]$. Then, since all branch trails of $\tilde{H}$'s underlying grid and vertical edges are in $V(G)-Y-N(Y)$, by Claim \ref{lem:splitting-off}, these branch trails only contain vertices in $T$. So by the 4-regularity of $G$, these branch trails contain no internal vertices. Hence, $P_1,\ldots,P_k$ must end at boundary branch vertices of $\tilde{H}$, where at most two of them end at any corner and otherwise their ends are distinct, without intersecting edges of the underlying grid and vertical edges of $\tilde{H}$.
\end{proof}
This concludes the proof of \Cref{prop:route-to-subgrid}.
\end{proof}

We apply \Cref{cor:allsignaturesroots} with
$$\psi(x) \coloneqq 2M(x+4\tau)\,(2x^2+5x+1), \qquad \text{where } M(x) \coloneqq 50x(1+|U|+q)+10,$$
which is non-decreasing and satisfies $\psi(x) > x$. The extra $4\tau$ makes room in the grid for the up to $\tau$ non-root branch vertices of $\mathcal{F}$. Let $(\mathcal{\tilde{H}}, \tilde\phi)$, $A$, $S$, $N$, $\lambda'_{U \setminus A}$ and $Z'$ be as given by \Cref{cor:allsignaturesroots}, let $X \coloneqq Z'$ and $k \coloneqq |\delta(X)|$, and let $t \coloneqq \psi(k)$. So $\mathcal{\tilde{H}}$ is an $(A, U \setminus A)$-clean $S$-saturated grid of height, width, and richness at least $t$. Note that if $\tilde{H}$ has height, width, and richness strictly greater than $t$, we can easily find a grid $(\mathcal{\tilde{H}}', \tilde\phi')$ of height, width, and richness exactly $t$, fully immersed by $\mathcal{G}$, such that the branch trails of $\tilde H'$'s underlying grid and vertical edges given by $\tilde{\phi}$   are contained in those of $\tilde\phi'$. So, we may assume $\tilde{H}$ has height, width, and richness exactly $t$. Let $T$ denote the branch vertices of $\tilde{H}$, so $|T| = t^2$.

We want to apply \Cref{prop:route-to-subgrid} to a superset of $X$ and $\tilde{H}$. Let $b \coloneqq 2M(k+4\tau)$.
% Note that by \Cref{cor:allsignaturesroots}(v), $|X \cap T| \le k^2$, and $X$ contains every root of $\mathcal{G}$.

\begin{lemma}\label{lem:route}
    There exist an $X \subseteq X' \subseteq V(G)$ with $|\delta(X')| = k' \le k$ and an $(A, U \setminus A)$-clean $S$-saturated grid $(\mathcal{H'}, \phi')$ in $\mathcal{G}$ with subspace $N$ and height, width, and richness at least $b$, such that \begin{enumerate}
        \item $X'$ is disjoint from $\phi'(V(H'))$,
        \item there exist $k'$ pairwise edge-disjoint trails $P_1,\ldots, P_{k'}$ in $G$ from $X'$ to boundary branch vertices of $H'$, and
        \item $P_1,\ldots,P_{k'}$ are pairwise edge-disjoint from the branch trails of the underlying grid and vertical-outside edges of $H'$.
    \end{enumerate}
\end{lemma}
\begin{proof}
We want to apply \Cref{prop:route-to-subgrid}, so we start with the following:

\begin{observation}
    We can obtain a set $X' \supseteq X$ that is $k'$-connected to $T$, where $k' \coloneqq |\delta(X')| \le k$, in $\Oh_{|T|,k}(n)$ time.
\end{observation}
\begin{proof}
    First note that we can verify if a candidate set $Y \supseteq X$ is $k$-connected to $T$ by testing for the value of a minimum $(W,R)$-cut in $G$, where $W$ is the union of $Y$ and every possible subset of $T$ of size at most $k^2 - |X \cap T|$, and $R$ is the rest of the vertices in $T$. $Y$ is $k$-connected to $T$ if and only if the minimum $(W,R)$-cut value across all of our tested sets $W$ and $R$ is at least $k$, given that $|\delta(Y)| \ge k$. This takes $\Oh(|T|^{k^2} k n)$ time, since each minimum cut test runs at most $k$ iterations of Ford-Fulkerson's algorithm, which each take $\Oh(n)$ time, since $G$ is 4-regular.
    So, we start by testing if $X$ itself is $k$-connected to $T$. If it is, we are done. Otherwise, we find an $(W,R)$-cut, $(Y, \bar{Y})$ such that $|\delta(\bar{Y})| = k' < k$; note that $Y \supseteq X$, since $W \supseteq X$. By \Cref{lem:connGrid}, it must be that $|Y \cap T| \le k'^2$. We repeat this process on $Y$, now testing whether $Y$ is $k'$-connected to $T$, noting that this process clearly terminates, and it can repeat at most $k$ times, since $k'$ strictly decreases, until we obtain a set $X' \supseteq X$ that is $k'$-connected to $T$, where $k' = |\delta(X')| \le k$.
\end{proof}

Since $t = \psi(k) = b(2k^2+5k+1) \ge b(2k'^2+5k'+1)$ and $b > k \ge k'$, we can apply \Cref{prop:route-to-subgrid} to $X'$ (with $k'$ in place of $k$) and $\tilde{H}$. This gives $k'$ edge-disjoint trails $P_1,\ldots, P_{k'}$ from $X'$ to boundary vertices of a grid $(\mathcal{H'}, \phi')$ in $\mathcal{G}$, such that $P_1,\ldots,P_{k'}$ are pairwise edge-disjoint from the edges of the underlying grid and vertical edges of $H'$, and $\mathcal{H'}$ is an $(A, U \setminus A)$-clean $S$-saturated grid in $\mathcal{G}$ with subspace $N$ and height, width, and richness at least $b$.

\end{proof}

From now on, we write $X$ for $X'$ and $k$ for $k'$; since $M$ is non-decreasing, we still have $b \ge 2M(k+4\tau)$. We want to further route to a slightly narrower subgrid, with the endpoints of our trails ``close to the middle'', so that we can see which connected components remain in the graph after removing trails more clearly.

\begin{lemma}\label{lem:route-to-diagonal}
    There exists a set of edge-disjoint trails $P'_1, \ldots, P'_k$ from $X$ to an $(A, U \setminus A)$-clean $S$-saturated grid in $\mathcal{G}$, $(\mathcal{H}'', \phi'')$, with subspace $N$, height at least $M(k+4\tau)$, and width and richness at least $M(k+4\tau)-2$ such that \begin{enumerate}
        \item $P'_1, \ldots, P'_k$ do not intersect the edges in the underlying grid of $H''$ or its outside-vertical edges, and
        \item the endpoints of $P'_1, \ldots, P'_k$ lie on distinct rows of $H''$, and are each horizontally in line with a vertex in the middle $k \times k$ subgrid of $H''$.
    \end{enumerate}
\end{lemma}
\begin{proof}
Order the batches of $\mathcal{H}'$ from left to right according to their subgrids $H_i$. Consider the grid $(\mathcal{H''}, \phi'')$ in $\mathcal{G}$ whose underlying grid has the same height as $\mathcal{H}'$ and consists of the columns from the first column of the subgrid of batch $\lceil b/4 \rceil + 1$ to the last column of the subgrid of batch $b - \lceil b/4 \rceil$, with these batches, and with outside edges routed as in \Cref{fig:growingA3}; i.e. the ``middle'' subgrid. So, $\mathcal{H''}$ is $(A, U \setminus A)$-clean and $S$-saturated with subspace $N$, it has height at least $b \ge M(k+4\tau)$, and it has width and richness at least $b - 2\lceil b/4 \rceil \ge b/2 - 2 \ge M(k+4\tau) - 2$. Moreover, on each side of $\mathcal{H}''$ there are at least $\lceil b/4 \rceil$ subgrids $H_i$, and hence at least $b/4 > k$ columns, of $\mathcal{H}'$.

Then, order the end vertices of $P_1,\ldots,P_k$ as $l_1,\ldots,l_{k_L}$ and $r_1,\ldots,r_{k_R}$, on the left and right boundaries of $\mathcal{H'}$ respectively, by their relative heights on the grid $\mathcal{H'}$, where $k_L+k_R=k$. I.e., an end vertex of $P_1,\ldots,P_k$ on the left boundary closest to the top left corner would be $l_1$. Then, consider the middle $k$ rows of the underlying grid of $\mathcal{H}''$. Label the left boundary vertices on the first $k_L$ of these rows, and the right boundary vertices on the last $k_R$ of these rows, by $l'_h$ and $r'_h$, respectively, where $h$ is their relative height among these vertices. For example, of these vertices on the left boundary, the one closest to the top-left corner of $\mathcal{H''}$ is denoted $l'_1$. It is easy to verify that since $b/4 > k$, there is a way to route edge-disjoint $l_il'_i$ and $r_jr'_j$ trails using only edges in the underlying subgrid of $\mathcal{H'}$, without intersecting the underlying subgrid of $\mathcal{H''}$, for $i \in [k_L], j \in [k_R]$, as in Figure \ref{fig:route-to-inner-grid}. Finally, let $P'_1,\ldots,P'_k$ be $P_1,\ldots,P_k$ with the relevant $l_il'_i$ or $r_jr'_j$ path appended to it.

\begin{figure}[ht]
\centering
\includegraphics[width=0.9\textwidth]{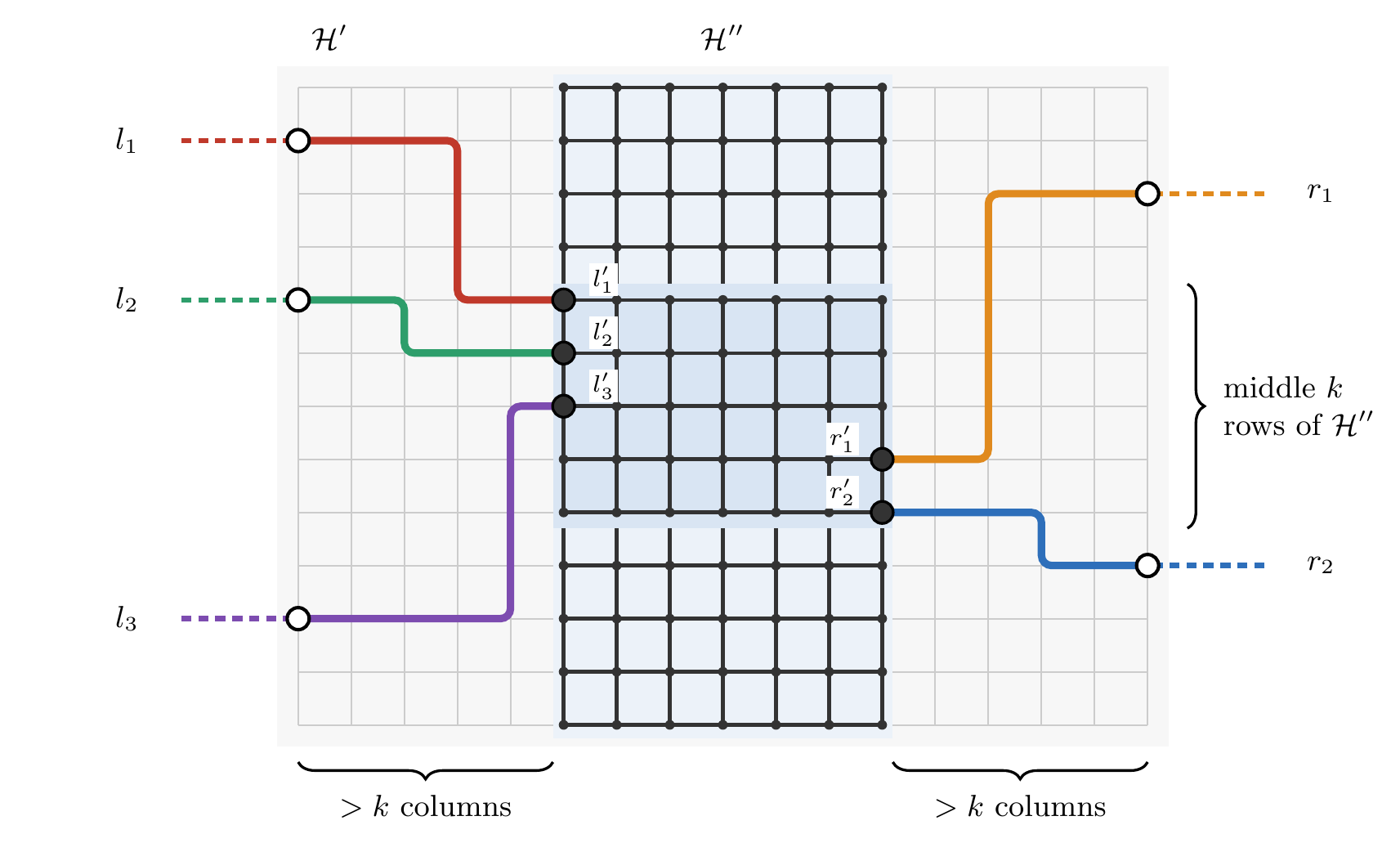}
\caption{Routing the ends of $P_1,\ldots,P_k$ to the middle $k$ rows of $\mathcal{H}''$. The ends $l_1,\ldots,l_{k_L}$ and $r_1,\ldots,r_{k_R}$ on the boundary of $\mathcal{H}'$ are joined, in order, to $l'_1,\ldots,l'_{k_L}$ and $r'_1,\ldots,r'_{k_R}$ by edge-disjoint trails that use only the more than $k$ columns of $\mathcal{H}'$ on either side of $\mathcal{H}''$.}
% \label{fig:diagonal}
\label{fig:route-to-inner-grid}
\end{figure}

\end{proof}

\subsection{Routing inside the subgrid} \label{subsec:routing}

Thus far, we have shown how to construct trails from a set of vertices $X$ containing the root vertices of $\mathcal{G}$ and at least one end of each of its ``problematic'' edges to a large $(A,U \setminus A)$-clean $S$-saturated grid in $\mathcal{G}$ with subspace $N$. Now, we show that for any way of pairing up the edges of $\delta(X)$ and up to $\tau$ new branch vertices, with any prescribed labelings in $N$ and colorings in $S$, we can find edge-disjoint trails realizing this pairing which together use every edge of $G$ not in $E(X)$, and that we can do so even after splitting off at a chosen vertex $v$.

\begin{proposition}\label{prop:universal-part}
Let $\mathcal{G} = (G, \lambda, R_G, \rho_G, \alpha_G)$ be a decorated graph, let $\tau \in \N$, let $A \subseteq U$ and $S \subseteq [q]$, and let $\lambda'_{U \setminus A}$ be a shifting of $\lambda_{U \setminus A}$. Let $X \subseteq V(G)$ and $k \coloneqq |\delta(X)|$, and suppose that
\begin{enumerate}
    \item $R_G \subseteq X$,
    \item $(\mathcal{H}'', \phi'')$ is an $(A, U \setminus A)$-clean $S$-saturated grid in $\mathcal{G}$ with subspace $N$ and height, width, and richness at least $M(k+4\tau)-2$, and $X$ is disjoint from $\phi''(V(H''))$,
    \item $\{e \in E(G): \lambda_A(e) \times \lambda'_{U \setminus A}(e)\notin N\} \subseteq E(X) \cup \delta(X)$,
    \item $\{e \in E(G): \alpha_i(e) = 1 \text{ for some } i \in [q] \setminus S\} \subseteq E(X) \cup \delta(X)$, and
    \item there exist edge-disjoint trails $P'_1, \ldots, P'_k$ such that each edge of $\delta(X)$ is the first edge of exactly one of them, and they end on the left and right boundaries of the underlying grid of $H''$ and satisfy (1) and (2) of \Cref{lem:route-to-diagonal}.
\end{enumerate}
Then there exists a vertex $v \in V(G) \setminus X$ such that the following holds for each of the three decorated graphs $\mathcal{G}_1, \mathcal{G}_2, \mathcal{G}_3$ obtained from $\mathcal{G}$ by splitting off at $v$. For every decorated graph $\mathcal{F}$ on at most $\tau$ vertices and every full immersion model $\phi$ of $\mathcal{F}$ in $\mathcal{G}$, there is a full immersion model $\phi_j$ of $\mathcal{F}$ in $\mathcal{G}_j$, for each $j \in [3]$, such that
\begin{enumerate}
    \item $\phi_j(u) = \phi(u)$ for every $u \in V(F)$ with $\phi(u) \in X$, and
    \item every subtrail of a branch trail of $\phi$ that is contained in $E(X) \cup \delta(X)$ is also a subtrail of the corresponding branch trail of $\phi_j$.
\end{enumerate}
\end{proposition}

\begin{proof}

Let $\mathcal{H''}$ and $P'_1, \ldots, P'_k$ be as in the statement. Then, let $D \coloneqq k + \tau$, and consider the middle $D \times D$ subgrid in the center of $H''$, which we denote by $\Delta$, and the $D$ vertices on its diagonal. We want to route each of $P'_1,\ldots,P'_k$ to one of these diagonal vertices, and use the remaining $\tau$ diagonal vertices to place non-root branch vertices of $\mathcal{F}$. So, we use concentric cycles, as in the proof of universality (\Cref{prop:universality}; see also~\cite[Proposition~5.2.1]{rosethesis}) to both extend $P'_1,\ldots,P'_k$ to a set of diagonal vertices, and construct trails between these diagonal vertices with any labelings in $N$ and colors in $S$.

\begin{figure}
\centering
    \begin{tikzpicture}[scale=1, every node/.style={MyNode, inner sep=.75}]
    \def \m {7}
    \def \n {10}
    \def \width {.75cm}
    \foreach \i in {0,...,\n} %Draw nodes
    	{\foreach \j in {0,...,\m}{
      	\node[draw,circle] (A\i\j) at (\i*\width,-\j*\width) {};
    }}
    \pgfmathparse{int(\n-1)}
    \foreach \i in {0,...,\pgfmathresult}%half-graph edges
    	{\foreach \j in {0,...,\m}{
        	\pgfmathparse{int(\i+1)}\edef\next{\pgfmathresult}
            \draw (A\i\j) -- (A\next\j);
    }}
    \pgfmathparse{int(\m-1)}
    \foreach \i in {0,...,\n}%half-graph edges
    	{\foreach \j in {0,...,\pgfmathresult}{
        	\pgfmathparse{int(\j+1)}\edef\next{\pgfmathresult}
            \draw (A\i\j) -- (A\i\next);
    }}
    \draw[MyRedArc] (A11) -- (A91) -- (A96) -- (A16) -- (A11);
    \draw[MyRedArc] (A22) -- (A82) -- (A85) -- (A25) -- (A22);
    \end{tikzpicture}
\caption{The underlying grid of $H''$, with the concentric cycles at distances~$1$ and~$2$ (bold red).}
\label{fig:rerouting1}
\end{figure}
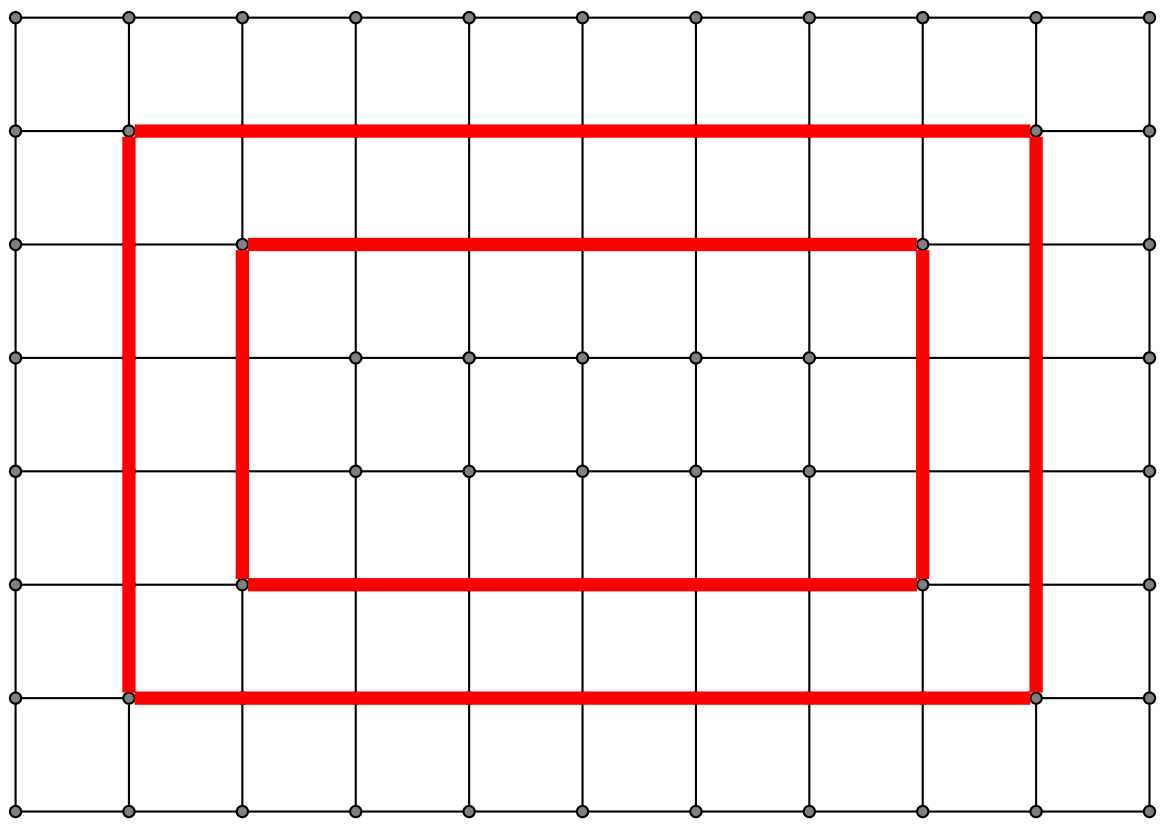

We would like to construct subtrails inside $H''$, where each subtrail will either join one of the end vertices $l'_1,\ldots,l'_{k_L}, r'_1,\ldots,r'_{k_R}$ of $P'_1,\ldots,P'_k$ to a diagonal vertex, or join two diagonal vertices. Each subtrail will have at least one end at a diagonal vertex, and each diagonal vertex will be the end of at most four subtrails, so there will be at most $4D$ subtrails.

First consider a subtrail $Q$ from an end vertex of one of $P'_1,\ldots,P'_k$, say $l'_i$. By \Cref{lem:route-to-diagonal}, $l'_i$ is horizontally in line with a unique vertex of the middle $k \times k$ subgrid of $H''$, and hence with a unique diagonal vertex $y$ of $\Delta$; we assign $y$ to $l'_i$. The subtrail $Q$ starts at $l'_i$, with its first edge being the horizontal edge going into the underlying grid of $\mathcal{H''}$, and continues horizontally until reaching $y$. Subtrails starting at a vertex in $\{r'_j : j \in [k_R]\}$ are constructed in the same fashion.

Then, call subtrails joining two diagonal vertices \textit{intermediate trails}.
To build our intermediate trails, we recall the definition of a concentric cycle in a grid. Let $(n_1,n_2)$ be the bottom-right vertex of $H''$. For each $i \ge 1$ such that the subgrid with top-left vertex $(1+i,1+i)$ and bottom-right vertex $(n_1-i,n_2-i)$ exists, the \emph{concentric cycle at distance $i$} is the cycle consisting of the boundary edges of this subgrid; see \Cref{fig:rerouting1}. Let
$$d \coloneqq 4D(|A|+2|S|+1) + 2, \qquad K \coloneqq k + 4\tau,$$
so $D \le K$, $d \le 8K(1+|U|+q)+2$, and $2d + D \le 17K(1+|U|+q) + 4 \le M(k+4\tau) - 2$. So $H''$ has height and width at least $2d + D$, and the concentric cycles at distances $1, \ldots, d-1$ exist and are disjoint from $\Delta$. We do not use the concentric cycle at distance $1$. We partition the concentric cycles at distances $2, \ldots, d-1$ into $4D$ parts of $|A|+2|S|+1$ consecutive cycles, so that the distances of each part are of the form $i, i+1, \ldots, i+|A|+2|S|$, and fix an injection from the intermediate trails to these parts.

Next we choose batches. Consider just those batches of $\mathcal{H''}$ whose edges all have their ends at distance at least $d$ from the corners of $H''$ and not on a column of $\Delta$. Since the batches of $\mathcal{H''}$ consist of vertical-outside edges with their ends on the top and bottom boundary of $H''$, at most $4d$ batches have an edge with an end too close to a corner of $H''$, and at most $D$ batches have an edge with an end on a column of $\Delta$. So, since $\mathcal{H''}$ has richness at least $M(k+4\tau) - 2 \ge 50K(1+|U|+q)+8 \ge 4D(2|S|+1) + 4d + D$, there are at least $4D(2|S|+1)$ such batches, and we fix an injection from the intermediate trails to $(2|S|+1)$-tuples of these batches. Each of these batches has every edge in a column which meets every concentric cycle at distances $1, \ldots, d-1$, so each such edge can be reached by going directly out from any of these concentric cycles along its column.

Now we show how to build the intermediate trails; see \Cref{fig:rerouting2} for an idea of the construction. Consider an intermediate trail $Q$ between two diagonal vertices $y_1, y_2$, with assigned part of distances $i', \ldots, i'+|A|+2|S|$ and assigned batches $F_{Q,1}, F_{Q,2}, \ldots, F_{Q,2|S|+1}$, and suppose we wish to give $Q$ a labeling $\mu \in N$ and a forward and backward coloring $\overrightarrow{\beta}, \overleftarrow{\beta}$ with $\overrightarrow{\beta} \cup \overleftarrow{\beta} \subseteq S$, where, starting at $y_1$, the colors in $\overrightarrow{\beta}$ must show up on the first half-edge, and the colors in $\overleftarrow{\beta}$ on the second half-edge, of an edge traversed by $Q$. We first pick the colors, then correct the label. List the required colors as $(\gamma_1, d_1), \ldots, (\gamma_{r_2}, d_{r_2})$, where $\gamma_j \in S$, $d_j \in \{\text{forward}, \text{backward}\}$, and $r_2 = |\overrightarrow{\beta}| + |\overleftarrow{\beta}| \le 2|S|$. Since $\mathcal{H''}$ is $S$-saturated, for each $j \in [r_2]$ there is an edge $e_j$ in $F_{Q,j+1}$ whose branch trail contains a half-edge of color $\gamma_j$. Traversed from one of its two ends, this branch trail meets $\gamma_j$ on the first half-edge of the corresponding edge, and traversed from the other end, on the second half-edge; we fix the direction in which $Q$ will traverse $e_j$ so that $\gamma_j$ shows up as required by $d_j$. Let $w$ be the sum of the $\lambda_A \times \lambda'_{U\setminus A}$-weights of $e_1, \ldots, e_{r_2}$. Since $\mathcal{H''}$ is $(A,U \setminus A)$-clean, the weights of the edges of $F_{Q,1}$ span $N$, and $\dim(N) = |A|$, there is a set of at most $|A|$ edges of $F_{Q,1}$ whose weights form a basis of $N$. So there exists a set of edges $f_1,\ldots,f_{r_1}$ in $F_{Q,1}$, with $r_1 \le |A|$, whose $\lambda_A \times \lambda'_{U\setminus A}$-weights sum to $\mu - w$. Write $g_1, \ldots, g_{r_1+r_2}$ for the edges $f_1,\ldots,f_{r_1}, e_1,\ldots,e_{r_2}$.

The intermediate trail $Q$ begins at $y_1$, and its first edge is an unused edge of $\Delta$ incident to $y_1$. It proceeds along this direction until hitting the concentric cycle at distance $i'+|A|+2|S|$. It then proceeds along this concentric cycle, then directly out to an end of $g_1$, along $g_1$, and back in to the next concentric cycle, at distance $i'+|A|+2|S|-1$. Here, since we may go directly out from the concentric cycle to either the top or the bottom end of the column of $g_1$, we can traverse $g_1$ in either direction, and we traverse each $e_j$ in the direction fixed above. This repeats until $Q$ has used all of $g_1, \ldots, g_{r_1+r_2}$ and returned to the concentric cycle at distance $i'+|A|+2|S|-(r_1+r_2)$. Finally, $Q$ proceeds along that concentric cycle, and then back to $y_2$ in the same manner, using an unused edge of $\Delta$ incident to $y_2$. Then $Q$ has $\lambda_A \times \lambda'_{U\setminus A}$-weight $\mu$, and, starting at $y_1$, every color in $\overrightarrow{\beta}$ shows up on the first half-edge, and every color in $\overleftarrow{\beta}$ on the second half-edge, of some edge traversed by $Q$.

Due to the choice of the parts and the batches, these intermediate trails exist and are pairwise edge-disjoint. Distinct subtrails use distinct concentric cycles and distinct batches; the segments going directly between a concentric cycle and a batch edge run along the column of that batch edge, which is not a column of $\Delta$; and the segments going directly between $\Delta$ and a concentric cycle run along the row or column of a diagonal vertex, and distinct diagonal vertices lie on distinct rows and columns. They are also edge-disjoint from the set of subtrails $P''$, joining ends of $P'_1, \ldots, P'_k$ to diagonal vertices. It is clear that upon deletion of the edges in intermediate trails and $P''$, the edges in $\mathcal{H''}$'s underlying subgrid and vertical-outside edges remain connected to $\mathcal{H''}$'s outer boundary.

Now that we have shown that we can build trails with arbitrary labelings and colorings, we must show that these trails can be built in a way such that every edge of $G$ belongs to a trail, even after arbitrarily splitting off at a chosen vertex. Consider splitting off a branch vertex $v$ of $H''$ that lies on a corner of the concentric cycle at distance 1. By construction, this vertex does not lie on any of the subtrails. Let the three potential decorated graphs obtained from this splitting off operation be $\mathcal{G}_1, \mathcal{G}_2, \mathcal{G}_3$. Then, consider the connected components left behind by our constructed trails so far, and splitting off at $v$. Let $G'_i$ be the subgraph of $G_i$ obtained by deleting the edges $E(X)$, $E(P'_1),\ldots,E(P'_k)$, and the edges of the subtrails, for $i \in [3]$.

\begin{claim}
    For $i \in [3]$, every connected component of $G'_i$ that is not an isolated vertex either \begin{enumerate}
        \item has weight zero on every edge, or
        \item is incident to the outer boundary of $\mathcal{H''}$'s underlying grid, or
        \item is incident to at least one of $P'_1,\ldots,P'_k$.
    \end{enumerate}
\end{claim}
\begin{proof}
    Let $i \in [3]$ be arbitrary. First, the branch trails of the underlying grid and vertical-outside edges of $\mathcal{H''}$ contain no edge of $P'_1,\ldots,P'_k$, by \Cref{lem:route-to-diagonal}, and no edge of $E(X)$, since every edge of $\delta(X)$ is in one of $P_1, \ldots, P_k$. Upon deletion of the subtrails, the components with edges in $\mathcal{H''}$'s underlying subgrid and vertical-outside edges are either connected to $\mathcal{H''}$'s outer boundary, or are only composed of edges in $\mathcal{H''}$'s underlying subgrid, in which case every edge has weight zero. Moreover, $v$ and its four incident edges are not used by any subtrail, and each neighbor of $v$ lies on the outer boundary of $H''$ or on the concentric cycle at distance $1$, which is not used by any subtrail. Then, all other edges of $G'_i$ belong to branch trails of a horizontal-outside edge of $\mathcal{H''}$ in $\mathcal{G}$, and the ends of these branch trails lie on the outer boundary of $\mathcal{H''}$'s underlying grid. So every remaining connected component $Q$ of $G'_i$ that is not an isolated vertex contains a subtrail of such a branch trail which either reaches an end of that branch trail, and so $Q$ is incident to the outer boundary of $\mathcal{H''}$'s underlying grid, or stops at a vertex incident to a deleted edge of that branch trail. Since the subtrails only use edges of the underlying grid and vertical-outside edges of $\mathcal{H''}$, such a deleted edge lies in one of $P'_1,\ldots,P'_k$. Hence $Q$ is incident to the outer boundary of $\mathcal{H''}$'s underlying grid or to one of $P'_1,\ldots,P'_k$.
\end{proof}

By this claim, we can construct $P'_1,\ldots,P'_k$, and then append an Eulerian circuit of each connected component of $G''_i \coloneqq G'_i$ with the underlying grid and vertical-outside edges of $\mathcal{H}''$ deleted, not containing any boundary vertices of $\mathcal{H}''$, intersecting one of $P'_1,\ldots,P'_k$ at an incident vertex. This may update the labeling and coloring of $P'_1,\ldots,P'_k$, and will inform which labeling and coloring to give the subtrail of each affected $P'_i$ going through the underlying subgrid and vertical-outside edges of $\mathcal{H''}$. Then, we may append every zero-weight component coming from the underlying grid of $\mathcal{H}''$ to an arbitrary trail. Finally, all remaining connected components of $G_i$, after the trails constructed so far have been deleted, are incident to the outer boundary of $\mathcal{H''}$'s underlying grid, and thus form a single connected component. So, this last component must have weight zero, since the total weight of the target full immersion, which is realized by the constructed trails, and $\mathcal{G}$ are equal. Hence, we may also append this component to an arbitrary trail.
\end{proof}

We have everything required to prove Proposition \ref{prop:inessential-vertex-decorated}, which we repeat here for convenience:
\irrelevantprop*
\begin{proof}
    Let $\psi$ and $M$ be as above, let $\psi_R(x) \coloneqq \psi(2x + 8r_G)$, and set
    $$f_{\ref{prop:inessential-vertex-decorated}}(\tau, r_G, |U|, q) \coloneqq f_{\ref{prop:allsignatures}}(\psi_R, |U|, q).$$
    Since $\psi$ depends only on $\tau$, $|U|$ and $q$, this is a function of $\tau$, $r_G$, $|U|$ and $q$. By \Cref{cor:allsignaturesroots}, we obtain $A$, $S$, $N$, $\lambda'_{U\setminus A}$, a set $X \coloneqq Z' \supseteq R_G$ and a clean grid of height, width and richness at least $t = \psi(|\delta(X)|)$. By \Cref{lem:route} and \Cref{lem:route-to-diagonal}, we obtain trails $P'_1, \ldots, P'_k$ and a clean grid $(\mathcal{H}'', \phi'')$ satisfying the hypotheses of \Cref{prop:universal-part}, which gives a vertex $v \in V(G) \setminus X$. In particular, $v$ is not a root.

    Let $\mathcal{G}'$ be obtained by splitting off at $v$. If $\mathcal{F}$ is in the $\tau$-color-folio of $\mathcal{G}'$, then it is in that of $\mathcal{G}$: replacing each new edge by the two edges it was formed from turns a full immersion model in $\mathcal{G}'$ into one in $\mathcal{G}$ with the same weights and forward and backward colors, by the definition of $\lambda'$ and $\alpha'$. Conversely, if $\mathcal{F}$ is in the $\tau$-color-folio of $\mathcal{G}$, then \Cref{prop:universal-part} gives a full immersion model of $\mathcal{F}$ in $\mathcal{G}'$.
    
    \emph{Running time.} Since $\psi_R$ depends only on $\tau$, $r_G$, $|U|$ and $q$, \Cref{cor:allsignaturesroots} takes $\Oh_{\tau, r_G, |U|, q}(n^4)$ time, and trimming $\tilde{H}$ to height, width and richness exactly $t$ takes $\Oh(n)$ time. Since $k \le g_{\ref{prop:allsignatures}}(\psi_R, |U|, q) + 4r_G$ and $|T| = \psi(k)^2$, both $k$ and $|T|$ are bounded in terms of $\tau$, $r_G$, $|U|$ and $q$. So the observation in \Cref{lem:route} takes $\Oh(n)$ time, \Cref{prop:route-to-subgrid} takes $\Oh(n^2)$ time, and \Cref{lem:route-to-diagonal} and choosing $v$ in \Cref{prop:universal-part} take $\Oh(n)$ time. In total, the running time is $\Oh_{\tau, r_G, |U|, q}(n^4)$.
\end{proof}

\section{Dealing with 2-edge cuts in \texorpdfstring{$G$}{G}}\label{sec:2-cuts}

By the above sections, we can test whether a decorated graph $\mathcal{H}$ is fully immersed by a 4-edge-connected decorated graph $\mathcal{G}$ in time $\Oh_{|V(H)|, |U|, q}(n^5)$. Since our goal is efficiently testing for a full immersion of an arbitrary colorless decorated graph $\mathcal{H}$ within an arbitrary colorless decorated graph $\mathcal{G}$, we will show a reduction from one problem to the other, and prove:

\begin{restatable}{proposition}{dynamicprop}\label{prop:dynamic}
    Let $\mathcal{G}$, $\mathcal{H}$ be colorless decorated graphs, with $|V(H)| = \delta_0$. Then, one can verify if $\mathcal{H}$ is fully immersed by $\mathcal{G}$ in $\Oh_{\delta_0, |U|}(n^5)$ time.
\end{restatable}

Note that since labeled tour graphs are colorless decorated graphs, this is equivalent to Theorem \ref{thm:main}. Throughout this section, we write $\delta_0 \coloneqq |V(H)|$. Let $\mathcal{G}= (G, \lambda, R, \rho, \alpha)$ be an arbitrary colorless decorated graph. Since the roots of $\mathcal{G}$ correspond to vertices of $\mathcal{H}$, we may assume that $\rho(R) \subseteq [\delta_0]$, and that every root of $\mathcal{G}$ corresponds to a root of $\mathcal{H}$ in any full immersion we are looking for; that is, $\rho(R) = \rho_H(R_H)$. If $G$ is 4-edge-connected, we are done by \Cref{prop:col-folio-alg}, since $\mathcal{H}$ is fully immersed by $\mathcal{G}$ if and only if $\mathcal{H}$ is in the $\delta_0$-color-folio of $\mathcal{G}$. Otherwise, $G$ must contain a 2-edge cut. The main idea is that we can split up our problem by looking at both sides of the cut and merging the solutions somehow. For this, we will define an auxiliary graph and explain how to process it.

\subsection{Graph processing and augmentation}\label{sec:cactus}

We will decompose $G$ using an auxiliary cactus graph $G_C$, as in~\cite{cactus}, where each vertex represents a set of vertices in $G$ that are pairwise 4-edge-connected in $G$. Formally,~$\cite{cactus}$ gives an algorithm, given a 2-edge-connected graph $G$, that in time linear in the size of $G$, returns an auxiliary graph $G_C=(V_C,E_C)$ such that \begin{enumerate}
    \item each edge $e \in E_C$ lies on exactly one simple cycle of $G_C$,
    \item there is a surjective mapping $\pi: V(G) \rightarrow V_C$ such that $\pi(u)=\pi(v)$ if and only if $G$ contains 4 edge-disjoint $uv$-paths,
    \item there exists a bijection $\iota : E_0 \rightarrow E_C$ where $E_0 \subseteq E(G)$ is the set of all edges belonging to some 2-edge cut of G, such that for every edge $e = uv \in E_0$, its image $\iota(e)$ is an edge of $G_C$ with endpoints $\pi(u)$ and $\pi(v)$, and
\item a pair of edges $e, f \in E_0$ forms a 2-edge cut in $G$ if and only if $\iota(e)$ and $\iota(f)$ belong to the same cycle of $G_C$.
\end{enumerate}

Note that~\cite{cactus} gives the above for 3-edge-connectivity, but since our given graph $G$ is Eulerian, 3-edge-connectivity implies 4-edge-connectivity. Further, since $G$ is 4-regular, this algorithm's runtime is $\Oh(n)$ where $n = |V(G)|$.

Root $G_C$ arbitrarily at a vertex $r$. We work our way up to $r$ from its ``leaf cycles''. By the cactus structure of $G_C$, every cycle $C$ of $G_C$ has a unique node closest to $r$, which we call the \textit{root} of $C$, and every node $v_i \neq r$ is a non-root node of exactly one cycle, which we call its \textit{parent cycle} $P_i$. The following is clear from the cactus structure of $G_C$:
\begin{observation}\label{obs:order}
    There exists an order in which to process the vertices $v_i \in V(G_C)$ such that, when $v_i$ is processed, all other vertices on cycles that $v_i$ lies on, other than $P_i$, have been processed.
\end{observation}

Now, we explain how to process each $v_i$, in the above order. Notice that for each $v_i \in V_C$, we can pair up its incident edges based on whether they appear in the same cycle in $G_C$. Suppose $\deg(v_i)=2k$, so $v_i$ is on $k$ cycles in $G_C$. Let $F_1,\ldots,F_{k}$ be a partition of $v_i$'s incident edges, each set of size two, based on their cycle membership, and write $F_j = (a_j, b_j)$ for $j \in [k]$. For ease of notation, let us always assume that the unique pair of edges belonging to $v_i$'s parent cycle is $(a_1,b_1)$. For every other cycle, $v_i$ is its root; we orient that cycle by traversing it from $v_i$ starting along $a_j$, and we choose the labels of each non-root node $v_p$ on it so that $a_1$ is the edge through which this traversal enters $v_p$. Then, we can focus on each 4-edge-connected component of $G$ somewhat separately using the following. We pair up the edges in $E_0$ corresponding to each pair $F_j$ in the same way: $F'_j = (a'_j,b'_j) :=(\iota^{-1}(a_j), \iota^{-1}(b_j))$ for $j \in [k]$. Let the endpoints of $a'_j, b'_j$ that are mapped to $v_i$ under $\pi$ be $s_j,t_j$, respectively, and write $a'_j = s_js'_j$ and $b'_j = t_jt'_j$. Consider the graph $G'_i$ obtained from $G$ by replacing each pair of edges $F'_j$, $j \in [k]$ by a single edge $e_j=s_jt_j$, and deleting all vertices not in the preimage of $v_i$ under $\pi$. We write $h^s_j$ and $h^t_j$ for the two half-edges of $e_j$, incident to $s_j$ and $t_j$, respectively.
Then, by the cactus structure of $G_C$, we have that $G_i'$ is 4-edge-connected. Informally, we will run our algorithm on each such $G'_i$. To properly combine signatures, we must also consider the connected component of $G - \{a'_j,b'_j\}$ not containing the preimage of $v_i$ under $\pi$. Let this component be denoted $L_{i,j}$. It is clear that the induced subgraph on the preimage of $v_i$, along with $L_{i,j}$ and $a'_j, b'_j$ for $j \in [k]$, partition $G$ into disjoint subgraphs.

\begin{figure}[ht]
\centering
\includegraphics[width=\textwidth]{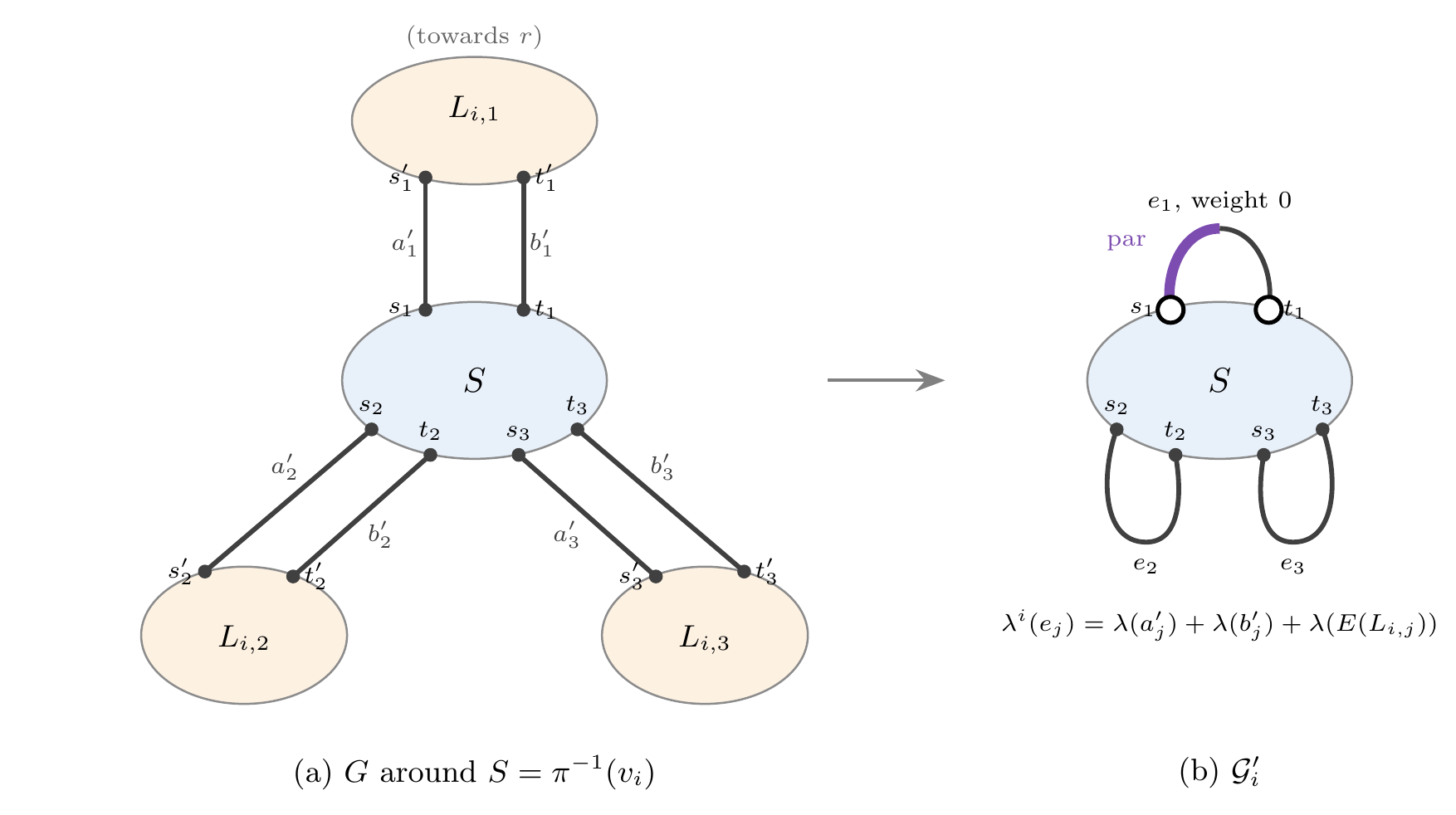}
\caption{A node $v_i$ of $G_C$ lying on its parent cycle and two further cycles. The set $S = \pi^{-1}(v_i)$ is separated from the parts $L_{i,1}, L_{i,2}, L_{i,3}$ of $G$ by the 2-edge cuts $\{a'_j, b'_j\}$, where $a'_j = s_js'_j$ and $b'_j = t_jt'_j$.}
\label{fig:twocut-G}
\end{figure}

See Figure \ref{fig:twocut-G} to see how $G'_i$ and $L_{i,j}$ relate. Formally, using the notation above, for each $v_i$ we consider an auxiliary decorated graph $\mathcal{G}'_i= (G'_i, \lambda^i, R^i, \rho^i, \alpha^i)$, where \begin{enumerate}
    \item $G'_i$ is the auxiliary graph obtained from $v_i$ as described above,
    \item $\lambda^i(e) = \lambda(e)$ for all $e \in E(G) \cap E(G'_i)$,
    \item $\lambda^i(e_j) = \lambda(a'_j) + \lambda(b'_j) + \sum_{e \in E(L_{i,j})}\lambda(e)$ for the virtual edges $e_j$ not corresponding to $v_i$'s parent cycle, and $\lambda^i(e_1) = 0$,
    \item $R^i = (R \cap \pi^{-1}(v_i)) \cup \{s_1,t_1\}$, and
    \item $\rho^i(v) = \rho(v)$ for $v \in R$, and $\rho^i(s_1) \coloneqq \rho^*_1$ and $\rho^i(t_1) \coloneqq \rho^*_2$ for those of $s_1, t_1$ not in $R$, where $\rho^*_1 \coloneqq \delta_0+1$ and $\rho^*_2 \coloneqq \delta_0+2$ are two fixed values, the same for every $v_i$ (if $s_1 = t_1 \notin R$, we only set $\rho^i(s_1) \coloneqq \rho^*_1$),
    \item $\alpha^i(h) = \emptyset$ for all half-edges $h$ of edges in $E(G) \cap E(G'_i)$, and
    \item for the pair of edges $(a'_1,b'_1)$ corresponding to $v_i$'s parent cycle, let $\alpha^i(h^s_1)=\{\mathrm{par}\}$ and $\alpha^i(h^t_1) = \emptyset$, where $\mathrm{par}$ is a fixed color.
\end{enumerate}
For the root $r$, which has no parent cycle, we instead set $R^r = R \cap \pi^{-1}(r)$ and $\rho^r(v) = \rho(v)$ for $v \in R^r$. We also record, for each $v_i$, the set
$$Q_i \coloneqq \rho\Big(R \cap \bigcup_{j \ge 2} V(L_{i,j})\Big) \subseteq [\delta_0]$$
of $\rho$-values of the roots of $\mathcal{G}$ that lie in the components replaced by the virtual edges $e_j$, $j \ge 2$ (for $r$, over all $j \in [k]$). This will help keep track of roots that are ``hidden'' by virtual edges replacing components.

Informally, for each $v_i \in V(G_C)$, we make the vertices $(s_1,t_1)$ (as defined above) roots of $G'_i$. Then, for each virtual edge we add that replaces a 2-edge cut separating $\pi^{-1}(v_i)$ from a connected component $W$ of $G - \{a'_j,b'_j\}$, we give it the labeling of an Eulerian trail of $W$ starting at $a'_j$ and ending at $b'_j$, except for $e_1$, which gets weight $0$.
Notice that we have not specified the coloring $\alpha^i$ of the virtual edges replacing pairs of outgoing edges $(a'_j,b'_j)$ not corresponding to $v_i$'s parent cycle. For this, we first need to define \textit{folios}.

\subsection{Folios}\label{sec:folio}
Note that our input graph is a labeled tour graph, and therefore does not have colored edges (equivalently, $\alpha(h)=\emptyset$ for all half-edges). Our purpose for adding colored edges is to be able to process the input graph in 4-edge-connected components, and use edge colorings to encode the folios of neighboring components. For this, each color will stand for a decorated graph that a neighboring component can realize, which we call a \textit{gadget}. For attaching gadgets, we first need to consider \textit{parent edges}.

\paragraph{Parent edges.} Some decorated graphs below have exactly one half-edge with a non-empty set of colors, namely $\{\mathrm{par}\}$; the edge containing it is the \emph{parent edge}, and the ordered pair of its ends, starting with the end at the colored half-edge, is its \emph{parent pair} (the pair may consist of the same vertex twice). Ignoring the color $\mathrm{par}$, these graphs are colorless.

\paragraph{Gadgets and colors.} A \emph{gadget} is a decorated graph $\mathcal{D}$ with a parent edge $z = s't'$ and parent pair $(s', t')$; we call $z$ its \emph{attachment edge} and $(s', t')$ its \emph{attachment pair} (so $s' = t'$ if $z$ is a loop). Let $\lambda^\circ(\mathcal{D})$ be the sum of the weights of the edges of $\mathcal{D}$ other than $z$. Let $\Gamma$ be the set of equivalence classes of gadgets (colorless apart from $\mathrm{par}$ on $z$) with \begin{enumerate}
    \item at most $\delta_0$ vertices other than $s',t'$,
    \item all $\rho$-values in $[\delta_0+2]$, and
    \item at most one root with each $\rho$-value.
\end{enumerate} So, $|\Gamma|$ is bounded by a function of $\delta_0$ and $|U|$. Importantly, we reserve one color for each equivalence class of gadgets. That is, we let the set of colors be $[q] \coloneqq \Gamma \cup \{\mathrm{par}\}$.

\begin{figure}[ht]
\centering
\includegraphics[width=0.3\textwidth]{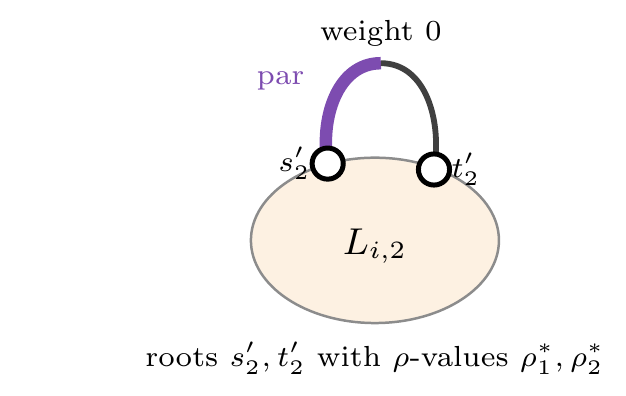}
\caption{The gadget $\mathcal{G}\langle L_{i,2}\rangle$: the part $L_{i,2}$ together with a zero-weight parent edge between its attachment roots $s'_2$ and $t'_2$.}
\label{fig:twocut-Gangle}
\end{figure}

Before seeing how to color virtual edges, we need the following additional auxiliary graphs. Let $X \subseteq V(G)$ with $r \notin \pi(X)$ be such that $\delta(X) = \{e, e'\}$ for two edges $e, e'$ of $E_0$ whose images lie on a common cycle $C$ of $G_C$, where $e$ comes before $e'$ when $C$ is traversed from its root as above. Let $u$ and $u'$ be the ends of $e$ and $e'$ in $X$. We write $\mathcal{G}\langle X\rangle$ for the decorated graph obtained from $G[X]$ by adding a zero-weight edge $uu'$ whose half-edge at $u$ has color $\{\mathrm{par}\}$, as parent edge with parent pair $(u, u')$, with roots $(R \cap X) \cup \{u, u'\}$, where $u, u'$ get $\rho$-values $\rho^*_1, \rho^*_2$ unless they are in $R$ (and only $\rho^*_1$ if $u = u' \notin R$). Informally, $\mathcal{G}\langle X\rangle$ is the part of $G$ inside $X$, with the rest of $G$ contracted to the single edge $uu'$; it is a gadget with attachment pair $(u, u')$. For a component $L$ of $G - \{e, e'\}$ we write $\mathcal{G}\langle L\rangle \coloneqq \mathcal{G}\langle V(L)\rangle$. In particular, for $j \ge 2$, $\mathcal{G}\langle L_{i,j}\rangle$ is the part of $G$ that the virtual edge $e_j$ replaces, with parent pair $(s'_j, t'_j)$; see Figure \ref{fig:twocut-Gangle}.

\begin{figure}[ht]
\centering
\includegraphics[width=0.5\textwidth]{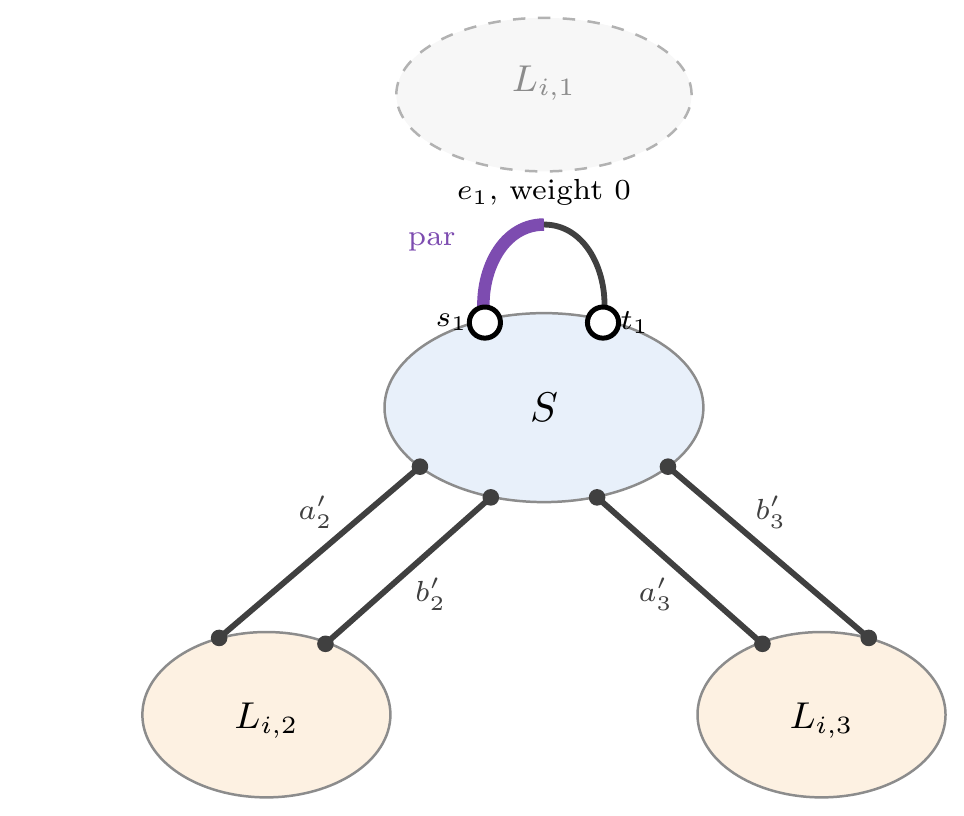}
\caption{The decorated graph $\mathcal{G}^*_i = \mathcal{G}\langle V(G) \setminus V(L_{i,1})\rangle$: the parts $L_{i,2}$ and $L_{i,3}$ and their cut edges are kept, and $L_{i,1}$ is replaced by the parent edge $e_1$.}
\label{fig:twocut-Gstar}
\end{figure}

For a node $v_i \neq r$, let $\mathcal{G}^*_i \coloneqq \mathcal{G}\langle V(G) \setminus V(L_{i,1})\rangle$, whose parent edge is $e_1 = s_1t_1$, and let $\mathcal{G}^*_r \coloneqq \mathcal{G}$, with no parent edge. Informally, $\mathcal{G}^*_i$ is the part of $G$ that $\mathcal{G}'_i$ represents: it agrees with $\mathcal{G}'_i$, except that each virtual edge $e_j$ with $j \ge 2$ is replaced by the part of $G$ it stands for, namely $a'_j$, $b'_j$ and $G[V(L_{i,j})]$. Its roots are $(R \setminus V(L_{i,1})) \cup \{s_1, t_1\}$, with $\rho$-values $\rho^i(R^i) \cup Q_i$; see Figure \ref{fig:twocut-Gstar}.

\paragraph{Folios. }For a decorated graph $\mathcal{G}$ that is colorless apart from possibly a parent edge, the \textit{folio} $\mathcal{P}(\mathcal{G})$ of $\mathcal{G}$ is the set of decorated graphs $\mathcal{F}$, up to equivalence, that are colorless apart from a parent edge if $\mathcal{G}$ has one (and colorless otherwise), with at most $\delta_0$ vertices other than the parent pair, that $\mathcal{G}$ fully immerses. In particular, $\mathcal{P}(\mathcal{G}\langle X\rangle) \subseteq \Gamma$. Since the parent pair of $\mathcal{G}\langle X\rangle$ consists of roots, every model maps the attachment edge of a member of $\mathcal{P}(\mathcal{G}\langle X\rangle)$ to the parent edge $uu'$ alone.

\begin{figure}[ht]
\centering
\includegraphics[width=0.5\textwidth]{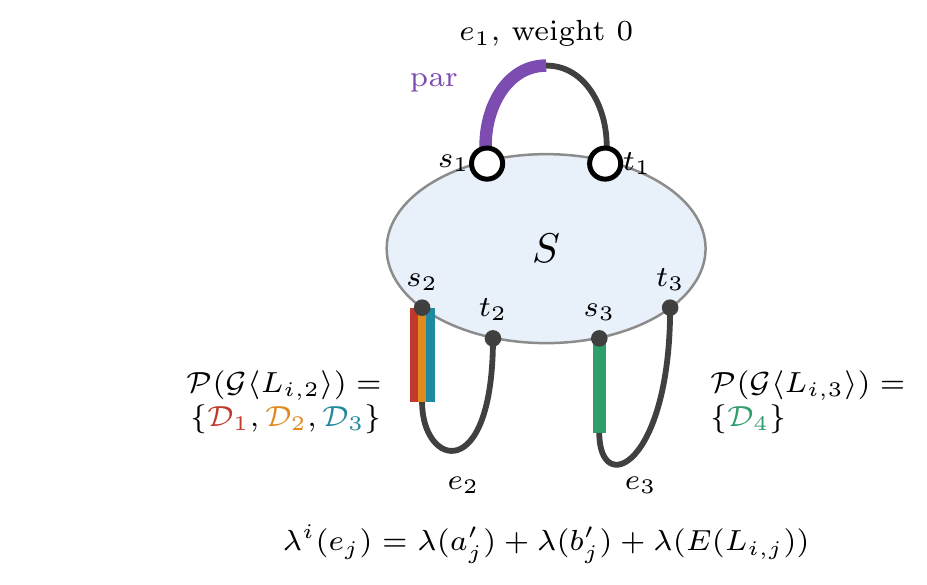}
\caption{The decorated graph $\mathcal{G}'_i$. Each cut $\{a'_j, b'_j\}$ is replaced by a virtual edge $e_j = s_jt_j$. The parent edge $e_1$ has weight $0$ and color $\mathrm{par}$ at $s_1$, and $s_1, t_1$ are roots (open circles). For $j \ge 2$, the half-edge of $e_j$ at $s_j$ is colored by the folio of $\mathcal{G}\langle L_{i,j}\rangle$.}
\label{fig:twocut-Gprime}
\end{figure}

We now explain how to color the virtual edges $e_j$ in the construction of $\mathcal{G}'_i$: for $j \ge 2$ (and for all $j \in [k]$ if $v_i = r$), we set
$$\alpha^i(h^s_j) \coloneqq \mathcal{P}(\mathcal{G}\langle L_{i,j}\rangle), \qquad \alpha^i(h^t_j) \coloneqq \emptyset.$$
Recall that the folio $\mathcal{P}(\mathcal{G}\langle L_{i,j}\rangle)$ is a set of gadget classes, each of which are a color, so $\alpha^i(h^s_j)$ is a set of colors; see Figure \ref{fig:twocut-Gprime}. Further, the colors of $e_j$ lie only on its half-edge at $s_j$; a trail traversing $e_j$ starting at $h^s_j$ enters the replaced component through $a'_j$ at $s'_j$, which is how we remember which end of a gadget is which. We will see in Section~\ref{sec:combine} how to compute $\mathcal{P}(\mathcal{G}\langle L_{i,j}\rangle)$, by assuming that nodes $v_i$ are processed in the order given by \Cref{obs:order}. We now formally define \textit{processing} a node $v_i$ as calculating the folio $\mathcal{P}(\mathcal{G}^*_i)$ of the part of $G$ that $\mathcal{G}'_i$ represents. Clearly, once we can color edges, we can use dynamic programming to obtain a decorated graph corresponding to the root vertex of $G_C$, which we will argue will be enough to obtain the folio of $\mathcal{G}$.

\paragraph{Insertion.} We are only concerned with $\delta'$-color-folios, where $\delta' \coloneqq 2(\delta_0+2)$, so we will abbreviate these to just color-folio, and write $\mathcal{C}(\mathcal{G}'_i) \coloneqq \mathcal{C}_{\delta'}(\mathcal{G}'_i)$. With this in mind, we will see that the folio of the part of $G$ that $\mathcal{G}'_i$ represents, $\mathcal{P}(\mathcal{G}^*_i)$, consists of the decorated graphs that are fully immersed by a decorated graph obtained from a member of the color-folio of $\mathcal{G}'_i$ by ``inserting'' the gadgets named by its colors. Formally, let $\mathcal{F}'' \in \mathcal{C}(\mathcal{G}'_i)$ with labeling $\mu$. The \emph{insertion} $\mathrm{Ins}(\mathcal{F}'')$ is obtained as follows. For each edge $f$ with a half-edge $h_f$ of color $\{\mathcal{D}\}$, where $\mathcal{D} \in \Gamma$ has attachment pair $(s', t')$ and attachment edge $z$, $a$ is the end of $f$ at $h_f$, and $b$ is the other end: \begin{enumerate}
    \item delete $f$,
    \item add a disjoint copy of $\mathcal{D} - z$,
    \item add an edge $as'$ of weight $\mu(f) + \lambda^\circ(\mathcal{D})$, and an edge $t'b$ of weight $0$, both colorless.
\end{enumerate}
The roots of $\mathrm{Ins}(\mathcal{F}'')$ are the roots of $\mathcal{F}''$ together with the roots of the inserted gadgets whose $\rho$-values lie in $Q_i$; so the attachment vertices $s', t'$ are not roots unless they are roots of $\mathcal{G}$. Notice that $\mathrm{Ins}(\mathcal{F}'')$ has at most one half-edge with color $\mathrm{par}$: the attachment edges of the gadgets are deleted, and $\mathcal{F}''$ has at most one.

\begin{lemma}\label{lem:insertion-shift}
    $\mathrm{Ins}(\mathcal{F}'')$ is well defined up to equivalence: replacing $\mathcal{F}''$ by an equivalent graph, or an inserted gadget by an equivalent one, gives an equivalent graph.
\end{lemma}
\begin{proof}
 Consider a gadget $\mathcal{D}$ inserted into $f$, with $a, b, s', t'$ as above. First, let $Y \subseteq V(F'')$. For a decorated graph $\mathcal{F}$, $Y \subseteq V(F)$ and $c \in \mathbb{F}_2^U$, we write $\mathcal{F}^{Y,c}$ for the decorated graph resulting from shifting at $Y$ by adding vector $c$ to $\delta(Y)$ in $\mathcal{F}$. In $\mathcal{F}''^{Y,c}$, the weight of $f$ changes by $c$ if and only if exactly one of $a, b$ is in $Y$. So $\mathrm{Ins}(\mathcal{F}''^{Y,c}) = \mathrm{Ins}(\mathcal{F}'')^{Y^+,c}$, where $Y^+$ is obtained from $Y$ by adding $V(D)$ if $b$ is in $Y$. Second, let $Y \subseteq V(D)$. Since $D$ is 4-regular, $|\delta_D(Y)|$ is even, so $\lambda^\circ(\mathcal{D})$, and hence the weight of $as'$, changes by $c$ if and only if $z \in \delta_D(Y)$, i.e.\ exactly one of $s', t'$ is in $Y$. So inserting $\mathcal{D}^{Y,c}$ instead of $\mathcal{D}$ gives $\mathrm{Ins}(\mathcal{F}'')^{Y,c}$ if $t' \notin Y$, and $\mathrm{Ins}(\mathcal{F}'')^{V(D) \setminus Y,c}$ if $t' \in Y$.
\end{proof}

We will show in Section~\ref{sec:folio-correct} how to compute $\mathcal{P}(\mathcal{G}^*_i)$ from $\mathcal{C}(\mathcal{G}'_i)$.

\subsection{Combining folios along cycles}\label{sec:combine}

To fully construct each $\mathcal{G}'_i$ by coloring virtual edges, we must know the folios of each component ``hanging off of'' $\mathcal{G}'_i$, that is not its parent cycle, meaning $\mathcal{P}(\mathcal{G}\langle L_{i,j} \rangle)$ for each $j \ge 2$. Let $C$ be the (non-parent) cycle at $v_i$ corresponding to the virtual edge $e_j$, so $v_i$ is the root of $C$. Let $v_1, \ldots, v_m$ be the other nodes of $C$, in order from $v_i$ starting along $a_j$, and set $v_0 = v_{m+1} \coloneqq v_i$. For $0 \le p \le m$, let $g_p$ be the edge of $G$ with $\iota(g_p) = v_pv_{p+1}$, and for $p \in [m]$ let $x_p, y_p \in \pi^{-1}(v_p)$ be the ends of $g_{p-1}$ and $g_p$. Then $C$ is the parent cycle of each $v_p$, and $(x_p, y_p) = (s_1, t_1)$ in the construction of $\mathcal{G}'_p$; also $g_0 = a'_j$, so $x_1 = s'_j$ and $y_m = t'_j$. Let $V_p \coloneqq V(\mathcal{G}^*_p)$ and $W_p \coloneqq V_1 \cup \cdots \cup V_p$. Then $\delta(W_p) = \{g_0, g_p\}$, so $\mathcal{G}\langle W_p\rangle$ is the part of $G$ formed by the first $p$ nodes of $C$, with parent pair $(x_1, y_p)$. In particular, $\mathcal{G}\langle W_1\rangle = \mathcal{G}^*_1$ and $\mathcal{G}\langle W_m\rangle = \mathcal{G}\langle L_{i,j}\rangle$.

Every member of $\mathcal{P}(\mathcal{G}\langle W_p\rangle)$ has its attachment pair mapped to $(x_1, y_p)$, and every member of $\mathcal{P}(\mathcal{G}^*_{p+1})$ has its parent pair mapped to $(x_{p+1}, y_{p+1})$; we name these pairs accordingly. For gadgets $\mathcal{D}_1$ with attachment edge $z_1 = x_1y_p$ and $\mathcal{D}_2$ with attachment edge $z_2 = x_{p+1}y_{p+1}$, and labelings $\mu_1, \mu_2$, the \emph{combination} $\mathrm{Comb}(\mathcal{D}_1, \mathcal{D}_2)$ is the gadget obtained by taking one copy of each, deleting $z_1, z_2$, adding the edge $g_p = y_px_{p+1}$ with weight $\lambda(g_p)$, and adding an attachment edge $x_1y_{p+1}$ of weight $\mu_1(z_1) + \mu_2(z_2)$ whose half-edge at $x_1$ has color $\{\mathrm{par}\}$. Its roots are those of $\mathcal{D}_1$ and $\mathcal{D}_2$, except that $y_p$ and $x_{p+1}$ are removed unless they are roots of $\mathcal{G}$ or equal to $x_1$ or $y_{p+1}$, respectively; $x_1$ and $y_{p+1}$ get $\rho$-values $\rho^*_1$ and $\rho^*_2$ unless they are roots of $\mathcal{G}$. Note that $\mathcal{G}\langle W_{p+1}\rangle = \mathrm{Comb}(\mathcal{G}\langle W_p\rangle, \mathcal{G}^*_{p+1})$.

\begin{lemma}\label{lem:comb-shift}
    $\mathrm{Comb}$ is well defined on equivalence classes.
\end{lemma}
\begin{proof}
    Let $Y \subseteq V(D_1)$. In $\mathrm{Comb}(\mathcal{D}_1^{Y,c}, \mathcal{D}_2)$, the edges of $\delta_{D_1}(Y) \setminus \{z_1\}$ change by $c$, and $x_1y_{p+1}$ changes by $c$ if and only if exactly one of $x_1, y_p$ is in $Y$. So $\mathrm{Comb}(\mathcal{D}_1^{Y,c}, \mathcal{D}_2)$ is the shifting of $\mathrm{Comb}(\mathcal{D}_1, \mathcal{D}_2)$ by $c$ at $Y$ if $y_p \notin Y$, and at $Y \cup V(D_2)$ if $y_p \in Y$. The case $Y \subseteq V(D_2)$ is symmetric.
\end{proof}

\begin{lemma}\label{lem:combine-folios}
    For $p \in [m-1]$, $\mathcal{P}(\mathcal{G}\langle W_{p+1}\rangle)$ is the set of gadgets $\mathcal{D}$, up to equivalence, with at most $\delta_0$ vertices other than the attachment pair, such that $\mathrm{Comb}(\mathcal{D}_1, \mathcal{D}_2)$ fully immerses $\mathcal{D}$, for some $\mathcal{D}_1 \in \mathcal{P}(\mathcal{G}\langle W_p\rangle)$ and $\mathcal{D}_2 \in \mathcal{P}(\mathcal{G}^*_{p+1})$. In particular, $\mathcal{P}(\mathcal{G}\langle W_{p+1}\rangle)$ can be computed from $\mathcal{P}(\mathcal{G}\langle W_p\rangle)$ and $\mathcal{P}(\mathcal{G}^*_{p+1})$ in time bounded by a function of $\delta_0$ and $|U|$.
\end{lemma}
\begin{proof}
    Let $\mathcal{D}_1, \mathcal{D}_2$ be as in the statement. By \Cref{lem:comb-shift}, we may assume that there are models $\chi_1, \chi_2$ of $\mathcal{D}_1, \mathcal{D}_2$ themselves in $\mathcal{G}\langle W_p\rangle, \mathcal{G}^*_{p+1}$. Since the ends of $z_1, z_2$ are roots, $\chi_1, \chi_2$ map them to the zero-weight attachment edges of $\mathcal{G}\langle W_p\rangle$ and $\mathcal{G}^*_{p+1}$, so $\mu_1(z_1) = \mu_2(z_2) = 0$. Mapping $g_p$ and $x_1y_{p+1}$ to themselves and all other edges by $\chi_1, \chi_2$ gives a model of $\mathrm{Comb}(\mathcal{D}_1, \mathcal{D}_2)$ in $\mathrm{Comb}(\mathcal{G}\langle W_p\rangle, \mathcal{G}^*_{p+1}) = \mathcal{G}\langle W_{p+1}\rangle$. By transitivity, $\mathcal{G}\langle W_{p+1}\rangle$ fully immerses $\mathcal{D}$, so $\mathcal{D} \in \mathcal{P}(\mathcal{G}\langle W_{p+1}\rangle)$.

    Conversely, let $\phi$ be a model of some $\mathcal{D} \in \mathcal{P}(\mathcal{G}\langle W_{p+1}\rangle)$ in $\mathcal{G}\langle W_{p+1}\rangle$. The only edges between $W_p$ and $V_{p+1}$ are $g_p$ and $x_1y_{p+1}$, and the latter is the trail of the attachment edge of $\mathcal{D}$. So cutting the trails of $\phi$ at each visit to $y_p$ and $x_{p+1}$ splits them into $g_p$, $x_1y_{p+1}$, and trails inside $G[W_p]$ or $G[V_{p+1}]$. Let $\mathcal{D}_1$ be the gadget whose vertices are the branch vertices of $\phi$ in $W_p$ together with $y_p$, whose edges are the trails inside $G[W_p]$ (with their weights) and a zero-weight attachment edge $z_1 = x_1y_p$ with $\mathrm{par}$ at $x_1$, and whose roots are its vertices that are roots of $\mathcal{G}\langle W_p\rangle$. It is 4-regular: each of its vertices has four trail-ends, where the ends at $g_p$ and $x_1y_{p+1}$ are replaced by the ends of $z_1$. Mapping each trail to itself and $z_1$ to $x_1y_p$ is a model of $\mathcal{D}_1$ in $\mathcal{G}\langle W_p\rangle$, and $\mathcal{D}_1$ has at most $\delta_0$ non-attachment vertices, since these are non-attachment vertices of $\mathcal{D}$. So $\mathcal{D}_1 \in \mathcal{P}(\mathcal{G}\langle W_p\rangle)$, and symmetrically we obtain $\mathcal{D}_2 \in \mathcal{P}(\mathcal{G}^*_{p+1})$. Finally, mapping each edge of $\mathcal{D}$ to the sequence of pieces of its trail is a model of $\mathcal{D}$ in $\mathrm{Comb}(\mathcal{D}_1, \mathcal{D}_2)$.

    For the computation, there are at most $|\Gamma|^2$ pairs, and full immersion can be tested by brute force on graphs with at most $2(\delta_0 + 2)$ vertices.
\end{proof}

So, by \Cref{lem:combine-folios}, we can calculate $\mathcal{P}(\mathcal{G}\langle W_m \rangle) = \mathcal{P}(\mathcal{G}\langle L_{i,j}\rangle)$, which tells us how to color the half-edge $h_j^s$ in $\mathcal{G}'_i$, using $m$ applications of \Cref{lem:combine-folios}. Then, once $\mathcal{G}'_i$ has been constructed, its color-folio can be calculated in time $\Oh_{\delta_0, |U|}(n_i^5)$, where $n_i \coloneqq |V(G'_i)|$, by \Cref{prop:col-folio-alg}, applied with $\tau = \delta'$; here $\mathcal{G}'_i$ has at most $\delta_0 + 2$ roots and $q = |\Gamma| + 1$ is bounded in terms of $\delta_0$ and $|U|$. Next, we see how to go from the color-folio of $\mathcal{G}'_i$ to the folio of $\mathcal{G}^*_i$.

\subsection{Converting color-folio to folio}\label{sec:folio-correct}

In this subsection we show that the folio of $\mathcal{G}^*_i$ can be read off from the color-folio of $\mathcal{G}'_i$.

\begin{lemma}\label{lem:folio-correct}
    Suppose $\alpha^i(h^s_j) = \mathcal{P}(\mathcal{G}\langle L_{i,j}\rangle)$ for all $j \ge 2$. Then $\mathcal{P}(\mathcal{G}^*_i)$ is the set of decorated graphs $\mathcal{F}$, up to equivalence, that are colorless apart from a parent edge (colorless if $v_i = r$), such that
    \begin{enumerate}
        \item $\mathcal{F}$ has at most $\delta_0$ vertices other than its parent pair,
        \item $\rho_F(R_F) = \rho^i(R^i) \cup Q_i$, and
        \item $\mathrm{Ins}(\mathcal{F}'')$ fully immerses $\mathcal{F}$, for some $\mathcal{F}'' \in \mathcal{C}(\mathcal{G}'_i)$.
    \end{enumerate}
    In particular, for $v_i = r$, this is $\mathcal{P}(\mathcal{G})$.
\end{lemma}
\begin{proof}
    \noindent\emph{Every $\mathcal{F}$ satisfying (1)--(3) is in $\mathcal{P}(\mathcal{G}^*_i)$.} By \Cref{lem:insertion-shift}, we may assume that there is a model $\psi$ of $\mathcal{F}''$ itself in $\mathcal{G}'_i$. Let $f$ be an edge of $F''$ with a half-edge of color $\{\mathcal{D}_f\}$, $\mathcal{D}_f \in \Gamma$, at its end $a$. The trail $\psi(f)$, read from $\psi(a)$, has $\mathcal{D}_f$ in its forward color, and only the half-edges $h^s_m$, $m \ge 2$, carry colors in $\Gamma$. So $\psi(f) = T_1 e_m T_2$ for some $m = m(f)$ with $\mathcal{D}_f \in \mathcal{P}(\mathcal{G}\langle L_{i,m}\rangle)$, where $e_m$ is traversed from $s_m$ to $t_m$, and subtrails $T_1,T_2$. Since the color of a half-edge on $e_{m(f)}$ dictates the color of $f$, call the edges $e_{m(f)}$ \emph{used}, and other virtual edges $e_j$ in $\mathcal{G}'_i$ \emph{unused}.

    Note that every $e_k$ such that $L_{i,k}$ contains a root of $\mathcal{G}$ is used. Indeed, let $\ell \in Q_i$ be the value of such a root. By (2), $\mathcal{F}$ has a root of value $\ell$, so by (3) so does $\mathrm{Ins}(\mathcal{F}'')$. The roots of $\mathcal{F}''$ have values in $\rho^i(R^i)$, which is disjoint from $Q_i$, so this root lies in some inserted gadget $\mathcal{D}_f$.

    We now build a model of $\mathrm{Ins}(\mathcal{F}'')$ in $\mathcal{G}^*_i$. For each $f$ as above with $m = m(f)$, by \Cref{lem:insertion-shift} we may assume there is a model $\chi_m$ of $\mathcal{D}_f$ itself in $\mathcal{G}\langle L_{i,m}\rangle$. It maps the attachment edge to $s'_mt'_m$, whose ends are roots, and the other edges of $\mathcal{D}_f$ to trails partitioning $E(L_{i,m})$; so $\lambda^\circ(\mathcal{D}_f) = \lambda(E(L_{i,m}))$. Map the copy of $\mathcal{D}_f - z$ by $\chi_m$, the edge $as'$ to $T_1a'_m$, and $t'b$ to $b'_mT_2$. Map every other edge $g$ of $F''$ to $\psi(g)$, and in every trail, replace each unused $e_k$ by an Eulerian trail of $\{a'_k, b'_k\} \cup E(L_{i,k})$ between its ends, which has weight $\lambda^i(e_k)$. These trails are edge-disjoint and cover $E(\mathcal{G}^*_i)$. They have the right weights, except possibly for $as'$ and $t'b$; for these, the trails $T_1a'_m$ and $b'_mT_2$ have total weight
$$w(T_1) + w(T_2) + \lambda(a'_m) + \lambda(b'_m) = \mu(f) + \lambda^i(e_m) + \lambda(a'_m) + \lambda(b'_m) = \mu(f) + \lambda^\circ(\mathcal{D}_f),$$
which is the total weight of $as'$ and $t'b$. Since $\delta(V(D_f)) = \{as', t'b\}$, a shifting at $V(D_f)$ fixes this. The roots of $\mathcal{F}''$ are mapped onto $R^i$, the roots of the inserted gadgets onto the roots of $\mathcal{G}$ in the used $L_{i,m}$, and the unused $L_{i,k}$ contain no roots; so every root of $\mathcal{G}^*_i$ is used. The parent edge of $\mathrm{Ins}(\mathcal{F}'')$ is mapped to a trail traversing $e_1$ from $s_1$, which has $\mathrm{par}$ in its forward color. So $\mathcal{G}^*_i$ fully immerses $\mathrm{Ins}(\mathcal{F}'')$, and by transitivity and (1), $\mathcal{F} \in \mathcal{P}(\mathcal{G}^*_i)$.

    \smallskip
    \noindent\emph{Every $\mathcal{F} \in \mathcal{P}(\mathcal{G}^*_i)$ satisfies (1)--(3).} Items (1) and (2) hold by definition. For (3), we may assume there is a model $\phi$ of $\mathcal{F}$ itself in $\mathcal{G}^*_i$. Let $J$ be the set of $j \ge 2$ such that $V(L_{i,j})$ contains a branch vertex of $\phi$. Since these sets are disjoint, $|J|$ plus the number of branch vertices in $\pi^{-1}(v_i)$ is at most $|V(F)|$. For $k \notin J$, $\phi$ traverses $a'_k$, $E(L_{i,k})$ and $b'_k$ as one segment of a single trail, of weight $\lambda^i(e_k)$. Cut every trail of $\phi$ at each visit to $s_j, t_j, s'_j, t'_j$ for $j \in J$.

    For $j \in J$, let $\mathcal{D}_j$ be the gadget whose vertices are the branch vertices of $\phi$ in $V(L_{i,j})$ together with $s'_j, t'_j$, whose edges are the pieces inside $L_{i,j}$ and a zero-weight attachment edge $s'_jt'_j$ with $\mathrm{par}$ at $s'_j$, and whose roots are its vertices that are roots of $\mathcal{G}\langle L_{i,j}\rangle$. See that this directly gives us a model of $\mathcal{D}_j$ in $\mathcal{G}\langle L_{i,j}\rangle$. Its non-attachment vertices are images of vertices of $F$ outside the parent pair (which maps to $s_1, t_1$), so $\mathcal{D}_j \in \mathcal{P}(\mathcal{G}\langle L_{i,j}\rangle) = \alpha^i(h^s_j)$.

    Let $\mathcal{F}''$ have as vertices the branch vertices of $\phi$ in $\pi^{-1}(v_i)$ and the vertices $s_j, t_j$ for $j \in J$, as roots those of its vertices in $R^i$, and as edges:
    \begin{itemize}
        \item for each $j \in J$, an edge $f_j = s_jt_j$ of weight $\lambda^i(e_j)$, whose half-edge at $s_j$ has color $\{\mathcal{D}_j\}$, with trail $e_j$;
        \item each remaining piece, with ends in $V(F'')$ and with every segment through some $L_{i,k}$, $k \notin J$, replaced by $e_k$; the piece whose trail is $e_1$ (the trail of the parent edge of $\mathcal{F}$) gets $\mathrm{par}$ on its half-edge at $s_1$.
    \end{itemize}
    These trails form a model of $\mathcal{F}''$ in $\mathcal{G}'_i$, and $|V(F'')| \le 2|V(F)| \le \delta'$. So $\mathcal{F}'' \in \mathcal{C}(\mathcal{G}'_i)$.

    Finally, mapping each vertex $u$ of $F$ to $\phi(u)$, and each edge of $F$ to the sequence of pieces of its trail, where $a'_j$ and $b'_j$ are read as the edges $s_js'_j$ and $t'_jt_j$ of $\mathrm{Ins}(\mathcal{F}'')$, is a model of $\mathcal{F}$ in $\mathrm{Ins}(\mathcal{F}'')$, which maps the parent edge of $\mathcal{F}$ to the parent edge of $\mathrm{Ins}(\mathcal{F}'')$. The weights agree after a shifting at each $V(D_j)$, since $\lambda(a'_j) + \lambda(b'_j) = \lambda^i(e_j) + \lambda^\circ(\mathcal{D}_j)$ is the total weight of $s_js'_j$ and $t'_jt_j$ in $\mathrm{Ins}(\mathcal{F}'')$. So (3) holds.

    For $v_i = r$ the same proof applies, with no parent edges and $\mathcal{G}^*_r = \mathcal{G}$.
\end{proof}

\begin{observation}\label{obs:folio-from-color-folio}
    Under the assumption of \Cref{lem:folio-correct}, $\mathcal{P}(\mathcal{G}^*_i)$ can be computed from $\mathcal{C}(\mathcal{G}'_i)$, $\rho^i(R^i)$ and $Q_i$ in time bounded by a function of $\delta_0$ and $|U|$.
\end{observation}
\begin{proof}
    See that each $\mathrm{Ins}(\mathcal{F}'')$ has at most $\delta' + 2\delta'(\delta_0+2)$ vertices. This follows because every decorated graph $\mathcal{F}''$ in $\mathcal{C}(\mathcal{G}'_i)$ has at most $\delta'$ vertices and $2\delta'$ edges, and each edge can be replaced by an inserted gadget with at most $\delta_0+2$ vertices. So, since $\delta' = 2(\delta_0 +2)$, whether it fully immerses a given $\mathcal{F}$ can be decided by brute force, and the number of candidates $\mathcal{F}''$ and $\mathcal{F}$ is bounded in terms of $\delta_0$ and $|U|$.
\end{proof}

\begin{proposition}\label{prop:folio-main}
    $\mathcal{P}(\mathcal{G})$ can be computed in time $\Oh_{\delta_0, |U|}(n^5)$.
\end{proposition}
\begin{proof}
    Process the nodes of $G_C$ in the order of Observation~\ref{obs:order}. For each node $v_i$ and each non-parent cycle $C$ at $v_i$, compute $\mathcal{P}(\mathcal{G}\langle L_{i,j}\rangle) = \mathcal{P}(\mathcal{G}\langle W_m\rangle)$ from the folios $\mathcal{P}(\mathcal{G}^*_1), \ldots, \mathcal{P}(\mathcal{G}^*_m)$ already computed, using $\mathcal{G}\langle W_1\rangle = \mathcal{G}^*_1$ and \Cref{lem:combine-folios}; construct $\mathcal{G}'_i$ with these colors; compute $\mathcal{C}(\mathcal{G}'_i)$ by \Cref{prop:col-folio-alg}; and compute $\mathcal{P}(\mathcal{G}^*_i)$ by Observation~\ref{obs:folio-from-color-folio}. This computes $\mathcal{P}(\mathcal{G}^*_i)$ correctly for every node, and in particular $\mathcal{P}(\mathcal{G})$ at $r$, in time $\Oh_{\delta_0, |U|}(n^5)$.

    Correctness follows by induction on the order, from Lemmas~\ref{lem:combine-folios} and~\ref{lem:folio-correct}. For the running time, $G_C$ is computed in $\Oh(n)$ time, and the labels $\lambda^i(e_j)$ and the sets $Q_i$ can be computed for all nodes in $\Oh(n)$ time, by accumulating them in the order of Observation~\ref{obs:order}. Every node other than $r$ is a non-root node of exactly one cycle, so there are at most $n$ applications of \Cref{lem:combine-folios} and at most $n$ applications of Observation~\ref{obs:folio-from-color-folio} in total, each taking $\Oh_{\delta_0,|U|}(1)$ time. Finally, each $\mathcal{G}'_i$ is 4-edge-connected with at most $\delta_0 + 2 \le \delta'$ roots and $q = |\Gamma| + 1$ colors, so by \Cref{prop:col-folio-alg} with $\tau = \delta'$, computing $\mathcal{C}(\mathcal{G}'_i)$ takes $\Oh_{\delta_0,|U|}(n_i^5)$ time, where $n_i \coloneqq |V(G'_i)|$. Since the sets $V(G'_i) = \pi^{-1}(v_i)$ partition $V(G)$, we have $\sum_i n_i^5 \le n^5$, so the total running time is $\Oh_{\delta_0,|U|}(n^5)$.
\end{proof}

Since $|V(H)| = \delta_0$ and $\mathcal{H}$ is colorless, $\mathcal{H}$ is fully immersed by $\mathcal{G}$ if and only if $\mathcal{H} \in \mathcal{P}(\mathcal{G})$, which gives the following:

\dynamicprop*

\section{Conclusion}
\label{sec:conclusion}

In this section we prove the remaining results mentioned in the extended abstract.

\subsection*{The sufficiency of irrelevant vertices for vertex-minor testing}

In the extended abstract we mentioned that, in order to find a non-uniform FPT algorithm for vertex-minor testing, it suffices to find irrelevant vertices. More specifically, fix $\ell\geq 2$ and suppose that we could find an $H$-irrelevant vertex in time $\mathcal{O}_H(n^\ell)$ when additionally given as input a sufficiently large (relative to $H$) comparability grid as an induced subgraph of our $n$-vertex host graph $G$. Then we can test for $H$ as follows.

By the grid theorem for vertex-minors \cite{GEELEN202393}, there exists a constant $k_H$ so that every graph of rank-width at least $k_H$ has a vertex-minor isomorphic to a comparability grid which is large enough to yield the irrelevant vertex. By \cite{almostLinearRw}, we can in $\Oh_{k_H}(n^2)$ time determine if the rank-width of $G$ is less than $k_H$ and find the desired rank-decomposition if so. In the latter case, we use \cite{Courcelle2000} and \cite{COURCELLE200791} to directly test if $G$ has a vertex-minor isomorphic to $H$ in $\Oh_{k_H}(n^2)$ time. In the former case, we find in $\Oh_{k_H}(n^3)$ time a graph $\widetilde{G}$ that is locally equivalent to $G$ and a set of vertices $T$ which induces in $\widetilde{G}$ the large comparability grid. (First delete vertices from $G$ until the rank-width is equal to the threshold of $k_H$. Then determine in which of the three ways each vertex should be removed by again applying \cite{Courcelle2000} and \cite{COURCELLE200791}.) Now, given $T$, we find an irrelevant vertex in $\mathcal{O}_H(n^\ell)$ time. We delete it and repeat the process, thus testing for $H$ in $\Oh_H(n^{\ell+1})$ time.

\subsection*{Apex-vertex-minor-testing}

Now we give a complete proof of \Cref{cor:apexVM}, which we restate below for convenience.

\apexTestingCor*
\begin{proof}
    Write $\mathcal{G}=(G, A)$ and $\mathcal{H}=(H, A)$; if $\mathcal{G}$ and $\mathcal{H}$ do not have the same apex set, then we immediately return no since isomorphisms are required to fix all the apex vertices. By \cite{recognizingCircleGraphs}, we can find a collection of chords representing the circle graph $G-A$ in $\Oh(n^3)$ time. (Also see \cite{Bouchet87} for an earlier algorithm and \cite{circleGraphLinear} for a faster algorithm.) This yields a corresponding tour graph and Eulerian circuit. Next, for each vertex $a \in A$, we obtain the corresponding signature $\Sigma_a$ on the tour graph in $\mathcal{O}(n)$ time. (Traverse the Eulerian circuit. The first time we see each neighbor $v$ of $a$ in $G$ we take the symmetric difference of the current signature with the two traversed edges incident to $v$; see \cite{BouchetCircleChar}.) Putting all of these signatures together, we obtain a labeled tour graph $\mathcal{T}_{\mathcal{G}}$ of $\mathcal{G}$.
    
    Now iterate over all labeled tour graphs with vertex-set $V(H)\setminus A$ and labelings in $\mathbb{F}_2^A$; the number of them is bounded as a computable function of $\mathcal{H}$. We check which of these labeled tour graphs $\mathcal{T}_\mathcal{H}$ represents $\mathcal{H}$. For each that does, we use \Cref{thm:main} to determine if $\mathcal{T}_{\mathcal{G}}$ has a full immersion isomorphic to $\mathcal{T}_\mathcal{H}$. If there exists such a $\mathcal{T}_\mathcal{H}$, we return yes, and otherwise we return no. Correctness follows from \Cref{thm:interpretation}.
\end{proof}

\subsection*{The rank-width of a circle graph}

Now we prove \Cref{lem:widthTourFuncEquiv}, which says that the rank-width of a circle graph is functionally equivalent to the treewidth of any of its tour graphs.

We begin by using a theorem of Bouchet \cite{connIsotropic} to prove the following lemma. A tour graph $\mathcal{T}$ is \emph{internally $6$-edge-connected} if it is $4$-edge-connected and for any set of vertices $X$ with $|\delta(X)|=4$, either $|X|=1$ or $|V(\mathcal{T})\setminus X|=1$. Given a set of vertices $X$ in a graph $G$, the \emph{cut-rank} $\rho(X)$ is the rank over the binary field of the submatrix of the adjacency matrix with rows $X$ and columns $V(G)\setminus X$.

\begin{lemma}
\label{lem:int6ConnCase}
    If $G$ is a circle graph and $\mathcal{T}$ is a tour graph of $G$, then for every non-empty set of vertices $X \subsetneq V(G)$, we have $2\rho_G(X)+2\leq |\delta_{\mathcal{T}}(X)|$. If moreover $\mathcal{T}$ is internally $6$-edge-connected, then we also have $|\delta_{\mathcal{T}}(X)|\leq 16\rho_G(X)$ for every set of vertices~$X$.
\end{lemma}
\begin{proof}
    We always refer to cut-rank in $G$ and edge cuts in $\mathcal{T}$. Bouchet \cite{connIsotropic} gave an exact formula for the cut-rank $\rho(X)$ in terms of the tour graph $\mathcal{T}$. Let $\mathcal{C}_X$ and $\mathcal{C}_Y$ be the collection of all vertex-sets of components of the subgraph of the tour graph induced on $X$ and $Y$, respectively. Let $A$ be the matrix over the binary field which has rows $\mathcal{C}_X$ and columns $\mathcal{C}_Y$: we put a $1$ between a pair of components if $\mathcal{T}$ has an odd number of edges joining them. Then Bouchet's formula is \begin{align*}
     \rho(X)=|\delta(X)|/2-|\mathcal{C}_X|-|\mathcal{C}_Y|+1+\text{rank}(A).
    \end{align*}

    Both sides have at least one component and the rank of $A$ is at most its number of rows/columns, so $\rho(X) \leq |\delta(X)|/2$. Moreover, equality cannot hold, since then we would have $|\mathcal{C}_X|=|\mathcal{C}_Y|=\text{rank}(A)=1$, but there would be an even number of edges between $X$ and $Y$ since $\mathcal{T}$ is Eulerian. This proves the desired lower bound on $|\delta(X)|$.

    For the upper bound, suppose that $\mathcal{T}$ is internally $6$-edge-connected. We call a component $C \in \mathcal{C}_X \cup \mathcal{C}_Y$ \emph{small} if there are exactly four edges in $\delta(C)$, and \emph{big} otherwise, in which case $|\delta(C)|\geq 6$. Let $\delta_b$ be the number of edges in $\delta(X)$ which are incident to a big component, and let $\delta_s$ be the number of edges in $\delta(X)$ whose ends are both in small components; so $|\delta(X)|=\delta_b+\delta_s$.

    First we claim that\begin{align*}
    \delta_b \leq 12 (|\delta(X)|/2-|\mathcal{C}_X|-|\mathcal{C}_Y|)=6|\delta(X)|-12|\mathcal{C}_X|-12|\mathcal{C}_Y|.
    \end{align*} To see this, first note that the right hand side equals the sum of the charges when each edge in $\delta(X)$ is given a charge of $6$ and each component in $\mathcal{C}_X \cup \mathcal{C}_Y$ is given a charge of $-12$. For each edge in $\delta(X)$ with both ends in small components, move $3$ charge to each of those components. For each edge in $\delta(X)$ with an end in a big component $C$, move $2$ charge to $C$, keep $1$ charge on the edge, and move $3$ charge to the other component. After moving the charge in this manner, each component has non-negative charge, and the sum of the charges of the edges equals $\delta_b$. This proves the desired inequality for $\delta_b$.

    Next we claim that \begin{align*}
    \delta_s \leq 16\cdot \text{rank}(A).
    \end{align*}This will complete the proof of the lemma; the three displayed formulas combined yield $|\delta(X)|=\delta_b+\delta_s \leq 16\rho(X)$. Let $M$ be the submatrix of $A$ obtained by restricting to rows and columns which are small components. Thus each row and column of $M$ has at most four $1$s. Moreover, since $\mathcal{T}$ is internally $6$-edge-connected, any two small components which are joined by an edge are joined by exactly one edge. Thus $M$ has at least $\delta_s$ many $1$s. Now consider a column basis of $M$, and let $k$ be the number of non-zero rows of $M$. Each of these $k$ rows has at least one $1$ among the column basis, and there are at most $4\cdot\text{rank}(A)$ many $1$s among these columns. So $k \leq 4\cdot\text{rank}(A)$. Since the number of $1$s is at most $4k$, this completes the proof that $\delta_s \leq 16\cdot \text{rank}(A)$.
\end{proof}

Now we prove the desired lemma, which is restated below for convenience.

\widthTour*
\begin{proof}
    First of all, the \emph{carving-width} of $\mathcal{T}$, denoted $\mathrm{cw}(\mathcal{T})$, is the minimum integer $c$ so that there is a subcubic tree $T$ whose leaves equal the vertex-set of $\mathcal{T}$ so that for each edge $e \in E(T)$, there are at most $c$ edges in the cut \emph{displayed} by $e$. This is the cut $\delta(X_e)$ where $X_e$ is the set of leaves of one of the components of $T-e$; we always consider cuts in the tour graph. The \emph{rank-width} of $G$ is defined in the same way, except that we consider the cut-rank $\rho(X_e)$ rather than the number of edges in the cut. We always consider cut-rank in the circle graph.

    We write $c$ for the carving-width of $\mathcal{T}$ and $b$ for the branch-width of $\mathcal{T}$. Thus, using well-known inequalities~\cite{bienstock1990embedding} which apply since $\mathcal{T}$ has maximum degree $4$,\begin{align*}
    (t+1)/3\leq b/2\leq c \leq 4b \leq 4(t+1).
    \end{align*}
    Consider a partition of the common vertex-set of $G$ and $\mathcal{T}$ into two parts $X$ and $Y$. By \Cref{lem:int6ConnCase}, $\rho(X) \leq (|\delta(X)|-2)/2$.
    Thus, if we consider a tree $T$ attaining the carving-width $c$ of the tour graph, the cut-rank of each set $X_e$ is at most $(c-2)/2$ in $G$. Thus $G$ has rank-width at most $2t+1$.

    For the other inequality we will show $c-2 \leq 16r$, which implies the desired bound. We need to reduce to the case that $\mathcal{T}$ is internally $6$-edge-connected. We proceed by induction on $|V(\mathcal{T})|$, and clearly the result holds if $c \leq 2$. Suppose that there is a partition of $V(\mathcal{T})$ into two non-empty parts $X$ and $Y$ so that $|\delta(X)|=2$. Then $G$ has no edges between $X$ and $Y$ by \Cref{lem:int6ConnCase}, and the rank-width of $G$ equals the maximum of the rank-width of its subgraphs induced on $X$ and $Y$. Similarly, if $\mathcal{T}_X$ and $\mathcal{T}_Y$ are obtained from the subgraphs of $\mathcal{T}$ induced on $X$ and $Y$ by adding one edge joining the ends of the edges in the cut $\delta(X)$ (that is, $\mathcal{T}_X$ is a tour graph of $G[X]$ and similarly for $Y$), then $\mathrm{cw}(\mathcal{T}) \leq \max\{\mathrm{cw}(\mathcal{T}_X), \mathrm{cw}(\mathcal{T}_Y), 2\}$. To see this, consider trees $T_X$ and $T_Y$ obtaining the carving-width of $\mathcal{T}_X$ and $\mathcal{T}_Y$ respectively, and let $x \in V(\mathcal{T}_X)$ and $y \in V(\mathcal{T}_Y)$ be ends of the edges in the cut $\delta(X)$. We form a tree $T$ from the disjoint union of $T_X$ and $T_Y$ by subdividing the edges incident to $x$ and $y$ and adding a single new edge between the new vertices. Every edge in the tree $T$ displays a cut of the same size as a cut in one of $T_X$ or $T_Y$, except for the new edge which displays a cut of size~2.

    So we may assume that $\mathcal{T}$ is $4$-edge-connected. Now suppose that there exists a partition of the vertex-set into two parts $X$ and $Y$ each of size at least two so that $|\delta(X)|=4$. Then \Cref{lem:int6ConnCase} gives that $\rho_G(X)=1$. Let $G_X$ and $G_Y$ be obtained from $G$ induced on $X$ and $Y$, respectively, by deleting all vertices on the other side except for one which has a neighbor in the kept side. Then the rank-width of $G$ is the maximum of the rank-widths of $G_X$ and $G_Y$. Moreover, $G_X$ and $G_Y$ have tour graphs $\mathcal{T}_X$ and $\mathcal{T}_Y$, respectively, which are obtained from $\mathcal{T}$ by identifying all the vertices in the other side to a single new vertex and removing the loops at that vertex. The carving-width of $\mathcal{T}$ equals the maximum of the carving-widths of $\mathcal{T}_X$ and $\mathcal{T}_Y$. In both cases one can see the bound on the width by taking an optimal subcubic tree of both sides, adding an edge between the leaves given by the special vertices, and suppressing degree 2 vertices.

    Thus we may assume that $\mathcal{T}$ is internally $6$-edge-connected. So by \Cref{lem:int6ConnCase}, the carving-width is at most $16r$, completing the proof.
\end{proof}

\subsection*{Irrelevant vertices for labeled tour graphs}

Finally, we use the theorems proven in this part to prove \Cref{thm:irrelevantLabeled}, which is restated below for convenience.

\irrTheorem*
\begin{proof}
    We fix a set $U$ of size $k$; for the rest of this proof, every graph is labeled over the binary vector space $\mathbb{F}_2^U$. Let $q$ be one more than the number of labeled tour graphs on at most $\ell+2$ vertices, considered up to isomorphism and shifting equivalence, and with at most one distinguished oriented edge. (The distinguished edge must be preserved by isomorphisms.) Let $f_{\ref{prop:inessential-vertex}}$ be the function from \Cref{prop:inessential-vertex}. We define $c=c(k, \ell)$ to be the maximum, over all non-negative integers $i \leq 2\ell+4$ and $j \leq 2$, of $f_{\ref{prop:inessential-vertex}}(i, j, k, q)$. If necessary, we also make $c$ slightly larger to ensure that $c \geq 4$.
    
    Let $\mathcal{G}$ be a labeled tour graph with treewidth at least $c$. If $\mathcal{G}$ is $4$-edge-connected, then by \Cref{prop:inessential-vertex}, $\mathcal{G}$ contains a vertex which is $\mathcal{H}$-irrelevant for every labeled tour graph $\mathcal{H}$ with $\ell$ vertices. So we may assume that $\mathcal{G}$ is not $4$-edge-connected. 
    
    Now let $(G_C, \pi, \iota)$ be a cactus decomposition of $\mathcal{G}$ as defined in \Cref{sec:cactus}. Consider a node $s$ of $G_C$ and the set $S$ of all vertices of $\mathcal{G}$ which are mapped to $s$. By considering the mapping $\pi$, we obtain a partition $W_1, W_2, \ldots, W_r$ of $V(\mathcal{G})\setminus S$ so that each part corresponds to a component of $G_C-s$. Then each cut $\delta(W_i)$ contains exactly two edges. We let $\mathcal{G}\langle s\rangle$ be the labeled tour graph obtained from the subgraph of $\mathcal{G}$ induced on $S$ by, for each set $W_i$, adding an edge between the ends in $S$ of the edges in $\delta(W_i)$. The label of this edge is the sum of the labels of all the edges incident to $W_i$ in $\mathcal{G}$. 

    Ignoring the edge labels for a moment, we note that $\mathcal{G}$ can be obtained from the graphs $(\mathcal{G}\langle s\rangle : s \in V(G_C))$ by repeatedly performing the following operation: take two different graphs $\mathcal{G}_1$ and $\mathcal{G}_2$ with edges $e_1$ and $e_2$, respectively, then delete $e_1$ and $e_2$ and add two new edges which join the ends of $e_1$ with the ends of $e_2$. If we write $t_1$ and $t_2$ for the treewidth of $\mathcal{G}_1$ and $\mathcal{G}_2$, respectively, then the treewidth of the resulting graph is at most $\max (t_1, t_2, 3)$. This is because we can combine the two tree decompositions using a single new bag of size at most $4$ containing the ends of $e_1$ and~$e_2$. Thus, since $\mathcal{G}$ has treewidth at least $c$ and $c \geq 4$, there exists a node $s$ of $G_C$ so that $\mathcal{G}\langle s\rangle$ has treewidth at least~$c$.

    We fix this node to be the root node of the cactus decomposition. Then, by \Cref{lem:folio-correct}, the $\ell$-folio of $\mathcal{G}$ is fully determined by the $(2\ell+4)$-color-folio of the decorated graph obtained from $\mathcal{G}\langle s\rangle$ with $q$ colors. Finally, by \Cref{prop:inessential-vertex}, this decorated graph contains a vertex $v$ such that any way to split off $v$ preserves its $(2\ell+4)$-color-folio; thus it also preserves the $\ell$-folio of $\mathcal{G}$. Hence, splitting off $v$ arbitrarily does not change the set of $\ell$-vertex labeled tour graphs that $\mathcal{G}$ immerses, as desired.
\end{proof}

\section*{AI Methodology}
We used GPT-5.6 to search the literature for a computable bound for \Cref{lem:gridImmersion}, prompting it to check the proofs in \cite{chudnovsky2013immersionfouredgeconnectedgraphs} and \cite{immersionEdgeConn}; ultimately this resulted in our proof using the latter. We also used Opus~5.5, Fable~5.1, GPT-5.6, and GPT‑6 Astra to make figures, to proofread, and to suggest localized improvements to the writing. The only substantive mathematical content affected by these tools was an earlier definition of folios: Opus~5.5 found a mistake in an earlier draft and suggested fixing it by coloring half-edges rather than edges. The authors verified the correctness and originality of all content, including references.

\section*{Acknowledgements} 
The authors thank Donggyu Kim for enlightening discussions about floodings. 

\section*{Funding}
Rose McCarty and Caleb McFarland were partially supported by NSF grant  DMS-2452111. Rose McCarty and Hadas Barabash were partially supported by AFOSR grant  FA9550271B013.

\bibliographystyle{alphaurl}
\bibliography{ref}

%\appendix

%\input{appendix1}

\end{document}